\documentclass[fleqn]{aa}
\usepackage{mathtools}
\usepackage{txfonts}
\usepackage{natbib}
\usepackage{graphicx}
\usepackage{stackengine}
\usepackage{epstopdf}
\usepackage{pgf}
\usepackage{longtable}
\usepackage{pdflscape}
\usepackage{geometry}
\usepackage{soul}
\usepackage{ulem}
\usepackage{orcidlink}
\usepackage{float}

\usepackage{soul}
\definecolor{lightgreen}{HTML}{B7F774}
\sethlcolor{yellow}

\providecommand{\mc}[1]{\multicolumn{1}{c}{#1}}

\usepackage{hyperref}
\hypersetup{colorlinks,linkcolor={blue},citecolor={blue},urlcolor={black}} 

\usepackage{color}
\usepackage{amstext}

\newcommand{\WHz}{$\,$W$\,$Hz$^{-1}$}

\begin{document}
\title{Core prominence as a signature of restarted jet activity in the LOFAR radio-galaxy population}

\author{
       {Dhanya~G.~Nair~\orcidlink{0000-0001-5357-7805}\inst{1,2}}\thanks{Corresponding author \email{dhanyagnair01@gmail.com}} 
        \and
        Raffaella~Morganti\inst{3,4}
        \and
        Marisa~Brienza~\orcidlink{0000-0003-4120-9970}\inst{5,6,7}
        \and
        Beatriz~Mingo~\orcidlink{0000-0001-5649-938X}\inst{8}
        \and
        Judith~H.~Croston\inst{8}
        \and
        Nika~Jurlin~\orcidlink{0000-0003-4270-5968}\inst{9}
        \and
        Timothy~W.~Shimwell\inst{3,10}
        \and
        Joseph~R.~Callingham~\orcidlink{0000-0002-7167-1819}\inst{3,10}
        \and
        Martin~J.~Hardcastle~\orcidlink{0000-0003-4223-1117}\inst{11}
}

\institute{Astronomy Department, Universidad de Concepción, Casilla 160-C,
Concepción, Chile 
    \and
  Joint Institute for VLBI ERIC, Oude Hoogeveensedijk 4, 7991 PD, Dwingeloo, The Netherlands  
    \and
ASTRON, Netherlands Institute for Radio Astronomy, Oude Hoogeveensedijk 4, 7991 PD, Dwingeloo, The Netherlands  
    \and
Kapteyn Astronomical Institute, University of Groningen, PO Box 800, 9700 AV, Groningen, The Netherlands 
    \and
INAF-Istituto di Radioastronomia, via Gobetti 101, Bologna 40129, Italy
\and
INAF – Osservatorio di Astrofisica e Scienza dello Spazio di Bologna, Via P. Gobetti 93/3, 40129 Bologna, Italy
\and
Dipartimento di Fisica e Astronomia, Universit. di Bologna, via P. Gobetti 93/2, 40129 Bologna, Italy
    \and   
School of Physical Sciences, The Open University, Walton Hall, Milton Keynes, MK7 6AA, UK
     \and
Department of Astronomy, The University of Texas at Austin, 2515 Speedway, Stop C1400, Austin, TX 78712-1205, USA
   \and
Leiden Observatory, Leiden University, PO Box 9513, 2300 RA Leiden, The Netherlands
   \and
Centre for Astrophysics Research, Department of Physics, Astronomy and Mathematics, University of Hertfordshire, College Lane, Hatfield AL10 9AB, UK
}

\authorrunning{D.~G.~Nair et al.}
\titlerunning{Core prominence as a signature of restarted jet activity in the LOFAR radio-galaxy population}

\date{Received 16 July 2024 / Accepted 23 August 2024}

\abstract
% context heading (optional) 
{Recurrent phases of dormancy and activity occur in the supermassive black holes in active galactic nuclei. Characterizing the duty cycles of this process is crucial in understanding the impact of the energy released on the host galaxies, and their evolution. However, it is challenging to identify sources in the quiescent and restarted phases.}
  % aims heading (mandatory)
{Our goal is to identify and characterize a substantial sample of radio galaxies in a restarted phase and explore the idea of core prominence as a signature of restarted activity. We expand our prior study of identifying restarted sources from a 
$30\, \mathrm{deg^2}$ area in the Lockman Hole to a larger 
$424\,\mathrm{deg^2}$ region in the \textit{Hobby-Eberly}
Telescope Dark Energy Experiment (HETDEX) extragalactic field using a sample of core-dominated radio galaxies selected visually.}
% methods heading (mandatory)
 {We used the 144 MHz LOFAR survey images of the HETDEX field to identify galaxies with restarting jets. By assessing diverse selection criteria including
 radio core dominance along with low surface brightness extended emission,
spectral index properties (e.g., steep or inverted spectra of
a core and an ultra-steep spectrum of extended emission), and morphology, we found 69 candidate restarted radio galaxies in the HETDEX field.}
 % results heading (mandatory)      
{The restarted candidates show a diverse intrinsic morphology, spanning from FRI, FRII, core-with-halo, to asymmetric forms, feasibly proposing different progenitors. 
Within these 69 restarted candidates, we identified a subset of nine galaxies 
characterized by ultra-steep spectrum extended emission combined with high radio core prominence, representing previous and current epochs of jet activity.
We interpret our findings for this small subset as support for a model in which the switch-on and switch-off mechanism happens with a relatively fast duty cycle in these sources.
We found a peculiar case, J131728.61+561544.8, which appears to have altered its jet orientation, possibly due to changes in the angular momentum or spin of its supermassive black hole, interactions with surrounding material, and/or variations in the magnetic flux threading its accretion disk.
}
 % conclusions
{The restarted candidates span a range of radio luminosities from log$_{10}$(L$_\mathrm{144 MHz}$/$\mathrm{WHz^{-1}}$) = 23.24
to log$_{10}$(L$_\mathrm{144 MHz}$/$\mathrm{WHz^{-1}}$) = 26.80,  and linear sizes between 88 and 1659 kpc at 144 MHz, with 16 sources identified as giant radio galaxies with sizes exceeding 0.7 Mpc.
The total stellar content of restarted sources is consistent with massive elliptical galaxies, with at least 17\% inhabiting cluster environments. 
Our findings at z$<$0.4 suggest that many restarting radio galaxies are not found in rich cluster environments, aligning with the environmental properties of the broader radio-galaxy population.
The present study confirms the core prominence as an effective parameter for selecting candidate restarted radio sources.
}

\keywords{galaxies: active – radio continuum: galaxies – surveys}

\maketitle

\section{Introduction}
\label{Introduction}

The fact that a supermassive black hole (SMBH) can cycle through periods of activity and quiescence is a crucial ingredient required by models of galaxy evolution to regulate the level of star formation in the host galaxy and self-regulate the growth of the black hole  \citep{shabala2009, gaspari2015,harrison2018,hardcastle2020,magliocchetti2022}. 
However, being able to get constraints on the duration and properties of this cycle is a challenging task. To make progress on this, SMBHs in active galactic nuclei (AGNs) in different phases of their evolution need to be found and characterized.

For radio AGNs, extensive work has been done to identify and characterize sources in their young (or recently born) phase for example, gigahertz peak spectrum (GPS) and compact steep spectrum (CSS) sources
\citep[for reviews,][]{anbaan2012,orienti2016,callingham2017,odeasaikia2021}.
Typically, radio cores have a flat or inverted spectral index\footnote{In this paper, the spectral index $\alpha$ is deﬁned through $S \propto \nu^\alpha$, where $S$ is flux density, $\nu$ the frequency} due to self-absorption \citep{blandfordkonigl1979,feretti1984}.  
An inverted spectrum with $\alpha_\mathrm{1400MHz}^\mathrm{144MHz} \gtrsim 0.7$ could indicate GPS or CSS sources which include restarted jets \citep{odea1998,orienti2016}.
It is more difficult to identify radio sources at the other extreme of the cycle: remnant and restarted radio sources. New possibilities for making progress on this are now provided by the low-frequency surveys. Low-frequency observations are essential because they allow us to perform a sort of `\textquoteleft archaeology\textquoteright' by tracing aged electrons and, therefore, the signatures of past nuclear activity \citep{kardashev1962}; readers can refer to \cite{morganti2017} for an overview.  In particular, the combination of a relatively high angular resolution and sensitivity to low surface brightness emission offered by the Low-Frequency Array (LOFAR) surveys has opened up new possibilities for selecting AGNs in their remnant and restarting phase.

Remnant radio galaxies represent dying sources where the jets have switched off
and the electrons are not replenished anymore into the radio lobes.  Radio sources entering this phase have been historically identified via an ultra-steep spectrum (USS), that is, $\alpha < -1.2$. 
However \cite{brienza2017} have shown that this selection is biased toward the oldest tail of the remnant populations, and so morphological selection is necessary.
As illustrated starting from the work of \cite{saripalli2012} and \cite{brienza2017},
they can also be identified using their amorphous morphology, the low surface brightness (SB) of their emission, and the lack (or very much dimmed emission) of the central core.
These parameters taken together have allowed a more complete census of the objects in this phase of their evolution (see \citealt{brienza2017,jurlin2021,shabala2020,quici2021,duttas2023}).
The relatively small remnant fraction of $<$ 10\% with respect to the entire radio galaxy population found by \cite{godfrey2017}, \cite{brienza2017}, \cite{mahatma2018}, and \cite{hardcastle2018} 
is thought to be a result of rapid expansion losses combined with synchrotron losses.

After this remnant phase, a restarted phase can occur, where the jet activity reignites. 
The group of restarted sources is particularly challenging to identify and characterize, and an attempt to make progress with this is the topic of the present study. 
The prototypes of restarted radio sources are the so-called double-double radio galaxies \citep[DDRGs;][]{shoenmakers2000a,mahatma2019} which have a pair of bright inner radio lobes and a pair of faint outer radio lobes. They likely represent a subset of restarted activity, occurring under very specific conditions.

However, other morphological and spectral signatures of restarted activity have been suggested by recent studies (e.g., \citealt{saripalli2012,jurlin2020,bruni2019,bruni2020}).
For example, as described by \cite{jurlin2020}, a high core prominence (CP), which is defined as the ratio between the flux density of the core and the total flux density of the source ($\mathrm{CP} = \mathrm{S_{core}} /  \mathrm{S_{total}} $) combined with low SB from the fading extended emission, can indicate recently restarted activity. 

Spectral properties can also help support the restarting scenario.
Restarted activity can be identified by extremely different radio spectral properties between the central and the diffuse emission (e.g., \citealt{roettiger1994,brienza2020,2021A&A...648A...9M}) or by steep spectrum cores \citep{fomalontmiley1979,bridle1981,barthel1984,akujor1996}. 
However, the interpretation of these features is not always straightforward, as discussed in, for instance, \cite{shulevski2012,mckinley2013, callingham2015, mckinley2018,frank2016,brienza2018,sridhar2020,kukreti2022}. 
For example, these studies have shown that even in objects appearing to host a recently restarted activity in their centers (e.g., traced by a peaked radio spectrum), the surrounding low surface brightness diffuse emission can still have a relatively standard spectral index.
As discussed in these studies, this can suggest an intermittent fueling with a relatively rapid duty cycle or, alternatively, it could be the result of intermittent flow due to a strong interaction between a recently created jet and the surrounding rich medium, with the subsequent temporary decollimation of the flow. The presence of self-absorption and mixing of different electron populations can further complicate the analysis. Thus, all this clearly shows the variety and complexity of the restarted phenomenon and the need for increasing the statistics to explore the large parameter space, for example, in the duty cycle and conditions.

\begin{table*}[ht!]
\caption{Properties of the data and images used in this work. }
\label{table-data-images}
\begin{center}
\begin{tabular}{llllll}
\hline\hline
\mc{Data/Images}  &  Field of view & Beam size & RMS noise & Frequency &  Telescope \\
\mc{}   & [$\mathrm{deg^{2}}$]  & [arcsec, deg] &   [$\mathrm{\mu Jy/beam}$] & [MHz]    \\
\mc{(1)}        & (2)  & (3) & (4)  & (5) & (6)   \\ \hline
LoTSS DR2 Image-I\,\, (LR) & 35 & $20\times20$ & 100--390 & 144 & LOFAR   \\
LoTSS DR2 Image-II (HR)  & 35 & $6\times6$ & 47--96 & 144 &  LOFAR  \\
NVSS & 16 &  $45\times45$ & 450 & 1400 &  VLA  \\
FIRST & 0.45 & $5\times5$ & 150 & 1400 &  VLA  \\
VLBA-C band  & 0.16 & $0.005\times0.002, \mathrm{PA=4.36}$ & 55 & 4300 &  VLBA   \\
VLBA-X band & 0.09 & $0.003\times0.0009, \mathrm{PA=3.93}$ & 33 & 7600 &  VLBA    \\
\hline  
\end{tabular}
\end{center}
\footnotesize{ {\bf Columns:} 1~--~Data; 2~--~Field of view; 3~--~Beam size; 4~--~RMS noise; 5~--~Frequency; 6~--~Telescope. The data/images and properties of LoTSS DR2 are obtained from \cite{shimwell2022}, NVSS from \cite{condon1998}, FIRST from \cite{becker1995}, and VLBA-C band and VLBA-X band from the Radio Fundamental Catalog.
}
\end{table*}

To understand more about the restarting phenomenon, its occurrence, and the time scales involved we need to go beyond single-object studies.
For instance, \cite{jurlin2020}  
found 23 restarting candidates in the Lockman Hole (LH) LOFAR Two-Meter Sky Survey (LoTSS) Deep field,
constituting a restarter fraction of 15\% (out of 158 sources $> 60\arcsec$). 
\cite{jurlin2020} used two main selection criteria for their sample, which we have also considered for the present study: i) the high CP combined with low SB extended emission; and ii) the presence of a steep spectrum inner region ($\alpha_\mathrm{1400~MHz}^\mathrm{144~MHz} \le -0.65$).
Complementary studies have also identified extremely different radio spectral properties between the central and the diffuse emission \citep{2021Galax...9...88M,2021A&A...648A...9M}.
An initial confirmation of the strengths of these selection criteria to identify restarted radio sources has been obtained by exploring the central regions of some of these sources in the LH area using high-resolution observations at 144~MHz with the LOFAR international baselines \citep{jurlin2024}.

Despite the amorphous, low surface brightness of the extended emission, small jets were seen in the prominent core regions when observed with high enough angular resolution. Building on this success, the search and characterization of restarted radio sources needs now to be expanded to larger samples.

In the present study, we aim at further expanding the available sample of candidate restarted galaxies, covering a much larger sky area (the \citealt{jurlin2020} study was limited to 30 $\deg^2$). We do this by starting with a  visually selected sample of core prominent radio galaxies identified by \cite{mingo2019}\, over an extragalactic region, the 
\textit{Hobby-Eberly} Telescope Dark Energy Experiment \citep[HETDEX:][]{hill2008} field, covering over 424 $\deg^2$ and covered by the LoTSS First Data Release 
(LoTSS DR1, \citealt{shimwell2019}) observations. 
In this paper, we characterize these core bright radio galaxies, using the criteria
described above, which will allow us to also test how reliably the core prominence parameter can be used to build large samples of restarted radio galaxies.

The paper is organized as follows: in Sect.~\ref{The Sample: Core-dominated AGN in LoTSS-DR1} we present the sample selection from LoTSS DR1 and \cite{mingo2019}, followed by an explanation of the optical and infrared identification. In Sect.~\ref{Identification of candidate restarted radio galaxies}, we present the various processes used to select restarted galaxies in the sample, and we present the final results and discuss our interpretation of results for the restarted candidates, including source morphology, radio luminosity, and optical host-galaxy properties in Sect.~\ref{Results and Discussion}, before presenting our summary and prospects in Sect.~\ref{Summary and prospects}. For this paper, we have used a Lambda cold dark matter ($\mathrm{\Lambda CDM}$) cosmology model for the Universe with Hubble constant ($ H_{0}) = 67.74\,\mathrm{km s^{-1} Mpc^{-1}}$, $\Omega_\mathrm{m} = 0.3089$ and $\Omega_\mathrm{\Lambda} = 0.6911$ based on \textit{Planck} measurements \citep{planck2016}. 

\section{The sample of core-dominated AGNs in the HETDEX region}
\label{The Sample: Core-dominated AGN in LoTSS-DR1}

LoTSS is a low-frequency radio survey at 144 MHz that aims to cover the entire northern sky. 
LoTSS DR1 \citep{shimwell2019}
covers the HETDEX field over 424 $\deg^2$ of the northern sky within 
$161^\circ$ < RA < $231^\circ$ and $45.5^\circ$ < Dec < $57^\circ$,
while a LoTSS Second Data Release (LoTSS DR2, \citealt{shimwell2022})
is now available covering a larger area of sky. LoTSS DR1 has provided images with median 
root mean square (RMS) noise of 71 $\mathrm{\mu Jy/beam}$ and $6\arcsec$ resolution, and value-added catalogs from the cross-identification of the radio sources with their optical counterparts. The LoTSS DR1 mosaic of the HETDEX region contains 318,520 sources among which 73 percent have a host galaxy identification \citep{williams2019} and 51 percent have spectroscopic or photometric redshifts \citep{duncan2019}.

\subsection{Selection of the sample} 
\label{Selection of the sample}

Within the LoTSS DR1 value-added catalog, \cite{mingo2019} investigated the Fanaroff \& Riley \citep[FR:][]{fanaroffriley1974} dichotomy using a well-resolved set of 5805 RLAGNs, selected from a larger sample of 23344 RLAGNs described by \cite{2019A&A...622A..12H}.
This well-resolved RLAGNs sample of \cite{mingo2019} contains only AGNs having an optical ID and reliable redshift ($z$). 
Full details on the completeness and the selection strategies of the well-resolved RLAGNs sample are discussed in \cite{mingo2019} and \cite{2019A&A...622A..12H}. 

In the morphological classification for the FR class (see Table 4 in \citealt{mingo2019}), a subset of 99 sources were classified as core-dominated radio sources (hereafter core-D AGNs). 
These sources presented a sharp drop and rise in brightness beyond the core, unlike traditional FRIs, and therefore they were excluded from the final FRI sample.
These sources have a high dynamic range\footnote{Dynamic range (DR) is the ratio of 
the brightest source in the field to the off-source RMS in the image, that is, DR $=$ peak/RMS} (75$\%$ have dynamic ranges > 4.5) and sizes mostly above $40\arcsec$.

A severe rise and drop in the brightness beyond the subkiloparsec nuclear region is also a characteristic of steep spectrum cores (SSC) found in early studies 
\citep{fomalontmiley1979,bridle1981,akujor1996}. 
These properties are also seen in well-studied objects where candidate restarted jets (or young jets temporarily disrupted in the interaction with the medium) are present, for example, \cite{frank2016,brienza2018,kukreti2022}. As described in Sect.~\ref{Introduction}, a restarting galaxy can be characterized by a bright core and low surface brightness, amorphous extended structures that are fading. Therefore, this 99 core-D AGNs sample with extended emission selected by \cite{mingo2019} from LoTSS DR1 shows morphological characteristics of restarted galaxies. Hence this sample forms an ideal starting point to identify the restarting galaxy population in the HETDEX field
and test the evolutionary scenario described in Sect.~\ref{Introduction}.

For the analysis presented in this study, we used the 6$\arcsec$ high-resolution maps, and the 20$\arcsec$ low-resolution maps, all taken from LoTSS DR2 \citep{shimwell2022}.
We refer to these images as LOFAR-HR and LOFAR-LR, respectively. The different resolutions were used for the identification 
and flux density measurement of the radio cores and that of the diffuse emission, respectively
at 144 MHz of the 99 core-D AGNs (Sect.~\ref{Identification of candidate restarted radio galaxies}). The properties of the images are summarized in  Table \ref{table-data-images}.

\subsection{Optical and infrared identification}
\label{Optical and infrared identification}
\begin{figure}[ht!]
\centerline{\includegraphics[width=0.5\textwidth]{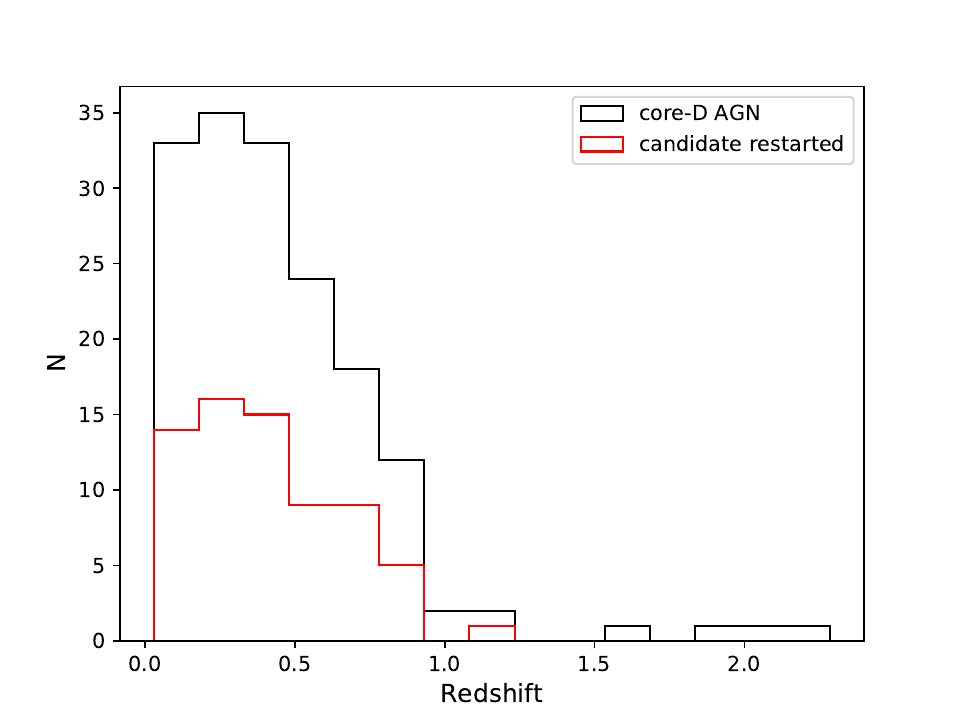}}
\caption{Redshift distribution of the core-D AGNs sample (black), and the final sample of 69 candidate restarted galaxies (red) identified in the HETDEX field as discussed in Sect.~\ref{Final sample}. For a description of candidate restarted galaxies, Sect.\ref{Identification of candidate restarted radio galaxies} provides the relevant details, with the final sample described in Sect.\ref{Final sample}.
}
\label{fig:redshift-restarted}
\end{figure}

As mentioned above, the 99 core-D AGNs all have an optical identification. Furthermore, 71 out of the 99 core-D sources have a spectroscopic redshift available. The spectroscopic redshift is obtained from the Sloan Digital Sky Survey Data Release 14 (SDSS DR14). For the rest, photometric redshifts estimated by \cite{duncan2019} are used. 
The redshifts are presented in Table~\ref{list of sources} and Table~\ref{list of sources-rejected}. The redshift distribution of the 99 core-D AGNs is shown in 
Fig.~\ref{fig:redshift-restarted} ranging from 0.03 to 2.3 with 86$\%$ of the sample lying at $z < 0.8$, and is discussed in detail in Sect.~\ref{Redshift distribution of restarted candidates}.

\begin{table}
\caption{Number of sources satisfying each selection criteria that resulted in the selection of candidate restarted galaxies.
}
\label{sample}
\begin{center}
\begin{tabular}{ccc}\hline\hline
    &  No.   & \%  \\ 
\hline 
FRI+FRII RLAGNs sample in BM19 &  1679  &    \\
core-D AGNs sample in BM19 &  99  &  \\
\hline 
small-unreliable (excluded) & 5  &   \\
core-D AGNs sample used  &  94  &   \\ 
\hline  
LOFAR $18\arcsec$ detections   &  94  & 100   \\
LOFAR $6\arcsec$ core detections   &  94  & 100   \\
FIRST core detections    &  81/94  &  86  \\
NVSS detections     &  79/94  & 84    \\
FIRST core and NVSS detections    &  77 &     \\
\hline 
$\mathrm{CP}_{\mathrm{1.4GHz}}$ \,$+$\,$\mathrm{SB}_{\mathrm{144\,MHz}}$  & 73 &  \\
$\mathrm{SSC}_{\mathrm{144\,MHz,1.4\,GHz}}$ \,$\le$\,-0.65 & 15  &  \\
$\mathrm{USS}_\mathrm{ext}$ \,($< -1.2$) $+$ bright core & 9 &   \\
Morphology     &   0 &   \\
\hline
Type-1 AGNs & 10  &   \\
Type-1 AGNs + USS & 2 &   \\
Type-1 AGNs (excluded) & 8  &   \\
One-sided jet morphology (excluded)  &  2   &  \\
\hline 
No. of candidate restarted galaxies   &  69  &   \\
\hline  
\end{tabular} 
\end{center}
\footnotesize{{\bf Notes:} For more details, readers can refer to  Sect.~\ref{Identification of candidate restarted radio galaxies}. Acronyms: CP~--~high radio core prominence at 1.4 GHz, SB~--~surface brightness at 144 MHz; SSC~--~steep spectrum core; USS~--~ultra-steep spectrum of the extended emission. BM19 stands for \cite{mingo2019}.}
\end{table}

\begin{figure}[ht!]
\centerline{\includegraphics[width=0.55\textwidth]{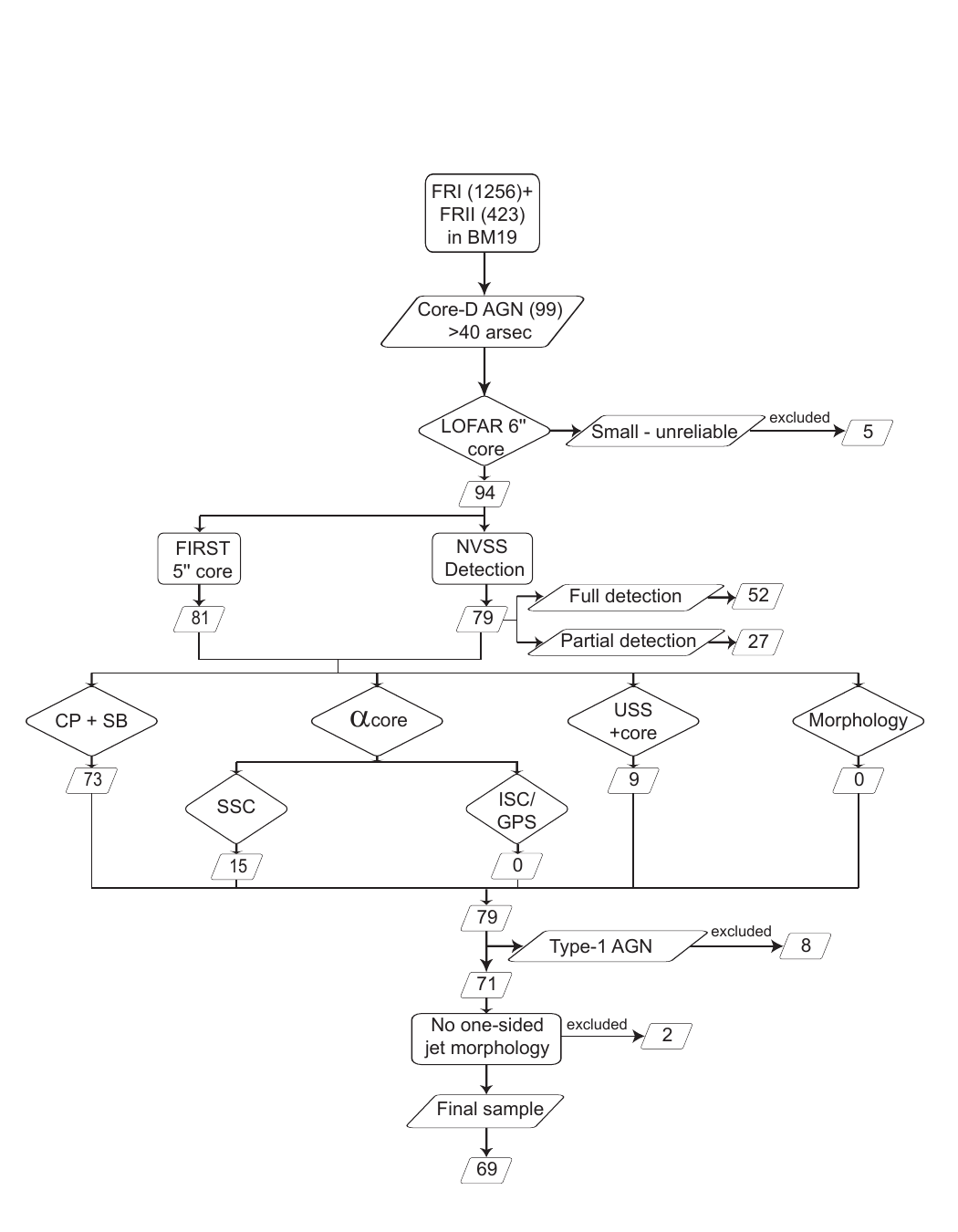}}
\caption{Flowchart illustrating the scheme followed to identify candidate restarted galaxies. The acronyms are the same as those in Table~\ref{sample} and are described in the text; details are provided in Sect.~\ref{Identification of candidate restarted radio galaxies}. BM19 stands for \cite{mingo2019}. 
}
\label{fig:flowchart}
\end{figure}

For the 71 sources with SDSS optical spectra, the wavelength coverage is $\approx$ 
3650--10400 $\AA$.
Given the redshift range of \, 0.03--2.3 of the objects in this study, the emission lines included are [OIII], H$\alpha$, [NII], Ly$\alpha$, Mg, CIV, and CIII. The permitted lines will be used in Sec.~\ref{Beaming effects contaminants} for identifying possible beamed sources. 

An infrared identification was also obtained for all the 99 core-D AGNs from the \textit{Wide-field Infrared Survey Explorer} `\textquoteleft AllWISE\textquoteright' catalog \citep{cutri2013} 
This will be used in Sec.~\ref{wise colors} to characterize the host galaxies of the restarted candidates. 
The AB magnitudes were converted to Vega magnitudes using the relation $m_{ab} = m_{vega} + \delta$m where $\delta$m is 2.699 for W1, 3.339 for W2, 5.174 for W3, and 6.62 for 
W4 taken from the AllWISE online documentation\footnote{\url{https://wise2.ipac.caltech.edu/docs/release/allsky/expsup/sec4_4h.html}}.

\section{Identification of candidate restarted radio galaxies}
\label{Identification of candidate restarted radio galaxies} 

In this section, we describe the process followed to characterize the 99 core-D AGNs sample presented in Sect.~\ref{The Sample: Core-dominated AGN in LoTSS-DR1} using the different selection criteria mentioned in Sect.~\ref{Introduction}. Fig.~\ref{fig:flowchart} depicts the steps we have followed to identify candidate restarted galaxies. 

Due to the resolution and dynamic range constraints of the LoTSS data, \cite{mingo2019} warn that some sources may have an uncertain morphology in the core-D AGNs sample. Therefore we did a detailed visual inspection of the  LOFAR LR and HR maps of 99 core-D AGNs. 
Five sources ($\mathrm{J}110501.70+472300.7,\, \mathrm{J}122459.65+554759.9, \, \mathrm{J}125556.65+520353.6, \, \mathrm{J}133128.47+475439.6, \, \mathrm{J}143644.70+515442.9$) are either very small, faint, or have an unreliable morphology in their LOFAR LR and HR maps, and so we removed them from the 99 core-D AGNs sample for any further analysis. The inspection of the LOFAR LR images also ensured that no blending with unrelated sources was present. We found distinct and reliable radio cores associated with the remaining 94 sources in the LOFAR HR maps at 144 MHz. 

\begin{figure*}[ht!]
\centerline{\includegraphics[width=0.5\textwidth]{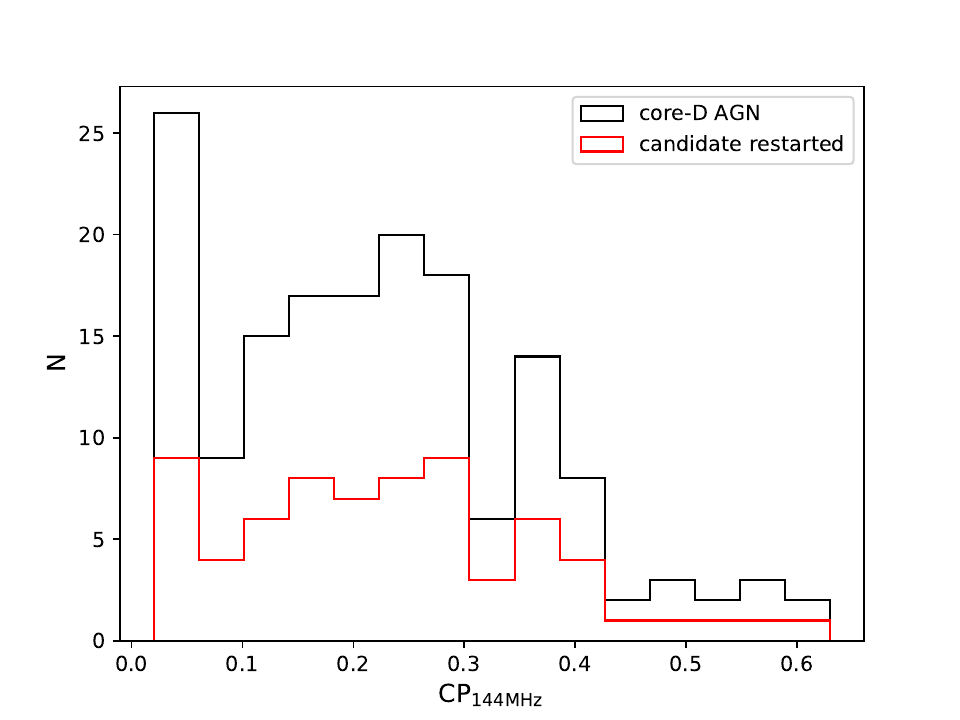}
           \includegraphics[width=0.5\textwidth]{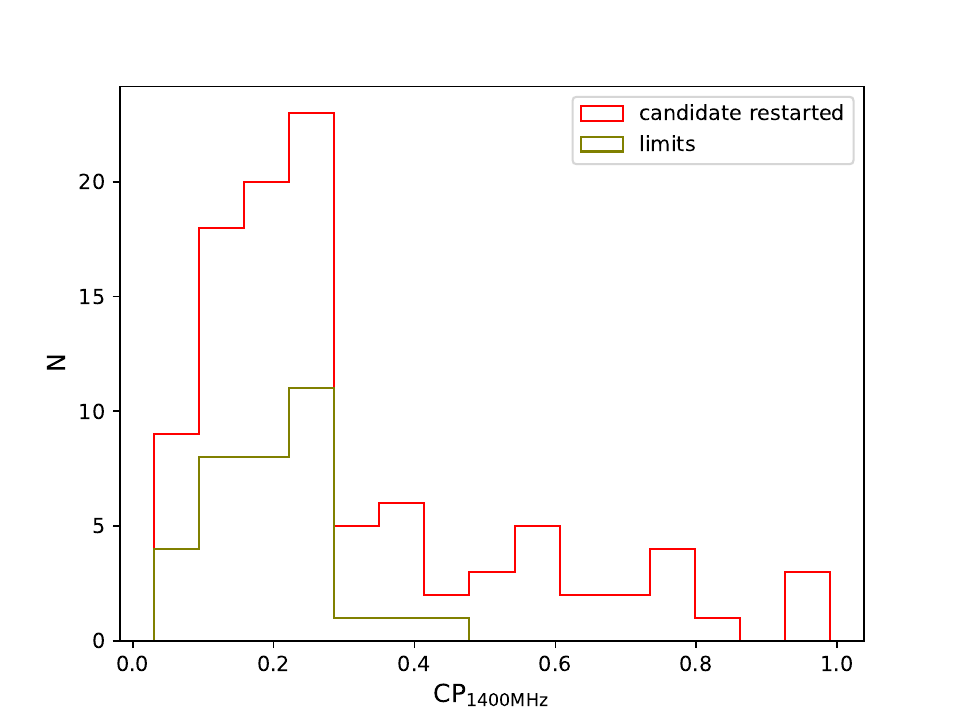}
           }
\caption{Distribution of radio core prominence at 144 MHz and 1.4 GHz.
\textit{\textbf{Left panel}}:\, The distribution of the radio core prominence at 144 MHz, $\mathrm{CP_{144}}$ of the core-D AGNs sample, and the final sample of 69 candidate restarted galaxies as discussed in Sect.~\ref{Final sample}. The $\mathrm{CP_{144}}$ was estimated using the peak core flux densities measured directly from the LOFAR $6\arcsec$ image and the total flux densities measured from the LOFAR $18\arcsec$ image. We note that J105922.02$+$523902.1 with $\mathrm{CP_{144}} = $ 5.0 is excluded in this plot for illustration purposes. However, this source is identified as a broad line AGNs in Sect.~\ref{Beaming effects contaminants} and excluded from the final sample of candidate restarted galaxies presented in Sect.~\ref{Final sample}. 
\textit{\textbf{Right panel}}:\, The distribution of the radio core prominence at 1.4 GHz, $\mathrm{CP_{1400}}$ of the final sample of 69 candidate restarted galaxies as discussed in Sect.~\ref{Final sample}. The distribution of limits (63 upper and six lower limits) is shown in olive color (see text for details). The mean and median values of the CP distribution are 0.35 and 0.26 respectively. We note that the candidate restarted galaxies selected based on steep spectrum core (SSC), inverted spectrum core (ISC), and ultra-steep spectrum (USS) of extended emission presented in Sect.~\ref{Final sample} are also included in both plots. 
}
\label{fig:cp}
\end{figure*}

\subsection{High radio core prominence and low surface brightness}
\label{High radio core prominence and low surface brightness}

As described in Sect.~\ref{Introduction}, the first criteria we use to identify the candidate restarted galaxies is based on their high CP combined with low-SB extended emission. 
To do this we quantify the radio core prominence at 144 MHz ($\mathrm{CP_{144}}$) and the surface brightness at 144 MHz ($\mathrm{SB_{144}}$) as described in the following text. 

$\mathrm{CP_{144}}$ is estimated as the ratio between the peak core flux densities measured from LOFAR HR map and the total flux densities measured from the LOFAR LR map ($\mathrm{CP_{144}} = \mathrm{S_{core,LOFAR-HR,peak}}/\mathrm{S_{total,LOFAR-LR}}$). Since LOFAR maps are sensitive to various scales of emission, we have used the peak core flux density from the LOFAR HR map to minimize any contamination from the extended emission. 
We measured the total flux densities at 144 MHz directly in the area following the 3$\times \sigma_\mathrm{rms,local}$ contours in their LOFAR-LR maps.

The distribution of $\mathrm{CP_{144}}$ for the core-D AGNs sample is shown in 
Fig.~\ref{fig:cp} in the left panel, spanning from 0.02 to 0.63, with mean and median values of $\mathrm{CP_{144}}$ at 0.22 and 0.21 respectively. As expected from the sample selection, $77\%$ of sources have $\mathrm{CP_{144}} \ge $ 0.1.

To compare the CP calculated in this work with other studies in the literature \citep{deRuiter1990,laing2014,baldi2015,jurlin2020}, we have also computed radio core prominence at 1.4~GHz ($\mathrm{CP_{1400}}$). 
In a similar way to what was done with the LOFAR HR image, to derive $\mathrm{CP_{1400}}$, we used the peak core flux densities measured directly from the Very Large Array faint Images of the Radio Sky at Twenty-cm survey \citep[VLA-FIRST,][]{becker1995} to avoid contamination from any extended emission. The total flux densities were instead measured from the VLA Sky Survey \citep[NVSS,][]{condon1998}, that is, $\mathrm{CP_{1400}} = \mathrm{S_{core,FIRST,peak}} /  \mathrm{S_{total,NVSS}} $. 
 
We have visually inspected the VLA-FIRST maps of 94 sources in the core-D AGNs sample. Out of the 94 core-D galaxies, 81 galaxies were detected in FIRST maps with $\ge$ 3$\times \sigma_\mathrm{rms,local}$. 
For the 13 sources without a detected core in FIRST, we measured the local RMS noise directly from the map and assumed a 3$\times \sigma_\mathrm{rms,local}$ as an upper limit of the core flux density, since the emission is unresolved. 
The local RMS noise is measured in a source-free box near the source location in the FIRST maps.

Out of the 94 core-D galaxies, 79 galaxies were detected with $\ge$ 3$\times \sigma_\mathrm{rms,local}$ in NVSS maps (52 galaxies fully detected, 27 galaxies partially detected and 15 galaxies not detected). 
There is a FIRST counterpart for 77 out of these 79 galaxies detected in NVSS as marked in column 11 in Table~ \ref{Core flux densities}. For the 52 sources with full detections, we measured their total flux densities directly in the area following the 3$\times \sigma_\mathrm{rms,local}$ contours in their NVSS images. 
For 15 sources with no detection in the NVSS maps, we computed a 3$\times \sigma_\mathrm{rms,local}$ upper limit to the undetected emission by measuring the standard deviation of the flux density in ten different boxes surrounding the source location. 
For 27 galaxies only part of the region detected by LOFAR is detected in the NVSS image.  
In these cases, the total flux density is derived by adding the flux density measured 
above 3$\times \sigma_\mathrm{rms,local}$ in the NVSS images and the upper limit
to the undetected emission. 
The latter is derived using the same strategy as described above for the undetected sources. The results are presented in column 5 in Table~\ref{Core flux densities}. The local RMS noise is measured in a source-free box near the source location in NVSS maps. 

As described in Sect.~\ref{Introduction}, a high CP is a relevant parameter in the process of identification of a restarted radio galaxy, and \cite{jurlin2020} applied a $ \mathrm{CP_{1400}}$ cut of 0.1 for the selection of their sample. A similar approach was also followed by \cite{saripalli2012}. Typically, FRI radio galaxies exhibit higher CP values, whereas FRII radio galaxies have faint cores and hence lower CP. A core prominence value $\ge$ 0.1 is much higher than the CP values found in FRI galaxies according to studies by \cite{deRuiter1990,laing2014,baldi2015,jurlin2020}. Based on the aforementioned articles, we also applied a $ \mathrm{CP_{1400}}$ cut $\ge$ 0.1 to the core-D AGNs sample to classify a source as restarted. 

Seventy-eight sources (83\%) in the core-D AGNs sample have a $ \mathrm{CP_{1400}}$ $\ge$ 0.1 and $51\%$ above 0.25. These high fractions are expected because of the way the sources are selected (see Table~\ref{list of sources}). We note that the $\mathrm{CP_{1400}}$ of 43 sources in the core-D AGNs sample are limit values (37 upper and six lower limits) in the case of no detection in FIRST, or partial/no detection in NVSS (32 out of these 43 sources, i.e., 75\%, have high $\mathrm{CP_{144}} > $ 0.1).

\begin{figure}[ht!]
\centerline{\includegraphics[width=0.5\textwidth]{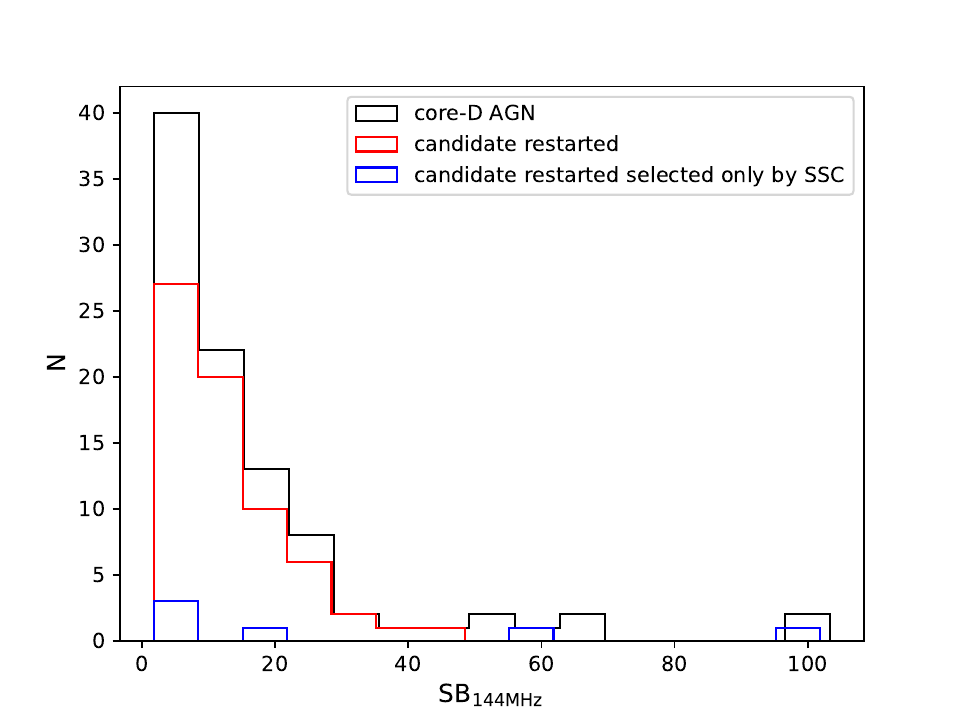}}
\caption{Distribution of SB at 144 MHz of the core-D AGNs sample (black), and the final sample of 69 candidate restarted galaxies (red) as discussed in Sect.~\ref{Final sample}. We note that the six candidate restarted galaxies selected only by SSC criteria presented in Sect.~\ref{Steep spectral index of the core} are indicated in blue color. The mean and median values of SB of candidate restarted galaxies are 14.6 and 10.5 $\mathrm{mJy\,arcmin^{-2}}$ respectively.
}   
\label{fig:sb}
\end{figure}

To estimate the SB of the extended emission of the 94 sources in the sample we first subtract the contribution from the core (the peak core flux density measured from the LOFAR HR image as described above) from the integrated flux density (measured from the LOFAR LR image as described above). Then, the flux density is divided by the source area measured directly in $\mathrm{arcmin^{2}}$.

A $\mathrm{SB_{144\,MHz}}$ \,$<$ \,50 \,$\mathrm{mJy\,arcmin^{-2}}$ is used as a cut-off for the low values of SB following works by \cite{jurlin2020,brienza2017,duttas2023} and references therein. 
A relaxed morphology with low SB $<$ 50 $\mathrm{mJy\,arcmin^{-2}}$ at 144 MHz criteria is used by \cite{brienza2017} to determine remnant galaxies. Recently, \cite{duttas2023} found that the SB is distributed with a median value of 63.4 $\pm$ 14.5 \,$\mathrm{mJy\,arcmin^{-2}}$ (range of 22.2 – 264 \,$\mathrm{mJy\,arcmin^{-2}}$) at 325 MHz for a sample of 21 remnant candidates in XMM-Newton Large-Scale Structure ﬁeld. 
The derived values of the SB for our sample are presented in Table \ref{list of sources}. The lowest SB measured is 1.92\,$\mathrm{mJy\,arcmin^{-2}}$ for J105922.02$+$523902.1. 

Out of the 78 galaxies with $\mathrm{CP_{1400}}$ $\ge$ 0.1, 73 sources have $\mathrm{SB_{144\,MHz}}$ \,$<$ \,50 \,$\mathrm{mJy\,arcmin^{-2}}$. Hence 73 objects satisfy both the high CP and low SB criteria.
The distribution of CP and SB of candidate restarted galaxies selected based on high CP and low SB are shown in Fig.~\ref{fig:cp} (right), and Fig.~\ref{fig:sb} respectively. We note that other candidate restarted radio galaxies will be selected also based on other criteria (i.e., steep spectrum core (SSC), inverted spectrum core (ISC), and ultra-steep spectrum (USS) of extended emission) as described in the next sections. The results combining all the selection criteria will be presented in Sect.~\ref{Final sample}.

\subsection{Steep spectral index of the radio core}
\label{Steep spectral index of the core}

The core of a radio galaxy is typically characterized by a much flatter spectrum ($\alpha \gtrsim -0.5$) compared to the large-scale lobes, due to the superposition of many synchrotron self-absorbed spectra from the base of the jet \citep[e.g.,][]{blandfordkonigl1979,miley1980}. 
However, in a newly started (or restarted) radio source where the jets are still embedded in the central dense medium, the radio spectrum is expected to appear peaked either due to synchrotron self-absorption or free-free absorption and optically thin with a relatively steep power law slope ($\alpha \lesssim -0.5$) above the peak frequency \citep{odea1998,odeasaikia2021}.

The resolution of our available radio data ($6\arcsec$) may not always allow us to resolve the jet base alone; instead, the central region that we observe likely comprises emission from the nucleus, the jet base, and/or inner lobes. Therefore, observing a steep spectrum core (SSC) potentially indicates the presence of compact, unresolved radio jets, which are possibly associated with a restarting activity, especially if embedded in old remnant plasma.

Following \cite{jurlin2020}, we use a large value of $\alpha_\mathrm{1400MHz}^\mathrm{144MHz} \le -0.65$ to identify a SSC \citep[e.g.,][]{readhead1979,marscher1988,bridle1981,schilizzi1981,jurlin2020}, this takes into account the uncertainties in the flux densities (and, as consequence, in the spectral indices). The spectral index of the core was computed between 144~MHz and 1400~MHz. We used the peak flux densities of the core detected in the LOFAR HR image at 144 MHz (6$^{\prime\prime}$) and the FIRST image at 1400 MHz (5$^{\prime\prime}$) to minimize any contamination from the diffuse emission surrounding the cores mainly at low frequencies. The error on the spectral index is calculated using: 
\begin{equation}
  \mathrm{\alpha_{err}} =  \frac {\mathrm{1}}  {\mathrm{ln \frac{\nu1} {\nu2}
  }}   
  \sqrt{  \frac {\mathrm{S_{1,err}}}  {\mathrm{S_1}}^2
  +  \frac {\mathrm{S_{2,err}}}  {\mathrm{S_2}}^2  }  
    \end{equation} 
where $\mathrm{S_{1}}$ and $\mathrm{S_{2}}$ are the flux densities at frequencies $\nu_{1}$ and $\nu_{2}$ and $\mathrm{S_{1,err}}$ and $\mathrm{S_{2,err}}$ are the errors on the respective flux densities.  

\begin{figure}[ht!]
\centerline{\includegraphics[width=0.5\textwidth]{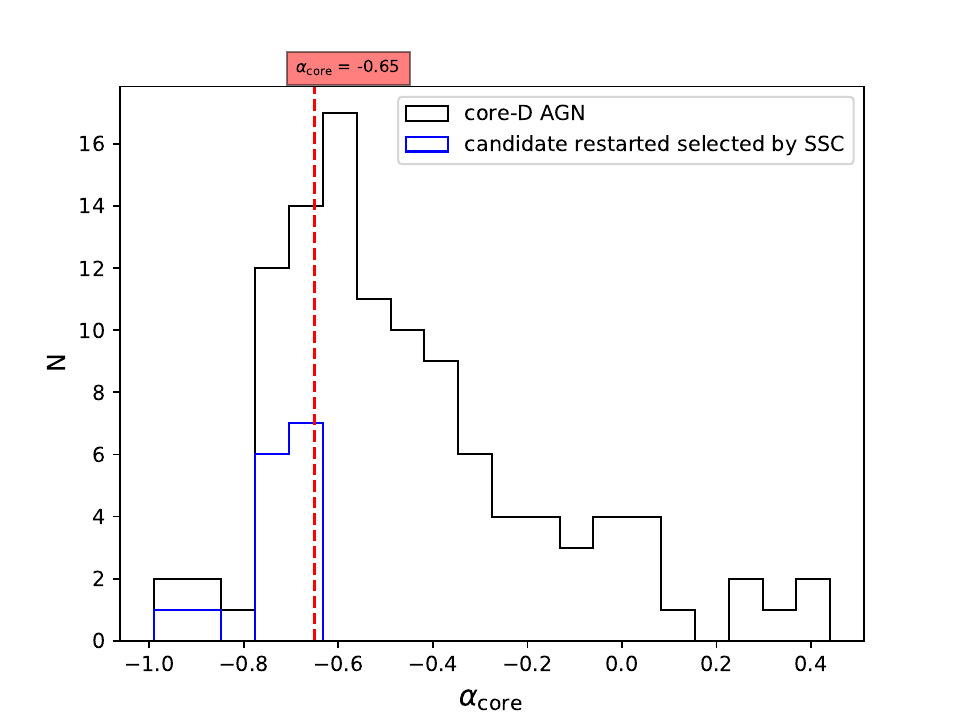}}
\caption{
Distribution of spectral indices of cores, $\alpha_\mathrm{1400MHz}^\mathrm{144MHz}$ of the core-D AGNs sample (black), and the 15 candidate restarted galaxies selected based on SSC criteria presented in Sect.~\ref{Steep spectral index of the core} (blue) in the final sample of 
69 candidate restarted galaxies presented in Sect.~\ref{Final sample}.
The dashed red line marks the value of SSC $=$ -0.65, and all sources to the left of the dashed line have SSC $\le$ -0.65.}
\label{fig:spec-inner}
\end{figure}

Using this criterion we selected 15 galaxies with SSC $\le$ -0.65, out of which three galaxies have an upper limit on their derived spectral indices as they were not detected in FIRST maps. 
To make sure that we minimize any contribution from the extended emission, we have also visually inspected all these 15 SSC galaxies and verified that the cores are isolated in the LOFAR HR and the FIRST images. The spectral index distribution of the cores of these 15 candidate restarted galaxies selected based on SSC criteria is shown in Fig.~\ref{fig:spec-inner}.

These candidate restarted sources are marked with `\textquoteleft SSC\textquoteright' on their respective maps in Figs.~\ref{fig:1}--\ref{fig:5} and their properties are given in Table~\ref{list of sources}. 
We note that as indicated in the table, six galaxies 
have been identified as candidate restarted galaxies based on only the SSC characteristics irrespective of their CP and SB values.

Because the peak of the radio spectrum can be at GHz frequencies \citep{odea1998,odeasaikia2021},  
we have also explored the presence of inverted spectrum cores (ISC, $\alpha_\mathrm{1400MHz}^\mathrm{144MHz}\geq 0.7$).  
Nonetheless, no ISC with $\alpha_\mathrm{1400MHz}^\mathrm{144MHz}\geq 0.7$ were detected within the sample. 

\begin{figure*}[ht!]
\centerline{\includegraphics[width=0.3\textwidth]{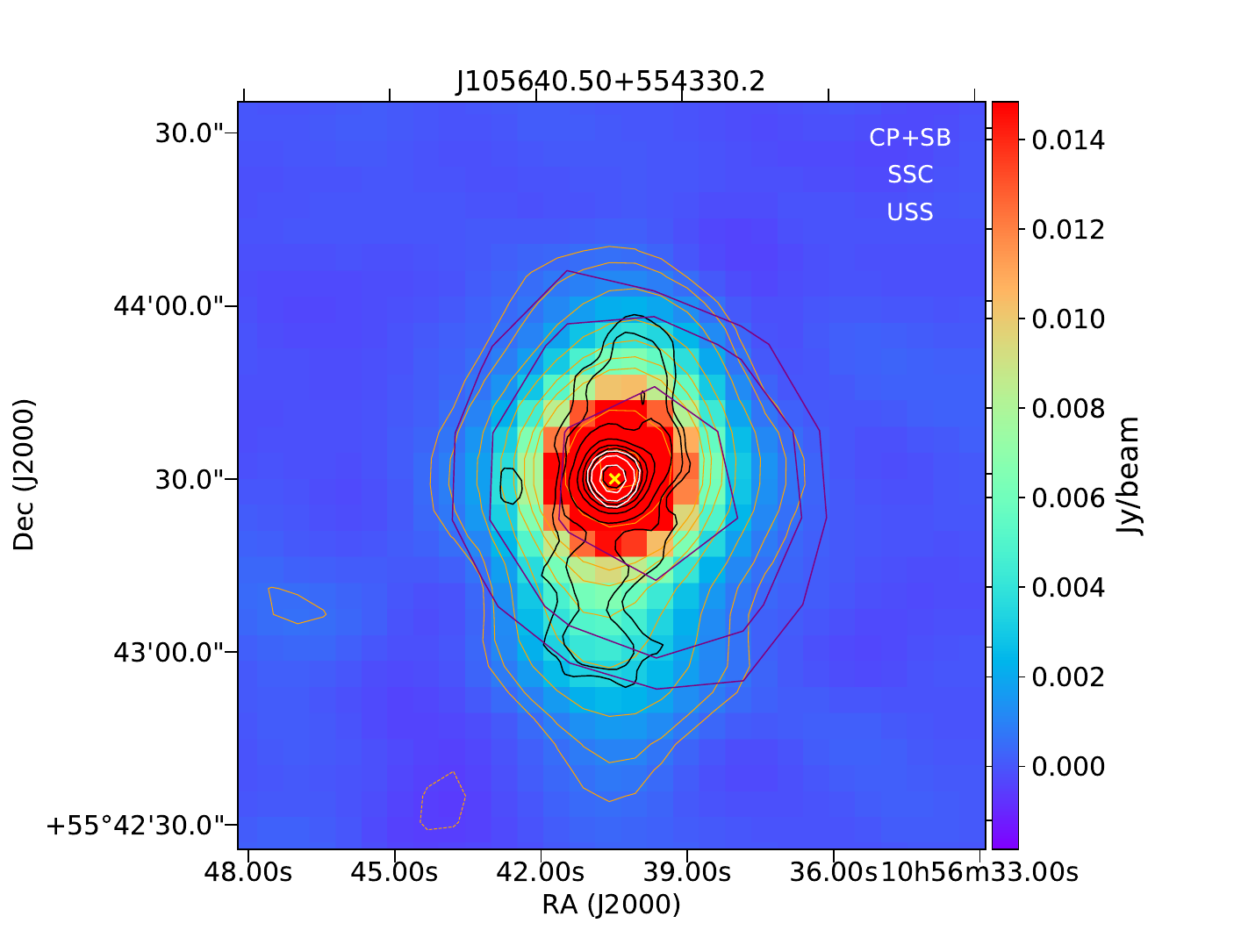}
            \includegraphics[width=0.3\textwidth]{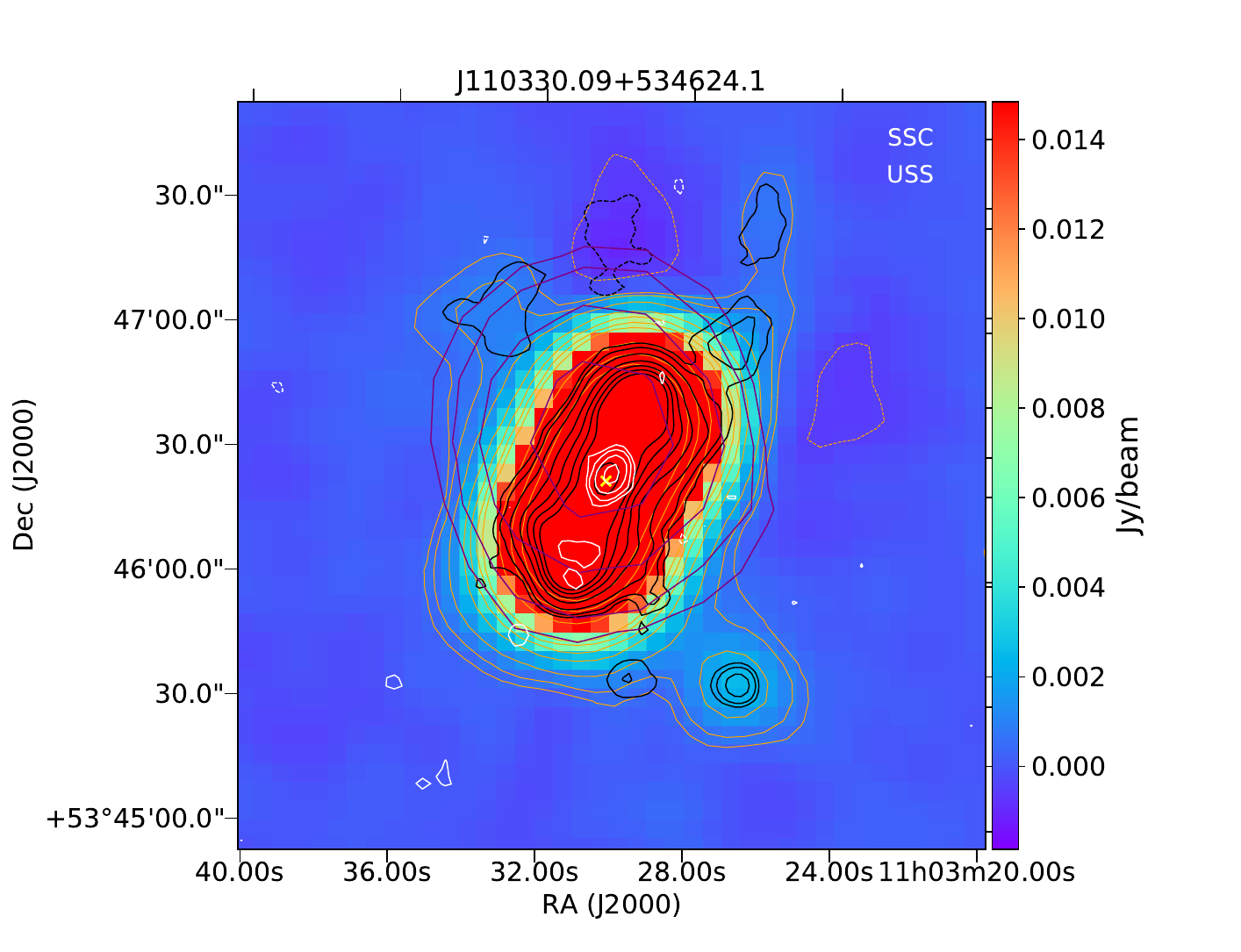}
            \includegraphics[width=0.3\textwidth]{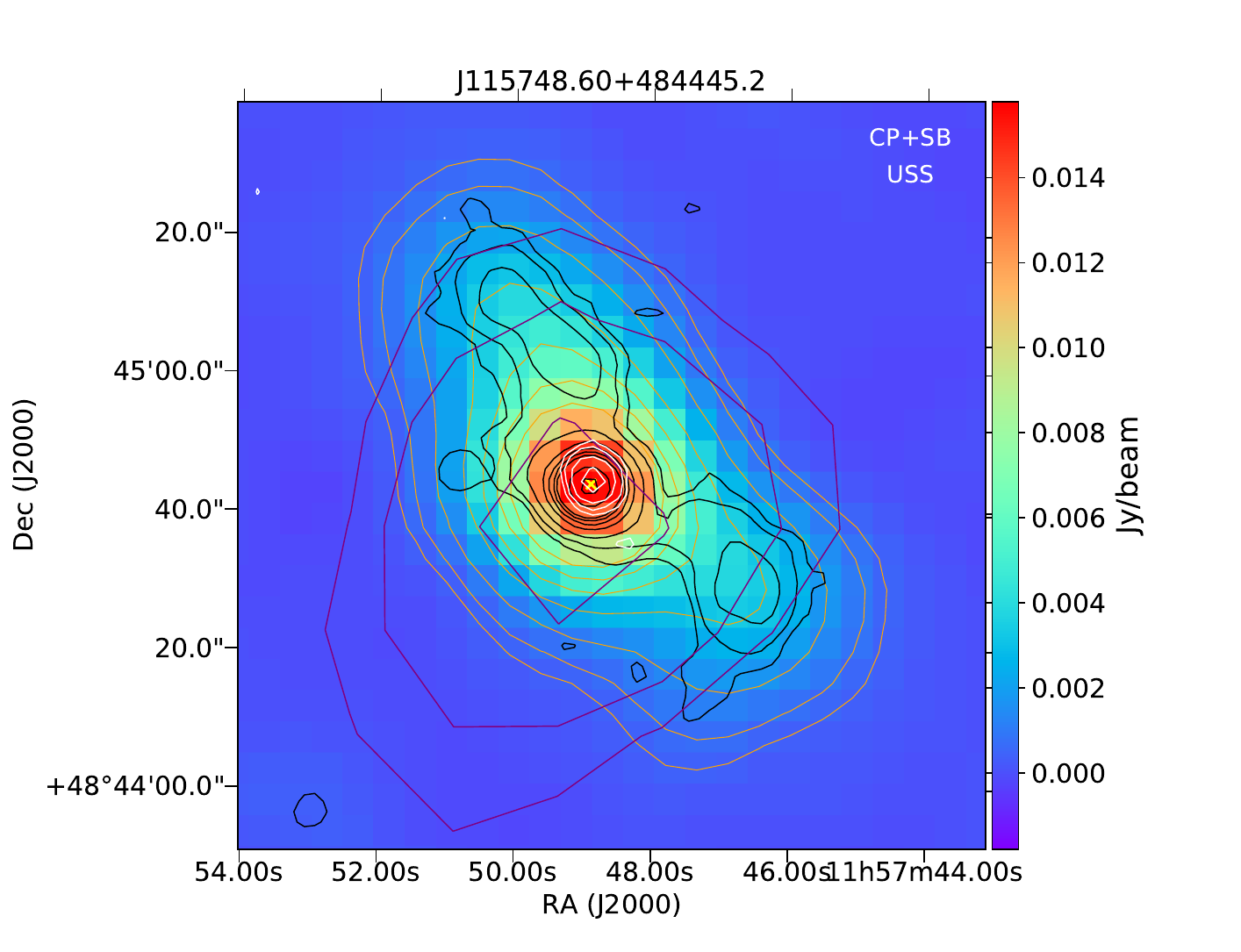}
           }         
\centerline{\includegraphics[width=0.3\textwidth]{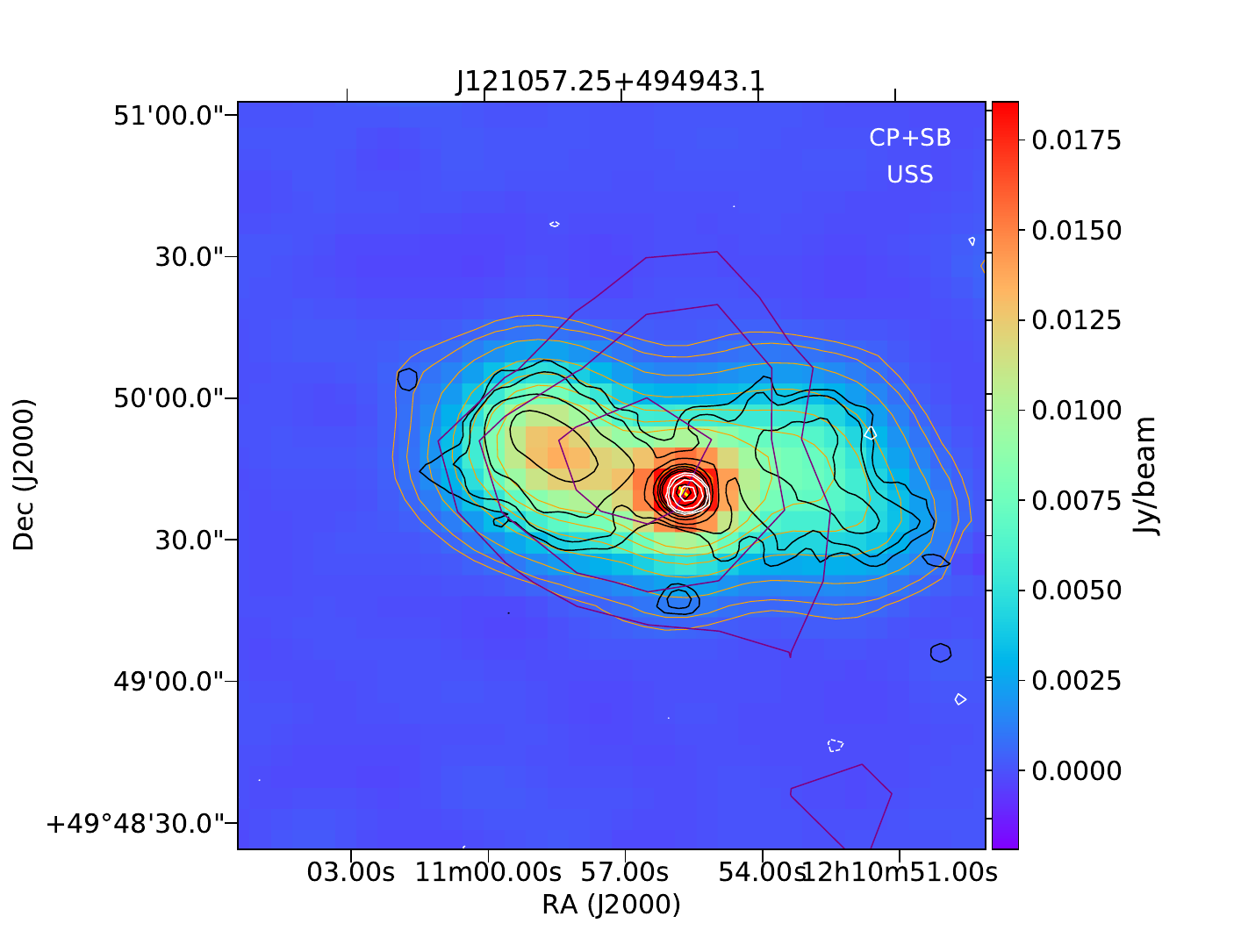}
            \includegraphics[width=0.3\textwidth]{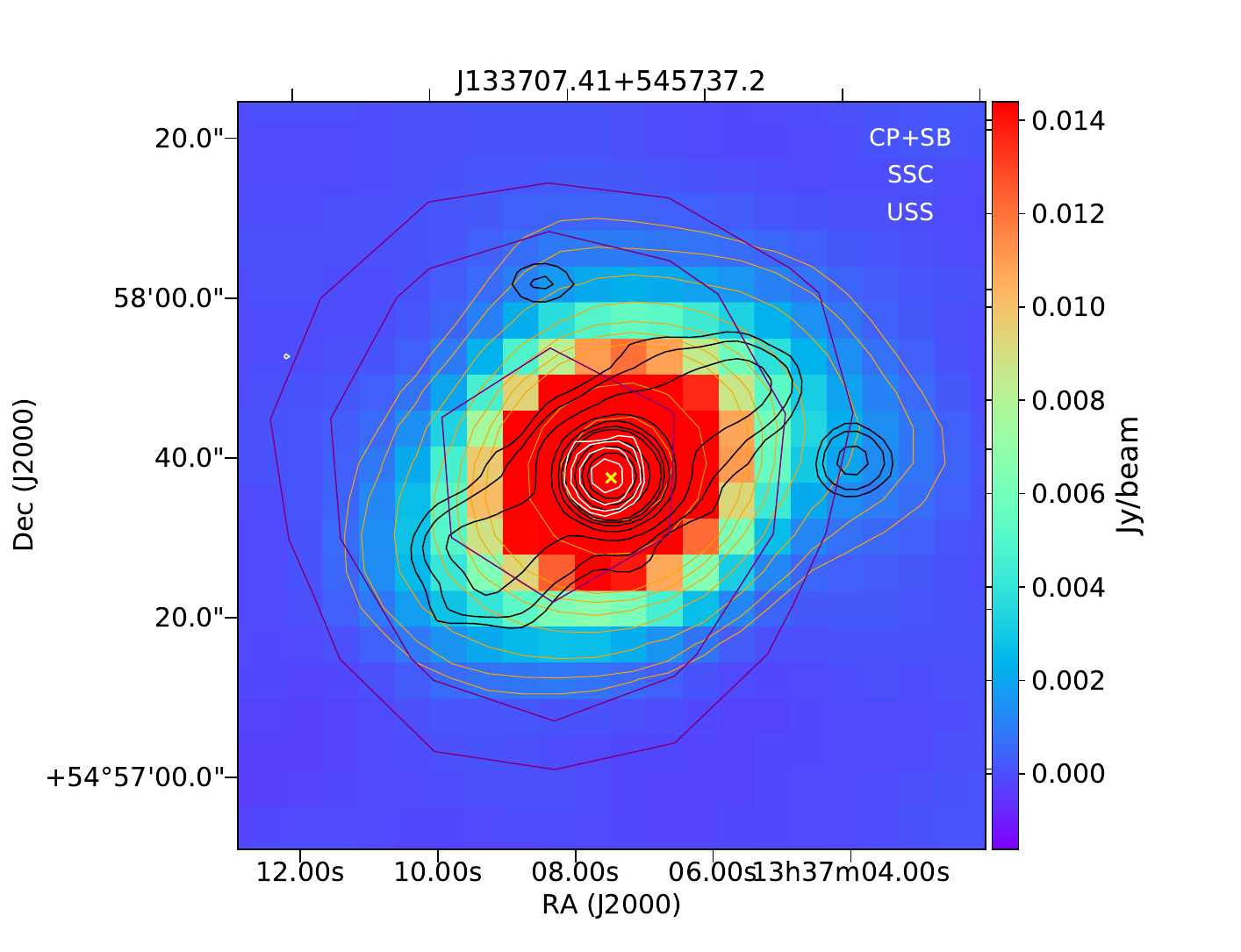}
            \includegraphics[width=0.3\textwidth]{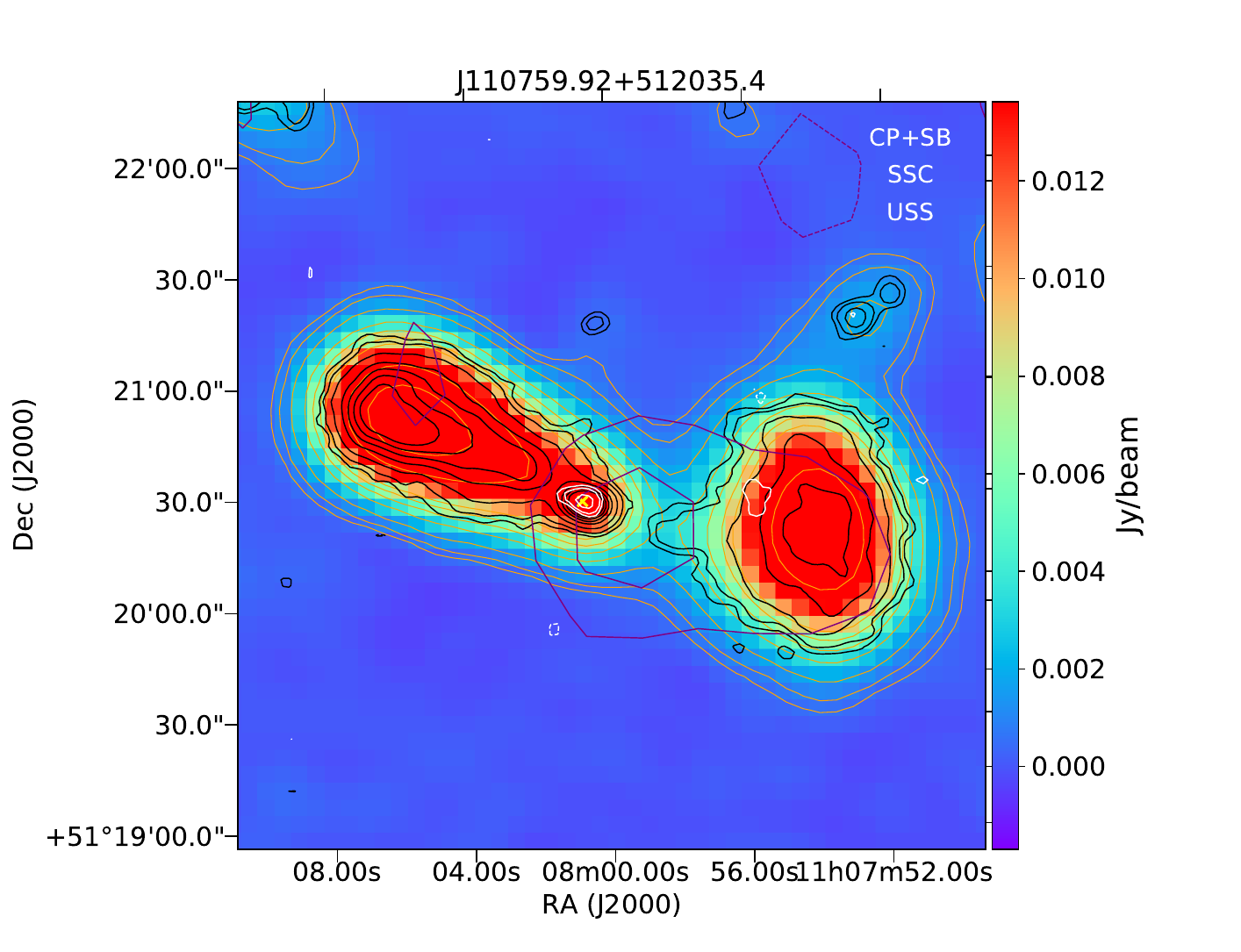}
           }
\centerline{ \includegraphics[width=0.3\textwidth]{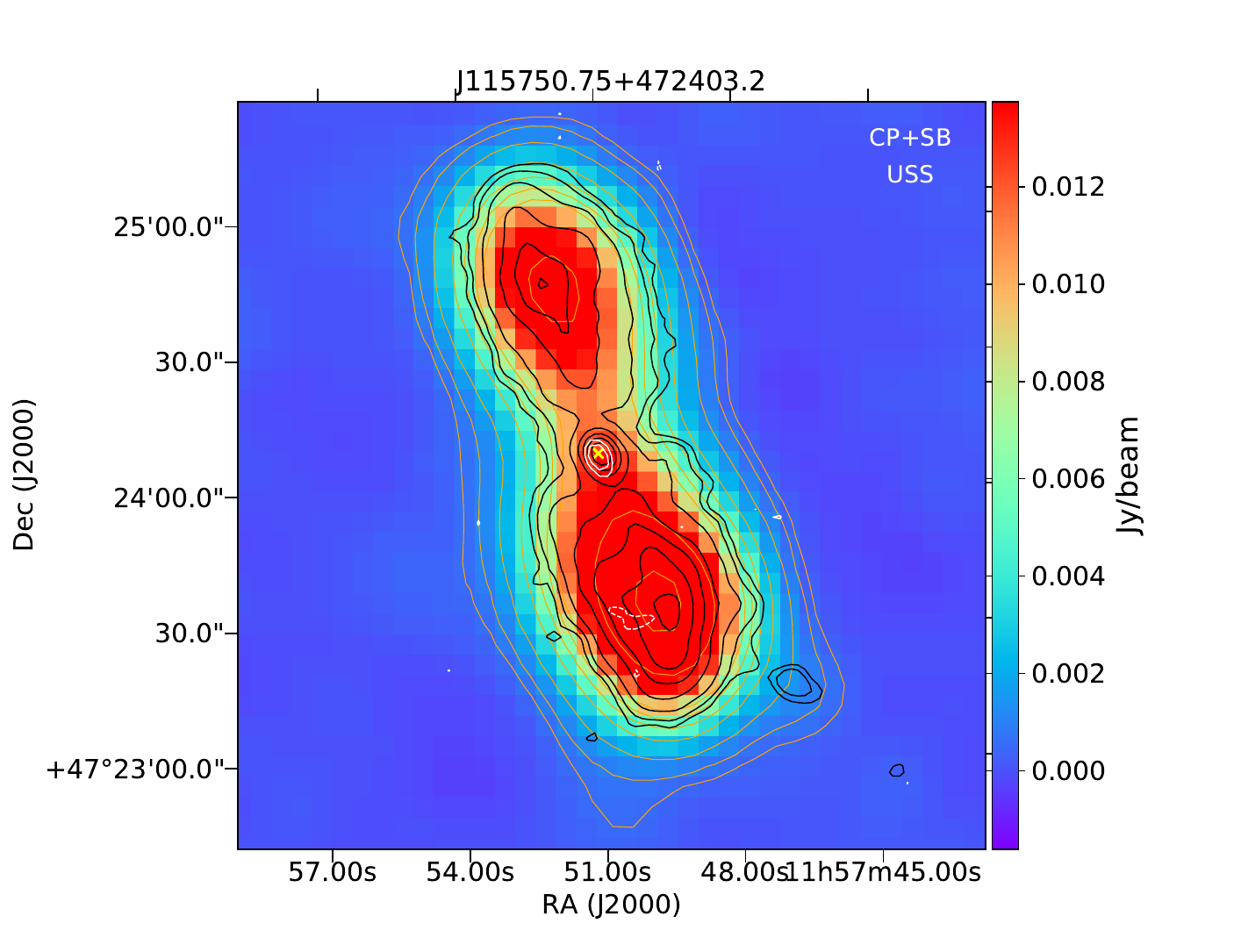}
            \includegraphics[width=0.3\textwidth]{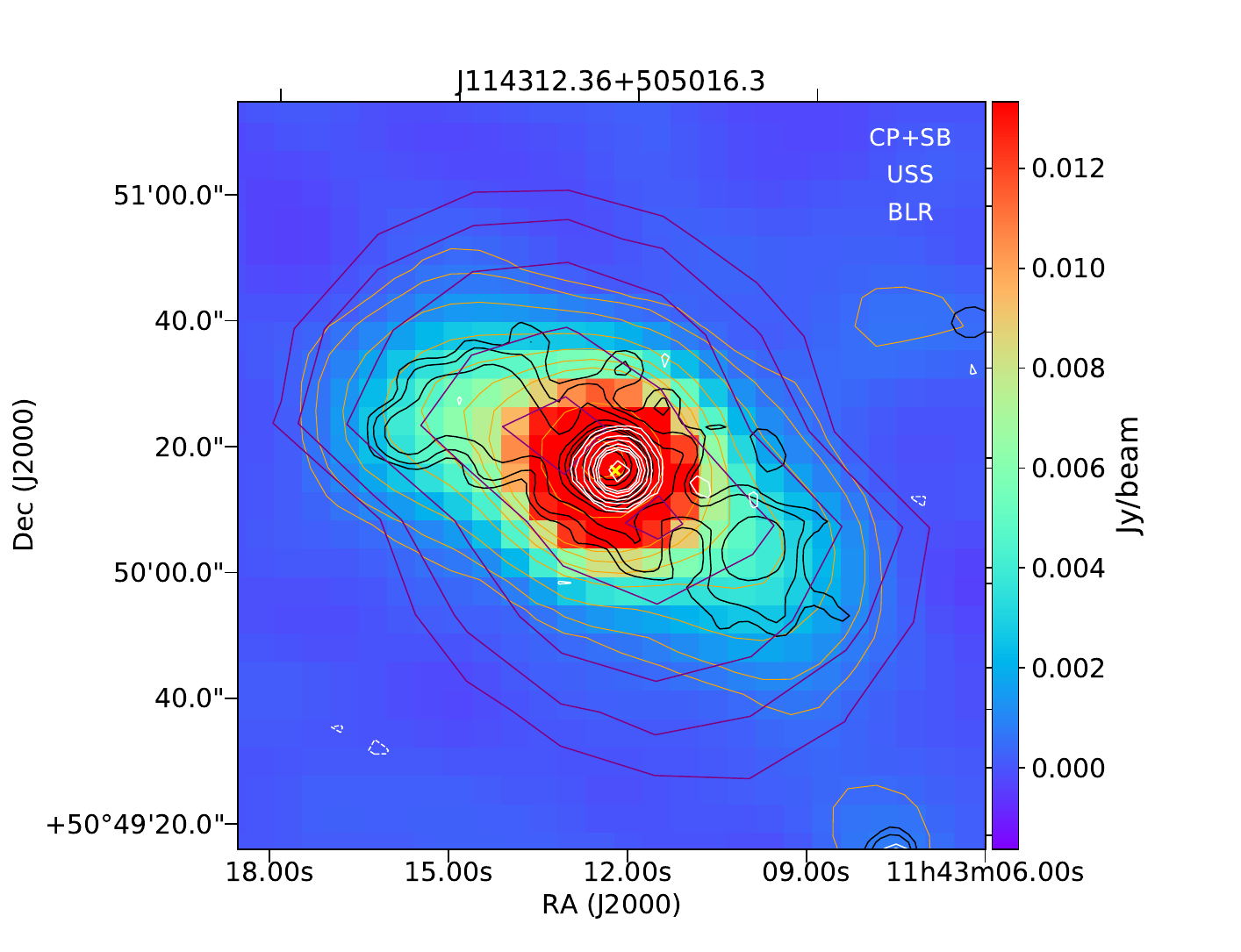}
            \includegraphics[width=0.3\textwidth]{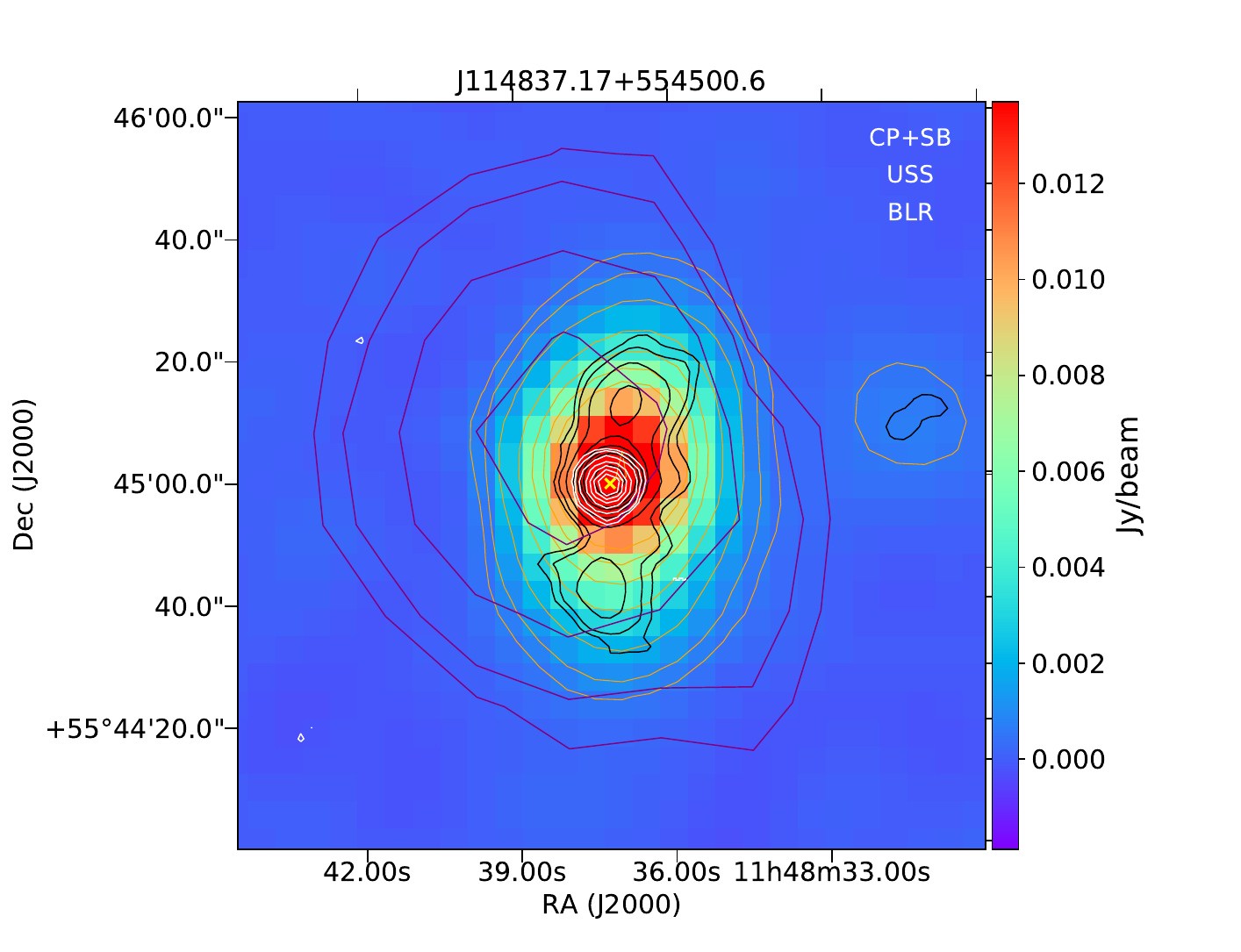}
           }
\caption{Sources with USS remnant emission and bright cores described in Sect.~\ref{Spectral index of the extended emission} in the candidate restarted galaxies. We note that the USS extended values calculated for J110759.92+512035.4 and J115750.75+472403.2 are limits due to partial or no NVSS detection as described in Sect.~\ref{Spectral index of the extended emission}. J114312.36+505016.3 and J114837.17+554500.6 with USS remnant lobes and broad Mg emission lines in their SDSS spectra are shown in the middle and right of the bottom panel. The contours represent: VLA FIRST radio contours (white), LOFAR radio contours of the high-resolution maps (black), and NVSS radio contours (purple) are overlaid on the LOFAR 20$^{\prime\prime}$ resolution maps (orange) for these sources. The host galaxy position is marked with a yellow cross.
}           
\label{fig:USS-morphology}
\end{figure*}

\subsection{Ultra-steep spectral index of the extended emission and a bright core}
\label{Spectral index of the extended emission}

Although the low SB indicates potential remnant emission, the most convincing evidence is obtained if the diffuse emission is also ultra-steep spectrum (USS). The integrated radio spectrum of active sources is characterized by a (broken) power law, where changes (steepening) in the spectral index (from the initial injection index typically in the range $\alpha_\mathrm{inj}$ = -0.5-0.7 \citep{blandfordostriker1978} are due to the synchrotron and inverse Compton losses of the electrons \citep{jaffeperola1973,carilli1991}. However, if the activity and the replenishment of fresh electrons stop, the steepening of the spectral index is expected to become more dramatic \citep{komissarovgubanov1994,brienza2017}. 
Thus, ultra-steep spectrum emission, typically $\alpha_\mathrm{ext} < -1.2$, can be considered as a good indicator for older remnant lobes \citep{komissarovgubanov1994,brienza2017}. Depending on the age of the remnant emission, this USS may appear only at high frequencies. Thus, the USS of the extended emission between 144~MHz and 1400~MHz, identified alongside a prominent core, serves as a distinct indicator of older remnant lobes from past activity, featuring a reactivated jet at the center. 

For computing $\alpha_\mathrm{ext,1400MHz}^\mathrm{ext,144MHz}$,
we used the flux density at 144 MHz of the extended emission calculated before for the SB.
In a similar way, we calculated the extended emission at 1400 MHz ($\mathrm{S}_\mathrm{ext,1400MHz}$) as the difference between the total flux densities measured from the NVSS image and the peak core flux density measured from the FIRST image. For sources partially or not detected in NVSS, we used the flux densities measured from the NVSS images that were calculated before for $\mathrm{CP_{1400}}$, as described in Sect.~\ref{High radio core prominence and low surface brightness}. Here, we matched (as closely as possible) the regions from which we extracted the flux for both frequencies, ensuring consistency in mapping the same structures.

Interestingly, we have found nine sources with ultra-steep spectral index $< -1.2$ (in the range from $-1.2$ to $-2.04$). These nine USS sources are shown in Fig.~\ref{fig:USS-morphology} and their properties are given in Table~\ref{list of sources}. These sources are noteworthy cases of restarted activity as their USS emission is the signature of the past, remnant activity, while they show the presence of new restarted activity in the bright central region. Four out of these nine USS sources are also characterized by a SSC. The USS values for two sources - J110759.92+512035.4 (< -1.40) and J115750.75+472403.2 (< -1.36) are limits due to partial or no NVSS detection. 
We return to these sources in Sect.~\ref{USS extended and steep spectrum core}.

\subsection{Beaming contamination}
\label{Beaming effects contaminants}

The high CP of the core-D sources can also be caused, at least in some cases, by the jet being strongly beamed 
(e.g., \citealt{marin2016}). If this happens, we cannot use the high CP as an indicator of a restarted source. 
In order to verify if beaming effects could be present in some sources, we have visually inspected their SDSS spectra to check for the presence of any broad permitted lines. Broad permitted emission lines are a signature of Type-1 AGNs, where the broad line regions are visible to the observer, due to the line of sight to the source being almost on the axis. 
Based on the SDSS spectra classification, we can confirm that there are ten sources categorized as `\textquoteleft QSO broad-line\textquoteright' (marked with `\textquoteleft BLR\textquoteright' on their respective maps in Figs.~\ref{fig:1}--\ref{fig:7}).
Thus, in these ten objects, the emission from the AGNs is likely beamed toward us and this likely affects also the radio emission. This suggests that also the emission associated with the core can be affected by beaming and the high CP can be a result of this. 

However, beaming alone does not exclude the presence of restarted activity. By looking at the other properties of these ten sources we find that two of them (J114312.36+505016.3 and J114837.17+554500.6) are also characterized by a USS extended emission as presented in Sect.~\ref{Spectral index of the extended emission}. Thus these two objects, despite being likely QSO affected by beaming, might be examples of restarted. Therefore, we keep these two as candidate restarted sources.
Following this step, we were left with a total of 71 sources identified as potential restarted galaxies. 

As a remark, we also note that while we primarily utilize broad emission lines to exclude beamed sources, it is important to consider the possibility of radiatively inefficient sources lacking such lines but still oriented toward us.

\subsubsection{VLBI morphology}
\label{VLBI morphology}

Another approach we adopted to check whether beaming effects are present in some of the sources is to look into their VLBI maps, if available. VLBI maps are the best tool to probe the parsec-scale morphology including the possible presence of objects with one-sided Doppler-beamed parsec-scale jets. 

We cross-matched the core-D AGNs sample with the Radio Fundamental Catalog (RFC\footnote{\url{https://astrogeo.org/rfc/}}) to obtain any VLBI counterpart and their high-angular resolution (arcseconds scale) images at C-band (4300 MHz) and X-band (7600 MHz). We obtained VLBI counterparts for five sources (see Table ~\ref{vlba detections}).
The VLBI morphology and spectral index properties of their VLBI cores are given in Table ~\ref{vlba detections}. 
The spectral indices of their VLBI cores are determined using the unresolved flux densities at 4300 MHz and 7600 MHz ($\alpha_\mathrm{4300\,MHz}^\mathrm{7600\,MHz}$). 

Three of the broad-line objects identified in Sect.~\ref{Beaming effects contaminants} have been identified with a VLBI counterpart (marked with $\dagger$ in Table~\ref{vlba detections}). Two among them, J105922.02$+$523902.1 and J110833.89$+$483202.9,
clearly show a flat spectrum at high frequencies (which is characteristic of quasars) with $\alpha_\mathrm{4300MHz}^\mathrm{7600MHz} = $ \,
-0.07 and 0.0 respectively.
Additionally, J105922.02+523902.1 shows a clear one-sided Doppler-beamed jet on parsec (pc) scales in its VLBI map, which confirms the likely presence of boosting. 
These further confirm these two objects as Type 1 AGNs reinforcing our findings in Sect.~\ref{Beaming effects contaminants}.

Interestingly, J114312.36+505016.3 (marked with $\star$$\dagger$ in Table~\ref{vlba detections}) identified with broad lines in the SDSS spectrum but with a USS extended emission as presented in Sect.~\ref{Beaming effects contaminants}, has a steep spectrum VLBI core at high frequencies ($\alpha_\mathrm{4300MHz}^\mathrm{7600MHz} = -1.33 $). This further confirms this object as a candidate restarted source (despite the CP likely being affected by beaming).

\subsubsection{One-sided jet morphology sources}
\label{One-sided jet morphology sources}

As a final check, we conducted a visual inspection of the LOFAR HR images to identify sources with one-sided jet morphology, which could also be indicative of underlying beaming effects.
We identified two sources, J113104.89+554530.7, and J140441.52+481100.6, with a clear one-sided jet morphology in the sample. Thus, it is possible that the core flux could be Doppler boosted and the high CP of these two sources is affected by orientation effects. 
In order to keep our final results free from any contamination on core prominence values, we remove these two one-sided jet morphology sources from restarted candidates. The maps of these two sources are marked with `\textquoteleft one-sided\textquoteright' in Fig.~\ref{fig:6}-\ref{fig:7}.

\begin{table*}[ht!]
\caption{Core-D sources with VLBA observations from literature (see Sect.~\ref{Beaming effects contaminants} for details) and the parameters derived. 
}
\label{vlba detections}
\begin{center}
\begin{tabular}{llllllllll}\hline\hline
\mc{Source}  &  $S_\mathrm{t,4300MHz}$ &  $S_\mathrm{un,4300MHz}$ & $S_\mathrm{t,7600MHz}$ & $S_\mathrm{un,7600MHz}$ & $\alpha_\mathrm{c/x,t}$ & $\alpha_\mathrm{c/x,un}$  & Type & Sep  & VLBI \\
\mc{}           & [Jy] & [Jy] & [Jy] & [Jy] & &  & & [\arcsec] & morph   \\
\mc{(1)}        & (2)  & (3)  & (4)  & (5) & (6) & (7) & (8) & (9) & (10)\\
\hline
J105922.02+523902.1$\dagger$ & 0.046 & 0.026 & 0.045 & 0.025 &  $-0.04$ & $-0.07$ & F & 1.36 & core-jet     \\
J110833.89+483202.9$\dagger$ & 0.018 & 0.016 & 0.017 & 0.016 & $-0.10$ & ~0.0 & F & 0.96 & core \\
J111300.30+494230.0 & 0.013 & 0.011 & 0.015 & 0.014 & +0.25 & +0.42 & F/I & 8.06 & core \\
J114312.36+505016.3$\star$$\dagger$ & 0.037 & 0.032 & 0.025 & 0.015 & $-0.69$ & $-1.33$ & S & 0.97 & core \\
J144841.08+465638.2 & 0.038 & 0.018 & 0.027 & 0.029 & $-0.60$ & +0.84 & I & 0.24 & core \\ 
\hline  
\end{tabular}
\end{center}
\footnotesize{ {\bf Columns:} 1~--~Source; 2~--~C-band total flux density integrated over entire map in Jy;
  3~--~C-band unresolved flux density at long VLBA baselines in Jy; 4~--~X-band total flux density integrated over the entire map in Jy; 5~--~X-band unresolved flux density at long VLBA baselines in Jy; 6~--~spectral index estimated using total flux at C and X bands from columns 2 and 4; 7~--~spectral index estimated using VLBI flux at C and X bands from columns 3 and 5; 8~--~core spectrum type (F: flat spectrum, I: inverted spectrum, S: steep spectrum, I: inverted spectrum); 9~--~angular separation between radio positions in the LoTSS DR2 catalog and RFC in arcsec; and 10~--~VLBI morphology. 
  The source names marked with $\dagger$ are the ones identified with broad lines in their SDSS spectra as discussed in Sect.~\ref{Beaming effects contaminants}, and $\star$ are the sources identified with USS in Sect.~\ref{Spectral index of the extended emission}.
  }
\end{table*}

\subsection{Final sample: Overview and considerations}
\label{Final sample}
Following the framework and the parameters outlined by \cite{jurlin2020} and various other literature, as appropriate for identifying restarted sources as detailed in Sect.~\ref{Identification of candidate restarted radio galaxies}, we applied a variety of criteria (high CP along with low SB of extended emission, SSC or ISC, USS extended emission coupled with a bright core and morphology) to identify candidate restarted radio galaxies within the core-D sample. As summarized in Fig.~\ref{fig:flowchart}, we excluded five sources with small-unreliable morphology through visual inspection of the sample. Eight Type-1 AGNs were excluded from the sample due to the presence of broad permitted lines in their SDSS spectra (suggesting the presence of beaming).  
Two objects with broad permitted lines have been retained in the final sample due to the presence of ultra-steep extended emission, indicative of old remnant lobes. 
Finally, two more sources were excluded from the final sample because they display one-sided jet morphology, a characteristic of Doppler-boosted AGNs.

As a result of these steps, we identified a final sample of 69 candidate restarted radio galaxies within the core-D AGNs sample. This final sample comprises 63 sources identified through the high CP combined with low-SB criteria, and six through the SSC criteria (nine sources satisfy both CP+SB and SSC criteria).
Among them, a subset of nine sources exhibited USS characteristics. 
These 69 sources are shown in Figs.~\ref{fig:1}--\ref{fig:5}
with the VLA FIRST radio contours (white), LOFAR radio contours of the high-resolution maps (black), and NVSS radio contours (purple) overlaid on the LOFAR 20$^{\prime\prime}$ resolution maps (orange) for these sources, and the optical ID from SDSS DR14 indicated by a yellow cross. These maps are marked with the selection criteria they satisfy as `\textquoteleft CP+SB\textquoteright', `\textquoteleft SSC\textquoteright', or `\textquoteleft USS\textquoteright'.
Their properties are listed in Tables~\ref{list of sources} and \ref{Core flux densities}. 

The density of restarted sources we found in this study is one per 
$6.2\,\mathrm{deg}^2$.
We emphasize here that the 69 sources we identified as restarted candidates represent a lower limit on the number of candidate restarted galaxies in the HETDEX field. This is because of the original selection of the sample (from \citealt{mingo2019}) which consists of one type of source, that is, core-dominated sources.
Thus, the sample considered in this study misses other types of candidate restarted radio galaxies which can be identified by properties different from the CP.
For example, we could potentially overlook cases such as 3C~388 \citep{brienza2020}, where the restarted nature requires investigation through resolved spectral index images. A number of such objects have been found in recent studies (see e.g., \citealt{2021Galax...9...88M}). Other cases such as the double-double radio galaxies, the classic example of restarted radio AGNs (\citealt{mahatma2019} and references therein) are also missed by the CP selection of the initial sample we used. 

\begin{figure*}[ht!]
\centerline{\includegraphics[width=0.45\textwidth]{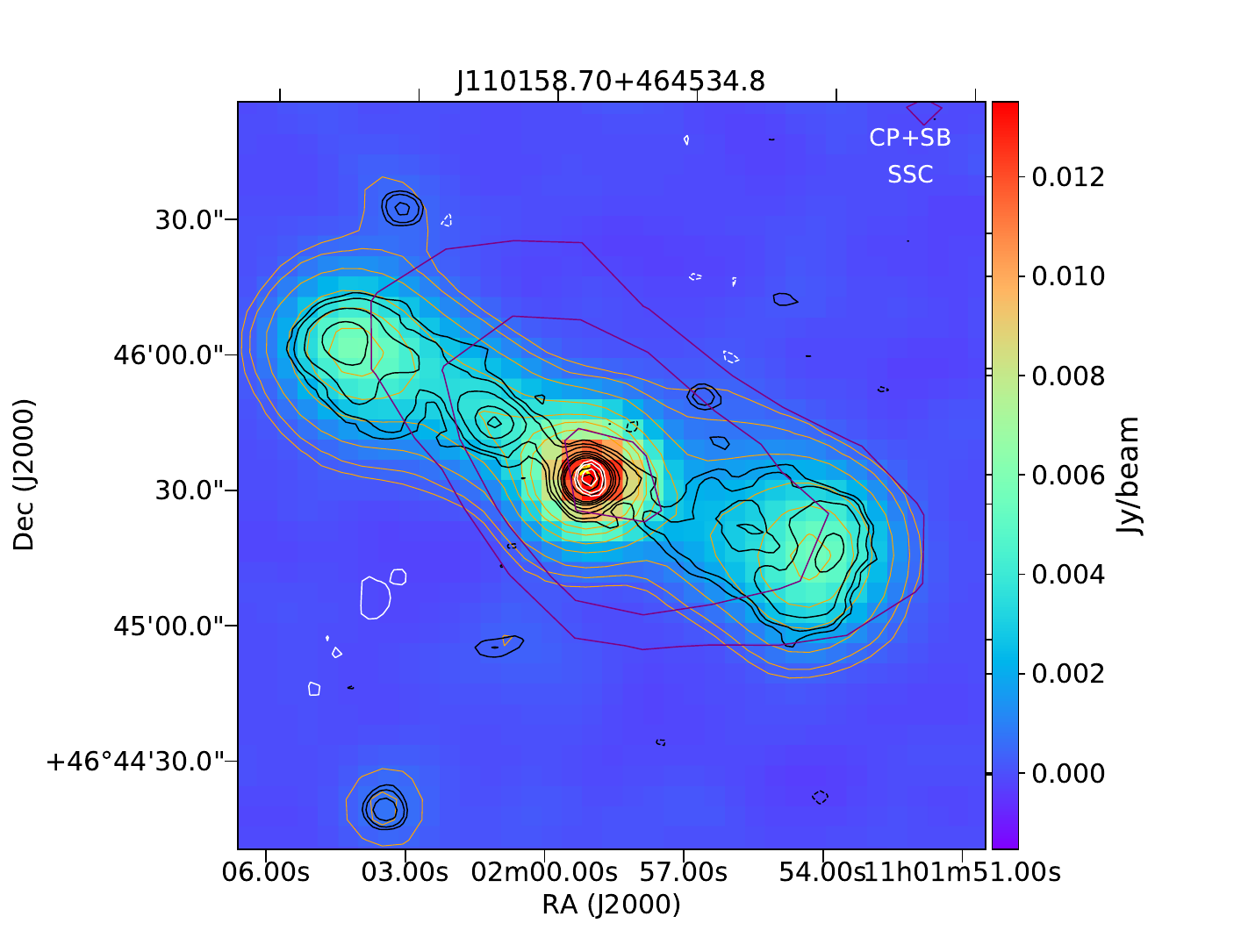}
            \includegraphics[width=0.45\textwidth]{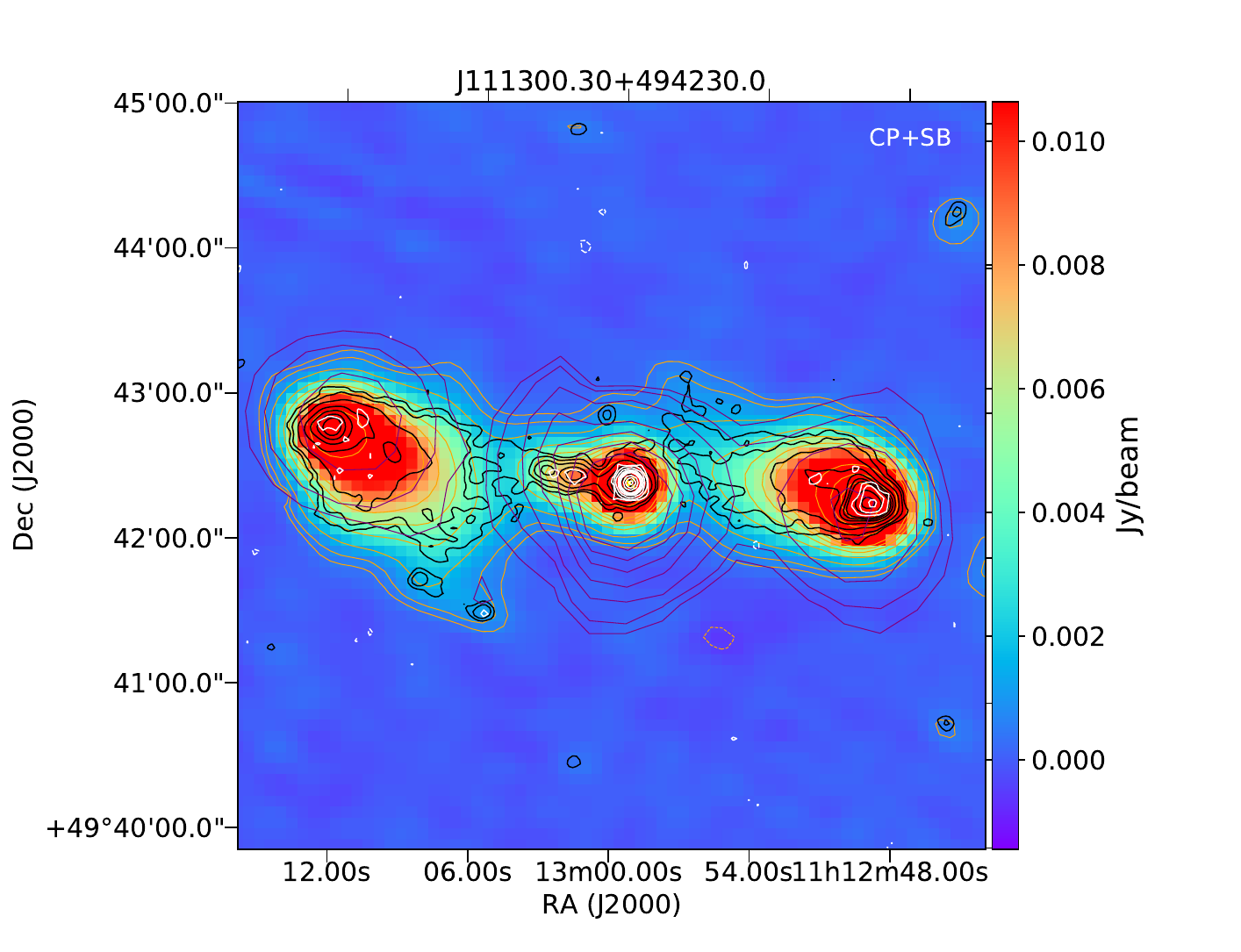}
           }       
\centerline{\includegraphics[width=0.45\textwidth]{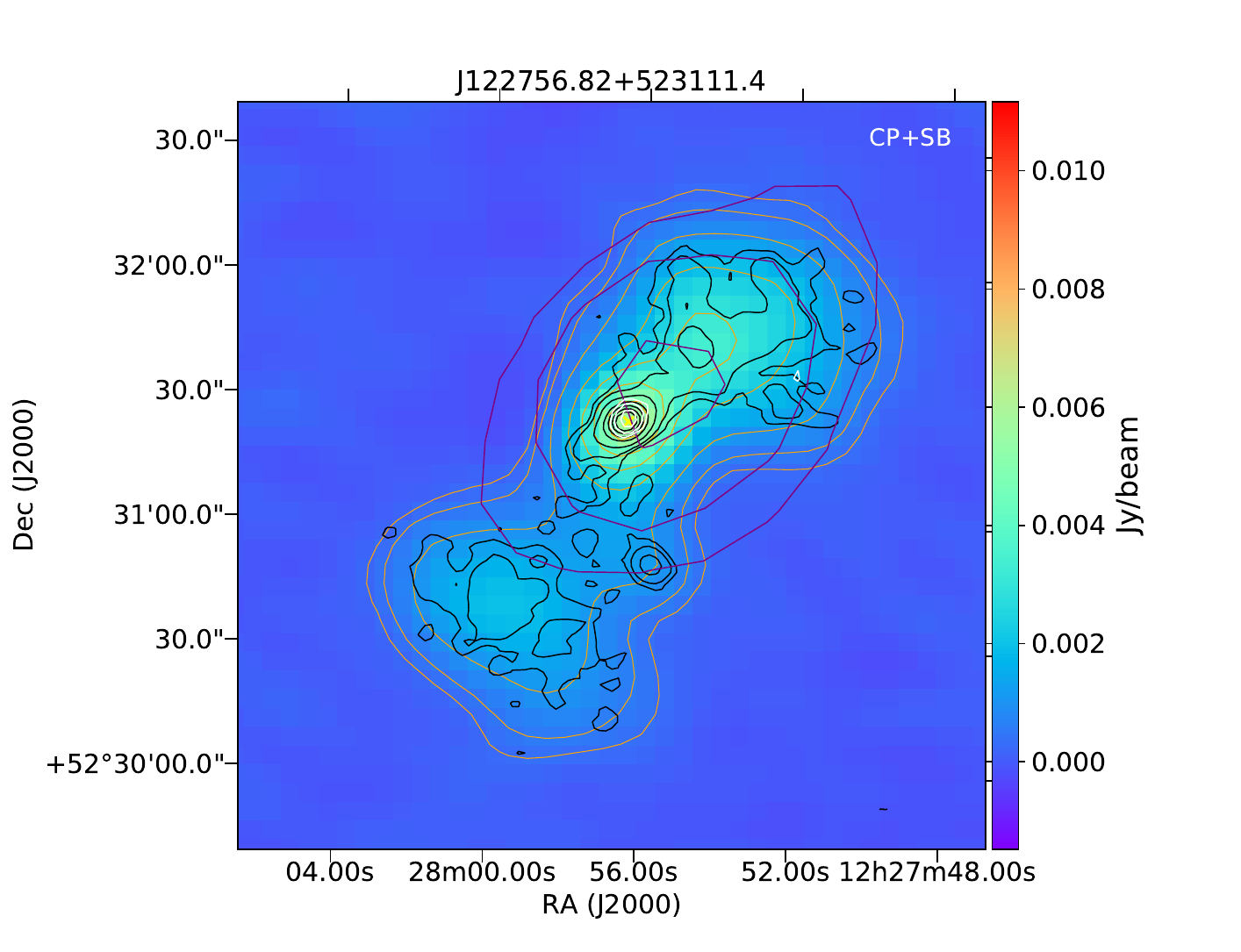}
            \includegraphics[width=0.45\textwidth]{Maps_final/tplot_40_3sig.pdf}
           }         
\centerline{\includegraphics[width=0.45\textwidth]{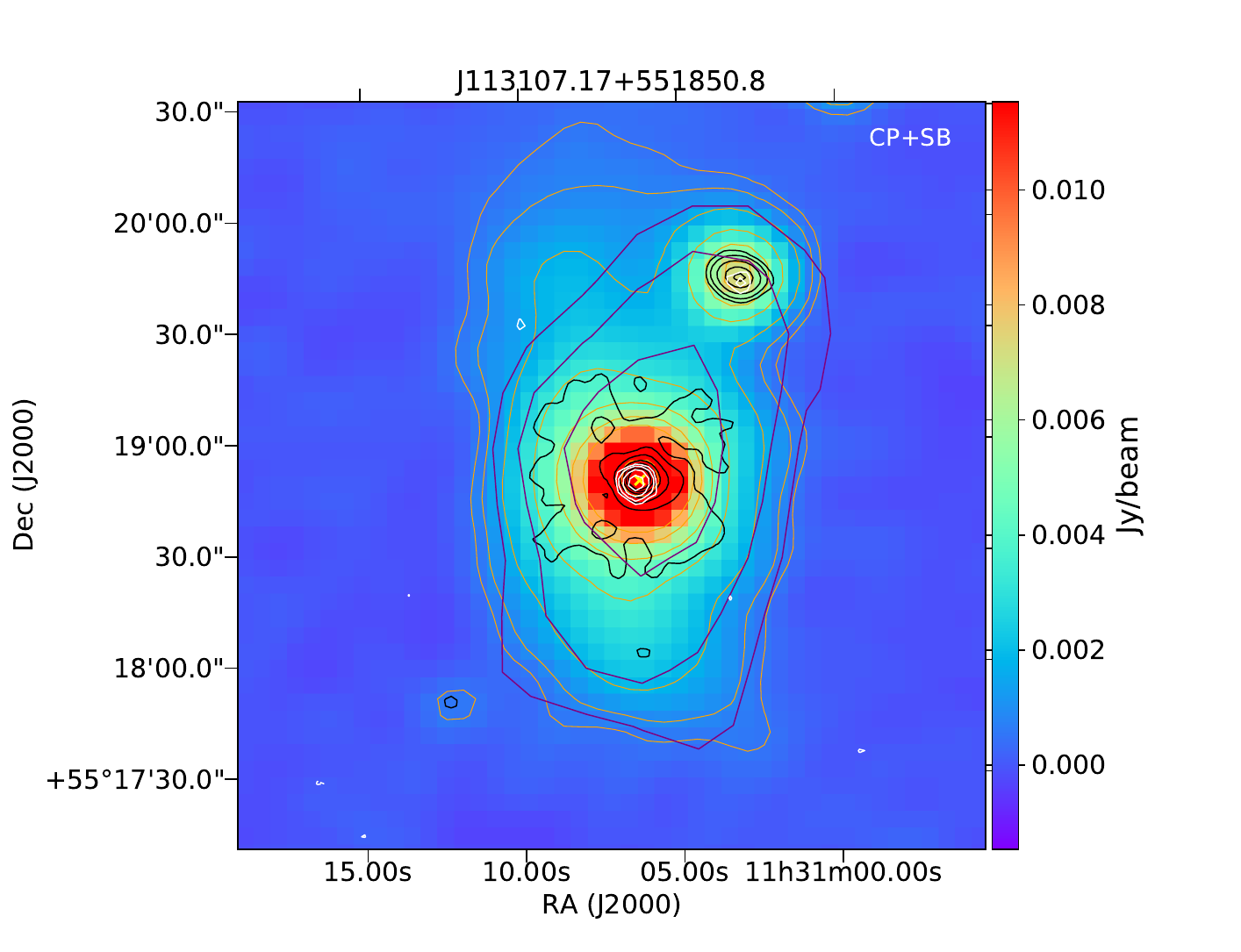}
            \includegraphics[width=0.45\textwidth]{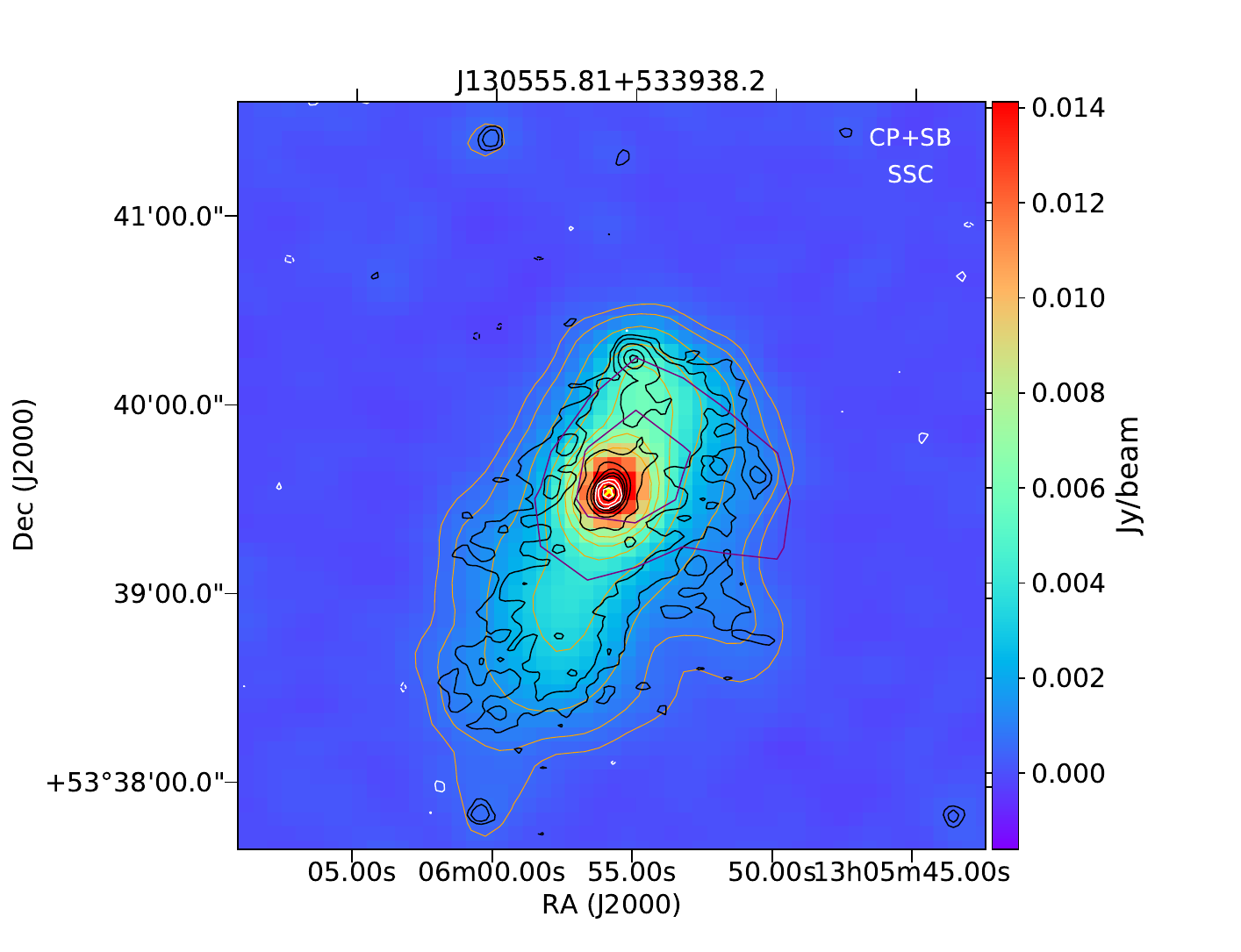}
}
\caption{Different morphologies of candidate restarted radio galaxies in the final sample presented in Sect.~\ref{Final sample}.
\textit{(Top panel)}: Examples of FRII-type candidate restarted radio galaxies (left: an example FRII-type with all structures connected, right: an example FRII-type with structures not connected). \textit{(Middle panel)}: Two FRI-type candidate restarted radio galaxies (left: an example FRI with fading lobes,  right: an example lobed FRI). 
\textit{(Bottom panel)}: Examples of core-with-halo type candidate restarted radio galaxies. The contours represent: VLA FIRST radio contours (white), LOFAR radio contours of the high-resolution maps (black), and NVSS radio contours (purple) are overlaid on the LOFAR 20$^{\prime\prime}$ resolution maps (orange) for these sources. The host galaxy position is marked with a yellow cross.
}             
\label{fig:maps-morphology}
\end{figure*}

\begin{figure}
\centerline{\includegraphics[width=0.55\textwidth]{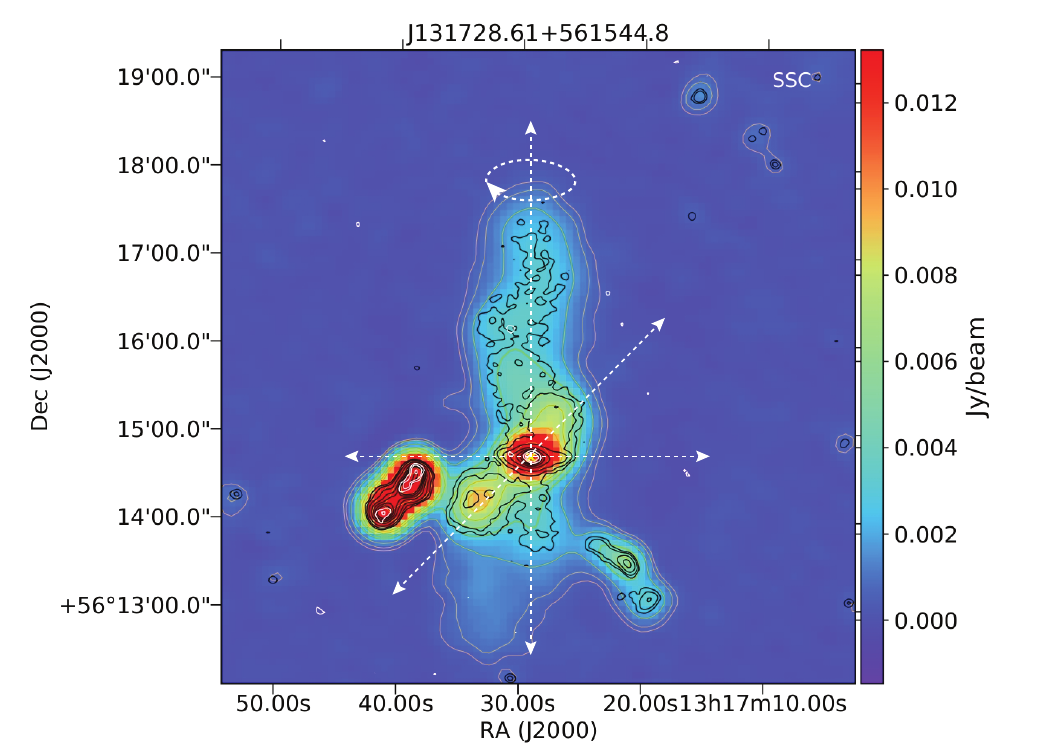}}
\caption{Map of J131728.61+561544.8 which is an interesting case of restarting activity in the sample presented. An ongoing jet realignment in this source is clearly seen in its morphology (see Sect.~\ref{Morphology of candidate restarted galaxies}).
The directions of jet precession are depicted by three axes, with an arrow indicating clockwise precession of the jet. The various contours represent: VLA FIRST radio contours (white), and LOFAR radio contours of the high-resolution maps (black) are overlaid on the LOFAR 20$^{\prime\prime}$ resolution maps (orange). The host galaxy position is marked with a yellow cross.}
\label{fig:J131728.61+561544.8}
\end{figure}

\section{Results and discussion}
\label{Results and Discussion}

In this work, we meticulously curated a sample of radio galaxies that are candidates to be in the restarted phase of their AGNs life cycle. We have started with an initial sample of core-D radio sources within the HETDEX region (from \citealt{mingo2019}), and then systematically applied criteria of
core dominance, surface brightness, spectral characteristics, and morphological features. 

The high fraction (69 out of 94) of the initial sample showing signatures of the presence of restarted activity supports the idea that CP can be a useful method for the selection of a restarted radio source, as also shown in \cite{jurlin2024} and references therein. 
The initial core-D sample, taken from \cite{mingo2019} was 
selected to include objects with a particularly prominent core (with respect to the extended emission). 
While it may not offer an exhaustive inventory of restarted radio source properties, our study marks the second endeavor after \cite{jurlin2020} to systematically identify restarted radio galaxies with diverse morphologies and characteristics based on predefined criteria previously employed in the literature.

Furthermore, compared to previous studies, we have expanded the search for candidate restarted sources to a larger area of the HETDEX region and have used multiple parameters to characterize these sources. 
Indeed, our study is complementary and expands the work of \cite{jurlin2020}.
While  \cite{jurlin2020}'s sample of restarted galaxies in the LH field had the advantage of deeper coverage (and better multiwavelength data to identify and characterize the hosts), it was limited to  23 sources of which only 13 were selected based on CP (i.e., 56\%). This makes it clear that by selecting a sample with high core prominence as we have done in this work we can explore only part (approximately half) of the population of candidate restarted sources. Therefore we cannot derive the full statistics of candidate restarted sources in the HETDEX field in this work.

Our study uncovered 69 sources, signifying a broader collection of potential restarted candidates. This is noteworthy considering our focus on a specific type of restarted source and acknowledging the inherent bias in the core-D initial sample.
Specifically, our study more deeply investigates core dominance as a criterion for selecting a restarted source. Because CP may provide an incomplete selection, we do not draw from the sample any statistical conclusion.

Below we discuss the interpretation of our findings in more detail:
specifically, we examine the morphology of restarted candidates (Sect.~\ref{Morphology of candidate restarted galaxies}), 
we discuss the USS class from our systematic analysis and 
their implications for future work (Sect.~\ref{USS extended and steep spectrum core}), 
we consider the radio properties (Sect.~\ref{The radio properties: radio luminosities and linear sizes}),
and finally, we examine their host galaxy and optical properties (Sect.~\ref{Host Galaxy properties of candidate restarted radio galaxies}).

\subsection{Morphology of candidate restarted galaxies}
\label{Morphology of candidate restarted galaxies}

Despite all sources being dominated by a bright radio core, the morphologies of the final sample of 69 candidate restarted galaxies are very diverse and not limited to DDRG. 
The evident diversity in their inherent morphologies, encompassing FRI, FRII, core-with-halo, and asymmetric forms (see examples in Fig.~\ref{fig:maps-morphology})
strongly indicates the existence of distinct progenitor types, suggesting that the progenitors are also a heterogeneous population \citep{2021A&A...648A...9M}.

We identify peculiar morphologies in some of the restarted candidates.
For example, J124308.33+521247.6 (first panel in the third row in Fig.~\ref{fig:3}) exhibits radio emission perpendicular to the direction of the main lobe, possibly indicating pressure gradients in the hot atmosphere that cause the lobes to bend toward the northeast \citep[e.g.,][]{leahy1984,hodgeskluck2012,2019MNRAS.488.3416H}. 
J151845.83+523707.9 (third panel in the third row in Fig.~\ref{fig:5}) is another example of an anisotropic pressure gradient in the environment, influencing the backflow of the northern radio lobe to expand toward the west perpendicular to the main lobe. 

The morphology of one of the restarted candidates, J131728.61+561544.8, depicted in 
Fig.~\ref{fig:J131728.61+561544.8}, reveals an ongoing extreme jet realignment. 
This source shows a clear example of quick intermittent activity with jet precession, characterized by three distinct phases that produced emissions oriented along three different axes: initially, faint emission in the north and south visible only at 3$\sigma$ in LOFAR HR in the N-S direction; subsequent clockwise movement of the jets resulting in brighter, tilted lobes in the NW-SE direction; and finally, another clockwise movement leading to jets oriented in the E-W direction.
Consequently, this source can be classified as an X-shaped radio galaxy \citep{mingo2011,kuzmicz2017,2019MNRAS.488.3416H}. 
This source exhibits a SSC ($\alpha_\mathrm{1400MHz}^\mathrm{144MHz} = -0.86$). 
The observed change in orientation of the jet could represent a signature of restarting activity: this could possibly be attributable to variations in the spin of its SMBH, or interactions with the ambient environment, and/or changes in the magnetic flux threading the accretion disk 
that powers the relativistic jets.
Such cases of variability in jet orientation and the mechanisms causing jet precession
have been studied, with detections reported in numerous individual jetted supermassive black hole systems (see e.g., \citealt{ekers1978,steenbrugge2008,kharb2014,krause2019}).
Recently, \cite{hernandez2017, hernandez2023} reported a similar case of extreme jet realignment in PBC~J2333.9-2343, exemplifying the transformation from a giant radio galaxy in the past to a blazar in the present epoch. 
Also, \cite{bruni2021} reported a similar case where they detected $\sim$300 kpc wide off-axis emission in B2 1144+35B, alongside an inner jet whose orientation does not align with the axis of the lobes. Furthermore, an orientation variability of the jet in Markarian 6 is reported by \cite{kharb2006} and \cite{mingo2011}. X-shaped radio sources (3C~136.1, 3C~223.1, 3C~403) are also reported by \cite{saripalli2012}. These types of sources are crucial to put constraints on the jet reorientation mechanisms and the timescales in this process.

Three restarted candidate galaxies (J104926.21+472859.0, J113107.17+551850.8, J130555.81+533938.2) exhibit a core-with-halo morphology, as illustrated in Fig.~\ref{fig:maps-morphology} (bottom panel). 
Remnant lobes are indeed expected to become amorphous in shape over time due to the plasma expansion in the surrounding medium if the source is over-pressured at the jets switch off \citep{blundell1999,brienza2017}.
\cite{saripalli2012} highlight significant asymmetry in the extended lobes for both restarting FRI and FRII sources, a characteristic we have also observed in some sources within our dataset (see e.g., J111113.93+500925.8, J113718.05+484715.2, J131417.95+473709.8, J141900.52+490308.0 in Figs.~\ref{fig:1}--\ref{fig:4}).
It is noteworthy that J141900.52+490308.0 resides in a cluster (see Sect.~\ref{Environment of the restarted candidate galaxies}) and hence a possible reason for the asymmetric shape is the influence of the cluster environment surrounding the sources.
J113718.05+484715.2 extends only in the southern direction relative to the core, potentially stemming from varied densities within the lobes, and 
is very similar to an asymmetric restarted source, J104912+575014, as reported by \cite{jurlin2020}, where extension occurs exclusively in one direction (see section 7.1 in \citealt{jurlin2020}).

We identified distinct hotspots (HS) in at least seven restarted sources in our sample 
(J110158.70$+$464534.8, J110921.45$+$481454.0 [HS in the S], J111300.30$+$494230.0, J114816.90$+$525038.4 [HS in NW side], J120529.19$+$561334.7, J131634.57$+$493226.4,
J151845.83$+$523707.9 in Fig.~\ref{fig:1}--\ref{fig:5}).
These HS at the end of the lobes may be fading due to the central activity switching off.
Interestingly, both \cite{mahatma2018} and \cite{quici2021} have found 
remnant sources showing the presence of HS. Although rare, they suggest the presence of sources where the central activity has only recently switched off. Modeling the synchrotron spectrum arising from a HS, \cite{quici2021} has also shown that these HS can persist for 5–10 Myr (at 5.5 GHz) after the jets switch oﬀ. 
\cite{jurlin2020} identified only one restarted source with HS.
On the other hand, HS are commonly identified in the DDRGs in the HETDEX field by \cite{mahatma2019} and in FRII type restarted radio galaxies by \cite{saripalli2012}. 

\subsection{USS extended sources}
\label{USS extended and steep spectrum core}

We have identified nine sources in the 69 candidate restarted galaxies that have two regions with diverse spectral index properties. 
Their central core region is characterized by a spectral index consistent with that of active galaxies and the outer regions are characterized by a USS ($\alpha_\mathrm{ext,1400MHz}^\mathrm{ext,144MHz}$ $< -1.2$) which is consistent with remnant structures. The two regions likely represent distinct populations of electrons. 
Four of the USS sources we found are also characterized by SSC.

Our results expand the group of objects with such properties, which used to be restricted to individual cases, such as 3C~388 \citep{roettiger1994,brienza2020} and has been recently extended by the study of the LH, combining LOFAR and Apertif images \citep{2021A&A...648A...9M, 2021Galax...9...88M}, and of the B\"ootes region \citep{kutkin2023}. 

All these studies underscore the existence of sources where USS emission (suggesting that the fueling of the outer lobes has stopped ) is coexistent with emission from active regions, as the one we are seeing in the present sample. 

For these two distinct phases to be detectable, and in particular for the USS emission being still visible, the off phase of activity must have been relatively short, approximately a few tens of mega years \citep[Myr;][]{2021A&A...648A...9M}.
We know from the recent work on this topic that a range in duty-cycle is present in radio galaxies. The work of \cite{sabater2019} has shown that in massive galaxies the low luminosity radio AGNs are `\textquoteleft switched on\textquoteright' most of the time, that is, have a very fast duty cycle. 
The occurrence of radio AGNs is lower for the luminosity of the sources in our sample  (above 10$^{24}$ \WHz, see next section). Thus, although the duty-cycle may not be so fast, we are still seeing a rapid on-and-off phase similar to other objects studied in detail.
For example, \cite{brienza2020} determined that the period of inactivity in 3C~388 is $\le$ 20 Myr, while the total age of the lobe is $\le$ 80 Myr. \cite{orienti2023} have very recently found sources with USS remnant emission with bright active cores. 
For USS regions observed at low frequencies, ages of the remnant emission between 160 Myr and 320 Myr have been estimated \citep{2021A&A...648A...9M}, suggesting these to be the older remnants still visible at LOFAR sensitivity.  

To determine the duration of the switch-off phase in the nine USS sources with bright cores we identified, it is essential to obtain follow-up, high-frequency observations. 

Within the USS sources, we have found two interesting cases, J114312.36+505016.3 and J114837.17+554500.6, both characterized by the presence of broad Mg emission lines in their SDSS spectra (see Sect.~\ref{Beaming effects contaminants}).
We interpret these two sources as possible cases with remnant emission from a past epoch of activity when the source had a different angle along the line of sight and was not as strongly beamed as in the present epoch. These two sources could be similar to cases discussed in detail in Sect.~\ref{Morphology of candidate restarted galaxies}, such as PBC J2333.9-2343 \citep{hernandez2023} and Markarian 6 \citep{mingo2011}.
Conducting a multi-frequency VLBI follow-up for these two USS sources could offer insights into their restarted nature, encompassing detailed spectra of their nuclear components and morphology.

All the other restarted candidates selected based on 
CP$+$SB criteria or SSC or morphology having a $\alpha$ of the extended emission in the range $-0.5$ to $-1.2$ (i.e., not ultra-steep, see Table~\ref{list of sources}), could represent sources only recently turned off and with relatively rapid restarted phase, preventing the outer lobes from fading away or becoming ultra-steep. If this is the case, the steeper part of the spectrum of these sources could be seen at higher frequencies: only follow-up observations can test this. \cite{brienza2017} predict that in the range 144-1400 MHz, only $\sim$ 50\% of the entire remnant lobe population is recovered, while about 100\% can be recovered if data up to 5 GHz are included.

Moreover, 17 sources in the restarted candidates have a 
$\alpha_\mathrm{1400MHz}^\mathrm{144MHz}$ $> -0.5$, implying that the available data at 1.4 GHz are not deep enough to derive a value for a steep spectrum emission or a tight upper limit.
Notably, all 17 sources show only partial or non-detection in NVSS. 
This results in a very poor constraint on the spectral index of the extended emission for these sources, likely due to their faintness and the limited sensitivity of NVSS.

As a final remark, it is worth noting that the interruption and the restarting events can also be due to temporary disruption of the jet flow, for example as a result of the interaction with a surrounding rich medium.  
A few examples are known where this could be happening (e.g., 3C~293 \citep{kukreti2022}, Centaurus A \citep{morganti1999, mckinley2013, mckinley2018}, B2~0258+35 \citep{brienza2018}, and NGC~3998 \citep{sridhar2020}). To trace the presence of these types of situations, a more detailed study of the gas in the host galaxy is needed.

\subsection{The radio properties: radio luminosities and linear sizes}
\label{The radio properties: radio luminosities and linear sizes}
 
The FRI+FRII parent sample of \cite{mingo2019} exhibits radio luminosities at 144 MHz (L$_\mathrm{144 MHz}$) spanning from log$_{10}$(L$_\mathrm{144 MHz}$/$\mathrm{WHz^{-1}}$) = 22.0 to 
log$_{10}$(L$_\mathrm{144 MHz}$/$\mathrm{WHz^{-1}}$) = 28.0 (see Fig.~5 in \citealt{mingo2019}), with median values of log$_{10}$(L$_\mathrm{144 MHz}$/$\mathrm{WHz^{-1}}$) = 25.30 for FRIs and log$_{10}$(L$_\mathrm{144 MHz}$/$\mathrm{WHz^{-1}}$) = 25.95 for FRIIs. 
The core-D AGNs sample exhibited a diverse range of radio luminosities, with a median value of log$_{10}$(L$_\mathrm{144 MHz}$/$\mathrm{WHz^{-1}}$) = 25.25 and a mean value of log$_{10}$(L$_\mathrm{144 MHz}$/$\mathrm{WHz^{-1}}$) = 26. This range covered regions predominantly occupied by FRIIs at the higher end to FRIs at the lower end, as depicted in Fig.~14 of \cite{mingo2019}. 

\begin{figure}[ht!]
\centerline{\includegraphics[width=0.45\textwidth]{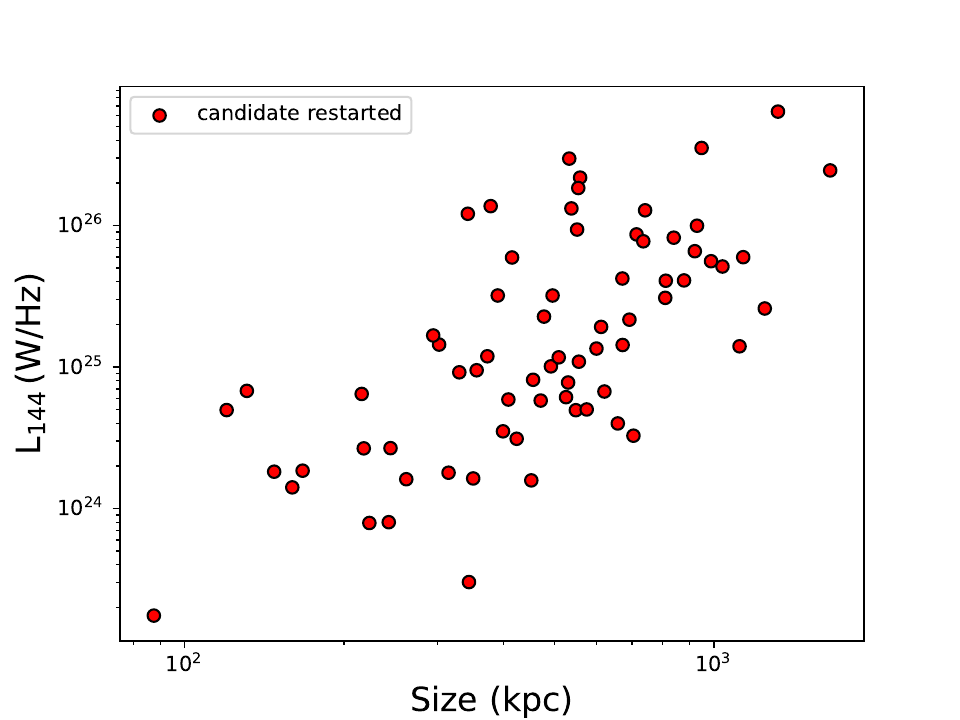}}
\caption{Radio luminosity at 144 MHz, $L_\mathrm{144}$ vs physical size for the final sample of 69 candidate restarted galaxies, Sect.~\ref{Final sample}. 
}
\label{fig:lm150-size}
\end{figure}

The restarted candidates presented in this work mostly occupy the region with high radio power as shown in Fig.~\ref{fig:lm150-size} with a median value of log$_{10}$(L$_\mathrm{144 MHz}$/$\mathrm{WHz^{-1}}$) = 25.13 and a mean value of log$_{10}$(L$_\mathrm{144 MHz}$/$\mathrm{WHz^{-1}}$) = 25.73. 
\cite{jurlin2020} also reported similar median values of radio luminosities, 
log$_{10}$(L$_\mathrm{144MHz}$/$\mathrm{WHz^{-1}}$) = 25.38 for the restarted candidates sample identified in the LH field.

Among the 69 restarted candidates, their linear sizes span between 
88--1659 kpc, exhibiting median and mean measurements of 526 kpc and 559 kpc, respectively. 
\cite{jurlin2020} observed a similar median value of linear sizes (487 kpc) among restarted candidates within the LH field, where only sources $>$ 60\arcsec were selected.
They did not detect any statistical difference in the linear sizes among three categories: remnant candidates, restarted candidates, and active samples, implying that the transition to the restarted stage occurs relatively soon after the remnant stage. Our findings, demonstrating similar linear sizes, further support this hypothesis. 

Finally, sixteen of the restarted candidate galaxies exhibit linear sizes surpassing 0.7 Mpc (Fig.~\ref{fig:lm150-size}), thus meeting the criteria for classification as giant radio galaxies \citep[GRGs,][]{ishwarachandra1999,shoenmakers2000b,machalski2001,machalski2008,dabhade2020,oei2023}. They are indicated with $\star$ alongside their source names in Table~\ref{Core flux densities}. Additionally, \cite{dabhade2020}, who presented the largest sample of GRGs in the HETDEX region, has also recognized ten out of these sixteen sources as GRGs. 
We recovered all the GRGs found by \cite{dabhade2020} so the discrepancy in number is likely due to the different ways in which the sizes are measured. The extra GRG objects we have which are not in the LoTSS DR1 catalog used by \cite{dabhade2020} are because of 
the flood-filling method used by \cite{mingo2019} that aims to capture extended emission from the source beyond what PyBDSF can fit.
Additionally \cite{mingo2019} used prerelease, LoTSS DR2 quality images, which have greater depth and better resolution than those used for the LoTSS DR1 catalog.

The median and mean values for the linear sizes of the GRGs we identified are 925 kpc and 975 kpc, respectively. Within our GRG sample, the largest reported size measures 1658 kpc, observed in J120856.45+464023.6. 
Additionally, the GRG restarted candidates exhibit notably high radio power, with median and mean values reaching $6.28 \times 10^{25}$ \WHz\ and $1.17 \times 10^{26}$ \WHz, respectively. 
This could be consistent with some findings which point to restarted GRGs having substantial accretion power (being luminous in the hard X-ray region), for example, \cite{bruni2019}.

\subsection{Host galaxy properties of candidate restarted radio galaxies}
\label{Host Galaxy properties of candidate restarted radio galaxies}

Both the study of \cite{jurlin2020} as well as the study of DDRG by \cite{mahatma2019} did not show any statistically significant difference between the host galaxies of restarted radio sources and comparison samples of radio AGNs.
This suggests that restarted candidates likely belong to the same parent population.

Below, we explore these findings further, analyzing the host galaxy properties of our sample of candidate restarted sources. 
In particular, we examine the distribution of their redshifts and environments in Sect.~\ref{Redshift distribution of restarted candidates} and Sect.~\ref{Environment of the restarted candidate galaxies}, evaluate their infrared characteristics in Sect.~\ref{wise colors}, and explore their stellar masses and associated implications in Sect.~\ref{Stellar masses}.

\subsubsection{Redshift distribution}
\label{Redshift distribution of restarted candidates}

The redshift distribution of the FRI+FRII parent sample shown in Fig.~4 in \cite{mingo2019} is predominantly concentrated at $z < 0.8$, with an extension toward higher redshifts (up to $z > 2$) due to the presence of quasars.
In fact, the observed redshift distribution is a consequence of the selection of AGNs with host-galaxy information and reliable redshifts, particularly below $z \sim 0.8$.

Among the 69 restarted sources, 50 have spectroscopic redshifts, while the rest rely on photometric redshifts.
The distribution of redshifts for the candidate restarted galaxies is depicted in Fig.~\ref{fig:redshift-restarted}, ranging from 0.03 to 1.14. 
The mean and median redshifts for these galaxies are 0.39 and 0.41, respectively.
Within the redshift range of $z < 0.8$, over 65\% of the restarted candidates lie at redshifts below 0.5. 
\cite{jurlin2020} also observed that 65\% of the restarted candidates in the LH field existed at redshifts below 0.5 (where median redshift was 0.394). 
However, it is important to note that we lack complete information for higher redshifts ($z > 0.8$); therefore, we cannot draw conclusions on whether restarting activity is rarer at higher redshifts based on the available information. 

\subsubsection{Environments of the candidate restarted galaxies}
\label{Environment of the restarted candidate galaxies}

The large-scale environment is found to be influential in the evolution of radio galaxies and driving AGNs jet activity and also in the remnant plasma evolution. 
Thus, understanding the physics of jet propagation and radio plasma evolution heavily relies on considering the environment as a crucial factor, providing valuable insights essential for accurate modeling \citep{parma2007,murgia2011,turner2015,
2019A&A...622A..12H,croston2019,shabala2020}.
To gain an initial understanding of the environments of candidate restarted galaxies, we conducted a cross-match with the two SDSS cluster catalogs outlined by \cite{wen2012} (using SDSS-III) and \cite{rykoff2014} (using SDSS DR8), via the process described by \cite{croston2019}. 

We found that 12 out of the 69 restarted candidates in our sample are located in cataloged SDSS clusters, denoted by $\dagger$ alongside their source names in Table.~\ref{Core flux densities}.
Specifically, we found twelve and four matches with the cluster catalogs in \cite{wen2012}, and \cite{rykoff2014} respectively. However, it is important to note that the SDSS cluster catalogs are incomplete at $z>0.4$.
Among the 69 candidate restarted galaxies, 35 have redshifts less than 0.4.
Consequently, the fraction of candidate restarted galaxies with $z<0.4$ present in SDSS clusters is $\sim$34\% (12 out of 35).
The main limitation of this analysis is incomplete information about environmental richness from the use of SDSS cluster catalogs.  Work is in progress to characterize group and cluster environments for LoTSS DR2 sources using a deeper wide area optical catalog, which will enable more complete information to lower cluster masses and higher redshifts (Croston et al. in prep).

It is noteworthy that among the candidate restarted sources located in clusters, 11 exhibit significant asymmetric morphology in their radio images, potentially influenced by the surrounding cluster environment. However, one source (J124818.86+500704.7), displays symmetric morphology. Notably, similar to our findings, all four restarted sources found by \cite{jurlin2020} in cluster environments also demonstrated asymmetric morphology.
Our findings at $z < 0.4$ indicate that many restarting radio galaxies are not located within a rich cluster environment, which is consistent with the environmental properties of the general radio-galaxy population. Based on the information we have available so far, there is no obvious difference between the large-scale environments of restarting radio galaxies and other radio galaxies.

\subsubsection{WISE colors}
\label{wise colors}

To characterize the host galaxies of the candidate restarted galaxies, we examined their infrared colors and how they are positioned within the WISE color-color diagram. This can help to distinguish between star-forming galaxies and AGNs at low $z$ \citep{yan2013,2019A&A...622A..12H}. 
The infrared emission at $3.4\, \mu m$ (W1) and $4.6\, \mu m$ (W2) in WISE trace continuum emission originating from stellar photospheres. Consequently, a higher W1-W2 value indicates dustier and/or increasing star-forming sources, whereas a lower value suggests the presence of older stellar populations. The infrared emission at $12\, \mu m$ (W3) and $22\, \mu m$ (W4) are more sensitive to warm dust emission heated by stars or the dusty torus surrounding some accreting black holes \citep{wright2010}. 

\begin{figure}[ht!]
\centerline{\includegraphics[width=0.5\textwidth]{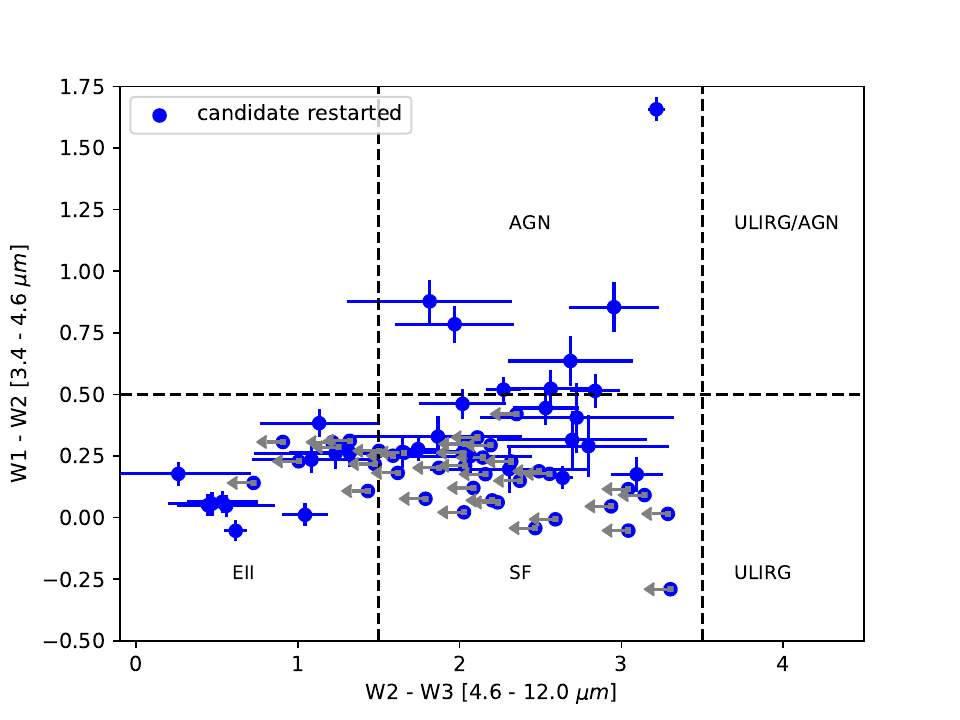}}
\caption{WISE color-color diagnostic plot for the final sample of 69 candidate restarted galaxies presented in Sect.~\ref{Final sample}, in Vega magnitudes. We follow the population division adopted by \cite{mingo2016} to identify the host galaxies to be elliptical galaxies (bottom-left), star-forming galaxies (bottom-center), 
starburst or ultra-luminous infrared galaxies (ULIRG, bottom-right and top-right), and AGNs-dominated (top-center and top-right). The host galaxy populations are divided by the lines, with the x-axis and y-axis as proxies of the prevalence of star formation and relative AGNs dominance respectively. W3 upper limits are represented with circles with gray arrows.}
\label{fig:wise}
\end{figure}

The AB magnitudes have been converted to Vega magnitudes and 
utilized in various studies of radio sources to make the WISE color–color plot, investigating the properties of host galaxies \citep[see e.g.,][]{mingo2019, mingo2022} and comparing characteristics across different classes of radio sources \citep[see e.g.,][]{mahatma2019, jurlin2020}.
We converted the AB magnitudes to Vega magnitudes using the relation described in Sect.~\ref{Optical and infrared identification} to make the WISE color-color plot shown in Fig.~\ref{fig:wise}.
For 38 restarted candidates, we have upper limits on W3, and for 11 restarted candidates, we have upper limits on W4. We followed the population classification adopted in \cite{mingo2016} for the WISE color-color plot which is the same classification followed by \cite{jurlin2020} also.

The analysis of the WISE color-color plot of the restarted candidates in Fig.~\ref{fig:wise} implies that the hosts are typical of the broader radio galaxy population. We also note that 38 sources with W3 upper limits marked with circles with arrows can further shift their position to the left in the horizontal direction in Fig.~\ref{fig:wise}. A more detailed classification of the hosts requires a detailed analysis of the stellar population, which is beyond the scope of the paper.

It's noteworthy that \cite{jurlin2020} also identified similar host types for restarted candidates within the LH field with the WISE color-color plot, and they claim a similar distribution of hosts to restarted sources in \cite{mahatma2019}. \cite{jurlin2020} reported analogous distributions for stellar population ages and dust content across restarted, remnant, and active comparison samples suggesting a similarity in the types of galaxies hosting both phases in the radio galaxy life cycle. \cite{mahatma2019} similarly reported that the restarted DDRGs sample and the active comparison sample within the HETDEX field exhibit comparable ranges and distributions in terms of dust content and emission from stellar populations.   
The source distribution we see in Fig.~\ref{fig:wise} is very similar to that of \cite{mingo2019, jurlin2020, mingo2022},
and this confirms the findings for restarted sources in \cite{jurlin2020} and \cite{mahatma2019}.

In a comprehensive modeling study of restarted, remnant, and active radio galaxy populations in the LH, \cite{shabala2020} found that the remnant radio lobes exhibit rapid fading. Consequently, any substantial observed proportion ($>$ 10\%) of remnant and restarted sources need a dominant population of short-lived jets which could feasibly be attributed to feedback-regulated accretion processes.

Earlier studies indicate that radiatively efficient sources typically have a higher prevalence of SF hosts compared to radiatively inefficient AGNs,
often associated with old elliptical galaxies \citep[e.g.,][]{janssen2012,gurkan2014,mingo2016,williams2018}.
The absence of ULIRGs in Fig.~\ref{fig:wise} is anticipated because of the selection criteria initially utilized by \cite{2019A&A...622A..12H} and \cite{gurkan2018}, specifically including sources where radio emissions significantly surpass those expected from star formation.

Very high W1-W2 and W2-W3 colors indicate the galaxy population hosted by AGNs (e.g., quasars or quasar-like hosts - high excitation radio galaxies (HERGs)). In the FRI+FRII parent sample in \cite{mingo2019}, 
the hosts of FRIIs are more commonly found to be star-forming galaxies on average than the FRIs, and notably encompass various HERGs (see Fig.~7 in \citealt{mingo2019}). Fig.~\ref{fig:wise} illustrates the presence of eight quasars or HERGs with W1-W2 $>$ 0.5 within the identified restarted candidates.
Our sample has a larger proportion of sources ($\sim$ 12\%) than \cite{mingo2019} ($\sim$ 4\%) with W1-W2 $>$ 0.5, where most sources are radiatively efficient AGNs (HERGs). This hints at the possibility that a higher proportion of our sources might be HERGs, compared to the parent FRI/FRII population of \cite{mingo2019}. 
This result would be consistent with recent studies (e.g., \citealt{bruni2019}) that suggest that restarting radio AGNs might have faster-accreting black holes compared to the broader radio AGNs population. 
We note, however, that studies of restarting sources, particularly at shorter wavelengths, tend to be biased toward brighter sources, which have higher accretion rates. In particular, \citealt{bruni2019}' sample of 15 sources was selected based on hard X-ray emission, which is linked to fast accretion. A comprehensive study of restarting radio AGNs is needed to establish the fraction of HERGs/LERGs in this population.
However, our sample is small, and, as discussed by several authors 
\citep[e.g.,][]{gurkan2014,mingo2016,mingo2019,mingo2022} the WISE color-color plot alone is not a good discriminator of radiatively efficient activity. We will investigate the accretion properties of this restarting population in more detail in a future study, with more reliable diagnostics similar to those presented by \cite{best2023} for the LOFAR Deep Fields.

\subsubsection{Stellar masses}
\label{Stellar masses}

The stellar mass ($\mathrm{M_{stellar}}$)  stands as a basic parameter in exploring a galaxy's evolution. The WISE W1 and W2 infrared (IR) photometry measurements track the prevailing stellar mass content within galaxies by capturing the Rayleigh-Jean's (RJ) limit blackbody emission for stars 2000 K and hotter \citep{jarrett2013}, allowing for an effective measurement of stellar masses. W1-W2 color value remains constant and unaffected by the age of the stellar population and the mass function \citep{pahre2004,jarrett2011,cluver2014}, though it can be affected by redshift as shown by examples \cite{assef2010,assef2013,donley2012}. 
We note that there are numerous methods to calculate stellar masses, and we utilize WISE 
short-wavelength bands here considering the available data. 
We estimate the cumulative baryonic stellar mass 
of the restarted candidates, using the W1 and W2 spectral luminosities and corresponding mass-to-light ratio (M/L), with the formula from \cite{pahre2004,jarrett2011,cluver2014}:
 \begin{equation} 
 \label{stellarmass-eqn}
  \mathrm{log_{10}} (M_{stellar}/L_{W1})  =  -1.96 \times (W1-W2)  - 0.03 
 \end{equation} 
in units of $(M_{\odot}/L_{\odot})$ where $M_{\odot}$ is solar mass ($\mathrm{1.989 \times 10^{30} kg}$) and $L_{\odot}$ is solar luminosity ($\mathrm{3.839 \times 10^{33} erg/sec}$). 

Since the IR photometry measurements can be influenced by AGNs-heated dust, we exclude the eight sources that belong to the region marked by AGNs in Fig.~\ref{fig:wise} (i.e., W1-W2 $>$ 0.5) in the calculation of stellar mass. This step left us with 61 sources for the calculation of the stellar masses.

\begin{figure}[ht!]
\centerline{\includegraphics[width=0.5\textwidth]{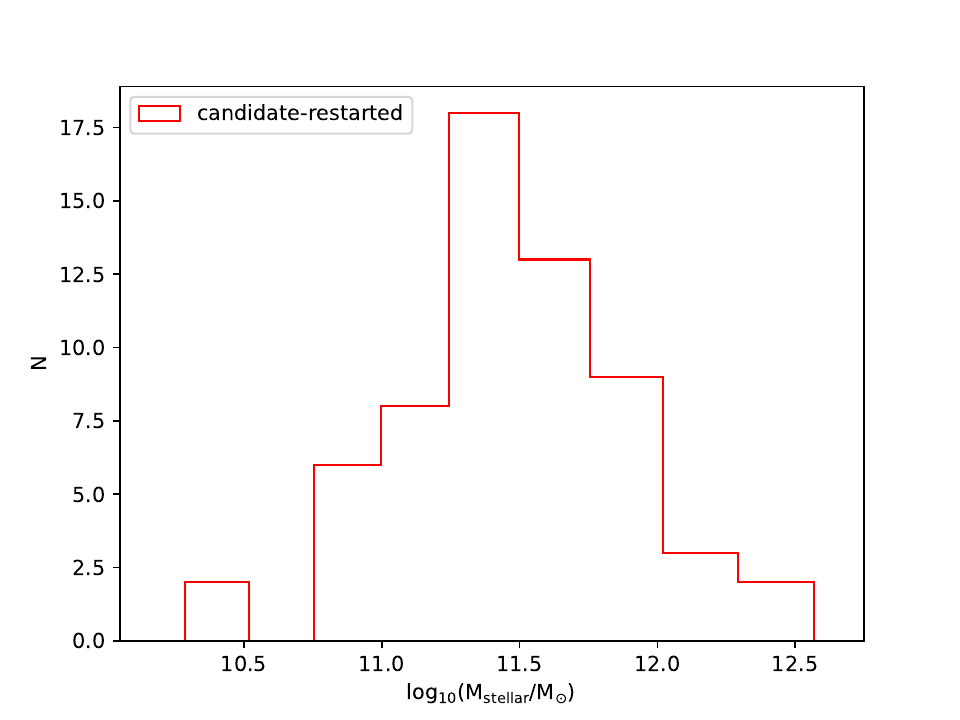}}
\caption{Distribution of the stellar masses of the final sample of candidate restarted galaxies presented in Sect.~\ref{Final sample}. Eight objects that belong to the region marked by AGNs in Fig.~\ref{fig:wise} (i.e., W1-W2 $>$ 0.5) are excluded in the calculation of stellar mass because their IR photometry measurements can be influenced by AGNs-heated dust.
} 
\label{fig:stellarmass}
\end{figure}

The WISE in-band luminosity for W1 ($L_{W1}$) in eqn.\ref{stellarmass-eqn} is derived using the formula:
 \begin{equation} 
  L_{W1} (L_{\odot}) =  10^{-0.4(M_{W1}-M_{\odot})}
 \end{equation}  
where $M_{W1}$ is the absolute in-band magnitude of the source at 3.4 $\mu m$ and $M_{\odot}$ is that of the Sun at 3.4 $\mu m$ which is 3.24 \citep{jarrett2013}.

We derived the absolute magnitude ($M$) of our sample at 3.4\,$\mu m$ using:
 \begin{equation} 
  M =  m - 5 (\mathrm{log_{10}}\,(d\,in\,pc) - 1.0 )
 \end{equation}  
where m is the apparent magnitudes at 3.4\,$\mu m$ obtained from the AllWISE catalog and 
d is the luminosity distance to the source. 

The stellar masses are given in Table ~\ref{Core flux densities}. 
The distribution of stellar masses for the restarted candidate galaxies depicted in Fig.~\ref{fig:stellarmass} is compatible with that of massive elliptical galaxies. Typical values of masses of elliptical galaxies span from $\sim$ $10^{9}$ to nearly $10^{12}$ solar masses \citep{best2006,sabater2019}.
The tail of the distribution in Fig.~\ref{fig:stellarmass} with high values of masses ($\mathrm{M} > 10^{12}\mathrm{M_{\odot}}$) is likely due to the influence of AGNs, potentially contaminating the IR emission due to AGNs-heated dust and consequently causing an overestimation of stellar masses.
Our results are compatible with the distribution of stellar masses among the restarted candidates in LH shown in Fig.~6 in \cite{jurlin2020}.

Some recent works have shown that radio emission tends to vary with the host mass \citep[e.g.,][]{capetti2022,capetti2023}.
In early-type galaxies, a positive correlation has been observed between median radio power and stellar luminosity \citep{capetti2023}. 
We conducted a preliminary investigation to explore potential correlations between radio properties and stellar masses. 
However, our analysis did not reveal any substantial correlation between radio luminosities and/or radio linear sizes with the host masses. The calculated Pearson correlation coefficients were 0.41 for radio luminosities and host mass p-value $= 0.001$), and 0.25 for radio linear sizes and host mass (p-value $= 0.05$), indicating an absence of significant correlation between the radio properties and the host mass in our sample.

\section{Conclusions and prospects}
\label{Summary and prospects}

In this paper we characterize a sample of core-dominated objects selected by \cite{mingo2019} within the HETDEX region. High CP has been used as a signature of the possible presence of restarted nuclear activity. Using various criteria, we have identified candidate restarted radio sources and characterized their properties, as well as those of their host galaxies.
In this way we have broadened the search for restarted candidates across a more extensive area, and, following previous studies, we employed diverse selection criteria encompassing 
high radio $\mathrm{CP_{1.4GHz}}$ combined with low SB of extended emission, 
a SSC or ISC, USS extended emission coupled with a bright core and morphological attributes
in identifying restarting radio galaxies, leveraging the comprehensive 144 MHz deep LOFAR observations, particularly within the LoTSS datasets.

The conclusions that we can draw from this study are:
\begin{itemize}
\item  A total of 69 candidate restarted radio galaxies were identified in the HETDEX field over $\sim$424 $\mathrm{deg^2}$ using LoTSS DR2 images; precisely 63 from the CP+SB criteria, six fulfilling SSC criteria, and none through ISC criteria (nine satisfying both CP+SB and SSC criteria).

\item We present the largest-to-date sample of core-bright restarted radio galaxies. This complements and expands the study by \cite{jurlin2020}. Our study also confirms that core prominence is a useful parameter for selecting restarted sources. 

\item The restarted candidates we found show a clear heterogeneity in intrinsic morphologies varying from FRI, FRII, core-with-halo, and asymmetric morphologies feasibly suggesting different kinds of progenitors. 

\item Our sample of 69 potential restarted galaxies spans the redshift range from 0.03 to 1.14, with over 65\% of the identified candidates located at redshifts below $z < 0.5$.

\item The restarted sample exhibited a diverse range of radio luminosities from log$_{10}$(L$_\mathrm{144 MHz}$/$\mathrm{WHz^{-1}}$) = 23.24 to log$_{10}$(L$_\mathrm{144 MHz}$/$\mathrm{WHz^{-1}}$) = 26.80
at 144 MHz, with a mean value of log$_{10}$(L$_\mathrm{144 MHz}$/$\mathrm{WHz^{-1}}$) = 25.73.

\item Our sample of restarted candidates spans a range of linear sizes between 88 and 1659 kpc, with an average measurement of 559 kpc.

\item We identified 16 GRGs with sizes exceeding 0.7 Mpc, constituting approximately 23\% of our sample.

\item We identified a unique case, J131728.61+561544.8, depicting notable variability in jet orientation and realignment across past and present epochs - an exceptionally 
rare occurrence. Furthermore, we identified two more sources [J114312.36+505016.3 and J114837.17+554500.6], with USS remnant emissions and broad emission lines in their SDSS spectra, resembling cases such as J131728.61+561544.8, indicating potential jet reorientation.

\item The analysis of the WISE color-color diagram indicates that the hosts of restarted candidates are typical of the broader radio galaxy population consistent with previous studies.

\item The stellar mass for the restarted candidates is found to be 
consistent with the total stellar content of massive elliptical galaxies.

\item We do not find any substantial correlation between radio 
properties (luminosities or linear sizes) with the host masses of the restarted sources.

\item  At $z < 0.4$, our findings suggest that many restarting radio galaxies are not located within a rich cluster environment. There is no indication that the large-scale environments of restarting radio galaxies are systematically different from the general radio galaxy population.

\item We found a subset of nine sources with a USS extended emission ($\alpha_\mathrm{ext,1400MHz}^{\mathrm{ext,144MHz}}$ $< -1.2$) surrounding a bright radio core. 
We associate these objects with a restarted phase characterized by a bright core exhibiting a spectral index consistent with that of active galaxies, along with remnant emission displaying USS indicative of aged plasma. 
The existence of these objects, tracing both past (remnant) and present (restarting) epochs, implies rapid switch-on and switch-off mechanisms occurring with a relatively fast duty cycle, suggesting that within a galaxy's life cycle, a restarted phase can swiftly follow a short remnant phase, 
allowing the continued detectability of USS emissions.

\end{itemize}

The results we found in this work have broadened our perception of galaxies with recurrent jet activity over the classical cases of DDRGs.
Our sample of 69 potential restarted sources represents a lower limit of the number of restarted sources in the HETDEX field since we explored only one of the possible ways for selecting candidates - those exhibiting core dominance and extension.
Going forward, we plan to apply our systematic approach to identify candidate restarted galaxies within the broader LoTSS DR2 sample, aiming to build a large significant group of such candidates over the entire HETDEX field. 

Identification of potential remnant galaxies in HETDEX is currently underway, utilizing the LoTSS datasets (Brienza et al. in prep.).
In the future, the collaborative outcomes merging their work on remnants with our identified restarted candidates will provide precise constraints on the duty cycle of radio galaxies's lifetime in the HETDEX region.
This would also enable us to make robust conclusions regarding the fractions and potential disparities in the radio properties, host galaxy classifications, stellar masses, and environments between restarted and remnant galaxies.

Follow-up observations with high-frequency and high-resolution capabilities
are required to confirm the restarted nature of the candidates and determine the timing of their duty cycles. As additional VLBI data becomes available, the central regions of more sources will be investigated in detail to further confirm their restarting nature.
The extension of LoTSS to include LOFAR international baselines, achieving even higher resolution, will contribute to confirming the restarted nature of our candidates and characterizing this specific population \citep[e.g.,][]{morabito2022,kukreti2022,sweijen2022,jurlin2024}.

The subclass of nine USS sources and J131728.61+561544.8 with a 
jet realignment morphology will be followed up by VLA observations with a broader frequency coverage at low and high resolutions to study their spatially resolved radio spectra. These observations, combined with LOFAR and GMRT data spanning 144-6000 MHz, would allow us to make spectral index maps and spectral curvature, facilitating the analysis of the radiative age of these sources and confining the duty cycle of the jet activity.
This approach is essential for estimating the duration of the switch-off phase \citep[e.g.,][]{brienza2020}. 
In the future, constructing resolved low-frequency radio spectra using the Apertif survey and LOFAR data, aimed at analyzing spectral index curvature \citep[e.g.,][]{2021Galax...9...88M}, will be crucial to confirm and broaden our perception of the evolutionary stages of a galaxy. 

%%%%%%%%%%%%%%%%%%%%%%%%%%%%%%%%%%%%%%%%%%%%%%%%%%%%%%%%%%%%%%%%%%%%%%%%
\begin{acknowledgements}
LOFAR is the Low Frequency Array, designed and constructed by ASTRON. It has observing, data processing, and data storage facilities in several countries, which are owned by various parties (each with their own funding sources), and which are collectively operated by the ILT foundation under a joint scientific policy. The ILT resources have benefited from the following recent major funding sources: CNRS-INSU, Observatoire de Paris and Université d’Orléans, France; BMBF, MIWF-NRW, MPG, Germany; Science Foundation Ireland (SFI), Department of Business, Enterprise and Innovation (DBEI), Ireland; NWO, The Netherlands; The Science and Technology Facilities Council, UK; Ministry of Science and Higher Education, Poland; The Istituto Nazionale di Astrofisica (INAF), Italy. 
DGN acknowledges funding from Conicyt through Fondecyt Postdoctorado (project code 3220195). MB acknowledges support from the Next Generation EU funds within the National Recovery and Resilience Plan (PNRR), Mission 4 - Education and Research, Component 2 - From Research to Business (M4C2), Investment Line 3.1 - Strengthening and creation of Research Infrastructures, Project IR0000034 – `\textquoteleft STILES - Strengthening the Italian Leadership in ELT and SKA\textquoteright'. BM acknowledges support from the UK Science and Technology Facilities Council (STFC) under grant ST/T000295/1. 
JHC acknowledges support from the UK Research and Innovation Science and Technology Facilities Council (UKRI STFC) under grants ST/X001164/1 and ST/T000295/1.
\end{acknowledgements}

%%%%%%%%%%%%%%%%%%%%%%%%%%%%%%%%%%%%%%%%%%%%%%%%%%%%%%%%%%%%%%%%%%%%%%%%
\bibliographystyle{aa}
\bibliography{restarted_hetdex}
\clearpage

\begin{onecolumn}

\newpage
%\clearpage

\begin{appendix}

\section{Tables}
\label{app:tables}

%%%%%%%%%%%%%%%%%%%%%%%%%%%%%%%%%%%%%%%%%%%
\centering
\tiny
\begin{longtable}{lllllllllll}
\caption{\label{list of sources} List of candidate restarted radio galaxies in the HETDEX region from LoTSS DR2. 
} \\
\hline\hline
\mc{No.}   & Source & $z$ & S$_\mathrm{int}$ & CP & SB & $\alpha_\mathrm{core}$ & $\alpha_\mathrm{ext}$ &  Selection & CP  \\
\mc{}       &  &  & 144MHz & 1.4GHz & 144MHz & & & criteria& 144MHz     \\
\mc{}       &  &  & [mJy]  &        &        & & &          &            \\
\mc{(1)}        & (2) & (3) & (4) & (5) & (6) & (7) & (8) & (9) & (10)   \\
\hline
\endfirsthead
\caption{List of candidate restarted radio galaxies in the HETDEX region from LoTSS DR2
\textit{(continued)}.}\\
\hline\hline
\mc{No.}   & Source & $z$ & S$_\mathrm{int}$ & CP & SB & $\alpha_\mathrm{core}$ & $\alpha_\mathrm{ext}$ &  Selection & CP  \\
\mc{}       &  &  & 144MHz & 1.4GHz & 144MHz & & & criteria& 144MHz       \\
\mc{}       &  &  & [mJy]  &        &        & & &          &             \\
\mc{(1)}        & (2) & (3) & (4) & (5) & (6) & (7) & (8) & (9) & (10)    \\
\hline
\endhead
\hline
\endfoot
1  & J104926.21+472859.0  & 0.761$^{\mathrm{s}}$ & 6.59   & >0.39 & 4.37   & -0.44 $\pm$  -0.05 & <-0.01    & CP+SB     & 0.63    \\
2  & J105611.37+472203.9  & 0.326$^{\mathrm{s}}$ & 7.80   & <0.23 & 4.08   & -0.47 $\pm$  -0.07 & <-0.08     & CP+SB     & 0.41                 \\
3$\star$  & J105640.50+554330.2$\star$  & 0.444$^{\mathrm{s}}$ & 46.71  & 0.76 & 16.19  & -0.68 $\pm$  -0.03 & -1.43   $\pm$  -0.25   & CP+SB,SSC,USS & 0.37                 \\
4  & J110158.70+464534.8  & 0.469$^{\mathrm{s}}$ & 43.28  & <0.24 & 7.10   & -0.65 $\pm$  -0.02 & <-0.56  & CP+SB,SSC & 0.28                \\
5$\star$  & J110330.09+534624.1$\star$  & 0.360$^{\mathrm{s}}$ & 277.97 & 0.24 & 101.84 & -0.71 $\pm$  -0.02 & -1.42   $\pm$  -0.02   & SSC,USS       & 0.06                 \\
6 & J110507.80+481110.4  & 0.213$^{\mathrm{s}}$ & 49.64  & <0.25 & 18.63  & -0.70 $\pm$  -0.03 & <-0.79     & CP+SB,SSC & 0.21                 \\
7 & J110533.20+525112.7  & 0.224$^{\mathrm{s}}$ & 22.07  & <0.19 & 7.94   & -0.59 $\pm$  -0.03 & <-0.18     & CP+SB     & 0.36                  \\
8 & J110609.86+495946.2  & 0.687$^{\mathrm{p}}$ & 17.83  & <0.23 & 6.15   & -0.44 $\pm$  -0.03 & <-0.13      & CP+SB     & 0.37                  \\
9$\star$  & J110759.92+512035.4$\star$  & 0.801$^{\mathrm{p}}$ & 223.57 & <0.17 & 33.22  & -0.66 $\pm$  -0.03 & <-1.40  & CP+SB,SSC,USS & 0.04                \\
10 & J110921.45+481454.0  & 0.392$^{\mathrm{s}}$ & 18.88  & <0.04 & 1.80   & -0.77 $\pm$  -0.06 & <0.14   & SSC       & 0.24      \\
11 & J111113.93+500925.8  & 0.596$^{\mathrm{s}}$ & 29.00  & 0.49 & 9.83   & -0.15 $\pm$  -0.02 & -0.81   $\pm$  -0.06   & CP+SB     & 0.18      \\
12 & J111300.30+494230.0  & 0.502$^{\mathrm{p}}$ & 237.88 & 0.61 & 27.25  & 0.23  $\pm$  0.003 & -0.99   $\pm$  -0.01   & CP+SB     & 0.09                \\
13 & J111853.71+523123.4  & 0.292$^{\mathrm{s}}$ & 8.06   & <0.12$\dagger$ & 2.96   & <-0.59  & <-0.24  & CP+SB     & 0.23                \\
14 & J112107.98+480217.8  & 0.126$^{\mathrm{s}}$ & 39.53  & > 0.26 & 5.28   & -0.61 $\pm$  -0.03 & <-0.75  & CP+SB     & 0.20                 \\
15 & J112113.61+542136.4  & 0.210$^{\mathrm{s}}$ & 66.39  & 0.35 & 18.96  & -0.50 $\pm$  -0.02 & -0.83   $\pm$  -0.05   & CP+SB     & 0.20              \\
16 & J112306.22+525750.2  & 0.470$^{\mathrm{s}}$ & 23.06  & 0.56 & 4.98   & -0.41 $\pm$  -0.01 & -0.53   $\pm$  -0.07   & CP+SB     & 0.49               \\
17 & J113017.57+474626.3  & 0.127$^{\mathrm{s}}$ & 13.60  & <0.03$\dagger$ & 3.05   & <-0.71  & <0.12  & SSC & 0.18                \\
18 & J113107.17+551850.8  & 0.244$^{\mathrm{p}}$ & 80.55  & 0.36 & 13.36  & -0.41 $\pm$  -0.01 & -0.92   $\pm$  -0.03   & CP+SB     & 0.15      \\
19 & J113718.05+484715.2  & 0.590$^{\mathrm{p}}$ & 34.33  & 0.59 & 11.33  & -0.34 $\pm$  -0.01 & -0.90   $\pm$  -0.08   & CP+SB     & 0.29                \\
20$\star$$\dagger$ & J114312.36+505016.3$\star$$\dagger$  & 0.915$^{\mathrm{s}}$ & 47.60  & 0.96 & 10.11  & -0.17 $\pm$  0.004 & -1.50   $\pm$  -0.37   & CP+SB,USS     & 0.54                 \\
21 & J114318.56+473601.8  & 0.654$^{\mathrm{p}}$ & 7.12   & <0.26$\dagger$ & 7.46   & <-0.50  & <-0.71  & CP+SB     & 0.18                 \\
22 & J114816.90+525038.4  & 0.050$^{\mathrm{s}}$ & 263.79 & 0.12 & 12.48  & -0.06 $\pm$  -0.01 & -0.65   $\pm$  0.005   & CP+SB     & 0.03                \\
23$\star$$\dagger$ & J114837.17+554500.6$\star$$\dagger$  & 0.421$^{\mathrm{s}}$ & 27.31  & 0.99 & 12.26  & -0.03 $\pm$  -0.01 & -2.04   $\pm$  -1.43   & CP+SB,USS     & 0.44                  \\
24 & J115646.56+551726.4  & 0.189$^{\mathrm{p}}$ & 39.57  & <0.14 & 15.06  & -0.61 $\pm$  -0.05 & <-0.53 & CP+SB     & 0.17                \\
25$\star$ & J115748.60+484445.2$\star$  & 0.343$^{\mathrm{s}}$ & 30.48  & 0.74 & 9.13   & -0.47 $\pm$  -0.02 & -1.16   $\pm$  -0.18   & CP+SB,USS     & 0.38                 \\
26$\star$  & J115750.75+472403.2$\star$  & 0.416$^{\mathrm{s}}$ & 144.22 & >0.17 & 30.44  & -0.57 $\pm$  -0.05 & <-1.36  & CP+SB,USS     & 0.03                 \\
27 & J120529.19+561334.7  & 0.681$^{\mathrm{p}}$ & 33.41   & <0.26 & 9.14   & -0.60	$\pm$	-0.02	& <-0.29    & CP+SB     & 0.42           \\
28 & J120856.45+464023.6  & 0.101$^{\mathrm{s}}$ & 734.09 & 0.36 & 22.32  & -0.36 $\pm$  0.002 & -0.76   $\pm$  0.003   & CP+SB     & 0.19                \\
29 & J120958.58+533419.2  & 0.431$^{\mathrm{s}}$ & 12.41  & >0.18 & 6.62   & -0.52 $\pm$  -0.06 & <-0.22 & CP+SB     & 0.30                 \\
30$\star$ & J121057.25+494943.1$\star$  & 0.580$^{\mathrm{p}}$ & 68.64  & 0.62 & 21.00  & -0.59 $\pm$  -0.02 & -1.38   $\pm$  -0.11   & CP+SB,USS     & 0.22          \\
31 & J122030.62+525126.8  & 0.346$^{\mathrm{s}}$ & 162.25  & <0.20 &  9.33  & -0.44	$\pm$	-0.02	  & <-1.03     & CP+SB    & 0.06           \\
32 & J122309.30+561327.7  & 0.507$^{\mathrm{s}}$ & 94.77  & 0.77 & 25.30  & -0.03 $\pm$  0.003 & -0.87   $\pm$  -0.03   & CP+SB     & 0.34                \\
33 & J122756.82+523111.4  & 0.084$^{\mathrm{s}}$ & 27.61  & <0.15 & 4.59   & -0.15 $\pm$  -0.03 & <-0.16   & CP+SB     & 0.14  \\
34 & J123230.22+554235.1  & 0.283$^{\mathrm{p}}$ & 19.25  & 0.32 & 7.03   & -0.57 $\pm$  -0.05 & -0.70   $\pm$  -0.09   & CP+SB     & 0.26                \\
35 & J123416.36+505425.3  & 0.172$^{\mathrm{s}}$ & 159.04 & 0.69 & 12.65  & -0.10 $\pm$  0.002 & -0.72   $\pm$  -0.01   & CP+SB     & 0.36                 \\
36 & J124300.29+483107.4  & 0.296$^{\mathrm{s}}$ & 14.25  & <0.28 & 3.76   & -0.07 $\pm$  -0.02 & <-0.05  & CP+SB     & 0.29               \\
37 & J124308.33+521247.6  & 0.200$^{\mathrm{s}}$ & 96.54  & 0.56 & 16.29  & -0.09 $\pm$  0.004 & -0.63   $\pm$  -0.02   & CP+SB     & 0.27                 \\
38 & J124729.10+552135.6  & 0.714$^{\mathrm{p}}$ & 14.05  & 0.80 & 7.38   & 0.09  $\pm$  -0.02 & -0.97   $\pm$  -0.24   & CP+SB     & 0.27                 \\
39 & J124818.86+500704.7  & 0.094$^{\mathrm{s}}$ & 78.20  & 0.31 & 21.92  & -0.52 $\pm$  -0.02 & -0.94   $\pm$  -0.04   & CP+SB     & 0.15                  \\
40 & J125140.01+561917.4  & 0.331$^{\mathrm{p}}$ & 22.53  & <0.16$\dagger$ & 6.97   & <-0.74  & <-0.52 & CP+SB,SSC & 0.24                 \\
41 & J130011.45+513806.2  & 0.461$^{\mathrm{s}}$ & 8.93   & >0.16 & 3.57   & -0.25 $\pm$  -0.08 & <-0.20   & CP+SB     & 0.17              \\
42 & J130044.40+543457.7  & 0.663$^{\mathrm{s}}$ & 105.76 & 0.48 & 37.40  & -0.46 $\pm$  -0.01 & -0.90   $\pm$  -0.03   & CP+SB     & 0.26               \\
43 & J130340.11+483050.1  & 0.319$^{\mathrm{s}}$ & 27.39  & 0.30 & 8.52   & -0.35 $\pm$  -0.04 & -0.83   $\pm$  -0.07   & CP+SB     & 0.13                 \\
44 & J130555.81+533938.2  & 0.029$^{\mathrm{s}}$ & 70.32  & <0.10 & 10.46  & -0.66 $\pm$  -0.03 & <-0.58   & CP+SB,SSC & 0.12     \\
45 & J131417.95+473709.8  & 0.225$^{\mathrm{s}}$ & 24.47  & 0.41 & 5.75   & -0.23 $\pm$  -0.02 & -0.58   $\pm$  -0.05   & CP+SB     & 0.24        \\
46 & J131634.57+493226.4 & 0.563$^{\mathrm{p}}$ & 88.75  & 0.51 & 16.19  & -0.51 $\pm$  -0.01 & -0.71   $\pm$  -0.02   & CP+SB     & 0.40   \\
47 & J131728.61+561544.8  & 0.108$^{\mathrm{s}}$ & 183.23 & <0.03 & 15.16  & -0.86 $\pm$  -0.06 & <-0.63  & SSC       & 0.05                \\
48$\star$ & J133707.41+545737.2$\star$  & 0.842$^{\mathrm{s}}$ & 51.69  & 0.95 & 17.15  & -0.77 $\pm$  -0.02 & -1.98   $\pm$  -1.02   & CP+SB,SSC,USS & 0.56  \\
49 & J133841.37+473547.9  & 0.119$^{\mathrm{s}}$ & 65.91  & 0.12 & 24.92  & -0.49 $\pm$  -0.07 & -0.86   $\pm$  -0.04   & CP+SB     & 0.05                \\
50 & J134015.53+563003.1  & 0.100$^{\mathrm{s}}$ & 42.35  & <0.10 & 9.80   & <-0.46  & <-1.05  & CP+SB     & 0.03                  \\
51 & J134606.47+485941.4  & 0.146$^{\mathrm{p}}$ & 23.68  & 0.20 & 12.87  & -0.65 $\pm$  -0.08 & -0.76   $\pm$  -0.07   & CP+SB,SSC & 0.16                \\
52 & J134928.74+462018.9  & 0.421$^{\mathrm{s}}$ & 60.71  & 0.21 & 12.32  & -0.71 $\pm$  -0.08 & -1.21   $\pm$  -0.06   & CP+SB,SSC & 0.08                \\
53 & J135635.66+560947.0  & 0.663$^{\mathrm{s}}$ & 26.28  & <0.11 & 8.57   & -0.37 $\pm$  -0.06 & <-0.34   & CP+SB     & 0.12               \\
54 & J135944.71+490552.6  & 0.165$^{\mathrm{p}}$ & 99.09  & 0.20 & 15.40  & -0.57 $\pm$  -0.03 & -0.95   $\pm$  -0.02   & CP+SB     & 0.09               \\
55 & J140014.67+472358.2  & 0.235$^{\mathrm{s}}$ & 20.19  & <0.24 & 5.49   & -0.55 $\pm$  -0.04 & <-0.34   & CP+SB     & 0.34                 \\
56 & J141027.26+465333.9  & 0.452$^{\mathrm{s}}$ & 6.47   & >0.14 & 4.64   & -0.37 $\pm$  -0.08 & <-0.01   & CP+SB     & 0.27                 \\
57 & J141900.52+490308.0  & 0.280$^{\mathrm{s}}$ & 22.67  & <0.09$\dagger$ & 7.12   & <-0.99  & <-0.57  & SSC       & 0.20                  \\
58 & J142742.37+471851.5  & 0.815$^{\mathrm{s}}$ & 25.65  & 0.22 & 6.03   & -0.51 $\pm$  -0.07 & -0.76   $\pm$  -0.06   & CP+SB     & 0.14       \\
59 & J143355.32+511749.9  & 0.193$^{\mathrm{p}}$ & 15.42  & <0.23 & 2.44   & 0.04  $\pm$  -0.02 & <0.04   & CP+SB    & 0.23                 \\
60 & J143423.51+544902.3  & 0.885$^{\mathrm{p}}$ & 38.07  & 0.60 & 14.92  & -0.59 $\pm$  -0.02 & -1.03   $\pm$  -0.12   & CP+SB     & 0.36                \\
61 & J143804.21+465050.0  & 0.745$^{\mathrm{s}}$ & 15.42  & <0.12 & 5.54   & -0.58 $\pm$  -0.09 & <-0.31  & CP+SB     & 0.20        \\
62 & J143912.24+541832.0  & 0.399$^{\mathrm{p}}$ & 233.08 & 0.56 & 43.58  & -0.33 $\pm$  0.002 & -0.75   $\pm$  -0.01   & CP+SB     & 0.33                \\
63 & J143951.79+533228.4  & 0.622$^{\mathrm{s}}$ & 52.13  & 0.69 & 14.70  & -0.50 $\pm$  -0.01 & -1.04   $\pm$  -0.09   & CP+SB     & 0.40                  \\
64 & J144841.08+465638.2  & 1.142$^{\mathrm{s}}$ & 43.04  & 0.78 & 20.02  & 0.33  $\pm$  0.004 & -0.65   $\pm$  -0.04   & CP+SB     & 0.29                  \\
65 & J145443.27+493358.8  & 0.668$^{\mathrm{p}}$ & 29.27  & <0.22 & 7.52   & -0.45 $\pm$  -0.05 & <-0.74  & CP+SB     & 0.13                \\
66 & J145833.56+472953.5  & 0.085$^{\mathrm{s}}$ & 77.05  & <0.28 & 24.93  & -0.51 $\pm$  -0.03 & <-1.08   &  CP+SB     & 0.10                \\
67 & J150240.77+510024.6  & 0.318$^{\mathrm{s}}$ & 12.63  & <0.31 & 5.15   & -0.40 $\pm$  -0.05 & <-0.43    & CP+SB     & 0.29                 \\
68 & J150917.60+535917.5  & 0.577$^{\mathrm{s}}$ & 60.74  & <0.42 & 11.86  & -0.33 $\pm$  -0.01 & <-0.71   & CP+SB     & 0.24           \\
69 & J151845.83+523707.9  & 0.521$^{\mathrm{s}}$ & 297.42 & 0.07 & 59.87  & -0.70 $\pm$  -0.04 & -1.14   $\pm$  -0.01   & SSC       & 0.03                \\
\hline
\end{longtable}
\footnotesize{{\bf Columns:} 1~--~Number; 2~--~Source; 3~--~redshift; 4~--~Total integrated flux densities at 144 MHz measured directly in the LOFAR $20\arcsec$ image in mJy; 5~--~radio core prominence at 1.4 GHz ($\dagger$ values poorly constrained); 6~--~surface brightness at 144 MHz in $\mathrm{mJy\,arcmin^{-2}}$; 7~--~spectral index of the core region; 8~--~spectral index of the extended emission; 9~--~selection criteria to identify a source as restarted (CP=high radio core prominence, SB= surface brightness at 144 MHz and SSC = steep spectrum core); 10~--~radio core prominence at 144 MHz. $\star$ in the source name denotes USS sources. $\dagger$ denotes sources with broad emission lines in their SDSS spectra. $^{\mathrm{s}}$ and $^{\mathrm{p}}$ in redshift values denote spectroscopic and photometric redshifts.
}

%%%%%%%%%%%%%%%%%%%%%%%%%%%%%%%%%%%%%%%
\centering
\tiny
\begin{longtable}{llllllllllll}
\caption{\label{Core flux densities} Flux densities at different frequencies for the candidate restarted radio galaxies in the HETDEX region from LoTSS DR2.
} \\
\hline\hline
\mc{No.}   & Source & S$_\mathrm{c,1400MHz}$ & S$_\mathrm{p,1400MHz}$ &   S$_\mathrm{tot,1400MHz}$  & S$_\mathrm{c,144MHz}$ & S$_\mathrm{p,144MHz}$ & L$_\mathrm{144MHz}$ & size &  $\mathrm{M_{stellar}}$ & NVSS,       \\
\mc{}       &  & [mJy] & [mJy] & [mJy] & [mJy] & [mJy]& [$\mathrm{W/Hz}$] & [kpc] & [$\times 10^{11} \mathrm{M_{\odot}}$] & FIRST \\
\mc{(1)}   & (2) & (3) & (4) & (5) & (6) & (7) & (8) & (9) & (10) & (11)\\
\hline
\endfirsthead
\caption{Flux densities at different frequencies for the candidate restarted radio galaxies in the HETDEX region from LoTSS DR2
\textit{(continued)}.} \\
\hline\hline
\mc{No.}   & Source & S$_\mathrm{c,1400MHz}$ & S$_\mathrm{p,1400MHz}$ &   S$_\mathrm{tot,1400MHz}$  & S$_\mathrm{c,144MHz}$ & S$_\mathrm{p,144MHz}$ & L$_\mathrm{144MHz}$ & size &  $\mathrm{M_{stellar}}$ & NVSS,       \\
\mc{}       &  & [mJy] & [mJy] & [mJy] & [mJy] & [mJy]& [$\mathrm{W/Hz}$] & [kpc] & [$\times 10^{11} \mathrm{M_{\odot}}$] & FIRST \\
\mc{(1)}   & (2) & (3) & (4) & (5) & (6) & (7) & (8) & (9) & (10) & (11)\\
\hline
\endhead
\hline
\endfoot
1  & J104926.21+472859.0  & 1.02  & 1.53  & 3.96$\dagger$   & 5.46   & 4.12 & 1.92E+25 & 612.24  & 6.83  & N,Y \\
2  & J105611.37+472203.9  & 0.99  & 1.14 & 4.99$\dagger$    & 4.63 & 3.23   & 3.99E+24 & 658.33  & 2.89  & P,Y \\
3  & J105640.50+554330.2  & 4.84 & 3.79 & 4.99   & 31.14  &  17.45 & 3.20E+25 & 390.51  & --  & Y,Y \\
4$\star$  & J110158.70+464534.8$\star$  & 2.47  & 2.89 & 11.84$\dagger$   & 14.63  & 12.20 & 3.08E+25 & 809.36  & 4.74  & P,Y \\
5${\star}$  & J110330.09+534624.1$\star$  & 6.97  & 3.54 & 14.51  & 23.83  & 17.36 & 1.28E+26 & 741.40  & 1.64  & Y,Y \\
6$\dagger$ & J110507.80+481110.4$\dagger$ & 3.67  & 2.24 & 8.91$\dagger$   & 18.59  & 10.63 & 6.45E+24 & 216.03  & 3.76  & P,Y \\
7 & J110533.20+525112.7  & 2.63  & 2.16 & 11.45$\dagger$  & 11.33  & 8.04 & 3.51E+24 & 399.50  & 1.54  & P,Y \\
8 & J110609.86+495946.2  & 2.81  & 2.49 & 10.77$\dagger$  & 14.20  & 6.64 & 4.22E+25 & 671.62  & 14.11 & P,Y \\
9${\star}$ & J110759.92+512035.4$\star$  & 2.93 & 1.97 & 11.39$\dagger$  & 14.28  & 8.64 & 6.38E+26 & 1321.64 & 28.34 & P,Y \\
10${\star}$ & J110921.45+481454.0$\star$  & 0.75  & 0.82 & 20.38$\dagger$  & 5.62   &  4.56 & 1.40E+25 & 1118.14 & 3.74  & P,Y \\
11${\star}$ & J111113.93+500925.8$\star$  & 5.32  & 3.71 & 7.59   & 7.91   & 5.19 & 5.13E+25 & 1038.11 & 13.10 & Y,Y \\
12${\star}$ & J111300.30+494230.0$\star$  & 45.78 & 37.26 & 61.13  & 29.56  & 22.10 & 2.45E+26 & 1658.79 & 0.74  & Y,Y \\
13 & J111853.71+523123.4  & 0.50$\dagger$  & 0.50$\dagger$ & 4.15$\dagger$   & 3.09   & 1.86 & 1.58E+24 & 451.90  & 2.68  & N,N \\ 
14$\dagger$ & J112107.98+480217.8$\dagger$  & 2.06  & 2.06 & 8.02$\dagger$   & 13.21  & 8.02 & 1.63E+24 & 350.95  & 2.35  & N,Y \\
15$\dagger$ & J112113.61+542136.4$\dagger$  & 8.36  & 4.36 & 12.63  & 25.43  & 13.26 & 9.48E+24 & 356.24  & 1.02  & Y,Y \\
16 & J112306.22+525750.2  & 5.11  & 4.53 & 8.12   & 17.22  & 11.23 & 2.16E+25 & 692.53  & 6.21  & Y,Y \\
17$\dagger$ & J113017.57+474626.3$\dagger$  & 0.51$\dagger$  & 0.51$\dagger$ & 14.95$\dagger$  & 4.02 & 2.51  & 3.02E+23 & 344.55  & 2.27  & N,N \\
18$\dagger$ & J113107.17+551850.8$\dagger$  & 6.05  & 4.94 & 13.74  & 18.49  & 12.45 & 1.35E+25 & 600.03  & 0.68  & Y,Y \\
19 & J113718.05+484715.2  & 5.23  & 4.70 & 7.93   & 13.94  & 9.98 &  2.27E+25 & 477.52  & 5.67  & Y,Y \\
20$\star$ & J114312.36+505016.3$\star$  & 21.89 & 17.62 & 18.39  & 35.13  & 25.86 & 2.18E+26 & 558.84  & --  & Y,Y \\
21 & J114318.56+473601.8  & 0.41$\dagger$  & 0.41$\dagger$ & 1.60$\dagger$   & 1.18   & 1.25 & 1.44E+25 & 302.58  & 9.29  & N,N \\
22 & J114816.90+525038.4  & 9.97  & 7.79 & 67.43  & 14.33  & 8.87 & 1.79E+24 & 315.25  & 1.61  & Y,Y \\
23$\star$ & J114837.17+554500.6$\star$  & 13.72 & 11.24 & 11.40  & 15.20  & 12.04 & 1.67E+25 & 294.91  & --  & Y,Y \\
24 & J115646.56+551726.4 & 1.41  &1.70 & 11.79$\dagger$  & 12.91  & 6.57 & 1.85E+24 & 167.06  & 0.25  & P,Y \\
25 & J115748.60+484445.2  & 4.42  & 4.08 & 5.50   & 16.45  & 11.56 & 1.19E+25 & 373.23  & 5.71  & Y,Y \\
26${\star}$ & J115750.75+472403.2$\star$  & 1.25  & 1.40 & 8.01$\dagger$   & 7.76  & 5.02 & 9.96E+25 & 929.26  & 1.88  & N,Y \\
27	& J120529.19+561334.7 & 4.00 & 3.63	& 13.9$\dagger$	& 19.22	& 13.97 & 6.58E+25	& 920.41&	7.90	& P,Y \\
28 & J120856.45+464023.6  & 73.74 & 62.10 & 170.49 & 190.94 & 139.86 & 4.96E+24 & 120.04  & 0.95  & Y,Y \\
29 & J120958.58+533419.2  & 1.27  & 1.16 &  6.51$\dagger$   & 4.92 & 3.71   & 1.43E+25 & 672.34  & 4.11  & N,Y \\
30${\star}$ & J121057.25+494943.1$\star$  & 4.05  & 4.0 & 6.44   & 18.24  & 14.82 & 8.65E+25 & 714.27  & 3.80  & Y,Y \\
31 &	J122030.62+525126.8 & 6.22	& 3.86 & 19.02$\dagger$	&15.98	&10.26 & 2.59E+25	&1247.98	& 3.17&	P,Y \\
32 & J122309.30+561327.7  & 38.53 & 29.96 & 38.82  & 45.30  & 32.34 & 9.36E+25 & 551.52  & 2.24  & Y,Y \\
33 & J122756.82+523111.4  &3.22  & 2.84 & 19.42$\dagger$  & 5.75   & 3.97 & 8.00E+23 & 243.03  & 2.98  & P,Y \\
34$\dagger$ & J123230.22+554235.1$\dagger$  & 1.71  & 1.40 & 4.37   & 7.05   & 5.03 & 5.89E+24 & 408.98  & 2.07  & Y,Y \\
35 & J123416.36+505425.3  & 60.43 & 45.74 & 66.10  & 76.38  & 57.76 & 6.78E+24 & 131.08  & --  & Y,Y \\
36$\dagger$ & J124300.29+483107.4$\dagger$  & 3.98  & 3.47 & 12.50$\dagger$  & 7.48   & 4.09 & 6.71E+24 & 621.16  & 2.46  & P,Y \\
37 & J124308.33+521247.6  & 30.52 & 21.39 & 38.54  & 35.93  & 26.42 & 9.17E+24 & 330.35  & 0.92  & Y,Y \\
38 & J124729.10+552135.6  & 5.09  & 4.63 & 5.80   & 3.97   & 3.80 & 3.20E+25 & 495.68  & 4.18  & Y,Y \\
39$\dagger$ & J124818.86+500704.7$\dagger$  & 4.53  & 3.72 &  11.91  & 13.10  & 11.79 & 1.82E+24 & 147.62  & 4.19  & Y,Y \\
40 & J125140.01+561917.4  & 1.04$\dagger$  & 1.04$\dagger$ & 6.38$\dagger$   & 8.26   & 5.40 & 1.01E+25 & 492.11  & 1.57  & P,N \\
41 & J130011.45+513806.2  & 0.69  & 0.88 & 5.61$\dagger$   & 2.42   & 1.52 & 5.79E+24 & 470.72  & 2.55  & N,Y \\
42 & J130044.40+543457.7  & 12.51 & 9.83 & 20.37  & 32.10  & 27.63 & 1.84E+26 & 554.34  & 8.29  & Y,Y \\
43 & J130340.11+483050.1  & 2.28  & 1.62 & 5.38   & 5.52   & 3.51 & 1.09E+25 & 555.77  & 2.23  & Y,Y \\
44 & J130555.81+533938.2  & 2.24  & 1.91 & 19.02$\dagger$  & 20.31  & 8.27 & 1.75E+23 & 87.45   & 2.63  & P,Y \\
45 & J131417.95+473709.8  & 3.30  & 3.59 & 8.71   & 8.21   & 5.94 & 4.95E+24 & 548.53  & 1.58  & Y,Y \\
46${\star}$ & J131634.57+493226.4$\star$ & 14.20 & 11.34 & 22.25  & 46.25  & 35.37 & 8.20E+25 & 839.80  & --  & Y,Y \\
47 & J131728.61+561544.8  &  3.72  & 1.41 & 43.57$\dagger$  & 38.79  & 9.60 &  5.01E+24 & 575.05  & 2.39  & P,Y \\
48 & J133707.41+545737.2  & 6.02  & 5.22 & 5.49   & 37.27  & 29.01 & 1.37E+26 & 378.65  & --  & Y,Y \\
49 & J133841.37+473547.9  &  0.83  & 1.19 & 10.25  & 3.42  & 3.55 & 2.66E+24 & 217.99  & 1.05  & Y,Y \\
50 & J134015.53+563003.1  &  0.46$\dagger$  & 0.46$\dagger$ & 4.39   & 1.61   & 1.28 & 7.90E+23 & 223.37  & 3.75  & Y,N \\
51$\dagger$ & J134606.47+485941.4$\dagger$  & 0.58  & 0.89 & 4.53   & 6.66   &  3.79 & 1.41E+24 & 159.73  & 0.19  & Y,Y \\
52${\star}$ & J134928.74+462018.9$\star$  & 0.66  & 0.97 &  4.68   & 7.45   & 4.74 & 4.09E+25 & 878.26  & 3.81  & Y,Y \\
53${\star}$ & J135635.66+560947.0$\star$ & 1.08  & 1.37 & 12.33$\dagger$  & 4.90   & 3.10 & 4.07E+25 & 811.25  & 11.08 & P,Y \\
54$\dagger$ & J135944.71+490552.6$\dagger$  & 3.76  & 2.59 & 13.25  & 17.68  & 9.17 & 8.12E+24 & 455.44  & 0.91  & Y,Y \\
55 & J140014.67+472358.2  & 2.88  & 2.0 & 8.26$\dagger$   & 12.87  & 6.84 & 3.11E+24 & 424.01  & 1.48  & P,Y \\
56 & J141027.26+465333.9  & 0.44  & 0.77 & 5.36$\dagger$   & 2.67   & 1.75 & 2.67E+24 & 244.89  & 2.60  & N,Y \\
57$\dagger$ & J141900.52+490308.0$\dagger$  & 0.49$\dagger$  & 0.49$\dagger$ & 5.53$\dagger$ & 10.06 & 4.50 & 7.77E+24 & 530.36  & 2.18  & N,N \\
58${\star}$ & J142742.37+471851.5$\star$   & 1.06  & 1.13 & 5.19   & 6.46   & 3.56 & 5.59E+25 & 987.77  & 3.03  & Y,Y \\
59${\star}$ & J143355.32+511749.9$\star$  & 4.82  & 3.86 & 16.86$\dagger$  & 6.29   & 3.53 & 3.27E+24 & 704.73  & 0.67  & P,Y \\
60 & J143423.51+544902.3  & 5.21  &  3.72 & 6.17   & 22.37  & 13.86 & 1.32E+26 & 538.03  & 3.60  & Y,Y \\
61 & J143804.21+465050.0  & 0.76  & 0.86 & 7.06$\dagger$   & 5.38   & 3.13 & 1.17E+25 & 509.19  & 37.02 & P,Y \\
62 & J143912.24+541832.0  & 33.41 & 37.17 & 66.64  & 91.17  & 77.44 & 1.21E+26 & 342.80  & --  & Y,Y \\
63${\star}$ & J143951.79+533228.4$\star$  & 5.17  & 6.86 & 9.90   & 30.17  & 21.06 & 7.73E+25 & 735.60  & 7.56  & Y,Y \\
64 & J144841.08+465638.2  & 21.42 & 25.85 & 33.05  & 15.06  & 12.40 & 2.97E+26 & 533.00  & --  & Y,Y \\
65${\star}$ & J145443.27+493358.8$\star$  & 1.09  & 1.36 & 6.28$\dagger$   & 4.05   & 3.75 & 5.97E+25 & 1136.23 & 8.06  & P,Y \\
66$\dagger$ & J145833.56+472953.5$\dagger$  & 2.37  & 2.48 & 8.76$\dagger$   & 11.31  & 7.66 & 1.61E+24 & 262.28  & 4.84  & P,Y \\
67 & J150240.77+510024.6  & 1.97  & 1.52 &  4.95$\dagger$   & 5.50   & 3.71 & 6.12E+24 & 525.60  & 2.27  & P,Y \\
68 & J150917.60+535917.5  & 7.85  & 6.94 & 10.34$\dagger$  & 22.63  & 14.51 & 5.94E+25 & 415.56  & 9.65  & P,Y \\
69${\star}$ & J151845.83+523707.9$\star$  & 2.09  & 1.72 &  24.20  & 16.77  & 8.15 & 3.53E+26 & 947.98  & 2.23  & Y,Y \\
\hline
\end{longtable}
\footnotesize{ {\bf Columns:} 1~--~Number; 2~--~Source; 3-4~--~the total and peak core flux densities at 1400 MHz measured directly in the FIRST $5\arcsec$ image in mJy; 5~--~the total flux densities at 1400 MHz measured directly in the NVSS $45\arcsec$ image in mJy; 6-7~--~the total and peak core flux densities at 144 MHz measured directly in the LOFAR $6\arcsec$ image in mJy; 8~--~radio luminosity of the source at 144 MHz; 9~--~linear size of the source at 144 MHz; 10~--~stellar mass; 11~--~detection in NVSS, detection in VLA-FIRST (Y=full detection, P= partial detection and N = no detection). $\star$ and $\dagger$ in source names denote GRGs, and galaxies residing in a cluster environment respectively. $\dagger$ in columns 3, 4, and 5, values represent upper limits (except for NVSS partial detection, where $\dagger$ in column 5 represents a lower limit).
}

%%%%%%%%%%%%%%%%%%%%%%%%%%%%%%%%%%%%%%%%%%%%%%%%%%%%%%%%%%%%%%%%%%%%%
%%%%% sources identified as "not restarted" below
%%%%%%%%%%%%%%%%%%%%%%%%%%%%%%%%%%%%%%%%%%%%%%%%%%%%%%%%%%%%%%%%%%%%%
\centering
\tiny
\begin{longtable}{llllllllll}
\caption{\label{list of sources-rejected} 
List of sources excluded from the sample of restarted candidates following the criteria discussed in Sect.~\ref{High radio core prominence and low surface brightness}, Sect.~\ref{Steep spectral index of the core} and Sect.~\ref{Spectral index of the extended emission}. 
} \\
\hline\hline
\mc{No.}   & Source & $z$ & S$_\mathrm{int}$ & CP & SB & $\alpha_\mathrm{core}$ & $\alpha_\mathrm{ext}$ & Reason for  \\
\mc{}       &  &  & 144MHz & 1.4GHz & 144MHz &   &    &  exclusion  \\
\mc{}       &  &  & [mJy]  &        &        &   &     &   \\
\mc{(1)}        & (2) & (3) & (4) & (5) & (6) & (7) & (8)  & (9)  \\
\hline
\endfirsthead
\caption{List of sources excluded from the sample of restarted candidates following the criteria discussed in Sect.~\ref{High radio core prominence and low surface brightness}, Sect.~\ref{Steep spectral index of the core} and Sect.~\ref{Spectral index of the extended emission}.
\textit{(continued)}. } \\
\hline\hline
\mc{No.}   & Source & $z$ & S$_\mathrm{int}$ & CP & SB & $\alpha_\mathrm{core}$ & $\alpha_\mathrm{ext}$ & Reason for  \\
\mc{}       &  &  & 144MHz & 1.4GHz & 144MHz &   &    &  exclusion  \\
\mc{}       &  &  & [mJy]  &        &        &   &     &   \\
\mc{(1)}        & (2) & (3) & (4) & (5) & (6) & (7) & (8)  & (9)  \\
\hline
\endhead
\hline
\endfoot
\sl{70$\dagger$   }  &  \sl{ J105922.02+523902.1$\dagger$   }  &  \sl{ 0.826$^{\mathrm{s}}$  }  &  \sl{ 5.57    }  &  \sl{ 0.86  }  &  \sl{ 1.92    }  &  \sl{ 0.37  $\pm$  0.003  }  &  \sl{ -  }  & \sl{Broad-line in SDSS spectrum}   \\
\sl{71$\dagger$   }  &  \sl{ J110150.26+554203.4$\dagger$   }  &  \sl{ 2.286$^{\mathrm{s}}$  }  &  \sl{ 45.31   }  &  \sl{ 1.09  }  &  \sl{ 14.09   }  &  \sl{ -0.57 $\pm$  -0.01  }  &  \sl{ -  } & \sl{Broad-line in SDSS spectrum}     \\
\sl{72   }  &  \sl{ J110229.18+501452.4   }  &  \sl{ 0.163$^{\mathrm{p}}$  }  &  \sl{ 100.47  }  &  \sl{ 0.05  }  &  \sl{ 15.55   }  &  \sl{ -0.33 $\pm$  -0.04  }  &  \sl{ -0.59   $\pm$  -0.01    } & \sl{Low $\mathrm{CP_{1.4GHz}}$} \\
\sl{73   }  &  \sl{ J110501.70+472300.7   }  &  \sl{ 0.269$^{\mathrm{p}}$  }  &  \sl{ 8.38    }  &  \sl{ <0.10$\dagger$  }  &  \sl{ 3.53    }  &  \sl{ <-0.42   }  &  \sl{ <-0.26   }   & \sl{Too small, removed from analysis}   \\
\sl{74$\dagger$  }  &  \sl{ J110833.89+483202.9$\dagger$   }  &  \sl{ 2.121$^{\mathrm{s}}$  }  &  \sl{ 33.68   }  &  \sl{ 1.17  }  &  \sl{ 5.54    }  &  \sl{ 0.07  $\pm$  0.004  }  &  \sl{ -  } & \sl{Broad-line in SDSS spectrum}    \\
\sl{75  }  &  \sl{ J111632.67+515613.2   }  &  \sl{ 0.357$^{\mathrm{p}}$  }  &  \sl{ 345.60  }  &  \sl{ 0.38  }  &  \sl{ 67.41   }  &  \sl{ -0.46 $\pm$  0.002  }  &  \sl{ -0.77   $\pm$  -0.01    }   & \sl{High SB}   \\
\sl{76$\dagger$  }  &  \sl{ J111818.90+510302.9$\dagger$   }  &  \sl{ 0.945$^{\mathrm{s}}$  }  &  \sl{ 53.89   }  &  \sl{ 0.59  }  &  \sl{ 16.43   }  &  \sl{ 0.03  $\pm$  -0.01  }  &  \sl{ -0.71   $\pm$  -0.03    }  & \sl{Broad-line in SDSS spectrum}      \\
\sl{77$\dagger$  }  &  \sl{ J111924.94+464728.1$\dagger$   }  &  \sl{ 1.547$^{\mathrm{s}}$  }  &  \sl{ 51.41   }  &  \sl{ 0.83  }  &  \sl{ 17.86   }  &  \sl{ 0.44  $\pm$  0.003  }  &  \sl{ -0.55   $\pm$  -0.03    }   & \sl{Broad-line in SDSS spectrum}     \\
\sl{78  }  &  \sl{ J112005.48+471400.8   }  &  \sl{ 0.110$^{\mathrm{p}}$  }  &  \sl{ 180.08  }  &  \sl{ 0.41  }  &  \sl{ 53.85   }  &  \sl{ -0.60 $\pm$  -0.01  }  &  \sl{-1.03  $\pm$  -0.018    }  & \sl{High SB}    \\
\sl{79$\dagger$  }  &  \sl{ J112429.66+463520.3$\dagger$   }  &  \sl{ 0.915$^{\mathrm{s}}$  }  &  \sl{ 37.60   }  &  \sl{ 0.26  }  &  \sl{ 13.29   }  &  \sl{ -0.29 $\pm$  -0.03  }  &  \sl{ -0.70   $\pm$  -0.03    }  & \sl{Broad-line in SDSS spectrum}    \\
\sl{80  }  &  \sl{ J113104.89+554530.7   }  &  \sl{ 0.192$^{\mathrm{s}}$  }  &  \sl{ 107.78  }  &  \sl{ 0.23  }  &  \sl{ 23.98   }  &  \sl{ -0.38 $\pm$  -0.01  }  &  \sl{ -0.66   $\pm$  -0.01    }   & \sl{One-sided morphology}   \\
\sl{81  }  &  \sl{ J113423.99+563353.6   }  &  \sl{ 0.626$^{\mathrm{p}}$  }  &  \sl{ 42.48   }  &  \sl{ <0.03$\dagger$   }  &  \sl{ 3.26    }  &  \sl{ <-0.61   }  &  \sl{ <-0.47    }   &  \sl{Low $\mathrm{CP_{1.4GHz}}$}   \\
\sl{82  }  &  \sl{ J114121.89+532614.7   }  &  \sl{ 0.300$^{\mathrm{p}}$  }  &  \sl{ 402.63  }  &  \sl{ 0.40  }  &  \sl{ 103.37  }  &  \sl{ -0.24 $\pm$  0.002  }  &  \sl{ -0.66   $\pm$  0.004    }  & \sl{High SB}     \\
\sl{83$\dagger$  }  &  \sl{ J115108.76+482400.6$\dagger$   }  &  \sl{ 1.064$^{\mathrm{s}}$  }  &  \sl{ 17.55   }  &  \sl{ 1.20  }  &  \sl{ 12.61   }  &  \sl{ 0.27  $\pm$  -0.01  }  &  \sl{ -  }  & \sl{Broad-line in SDSS spectrum}     \\
\sl{84  }  &  \sl{ J121708.17+485812.8   }  &  \sl{ 0.484$^{\mathrm{p}}$  }  &  \sl{ 21.68   }  &  \sl{ <0.04$\dagger$  }  &  \sl{ 4.86    }  &  \sl{ <-0.46   }  &  \sl{ <-0.29   } & \sl{Low $\mathrm{CP_{1.4GHz}}$}     \\
\sl{85  }  &  \sl{ J121920.21+481554.1   }  &  \sl{ 0.569$^{\mathrm{p}}$  }  &  \sl{ 7.33    }  &  \sl{ <0.09$\dagger$  }  &  \sl{ 4.82    }  &  \sl{ <-0.61   }  &  \sl{ <-0.11    }  & \sl{Low $\mathrm{CP_{1.4GHz}}$}  \\
\sl{86$\dagger$  }  &  \sl{ J121952.36+465426.1$\dagger$   }  &  \sl{ 1.850$^{\mathrm{s}}$  }  &  \sl{ 67.20   }  &  \sl{ 0.78  }  &  \sl{ 22.86   }  &  \sl{ 0.01  $\pm$  0.004  }  &  \sl{ -1.05   $\pm$  -0.05    }    & \sl{Broad-line in SDSS spectrum}  \\
\sl{87  }  &  \sl{ J122459.65+554759.9   }  &  \sl{ 0.537$^{\mathrm{p}}$  }  &  \sl{ 4.36    }  &  \sl{ <0.09$\dagger$  }  &  \sl{ 4.75    }  &  \sl{ <-0.23  }  &  \sl{ <0.11    }    & \sl{Too small, removed from analysis}  \\
\sl{88}  &  \sl{J123014.92+541141.4}  &  \sl{ 0.610$^{\mathrm{p}}$  }  &  \sl{ 12.03  }  &  \sl{ <0.07$\dagger$  }  &  \sl{  4.87   }  &  \sl{ <-0.34  }  &  \sl{ <-0.35    } & \sl{Low $\mathrm{CP_{1.4GHz}}$}    \\
\sl{89  }  &  \sl{ J123640.13+491141.4   }  &  \sl{ 0.135$^{\mathrm{p}}$  }  &  \sl{ 608.31  }  &  \sl{ 0.05  }  &  \sl{ 50.22   }  &  \sl{ -0.40 $\pm$  -0.01  }  &  \sl{ -0.82   $\pm$  -0.003   } &   \sl{Low $\mathrm{CP_{1.4GHz}}$ \& High SB}    \\
\sl{90  }  &  \sl{ J123648.05+552508.8   }  &  \sl{ 0.324$^{\mathrm{p}}$  }  &  \sl{ 220.69  }  &  \sl{ 0.23  }  &  \sl{ 62.77   }  &  \sl{ -0.57 $\pm$  -0.01  }  &  \sl{ -0.70   $\pm$  -0.01    }   & \sl{High SB}   \\
\sl{91  }  &  \sl{ J125556.65+520353.6   }  &  \sl{ 0.459$^{\mathrm{p}}$  }  &  \sl{ 6.34    }  &  \sl{ <0.07$\dagger$  }  &  \sl{ 2.40    }  &  \sl{ <-0.27   }  &  \sl{ <-0.04   }    & \sl{Too small, removed from analysis}  \\
\sl{92  }  &  \sl{ J131217.45+484754.9   }  &  \sl{ 0.364$^{\mathrm{p}}$  }  &  \sl{ 123.62  }  &  \sl{ 0.07  }  &  \sl{ 8.16    }  &  \sl{ -0.51 $\pm$  -0.04  }  &  \sl{ -0.67   $\pm$  -0.01    }  & \sl{Low $\mathrm{CP_{1.4GHz}}$}    \\
\sl{93  }  &  \sl{ J133128.47+475439.6   }  &  \sl{ 0.248$^{\mathrm{s}}$  }  &  \sl{ 3.28    }  &  \sl{ 0.74  }  &  \sl{ 2.01    }  &  \sl{ -0.20 $\pm$  -0.08  }  &  \sl{ -0.81   $\pm$  -1.03    }   & \sl{Too small, removed from analysis}   \\
\sl{94 }  &  \sl{ J134452.94+485816.2   }  &  \sl{ 0.340$^{\mathrm{s}}$  }  &  \sl{ 54.69
    }  &  \sl{ <0.03$\dagger$    }  &  \sl{ 6.21    }  &  \sl{ <-0.83	
   }  &  \sl{ <-0.54      }   & \sl{Low $\mathrm{CP_{1.4GHz}}$}   \\
\sl{95  }  &  \sl{ J140441.52+481100.6   }  &  \sl{ 0.162$^{\mathrm{s}}$  }  &  \sl{ 8.29    }  &  \sl{ <0.34  }  &  \sl{ 2.24    }  &  \sl{ -0.15 $\pm$  -0.03  }  &  \sl{ <0.14     }  & \sl{One-sided morphology}     \\
\sl{96  }  &  \sl{ J141806.69+483601.4   }  &  \sl{ 0.514$^{\mathrm{s}}$  }  &  \sl{ 7.94  }  &  \sl{ <0.08$\dagger$  }  &  \sl{ 2.11    }  &  \sl{ <0.02}	& \sl{<-0.09      }  & \sl{Low $\mathrm{CP_{1.4GHz}}$}    \\
\sl{97  }  &  \sl{ J142436.47+554938.5   }  &  \sl{ 0.485$^{\mathrm{p}}$  }  &  \sl{ 6.53    }  &  \sl{ <0.09$\dagger$  }  &  \sl{ 3.90    }  &  \sl{ <-0.24   }  &  \sl{ <0.01       }   & \sl{Low $\mathrm{CP_{1.4GHz}}$}   \\
\sl{98  }  &  \sl{ J143644.70+515442.9   }  &  \sl{ 0.204$^{\mathrm{p}}$  }  &  \sl{ 1.88    }  &  \sl{ <0.08$\dagger$  }  &  \sl{ 2.59    }  &  \sl{ <-0.18   }  &  \sl{ <0.65      }    & \sl{Too small, removed from analysis}   \\
\sl{99  }  &  \sl{ J145935.07+533350.1   }  &  \sl{ 0.077$^{\mathrm{p}}$  }  &  \sl{ 272.18  }  &  \sl{ <0.08  }  &  \sl{ 26.82   }  &  \sl{ -0.59 $\pm$  -0.03  }  &  \sl{ <-0.96    }  & \sl{Low $\mathrm{CP_{1.4GHz}}$}   \\
\hline
\end{longtable}
\footnotesize{{\bf Columns:} 1~--~Number; 2~--~Source; 3~--~redshift; 4~--~Total integrated flux densities at 144 MHz measured directly in the LOFAR $20\arcsec$ image in mJy; 5~--~radio core prominence at 1.4 GHz ($\dagger$ values poorly constrained); 6~--~surface brightness at 144 MHz in $\mathrm{mJy\,arcmin^{-2}}$; 7~--~spectral index of the core region; 8~--~spectral index of the extended emission; 9~--~Reason to exclude from the sample of restarted candidates. $\dagger$ in the source name denotes galaxies with broad emission lines in their SDSS spectra.
}

%%%%%%%%%%%%%%%%%%%%%%%%%%%%%%%%%%%%%%%%%%%%%%%%%%%%%%%%%%%%%%%%%%%%%
%%%%% sources identified as "not restarted" below
%%%%%%%%%%%%%%%%%%%%%%%%%%%%%%%%%%%%%%%%%%%%%%%%%%%%%%%%%%%%%%%%%%%%%
\centering
\tiny
\begin{longtable}{llllllllllll}
\caption{\label{Core flux densities-rejected} 
Flux densities at different frequencies for sources excluded from the sample of restarted candidates following the criteria discussed in Sect.~\ref{High radio core prominence and low surface brightness}, Sect.~\ref{Steep spectral index of the core} and Sect.~\ref{Spectral index of the extended emission}.
} \\
\hline\hline
\mc{No.}   & Source & S$_\mathrm{c,1400MHz}$ & S$_\mathrm{p,1400MHz}$ &   S$_\mathrm{tot,1400MHz}$  & S$_\mathrm{c,144MHz}$ & S$_\mathrm{p,144MHz}$ & L$_\mathrm{144MHz}$ & size &  $\mathrm{M_{stellar}}$ & NVSS,       \\
\mc{}       &  & [mJy] & [mJy] & [mJy] & [mJy] & [mJy]& [$\mathrm{W/Hz}$] & [kpc] & [$\times 10^{11} \mathrm{M_{\odot}}$] & FIRST \\
\mc{(1)}   & (2) & (3) & (4) & (5) & (6) & (7) & (8) & (9) & (10) & (11)\\
\hline
\endfirsthead
\caption{Flux densities at different frequencies for the candidate restarted radio galaxies in the HETDEX region from LoTSS DR2
\textit{(continued)}.} \\
\hline\hline
\mc{No.}   & Source & S$_\mathrm{c,1400MHz}$ & S$_\mathrm{p,1400MHz}$ &   S$_\mathrm{tot,1400MHz}$  & S$_\mathrm{c,144MHz}$ & S$_\mathrm{p,144MHz}$ & L$_\mathrm{144MHz}$ & size &  $\mathrm{M_{stellar}}$ & NVSS,       \\
\mc{}       &  & [mJy] & [mJy] & [mJy] & [mJy] & [mJy]& [$\mathrm{W/Hz}$] & [kpc] & [$\times 10^{11} \mathrm{M_{\odot}}$] & FIRST \\
\mc{(1)}   & (2) & (3) & (4) & (5) & (6) & (7) & (8) & (9) & (10) & (11)\\
\hline
\endhead
\hline
\endfoot
\sl{70  } &  \sl{ J105922.02+523902.1  } & \sl{ 77.97 } &  \sl{64.22} & \sl{ 74.51  } &  \sl{ 31.69  } &  \sl{28.36} & \sl{ 1.20E+26 } &  \sl{ 324.32  } &  -- &  \sl{ Y,Y} \\
\sl{71  } &  \sl{ J110150.26+554203.4  } &  \sl{ 5.91  } & \sl{4.66} &  \sl{ 4.28   } &  \sl{ 21.52  } & \sl{16.51} & \sl{ 1.56E+27 } &  \sl{ 539.35  } &  -- &  \sl{ Y,Y} \\
\sl{72 } &  \sl{ J110229.18+501452.4  } & \sl{ 1.92  } & \sl{1.46} & \sl{ 27.79  } &  \sl{ 49.19  } &  \sl{3.08} & \sl{ 7.20E+25 } &  \sl{ 423.49  } &  \sl{ 0.14  } &  \sl{ Y,Y} \\
\sl{73${\dagger}$ } &  \sl{ J110501.70+472300.7$\dagger$  } &  \sl{ 0.46$\dagger$  } &  \sl{0.46$\dagger$} & \sl{ 4.47$\dagger$  } &  \sl{ 1.48   } &  \sl{1.18} & \sl{ 5.13E+23 } &  \sl{ 186.33  } &  \sl{ 0.15  } &  \sl{ N,N} \\
\sl{74${\star}$ } &  \sl{ J110833.89+483202.9$\star$  } & \sl{ 26.95 } &  \sl{21.77} & \sl{ 18.54  } &  \sl{ 26.48  } & \sl{18.78} & \sl{ 8.48E+26 } &  \sl{ 768.52  } &  -- &  \sl{ Y,Y} \\
\sl{75 } &  \sl{ J111632.67+515613.2  } & \sl{ 37.62 } & \sl{28.71} &  \sl{ 76.48  } &  \sl{ 97.63  } &  \sl{80.84} & \sl{ 1.51E+26 } &  \sl{ 406.32  } &  \sl{ 2.83  } &  \sl{ Y,Y} \\
\sl{76${\star}$ } &  \sl{ J111818.90+510302.9$\star$  } &  \sl{ 16.19 } &  \sl{12.32} & \sl{ 20.90  } &  \sl{ 17.77  } &  \sl{11.64} & \sl{ 2.49E+26 } &  \sl{ 784.29  } &  -- &  \sl{ Y,Y} \\
\sl{77 } &  \sl{ J111924.94+464728.1  } &  \sl{ 59.94 } &  \sl{48.96} & \sl{ 58.74  } &  \sl{ 21.97  } &  \sl{18.35} & \sl{ 6.51E+26 } &  \sl{ 634.22  } &  -- &  \sl{ Y,Y} \\
\sl{78${\dagger}$ } &  \sl{ J112005.48+471400.8$\dagger$  } &  \sl{ 26.16 } &  \sl{10.01} & \sl{ 24.12  } &  \sl{ 86.45  } & \sl{38.61} &  \sl{ 5.84E+24 } &  \sl{ 119.79  } &  \sl{ 2.17  } &  \sl{ Y,Y} \\
\sl{79${\star}$ } &  \sl{ J112429.66+463520.3$\star$  } &  \sl{ 2.42  } & \sl{2.46} & \sl{ 9.33   } &  \sl{ 6.33   } & \sl{4.67} & \sl{ 1.58E+26 } &  \sl{ 1015.35 } & -- &  \sl{ Y,Y} \\
\sl{80 } &  \sl{ J113104.89+554530.7  } &  \sl{ 11.87 } & \sl{6.48} & \sl{ 27.85  } &  \sl{ 27.16  } & \sl{15.12} & \sl{ 1.11E+25 } &  \sl{ 350.27  } &  -- &  \sl{ Y,Y} \\
\sl{81${\star}$ } &  \sl{ J113423.99+563353.6$\star$ } &  \sl{ 0.38$\dagger$   } &  \sl{0.38$\dagger$ } & \sl{ 14.64$\dagger$  } &  \sl{ 1.74   } &  \sl{1.50} & \sl{ 4.86E+25 } &  \sl{ 1622.51 } &  \sl{ 5.19  } &  \sl{ P,N} \\
\sl{82${\dagger}$ } &  \sl{ J114121.89+532614.7$\dagger$  } &  \sl{ 72.80 } & \sl{ 48.71 } &  \sl{ 121.72 } &  \sl{ 126.75 } &  \sl{84.10} & \sl{ 1.17E+26 } &  \sl{ 304.11  } &  \sl{ 1.52  } &  \sl{ Y,Y} \\
\sl{83 } &  \sl{ J115108.76+482400.6  } &  \sl{ 9.43  } &  \sl{8.07} & \sl{ 6.73   } &  \sl{ 3.26   } &  \sl{4.44} & \sl{ 1.18E+26 } &  \sl{ 437.05  } &  -- &  \sl{ Y,Y} \\
\sl{84${\star}$ } &  \sl{ J121708.17+485812.8$\star$  } &  \sl{ 0.47$\dagger$} & \sl{0.47$\dagger$} & \sl{ 11.23$\dagger$} &  \sl{ 1.26   } &  \sl{1.30} & \sl{ 2.31E+25 } &  \sl{ 881.25  } &  \sl{ 9.50  } &  \sl{ N,N} \\
\sl{85 } &  \sl{ J121920.21+481554.1  } &  \sl{ 0.43$\dagger$  } &  \sl{ 0.43$\dagger$  } & \sl{ 4.88$\dagger$   } &  \sl{ 2.01   } & \sl{ 1.67} & \sl{ 8.37E+24 } &  \sl{ 481.29  } &  \sl{ 1.86  } &  \sl{ N,N} \\
\sl{86 } &  \sl{ J121952.36+465426.1  } &  \sl{ 21.68 } &  \sl{17.07} & \sl{ 21.90  } &  \sl{ 20.06  } &  \sl{16.88} & \sl{ 1.30E+27 } &  \sl{ 599.74  } &  -- &  \sl{ Y,Y} \\
\sl{87 } &  \sl{ J122459.65+554759.9  } &  \sl{ 0.45$\dagger$  } & \sl{0.45$\dagger$} & \sl{ 5.03$\dagger$   } &  \sl{ 0.94   } &  \sl{0.74} & \sl{ 2.39E+24 } &  \sl{ 289.33  } &  \sl{ 8.37  } &  \sl{ N,N} \\
\sl{88	} &  \sl{J123014.92+541141.4 } & \sl{0.41$\dagger$	} &  \sl{0.41$\dagger$	} & \sl{5.53$\dagger$	} &  \sl{0.90	} &  \sl{0.87} & \sl{6.18E+24	} &  \sl{1378.62	} &  \sl{2.03	} &  \sl{N,N} \\
\sl{89${\dagger}$ } &  \sl{ J123640.13+491141.4$\dagger$  } & \sl{ 6.67  } & \sl{4.98} & \sl{ 100.00 } &  \sl{ 15.86  }  & \sl{12.20} &  \sl{ 3.07E+25 } &  \sl{ 476.94  } &  \sl{ 0.64  } &  \sl{ Y,Y} \\
\sl{90${\dagger}$ } &  \sl{ J123648.05+552508.8$\dagger$  } &  \sl{ 14.65 }  & \sl{10.98} &  \sl{ 48.76  } &  \sl{ 52.60  }  & \sl{39.09} &  \sl{ 7.75E+25 } &  \sl{ 362.75  } &  \sl{ 3.03  } &  \sl{ Y,Y} \\
\sl{91 } &  \sl{ J125556.65+520353.6  } &  \sl{ 0.37$\dagger$  }  & \sl{0.37$\dagger$} &  \sl{ 5.52$\dagger$   } &  \sl{ 1.05   }  & \sl{0.67} &  \sl{ 1.83E+24 } &  \sl{ 314.64  } &  \sl{ 7.27  } &  \sl{ N,N} \\
\sl{92${\star}$${\dagger}$} &  \sl{ J131217.45+484754.9$\star$$\dagger$ } &  \sl{ 1.19  } &  \sl{1.89} & \sl{ 28.30  } &  \sl{ 9.22   } & \sl{5.91} &  \sl{ 6.88E+25 } &  \sl{ 1306.84 } &  \sl{ 1.00  } &  \sl{ Y,Y} \\
\sl{93 } &  \sl{ J133128.47+475439.6  } &  \sl{ 0.57  } &  \sl{0.89} &  \sl{ 1.20   } &  \sl{ 2.24   } &  \sl{1.40} &  \sl{ 5.45E+23 } &  \sl{ 292.69  } &  \sl{ 1.14  } &  \sl{ Y,Y} \\
\sl{94 } &  \sl{ J134452.94+485816.2  } &  \sl{ 0.45$\dagger$    } &  \sl{0.45$\dagger$  } &  \sl{ 15.87$\dagger$  } &  \sl{  10.28} &  \sl{2.89} &  \sl{ 3.17E+25 } &  \sl{ 1461.99  } &  \sl{ 2.60  } &  \sl{ P,N} \\
\sl{95 } &  \sl{ J140441.52+481100.6  } & \sl{ 3.23  } &  \sl{2.94} & \sl{ 8.66$\dagger$   } &  \sl{ 6.60   } & \sl{4.07} & \sl{ 7.59E+23 } &  \sl{ 333.25  } &  -- &  \sl{ P,Y} \\
\sl{96	} &  \sl{ J141806.69+483601.4 } & \sl{0.50$\dagger$	} & \sl{0.50$\dagger$	} &  \sl{6.65$\dagger$	} &  \sl{0.80	} &  \sl{0.0.48	} & \sl{5.86E+24	} &  \sl{787.15	} &  \sl{1.92} &  \sl{N,N} \\
\sl{97 } &  \sl{ J142436.47+554938.5  } &  \sl{ 0.53$\dagger$  } &  \sl{ 0.53$\dagger$  } &  \sl{ 6.23$\dagger$   } &  \sl{ 1.38   } &  \sl{0.92 } &  \sl{ 3.28E+24 } &  \sl{ 399.62  } &  \sl{ 6.83  } &  \sl{ N,N} \\
\sl{98 } &  \sl{ J143644.70+515442.9  } &  \sl{ 0.47$\dagger$  } &  \sl{ 0.47$\dagger$} & \sl{ 5.56$\dagger$} &  \sl{ 1.30   } &  \sl{0.70} & \sl{ 1.83E+23 } &  \sl{135.89} &  
\sl{ 0.27  } &  \sl{ N,N} \\
\sl{99 } &  \sl{ J145935.07+533350.1  } &   \sl{ 2.75  } &   \sl{ 2.80} & \sl{33.35$\dagger$} &  \sl{ 14.65  } &   \sl{10.40} & \sl{ 4.92E+24 } &  \sl{ 317.63  } &  \sl{ 2.27  } &  \sl{ P,Y} \\
\hline
\end{longtable}
\footnotesize{ {\bf Columns:} 1~--~Number; 2~--~Source; 3-4~--~the total and peak core flux densities at 1400 MHz measured directly in the FIRST $5\arcsec$ image in mJy; 5~--~the total flux densities at 1400 MHz measured directly in the NVSS $45\arcsec$ image in mJy; 6-7~--~the total and peak core flux densities at 144 MHz measured directly in the LOFAR $6\arcsec$ image in mJy; 8~--~radio luminosity of the source at 144 MHz; 9~--~linear size of the source at 144 MHz; 10~--~stellar mass; 11~--~detection in NVSS, detection in VLA-FIRST (Y=full detection, P= partial detection and N = no detection). $\star$ and $\dagger$ in source names denote GRGs, and galaxies residing in a cluster environment respectively. $\dagger$ in columns 3, 4, and 5, values represent upper limits (except for NVSS partial detection, where $\dagger$ in column 5 represents a lower limit).
}
%%%%%%%%%%%%%%%%%%%%%%%%%%%%%%%%%%%%%%%%%%%%%%%%%%%%%%%%%%%%%%%%%%%%%%%

\newpage
\section{Images}
\label{app:images}
%%%%%%%%%%%%%%%%%%%%%%%%%%%%%%%%%%%%%%%%%%%%%%%%%%%%%%%%%%%%%%%%%%%%%%%
\begin{figure*}[ht!]
\centerline{\includegraphics[width=0.3\textwidth]{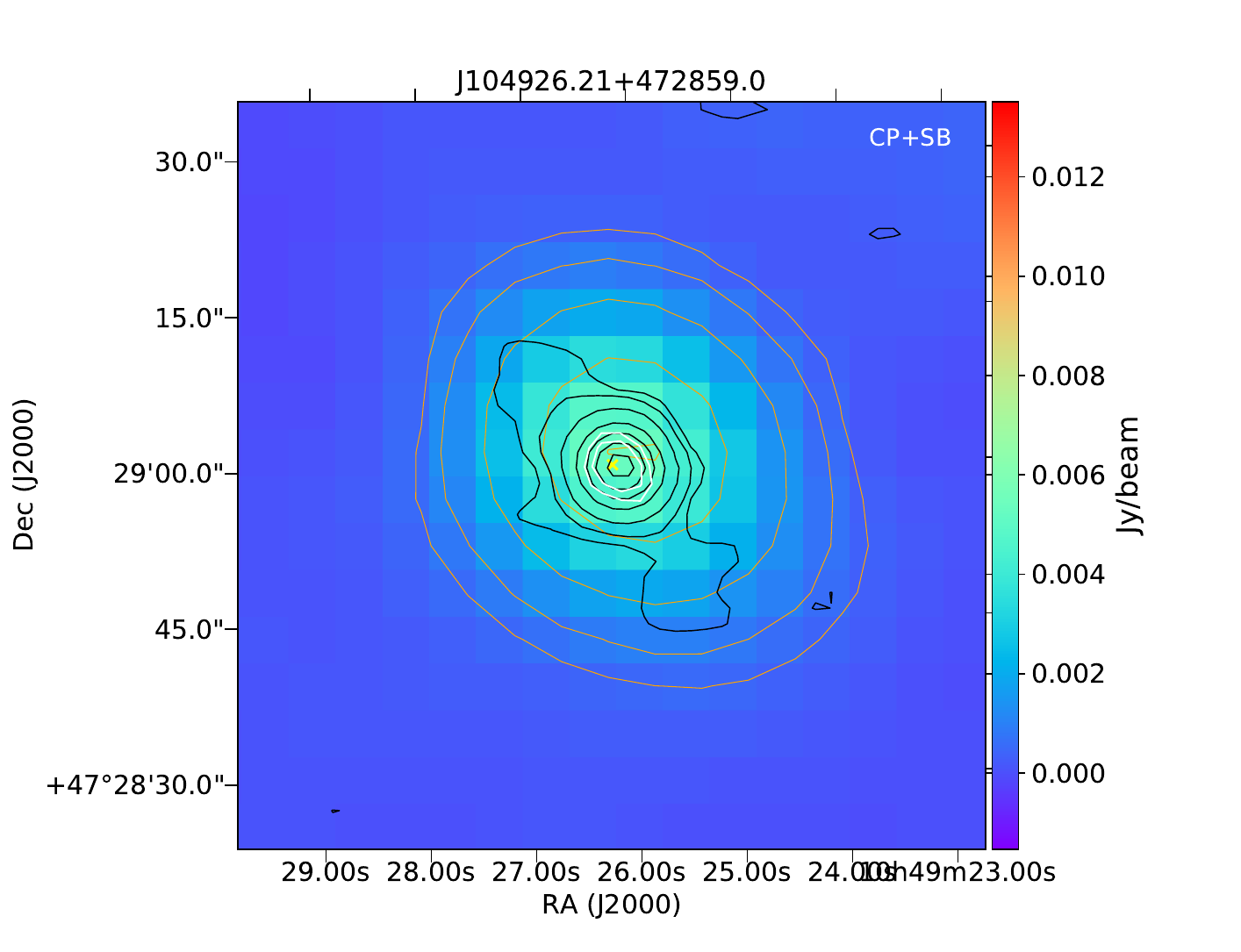}
            \includegraphics[width=0.3\textwidth]{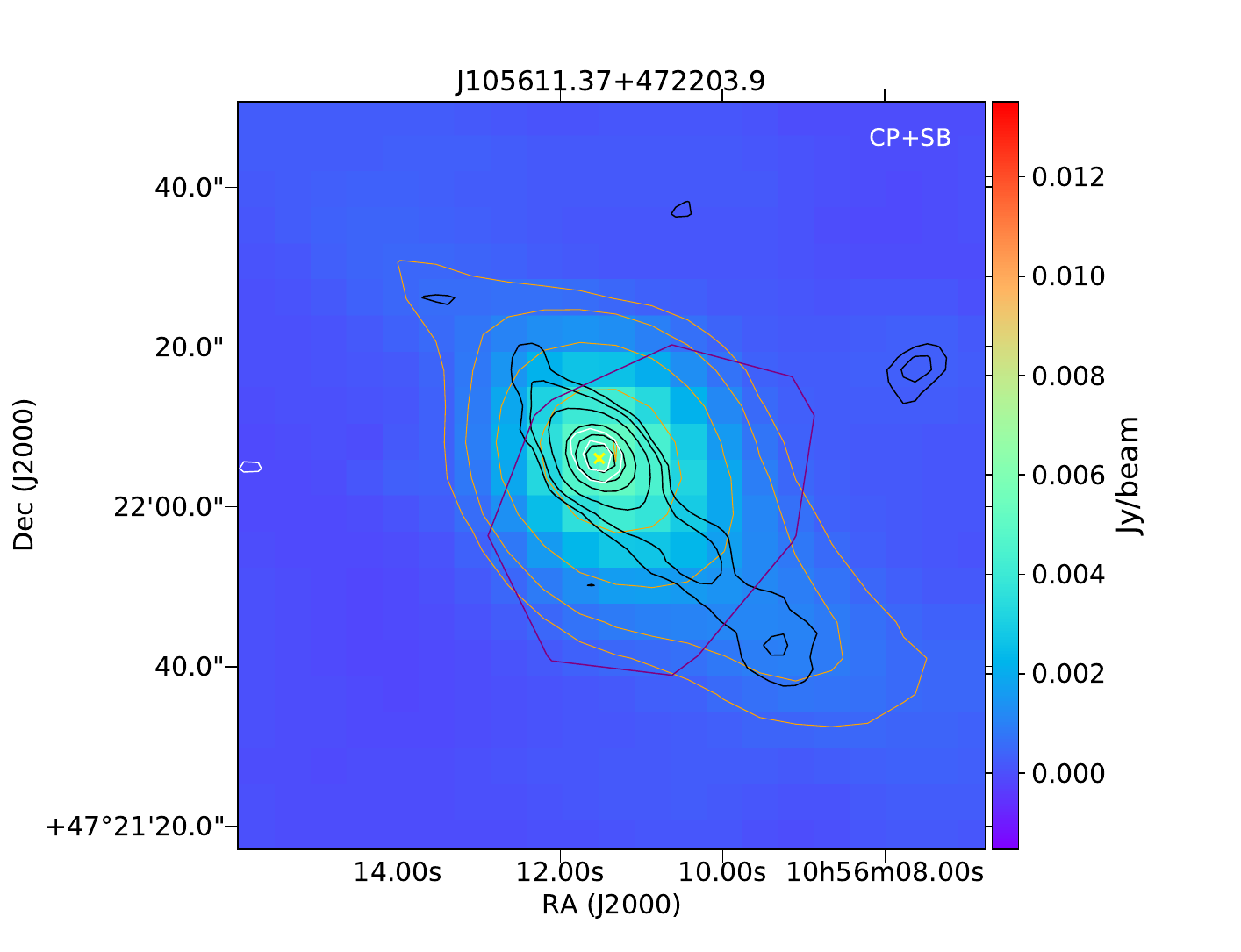}
            \includegraphics[width=0.3\textwidth]{Maps_final/tplot_3_3sig.pdf}     
            }        
\centerline{\includegraphics[width=0.3\textwidth]{Maps_final/tplot_6_3sig.pdf}
            \includegraphics[width=0.3\textwidth]{Maps_final/tplot_8_3sig.pdf}
            \includegraphics[width=0.3\textwidth]{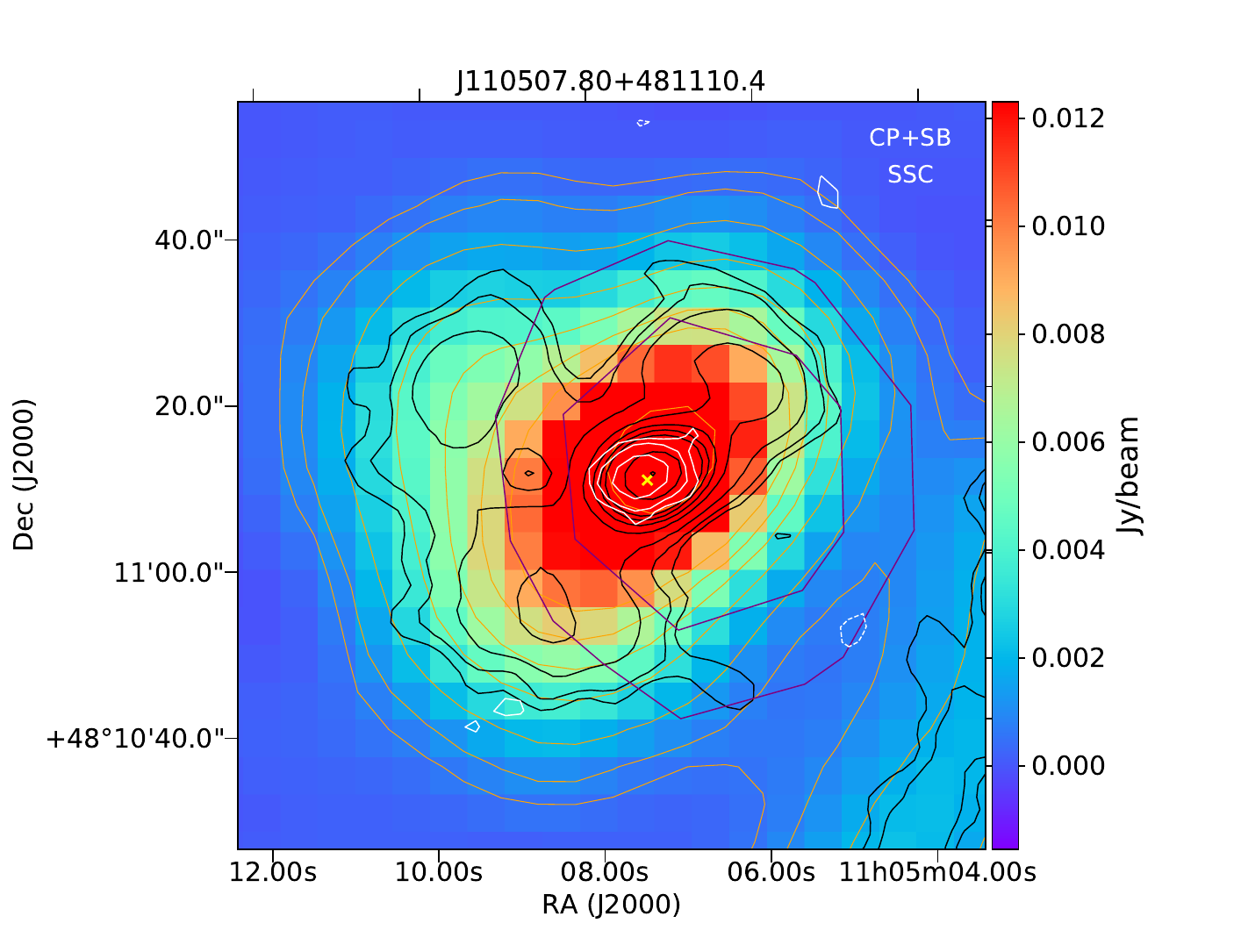}           
           }         
\centerline{\includegraphics[width=0.3\textwidth]{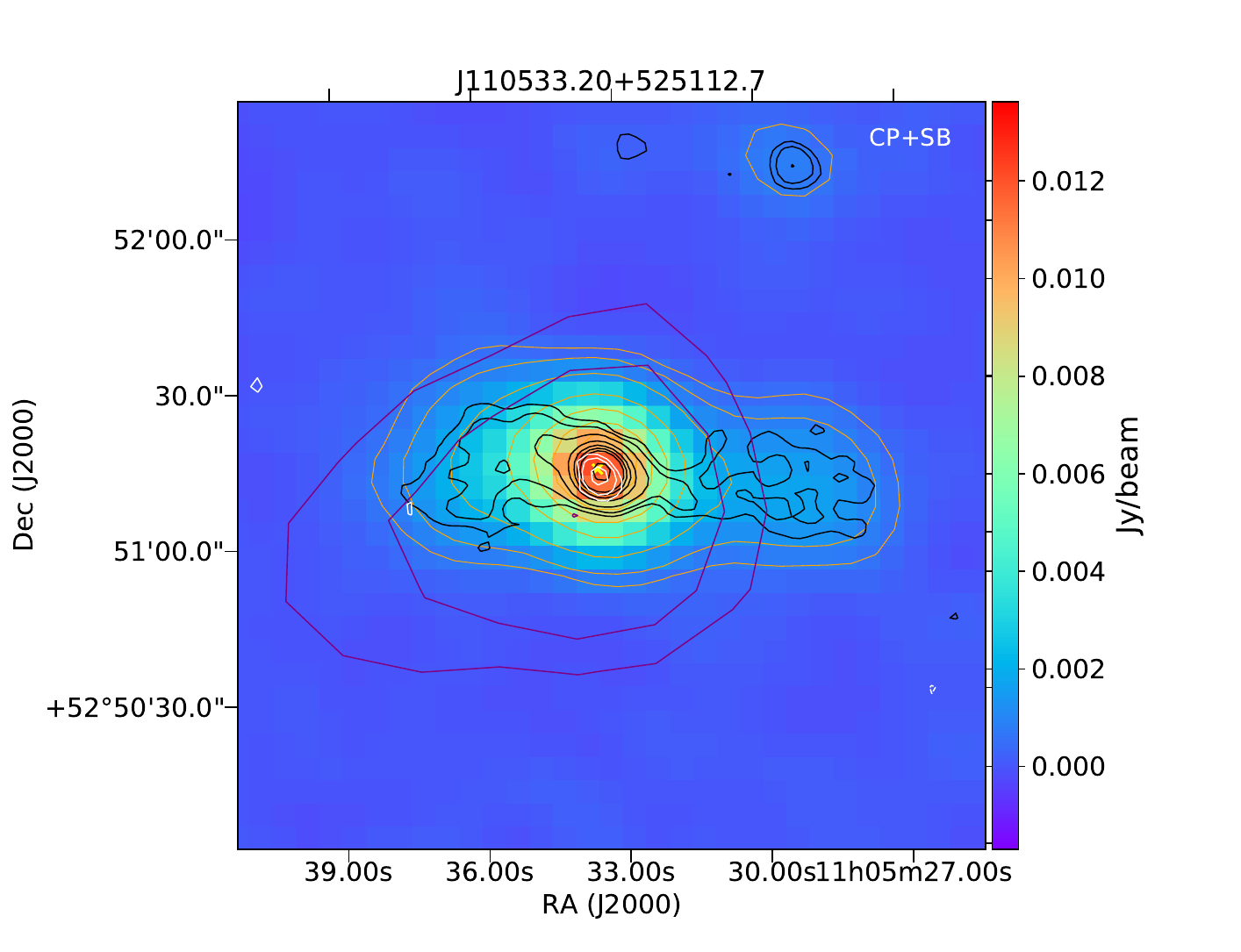}
            \includegraphics[width=0.3\textwidth]{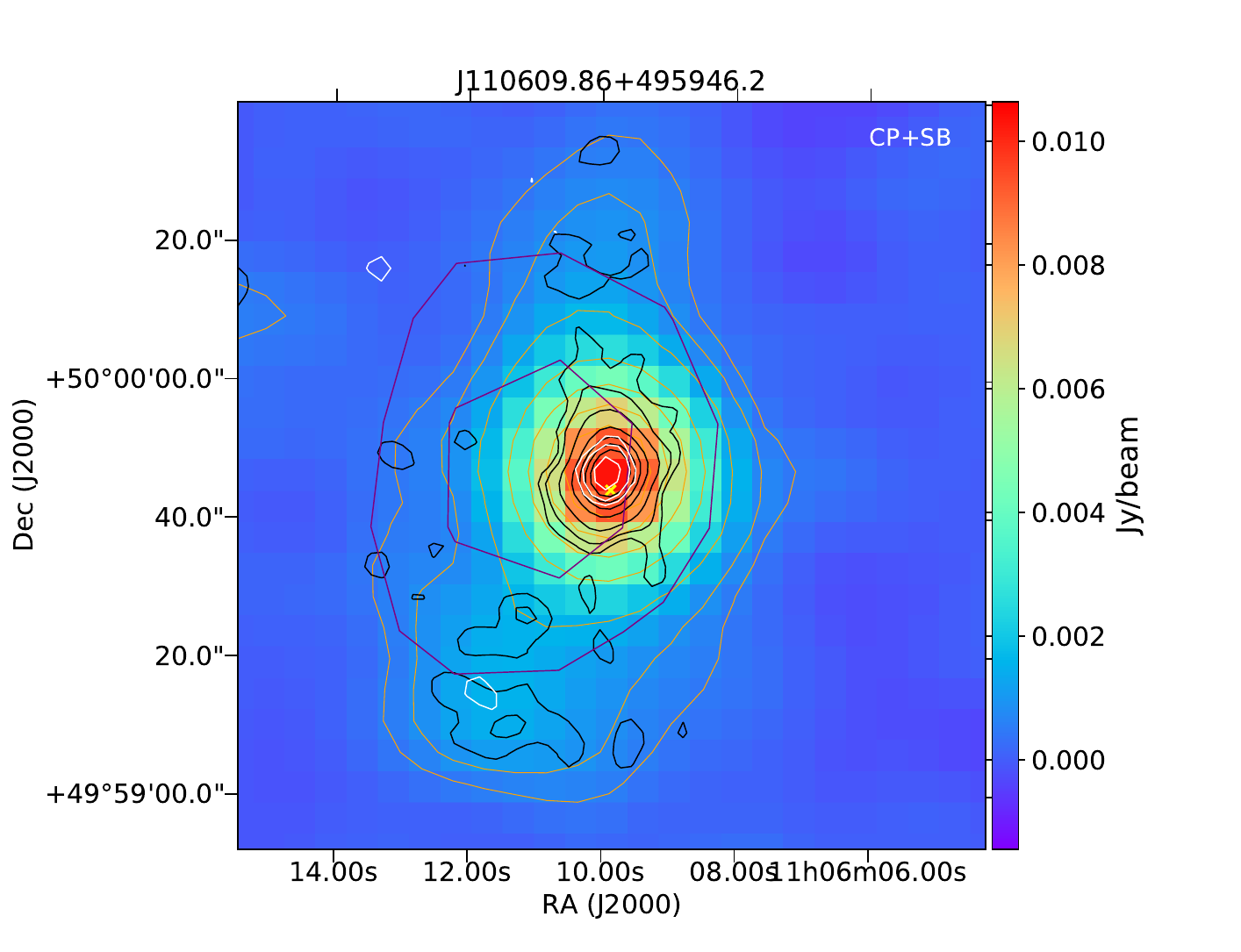}
            \includegraphics[width=0.3\textwidth]{Maps_final/tplot_13_3sig.pdf}            
}
\centerline{\includegraphics[width=0.3\textwidth]{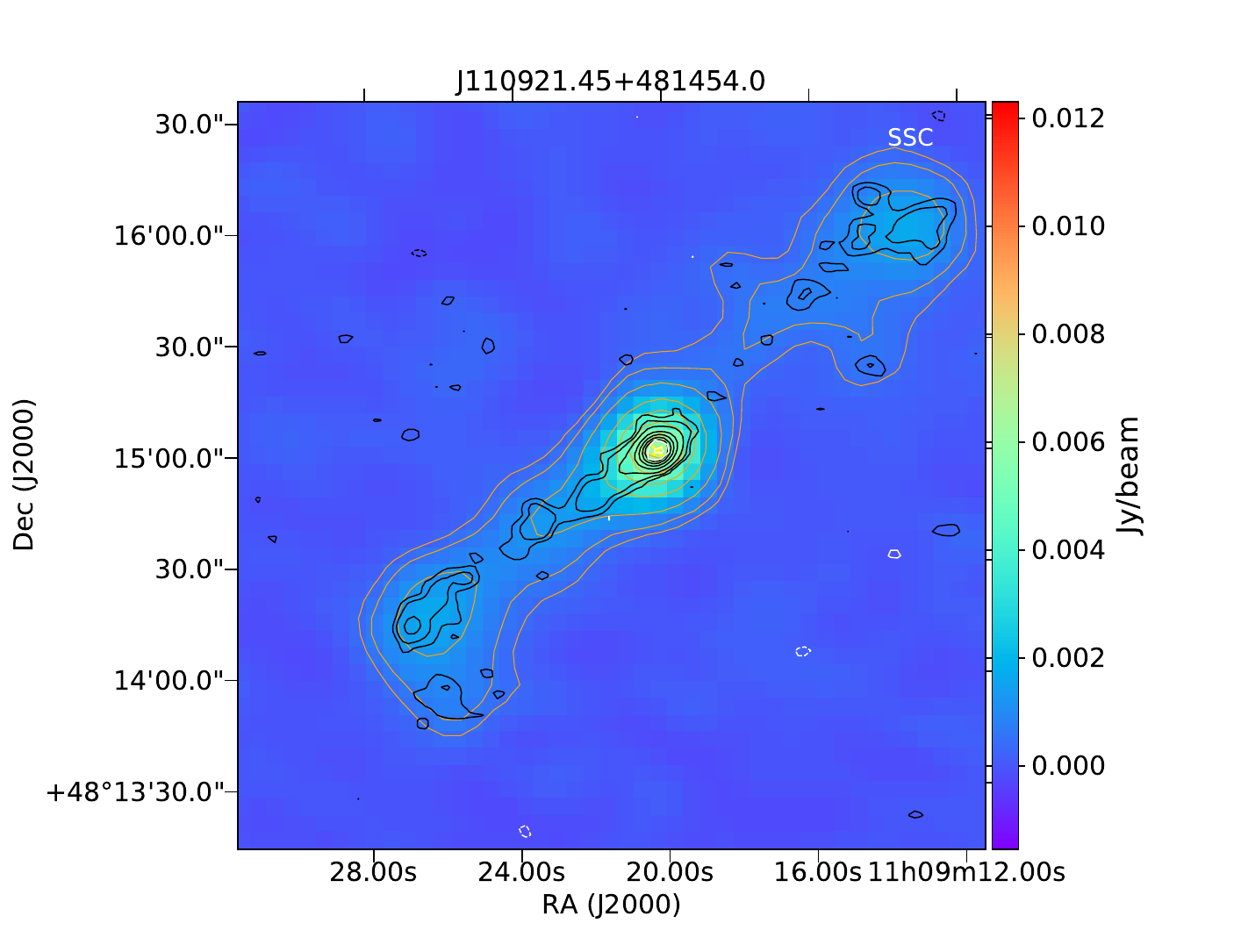}
            \includegraphics[width=0.3\textwidth]{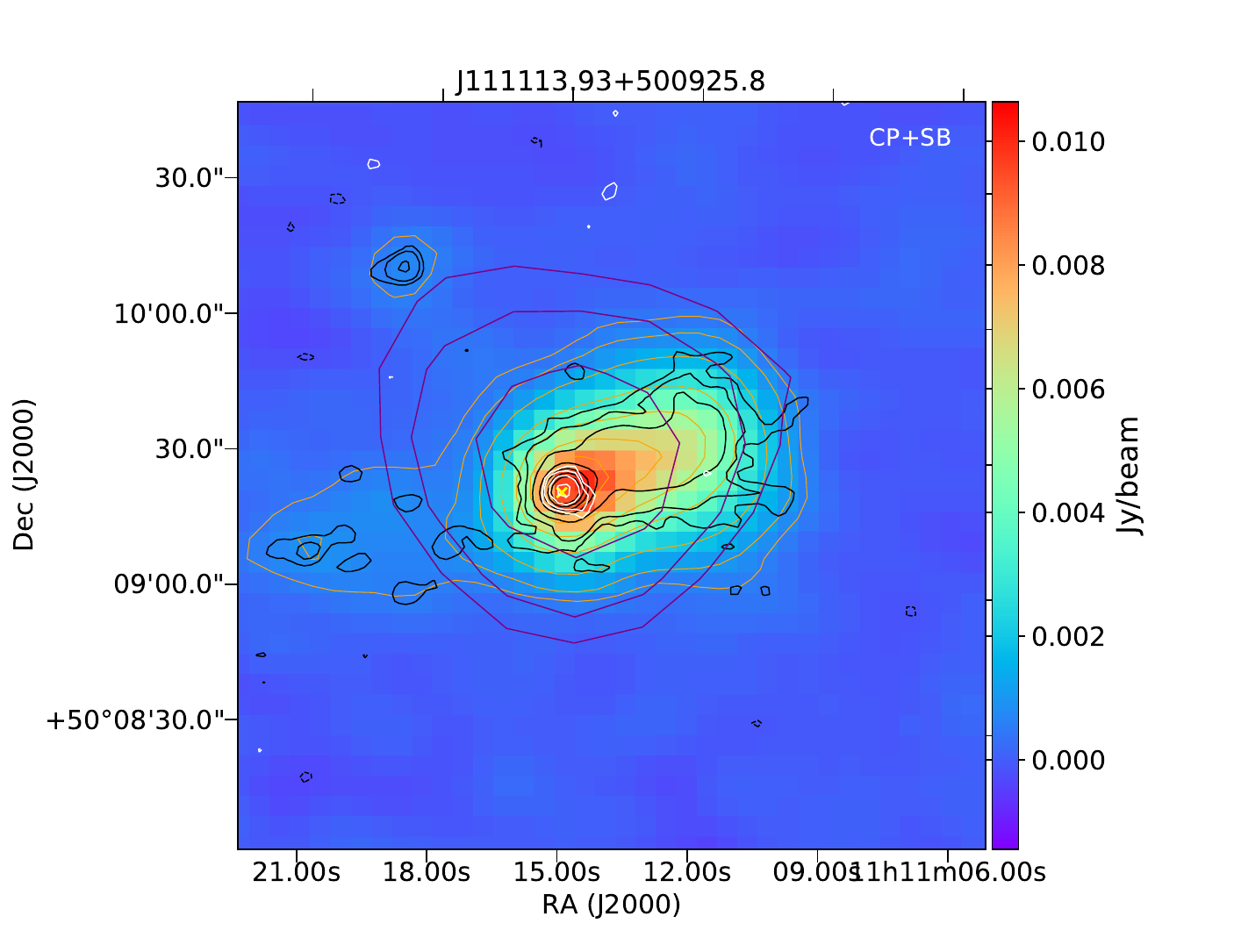}
            \includegraphics[width=0.3\textwidth]{Maps_final/tplot_17_3sig.pdf}            
}
\centerline{\includegraphics[width=0.3\textwidth]{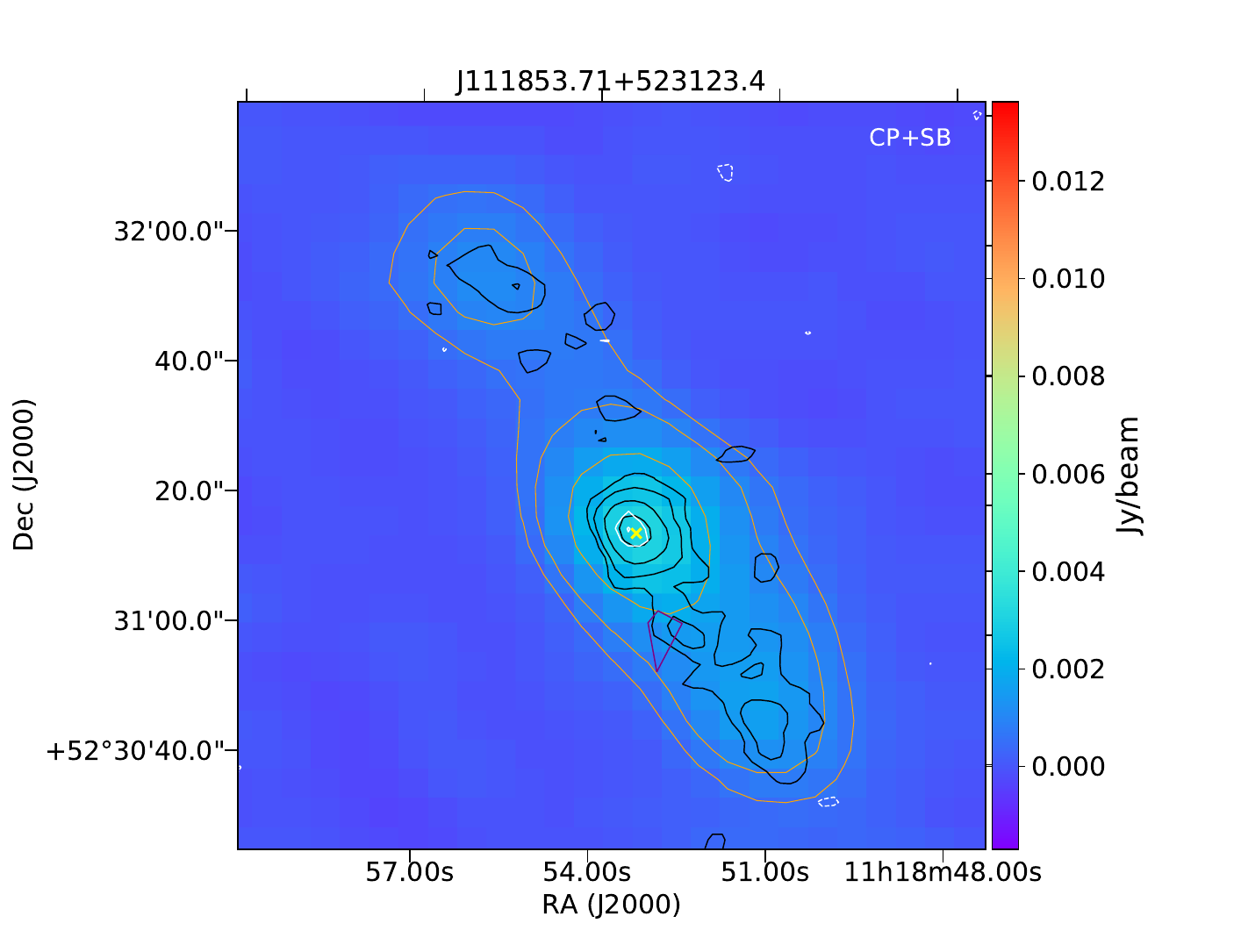}
            \includegraphics[width=0.3\textwidth]{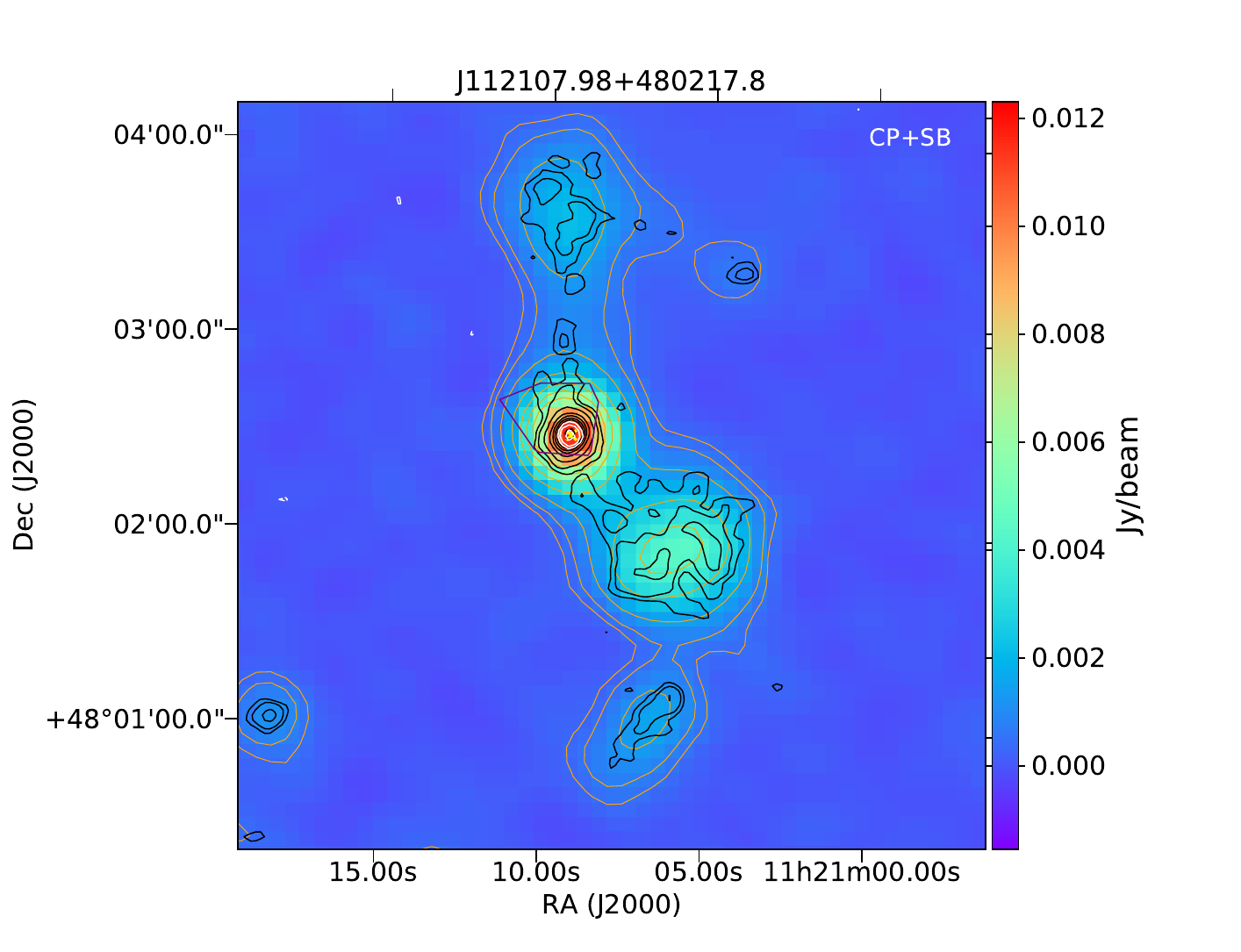}
            \includegraphics[width=0.3\textwidth]{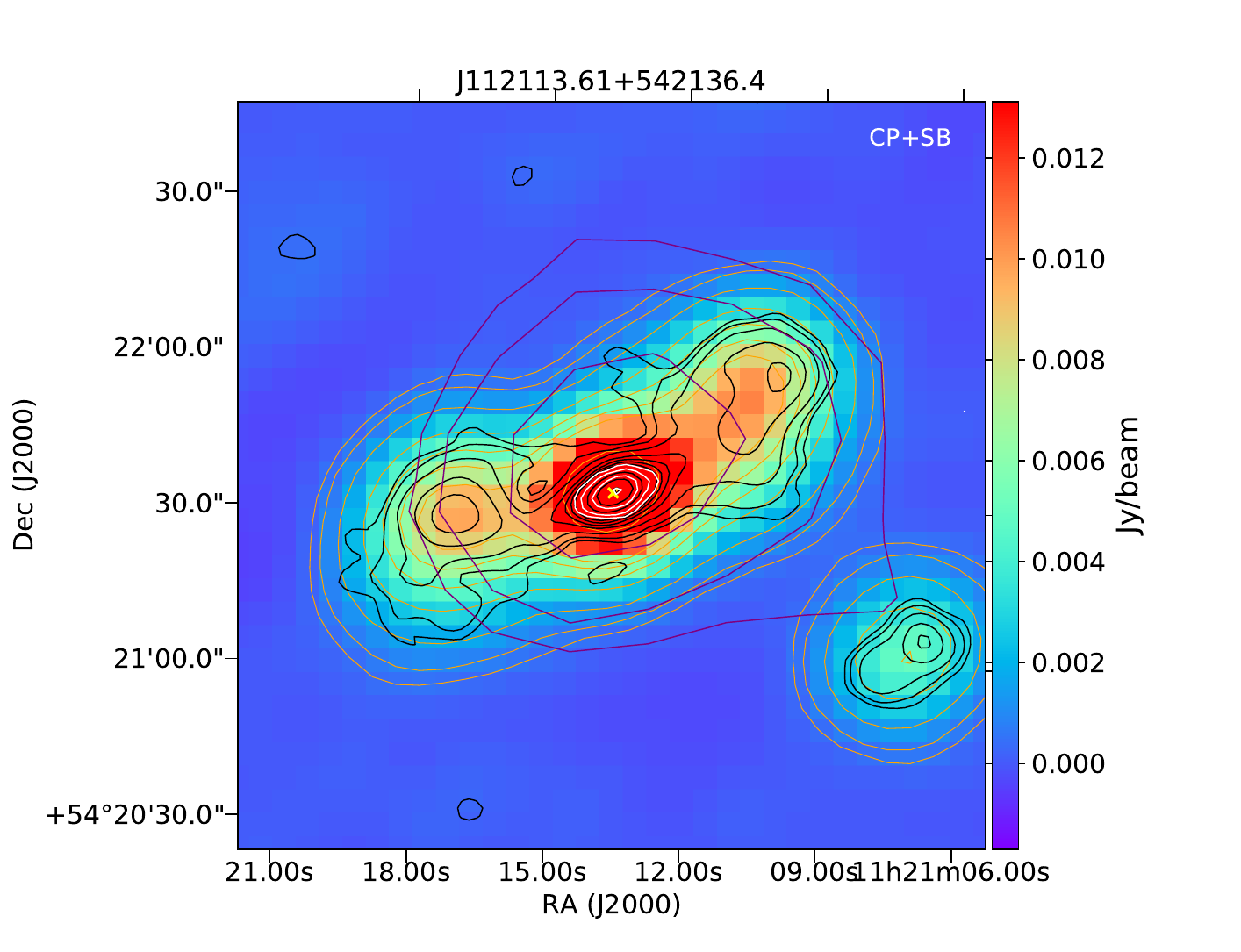}            
}
\caption{\footnotesize{
Images of 69 candidate restarted galaxies selected based on high radio $\mathrm{CP_{1.4GHz}}$ combined with low SB of extended emission, a steep spectrum of the core, and USS extended emission coupled with a bright core and summarised in Tables.~\ref{list of sources} and ~\ref{Core flux densities}. Radio contours from VLA FIRST maps (white, 5$\arcsec$), LOFAR high-resolution maps (black, 6$\arcsec$), and NVSS maps (purple, 45$\arcsec$) are overlaid on the LOFAR low-resolution resolution maps (orange, 20$\arcsec$).
The contouring of all the maps is made at $\,\sigma_\mathrm{local}\times(-3,3,5,10,20,30,40,50,100,150,200)$ levels, with $\sigma_\mathrm{local}$ representing the local RMS noise of the corresponding maps.
The host galaxy position is marked with a yellow cross.
}}
\label{fig:1}
\end{figure*}

\begin{figure*}[ht!]
\centerline{\includegraphics[width=0.3\textwidth]{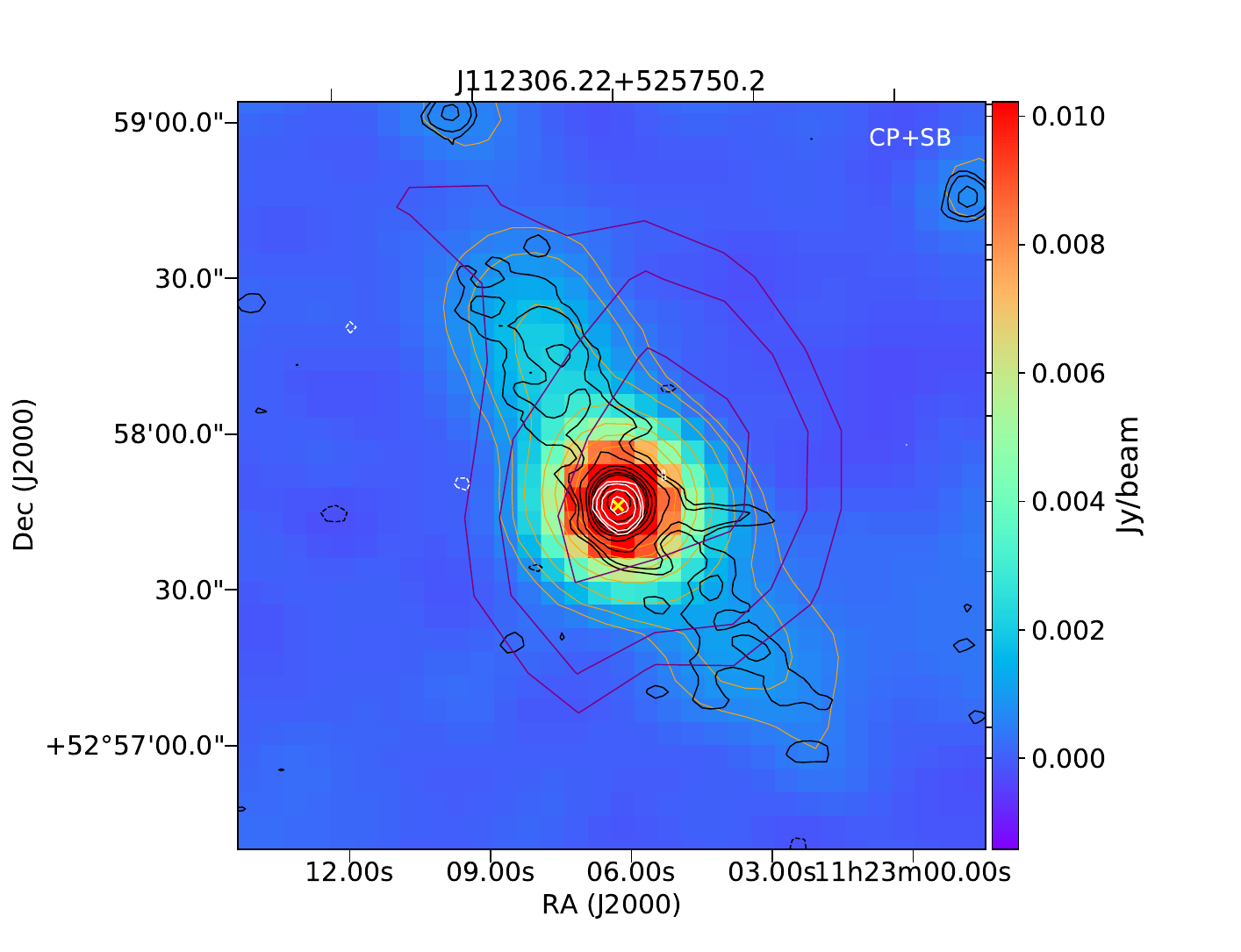}
            \includegraphics[width=0.3\textwidth]{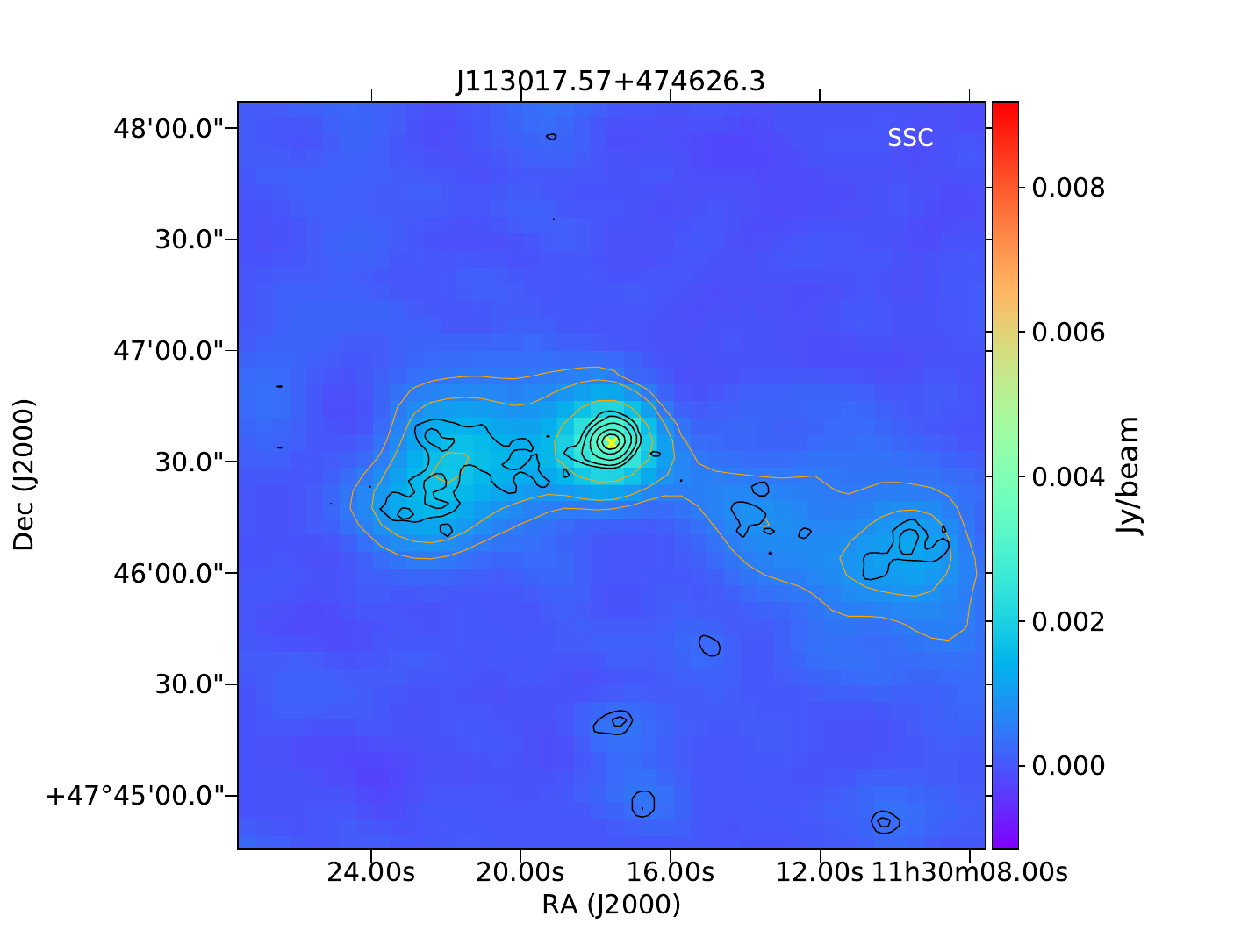}
            \includegraphics[width=0.3\textwidth]{Maps_final/tplot_29_3sig.pdf}                   }        
\centerline{\includegraphics[width=0.3\textwidth]{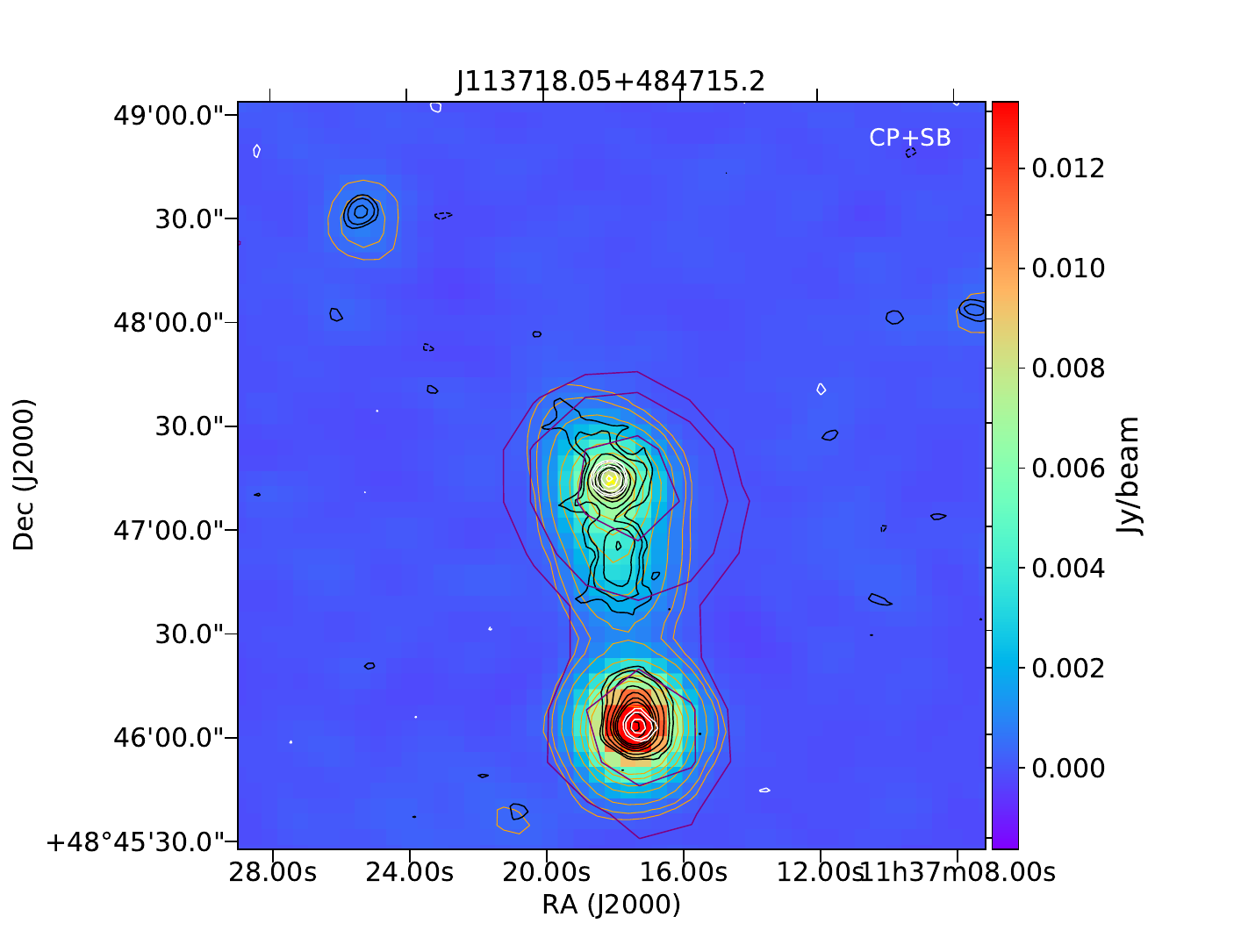}
            \includegraphics[width=0.3\textwidth]{Maps_final/tplot_33_3sig.pdf}
            \includegraphics[width=0.3\textwidth]{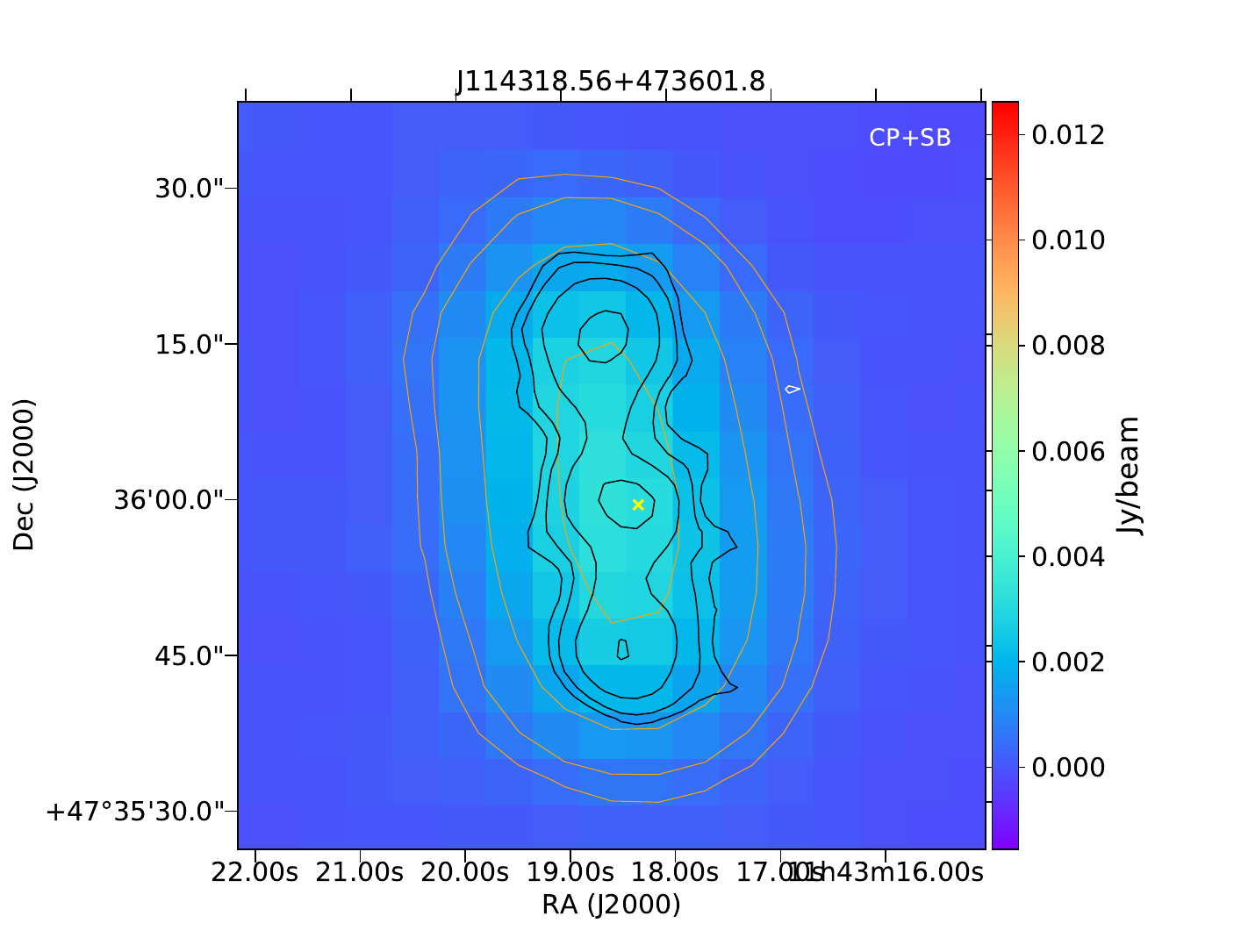}          
           }         
\centerline{\includegraphics[width=0.3\textwidth]{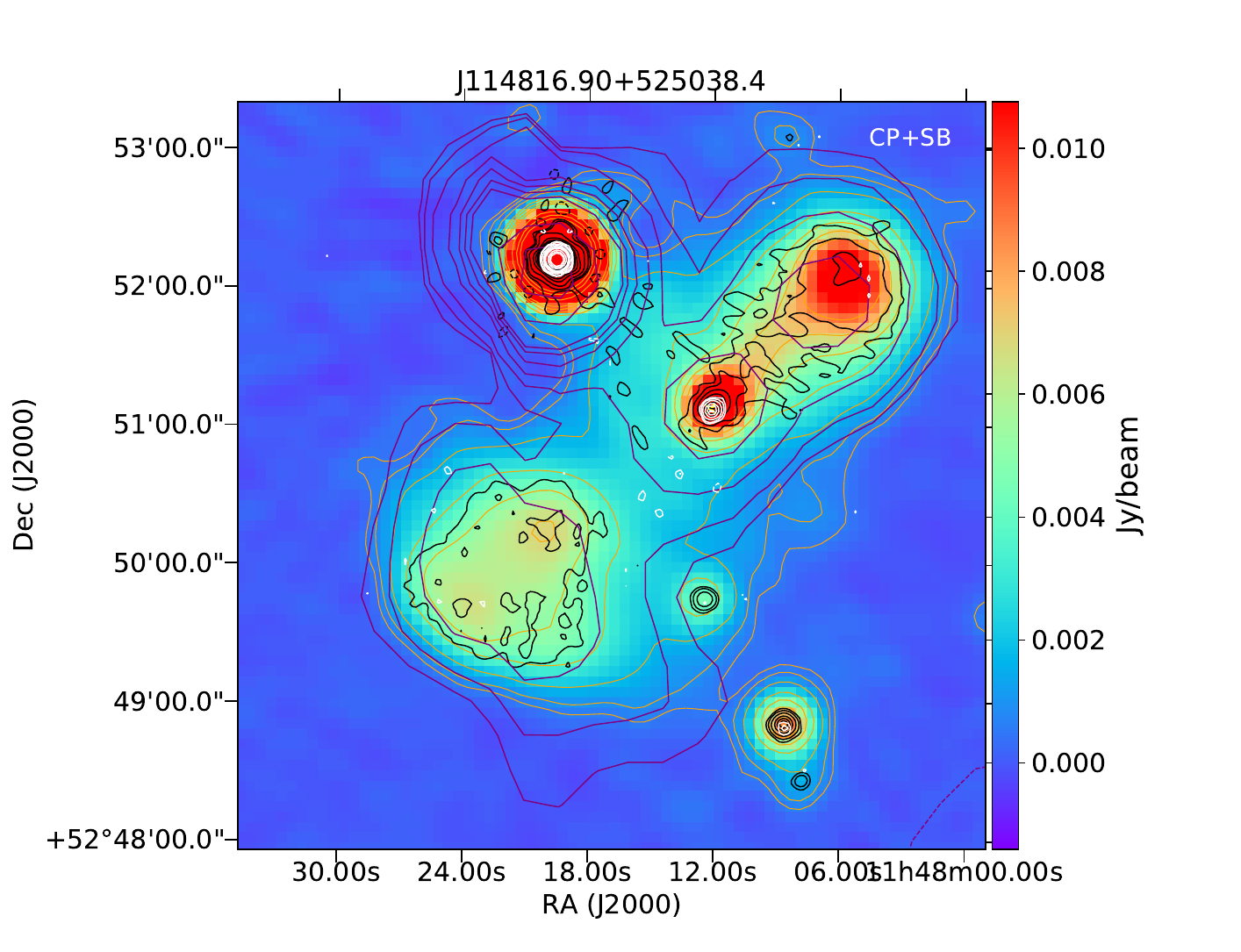}
            \includegraphics[width=0.3\textwidth]{Maps_final/tplot_36_3sig.pdf}
            \includegraphics[width=0.3\textwidth]{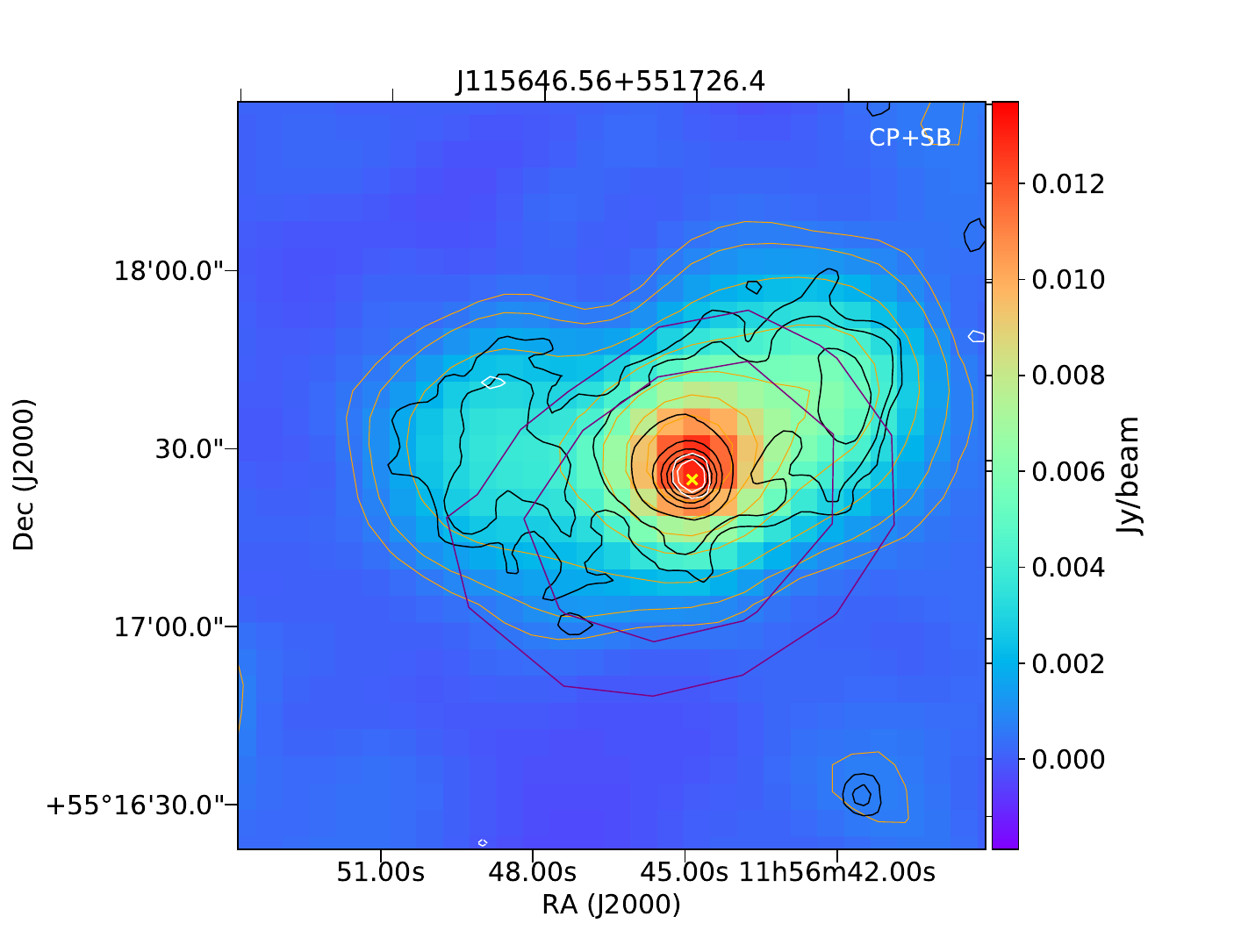}            
}
\centerline{\includegraphics[width=0.3\textwidth]{Maps_final/tplot_39_3sig.pdf}
            \includegraphics[width=0.3\textwidth]{Maps_final/tplot_40_3sig.pdf}
            \includegraphics[width=0.3\textwidth]{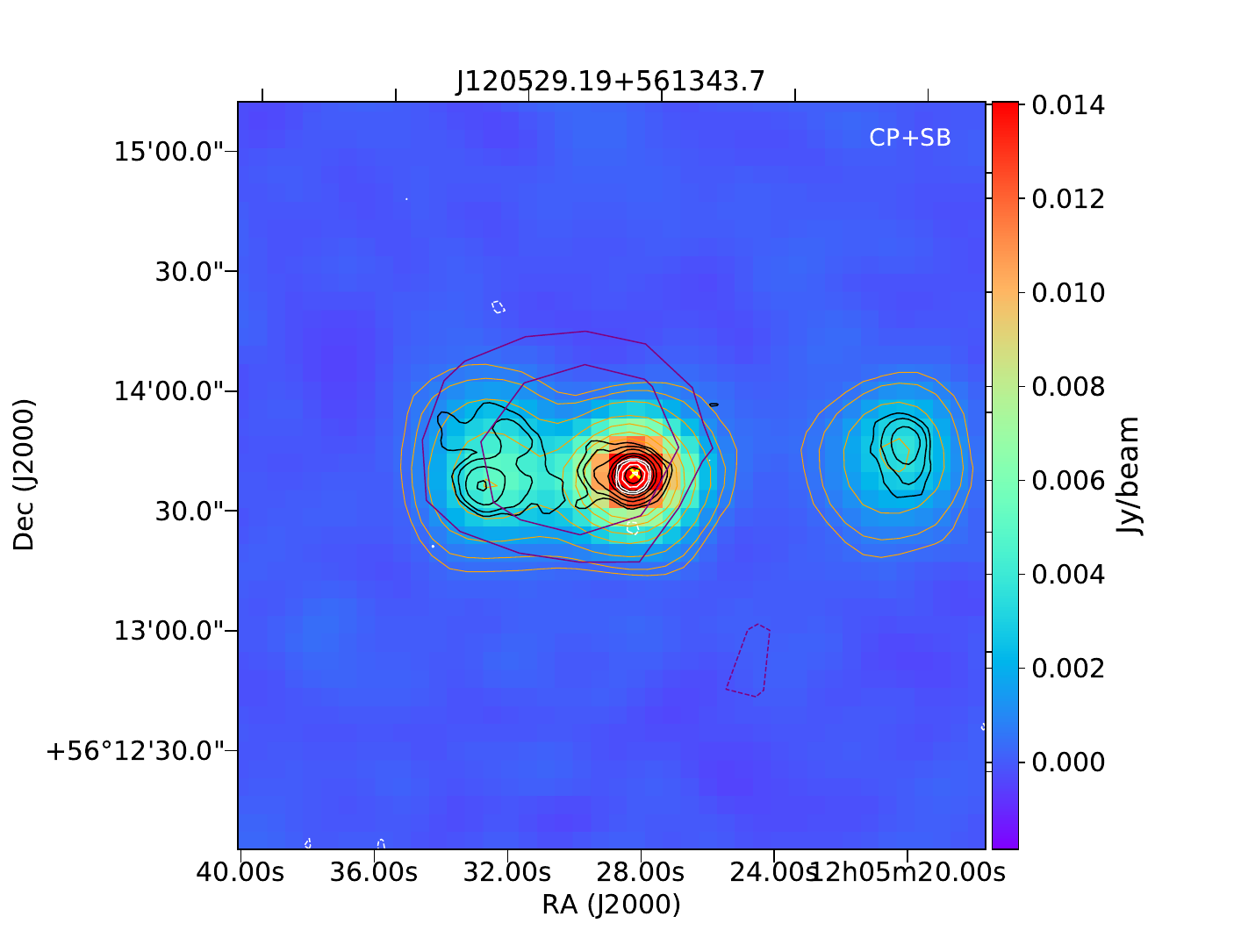}            
}
\centerline{\includegraphics[width=0.3\textwidth]{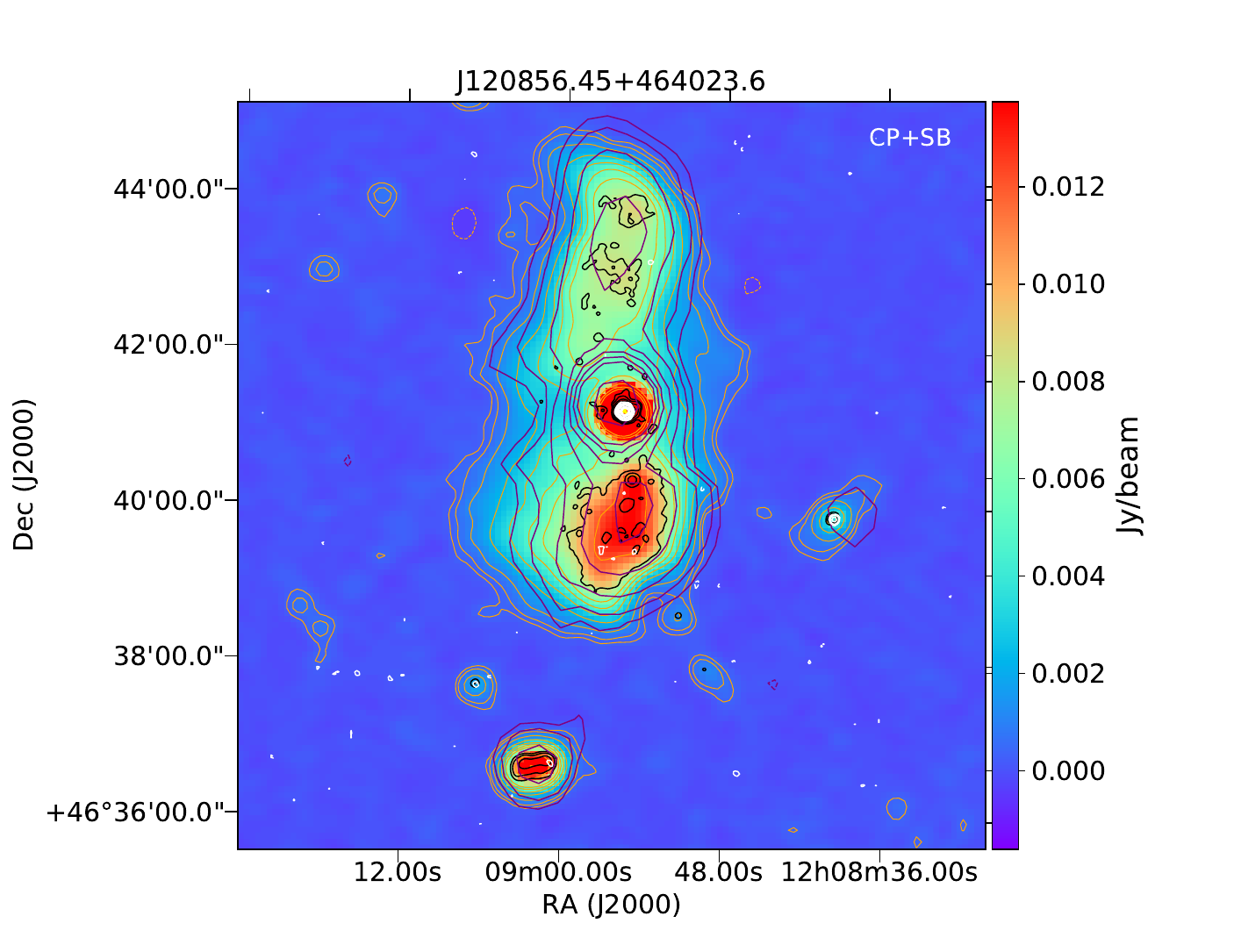}
            \includegraphics[width=0.3\textwidth]{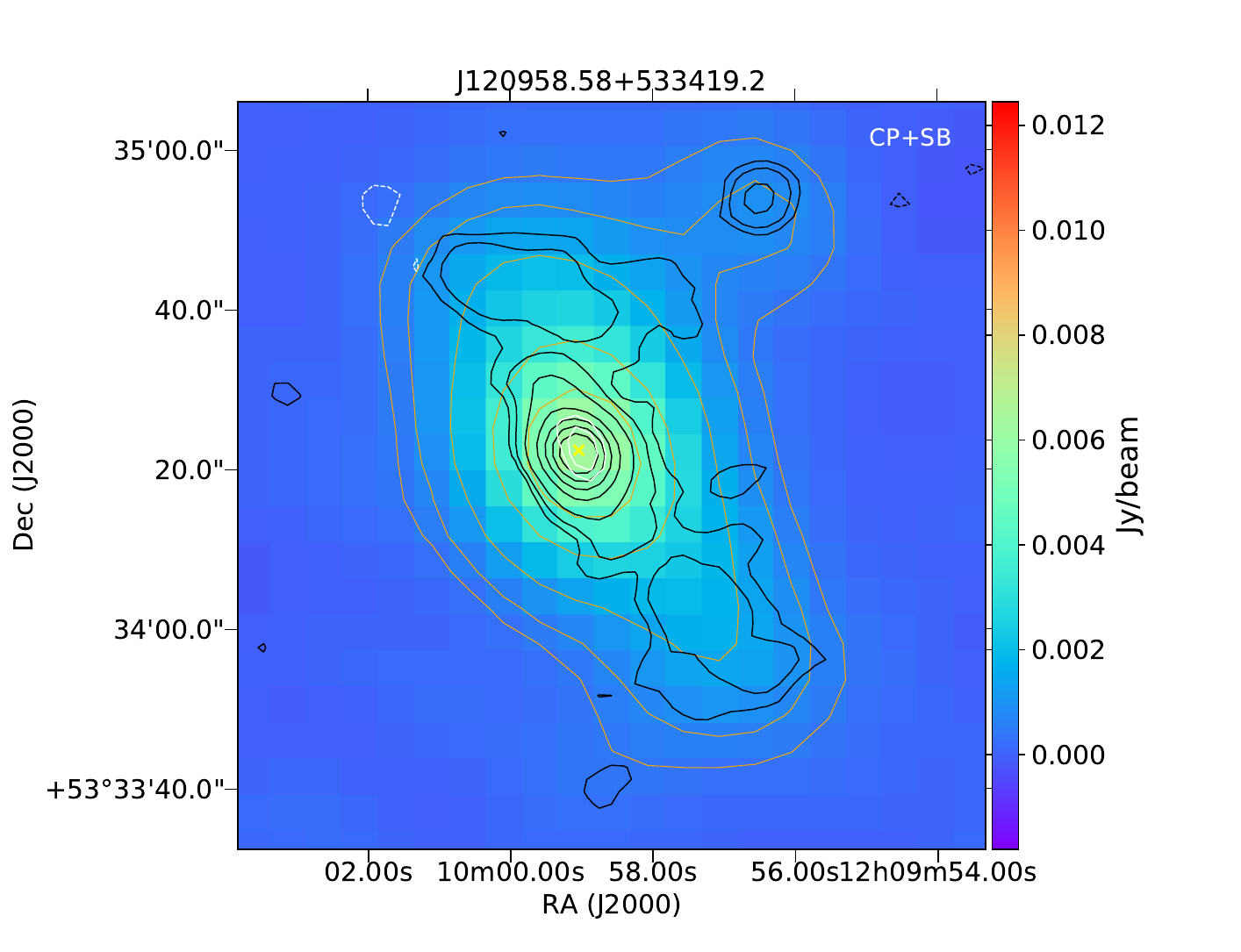}
            \includegraphics[width=0.3\textwidth]{Maps_final/tplot_44_3sig.pdf}            
}
\caption{\footnotesize{Images of 69 candidate restarted galaxies selected based on
high radio $\mathrm{CP_{1.4GHz}}$ combined with low SB of extended emission, a steep spectrum of the core, and USS extended emission coupled with a bright core and summarised in Tables.~\ref{list of sources} and ~\ref{Core flux densities}. 
Radio contours from VLA FIRST maps (white, 5$\arcsec$), LOFAR high-resolution maps (black, 6$\arcsec$), and NVSS maps (purple, 45$\arcsec$) are overlaid on the LOFAR low-resolution resolution maps (orange, 20$\arcsec$).
The contouring of all the maps is made at $\,\sigma_\mathrm{local}\times(-3,3,5,10,20,30,40,50,100,150,200)$ levels, with $\sigma_\mathrm{local}$ representing the local RMS noise of the corresponding maps.
The host galaxy position is marked with a yellow cross. \\ 
}}
\label{fig:2}
\end{figure*}

\begin{figure*}[ht!]
\centerline{\includegraphics[width=0.3\textwidth]{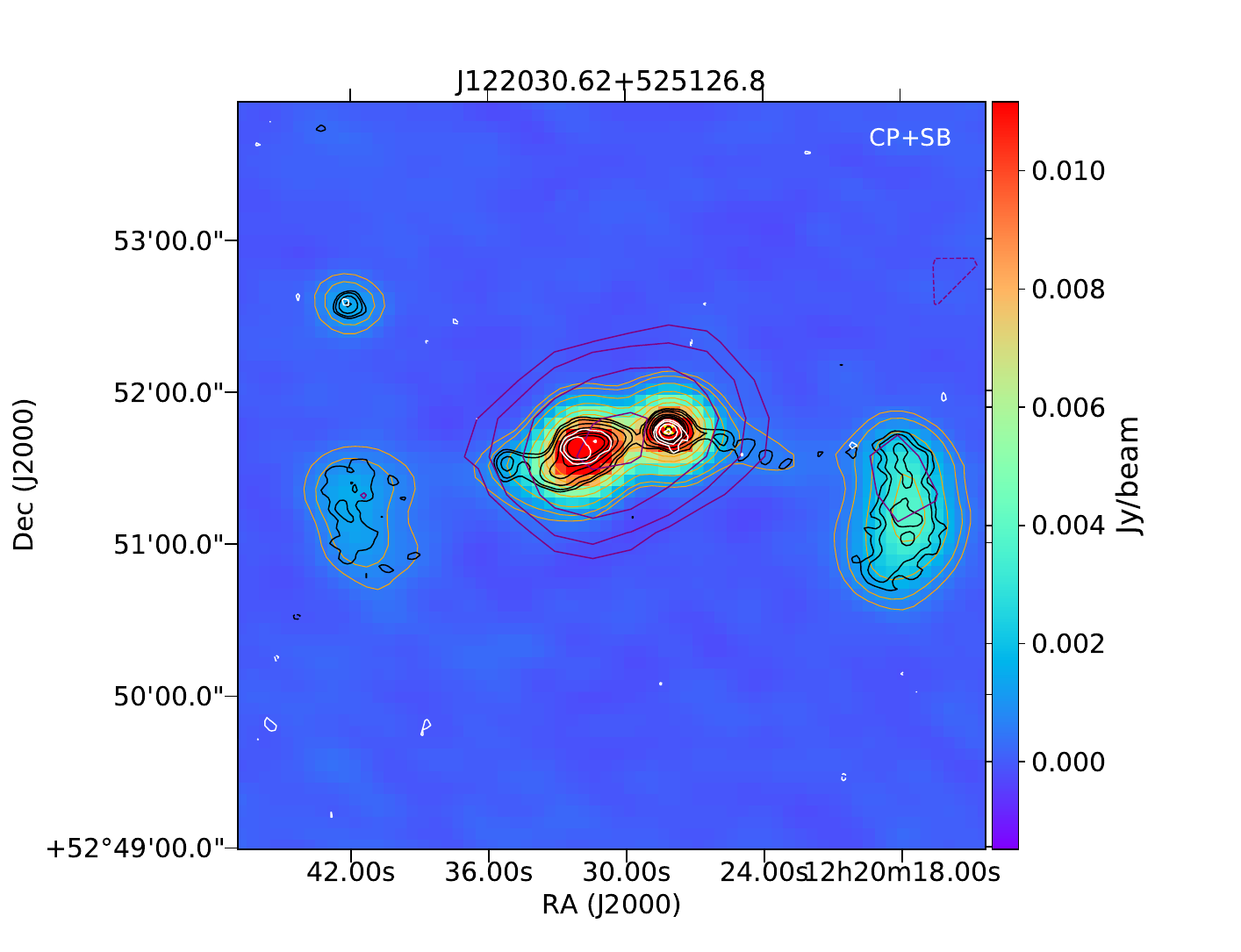}
            \includegraphics[width=0.3\textwidth]{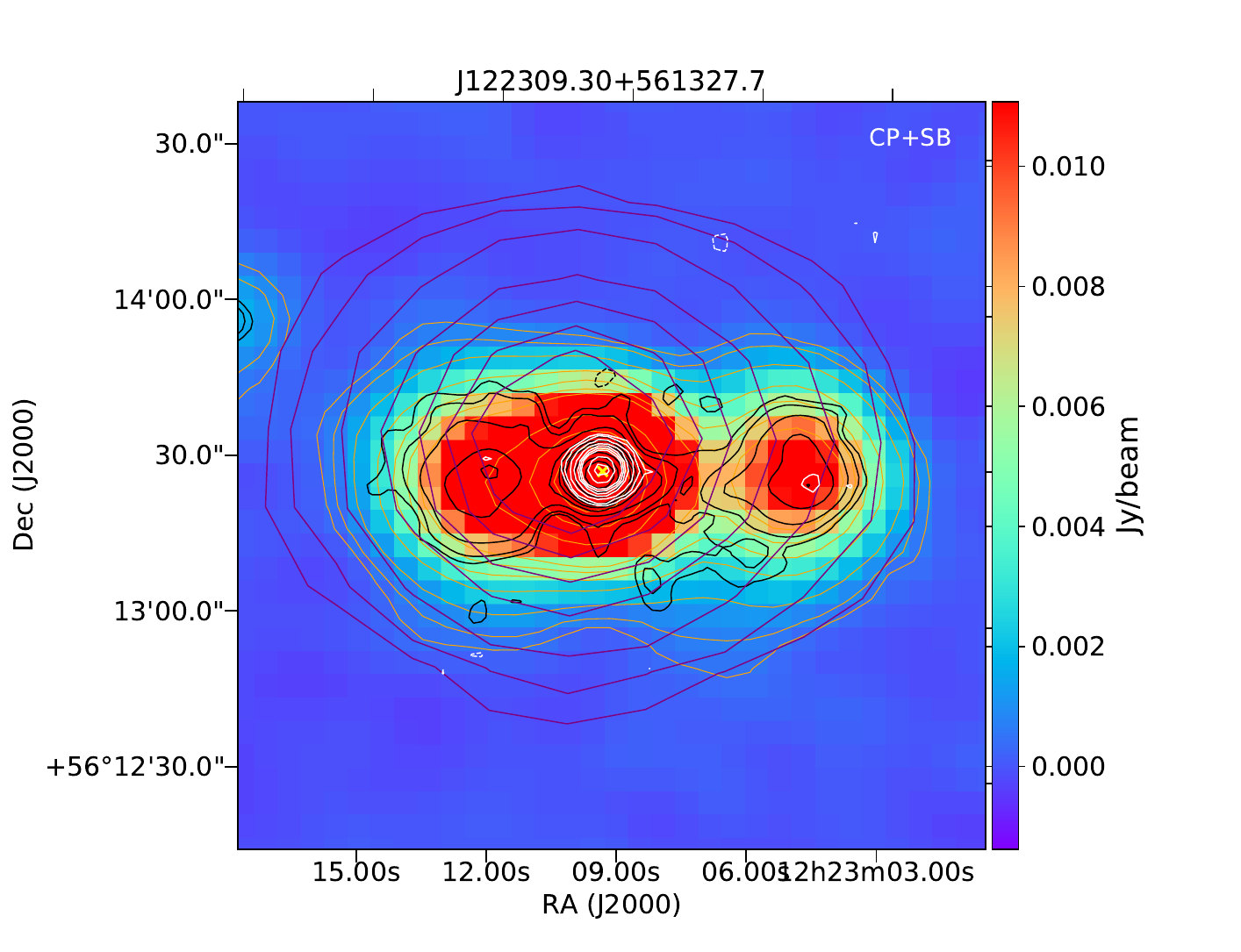}
            \includegraphics[width=0.3\textwidth]{Maps_final/tplot_51_3sig.pdf}                   }        
\centerline{\includegraphics[width=0.3\textwidth]{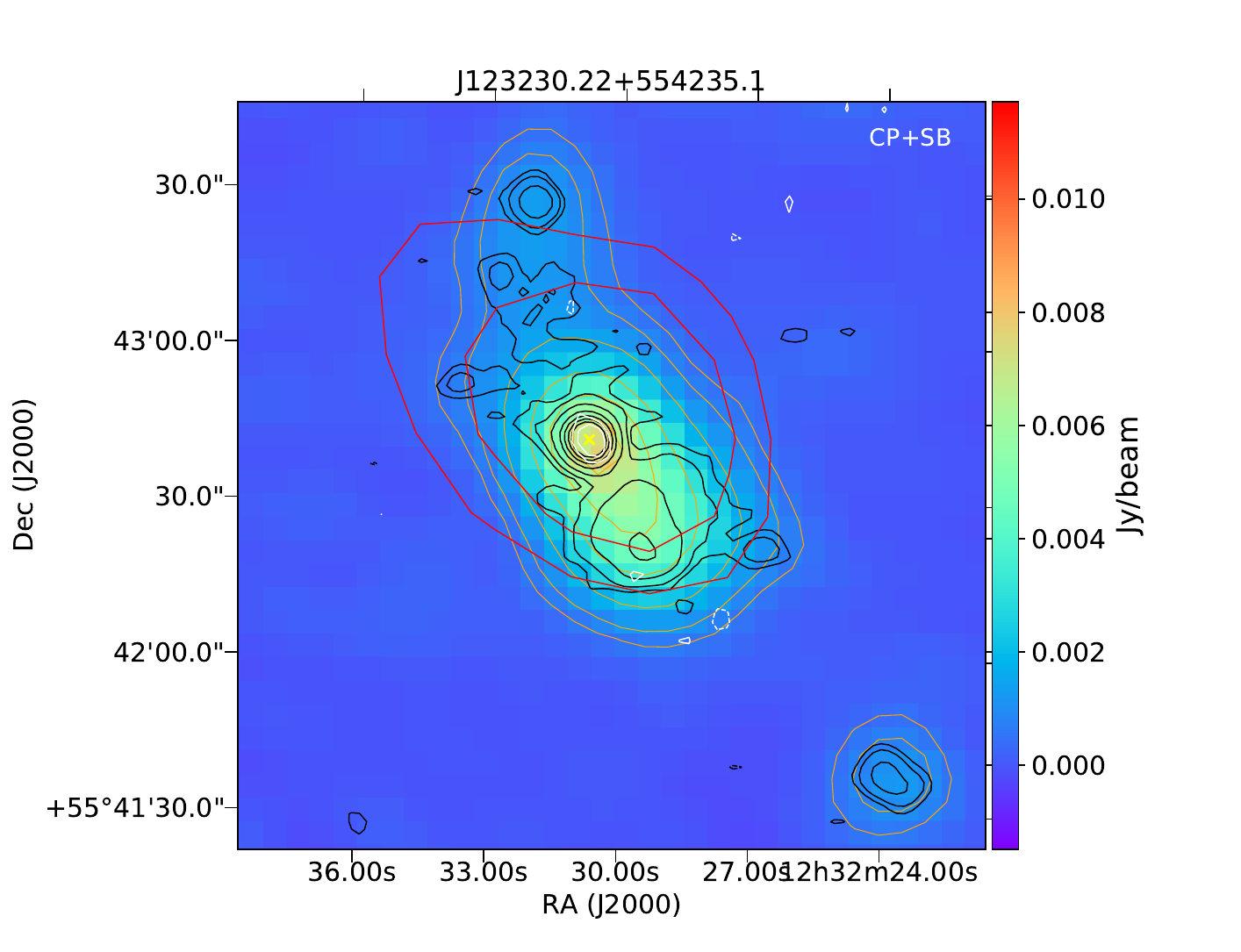}
            \includegraphics[width=0.3\textwidth]{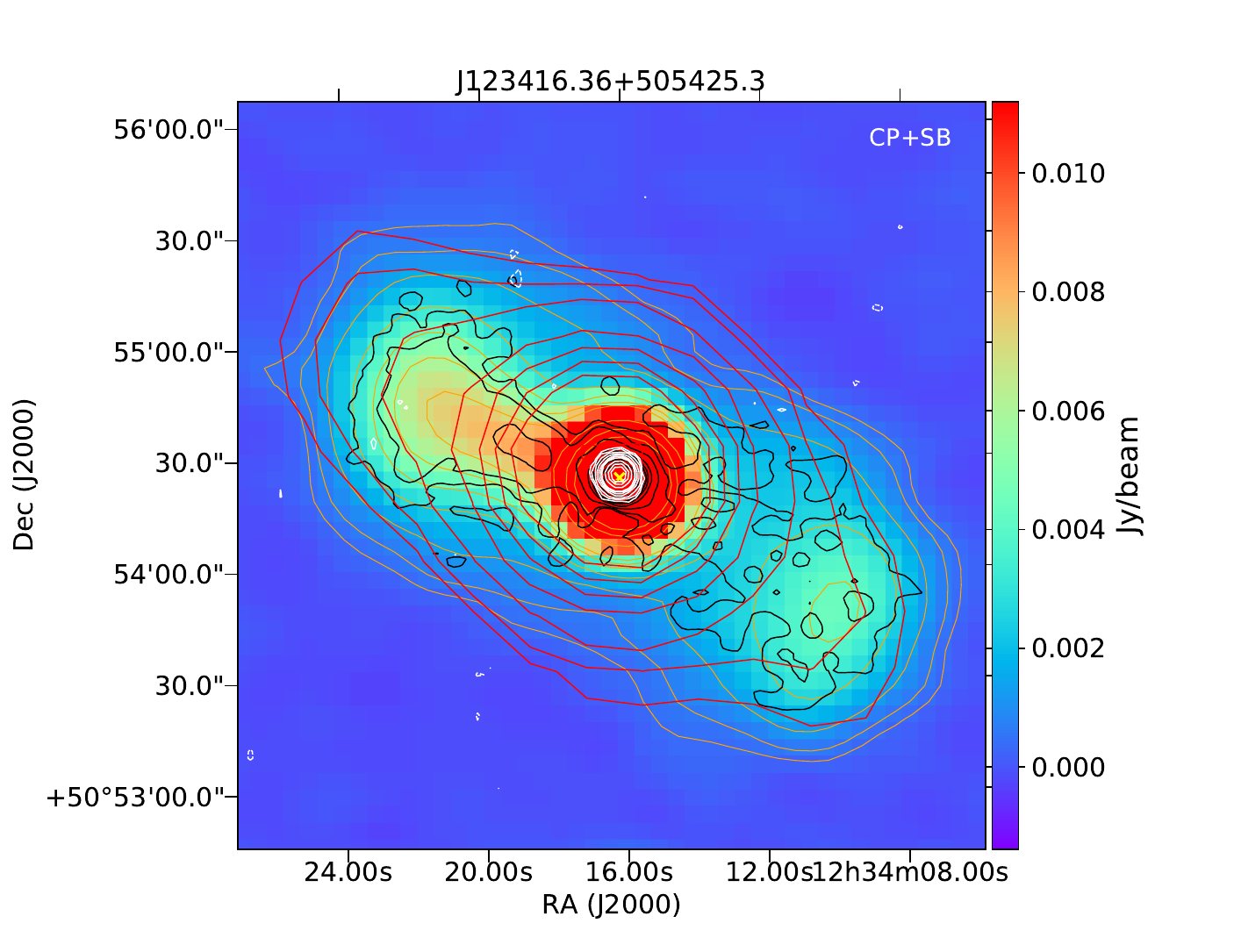}
            \includegraphics[width=0.3\textwidth]{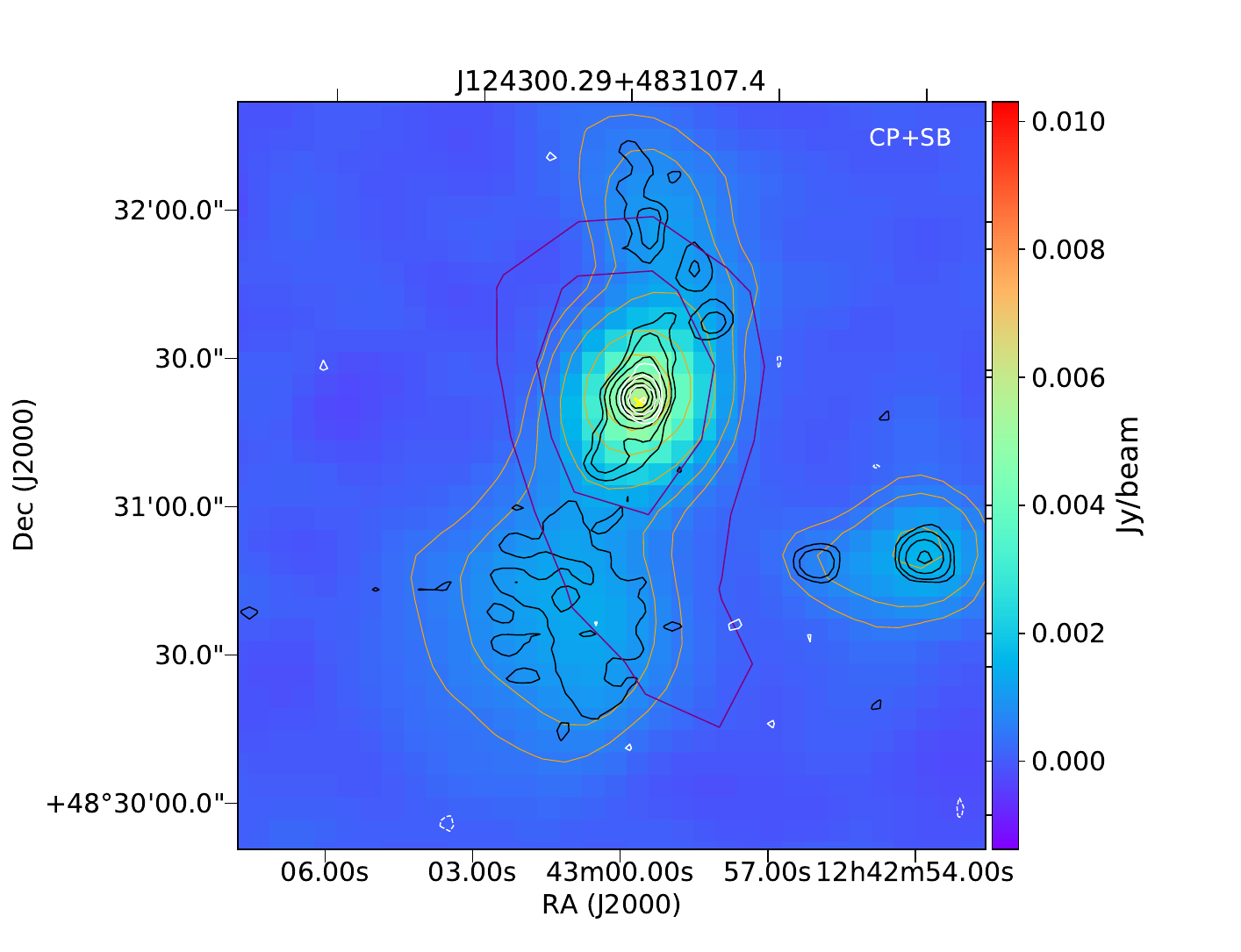}          
           }         
\centerline{\includegraphics[width=0.3\textwidth]{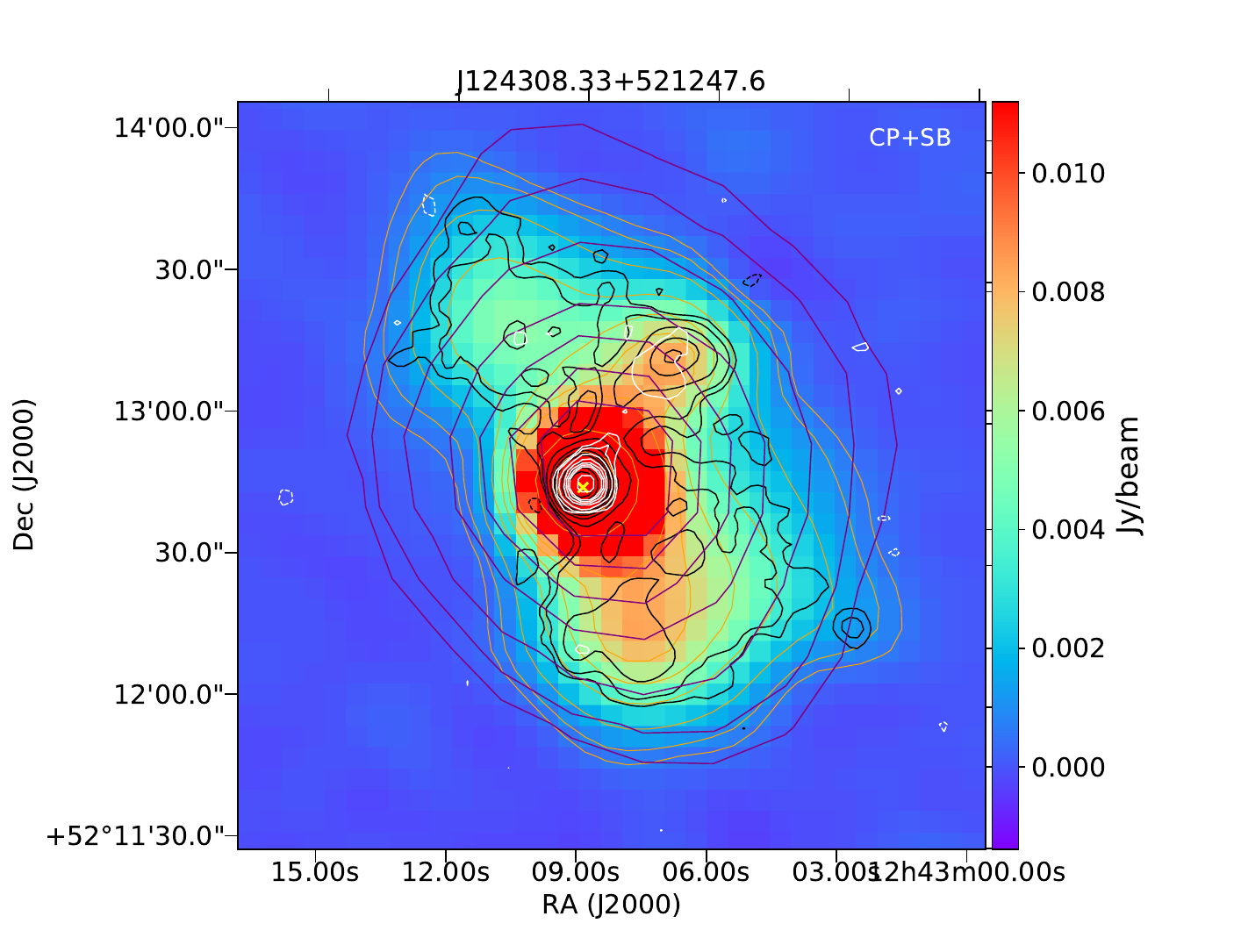}
            \includegraphics[width=0.3\textwidth]{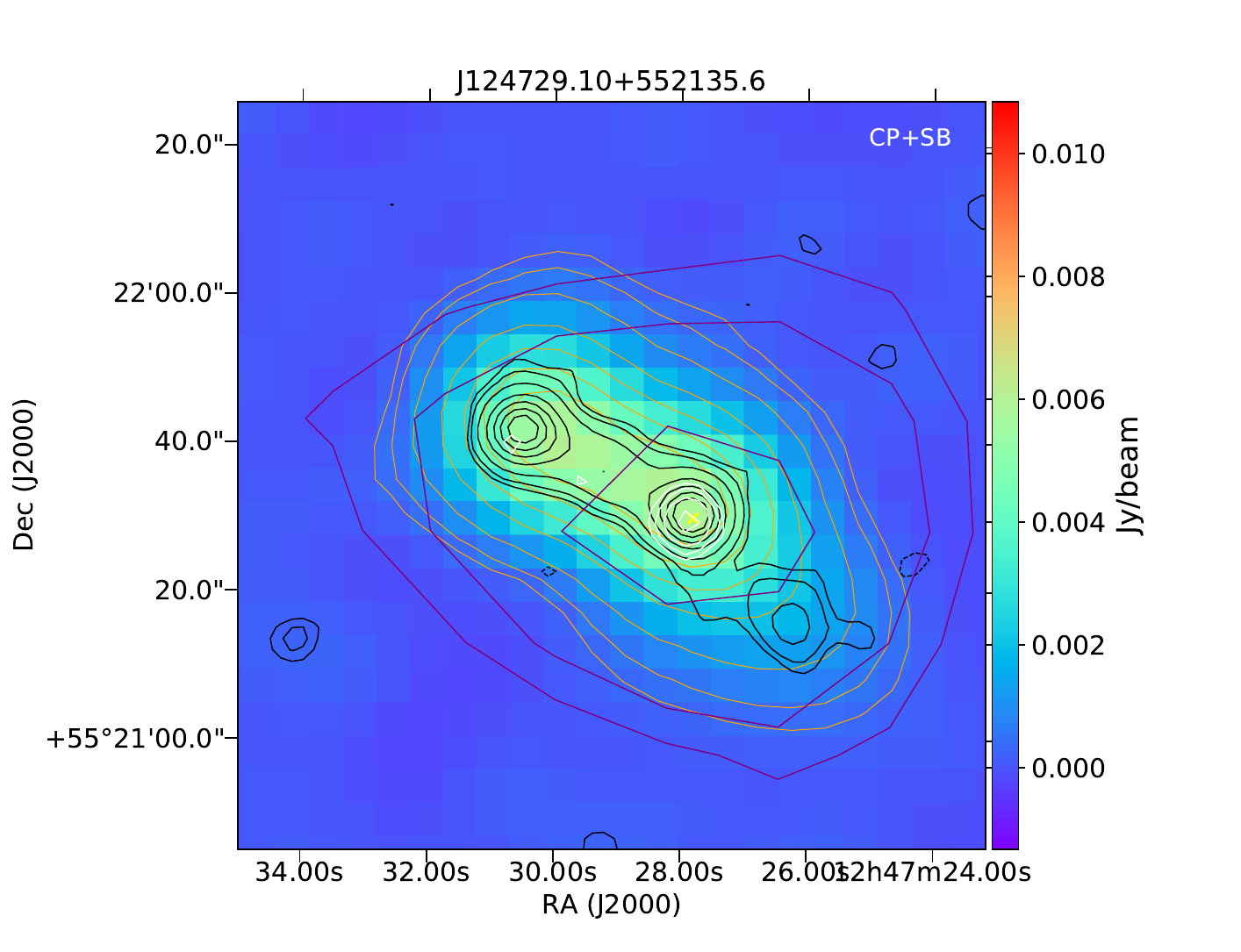}
            \includegraphics[width=0.3\textwidth]{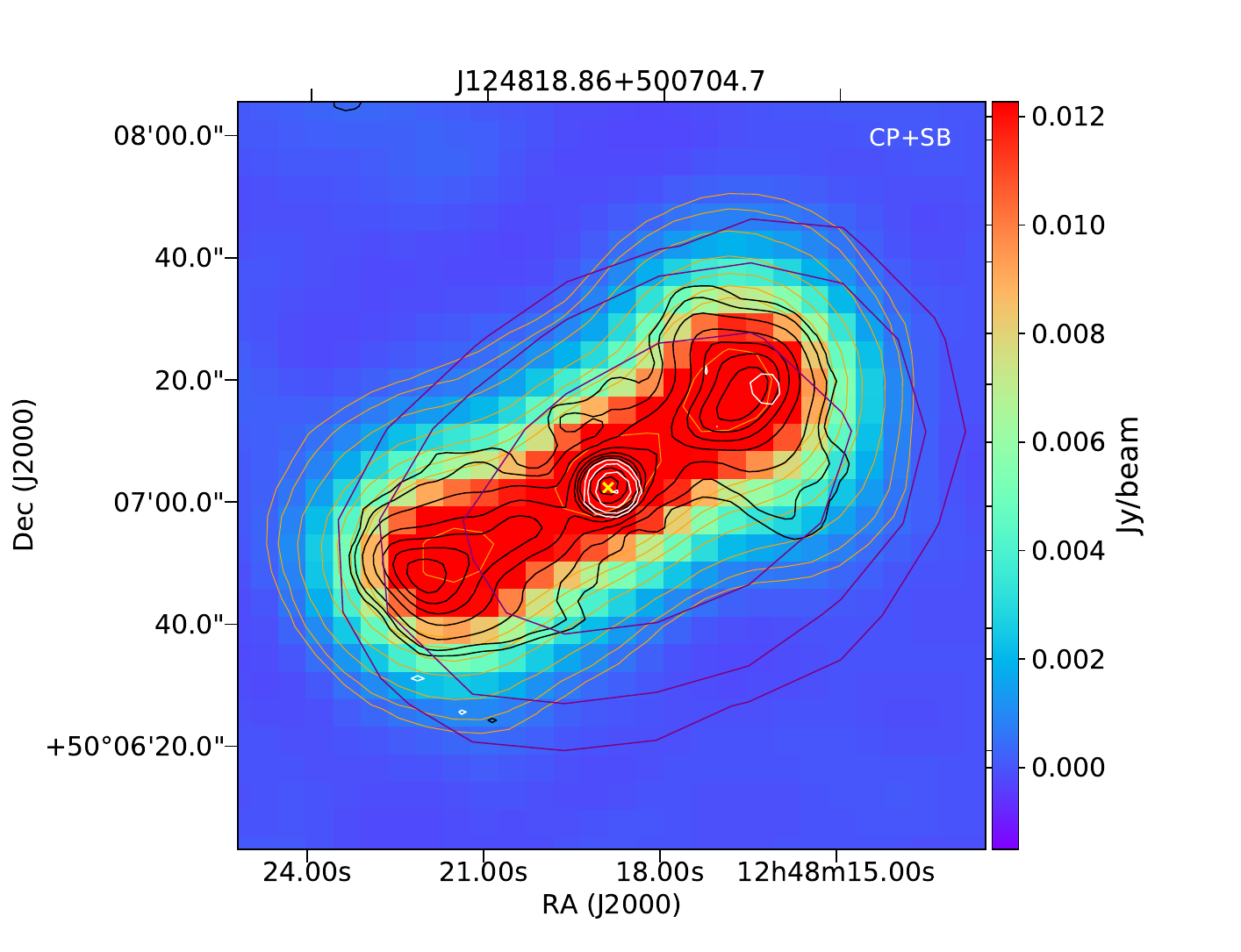}            
}
\centerline{\includegraphics[width=0.3\textwidth]{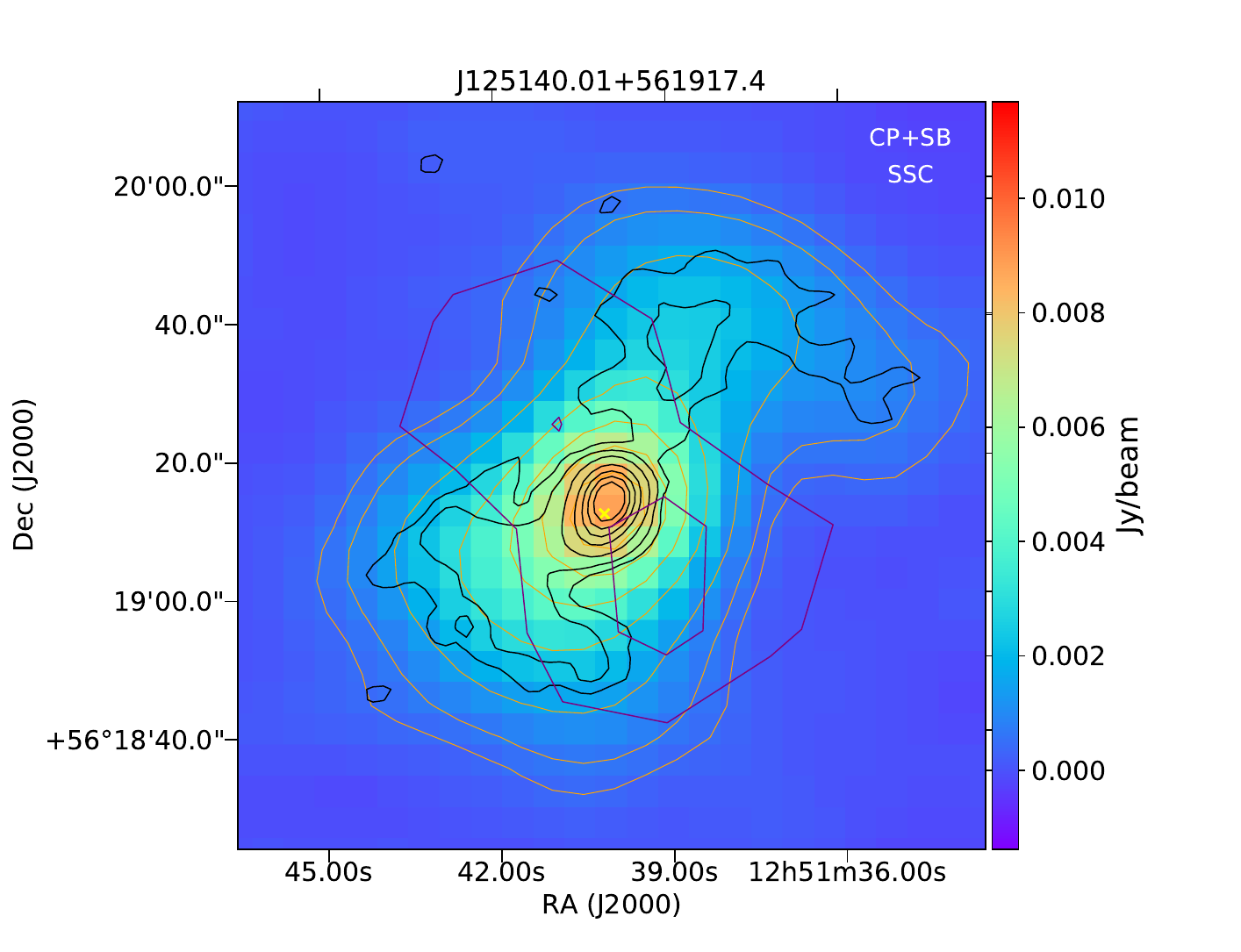}
            \includegraphics[width=0.3\textwidth]{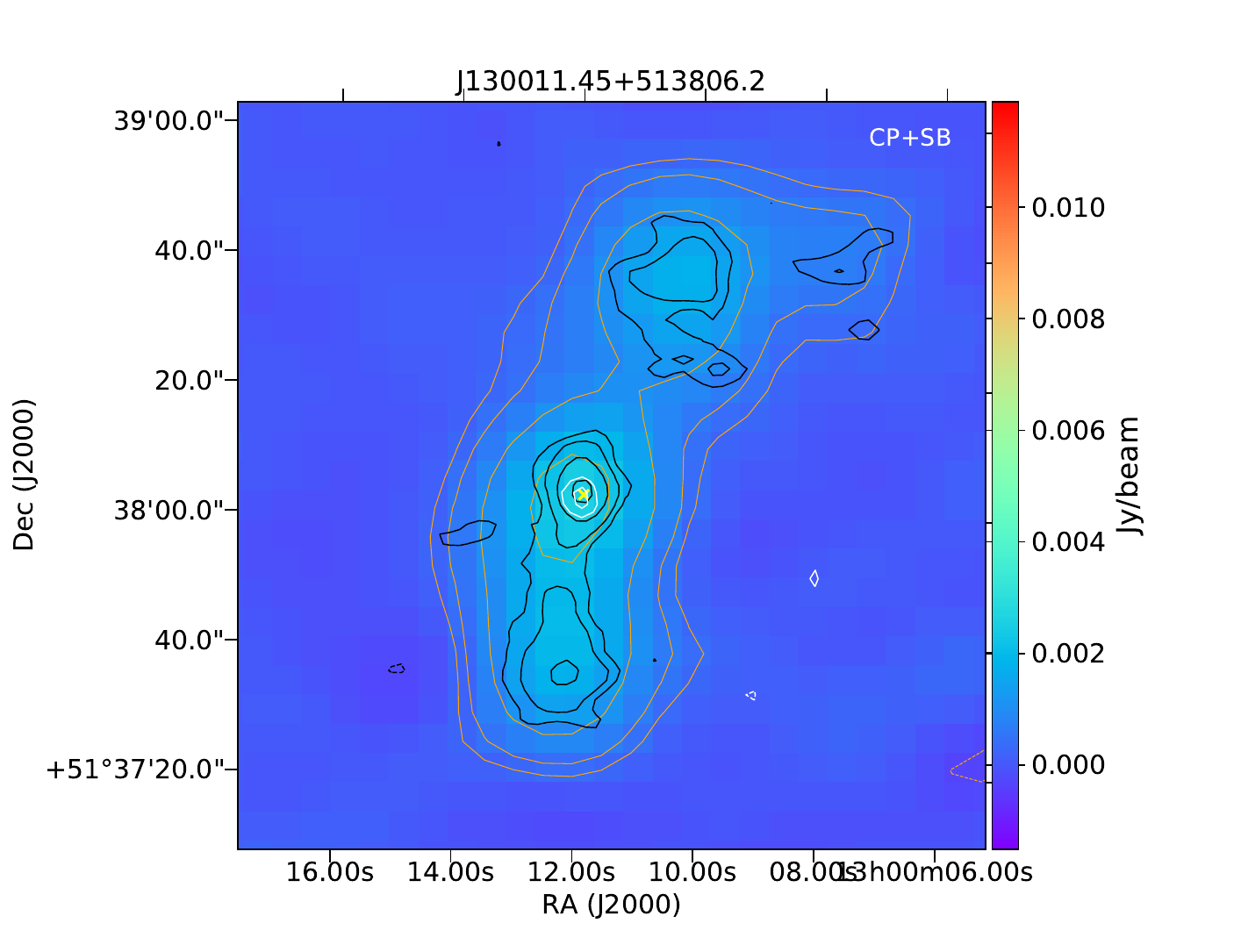}
            \includegraphics[width=0.3\textwidth]{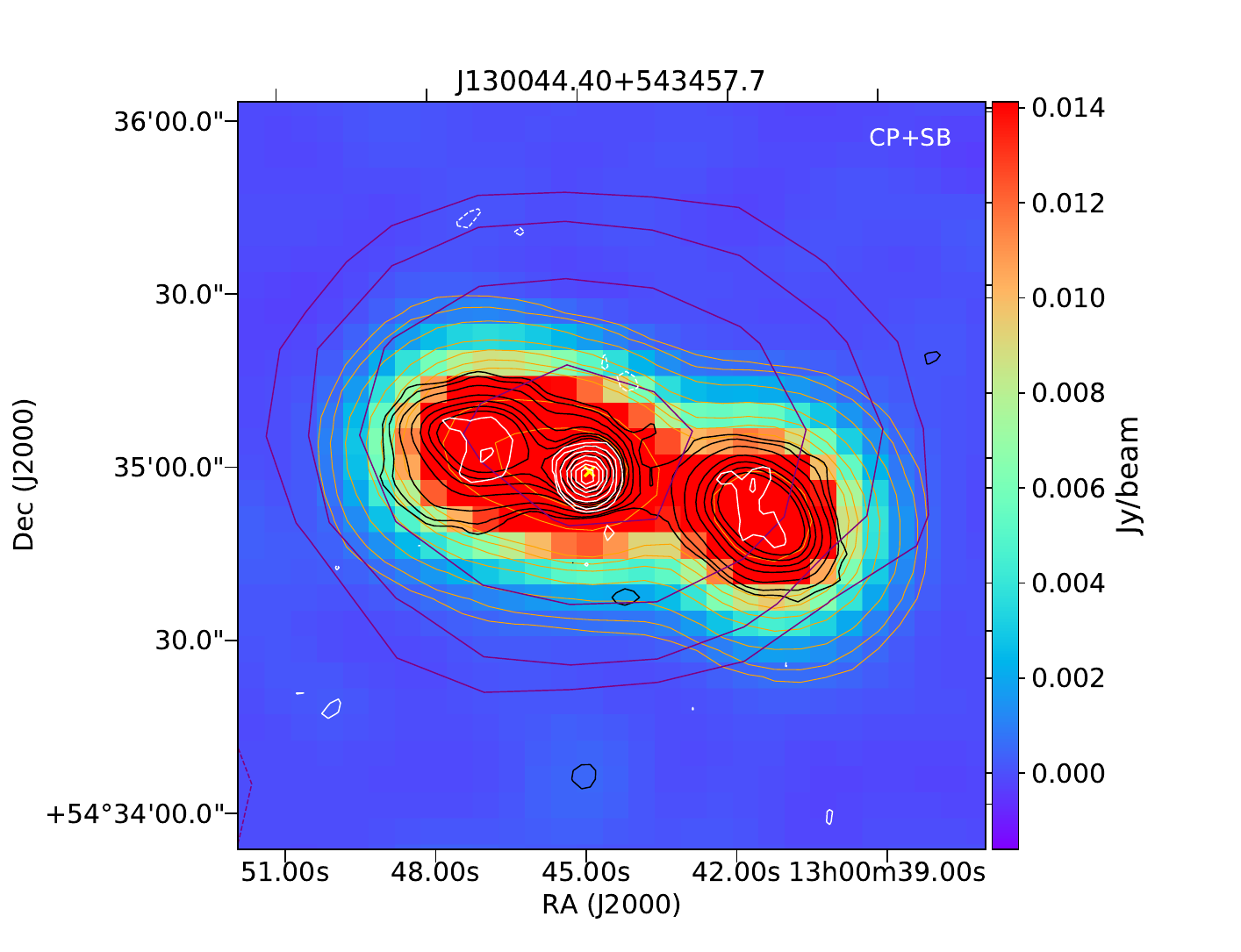}            
}
\centerline{\includegraphics[width=0.3\textwidth]{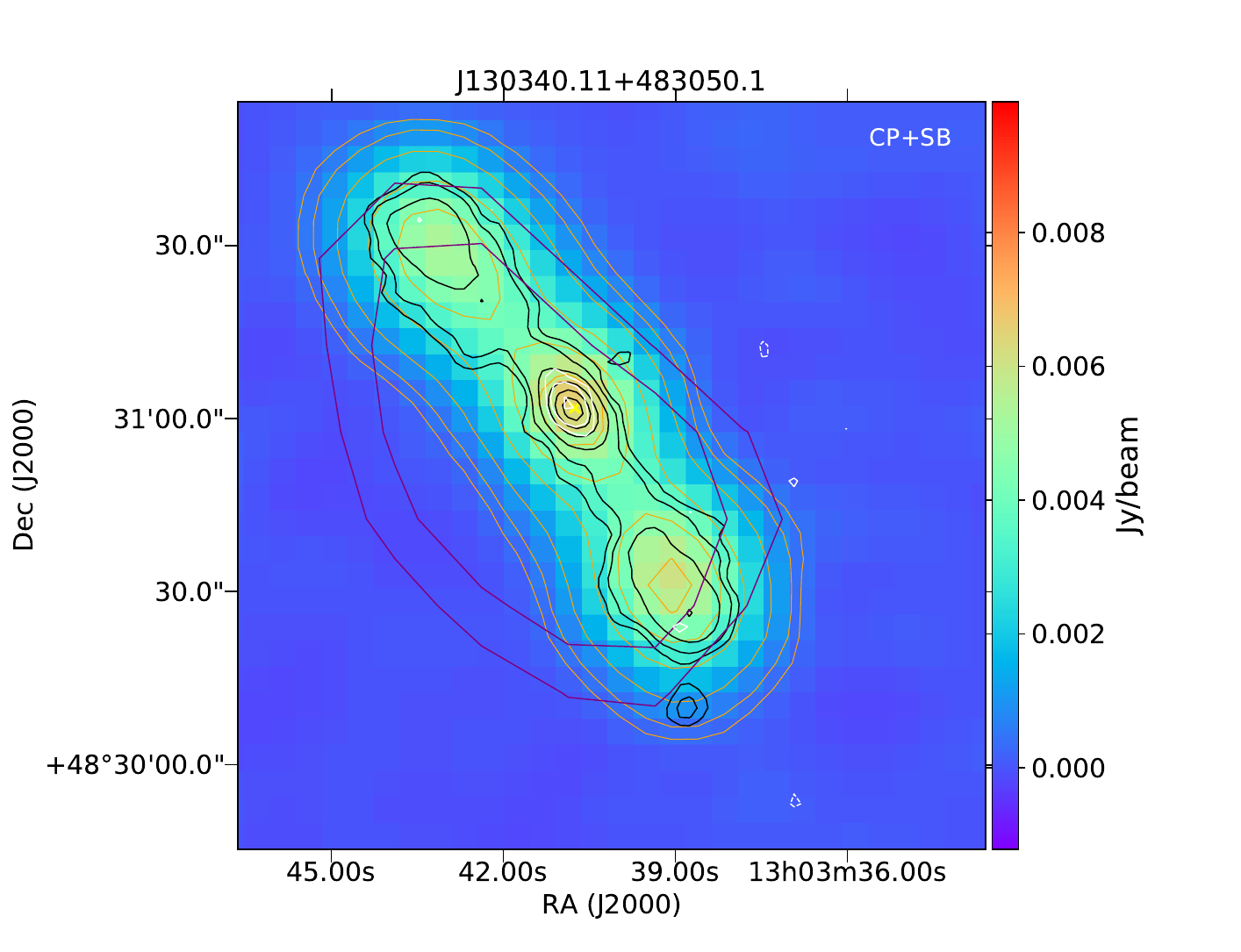}
            \includegraphics[width=0.3\textwidth]{Maps_final/tplot_66_3sig.pdf}
            \includegraphics[width=0.3\textwidth]{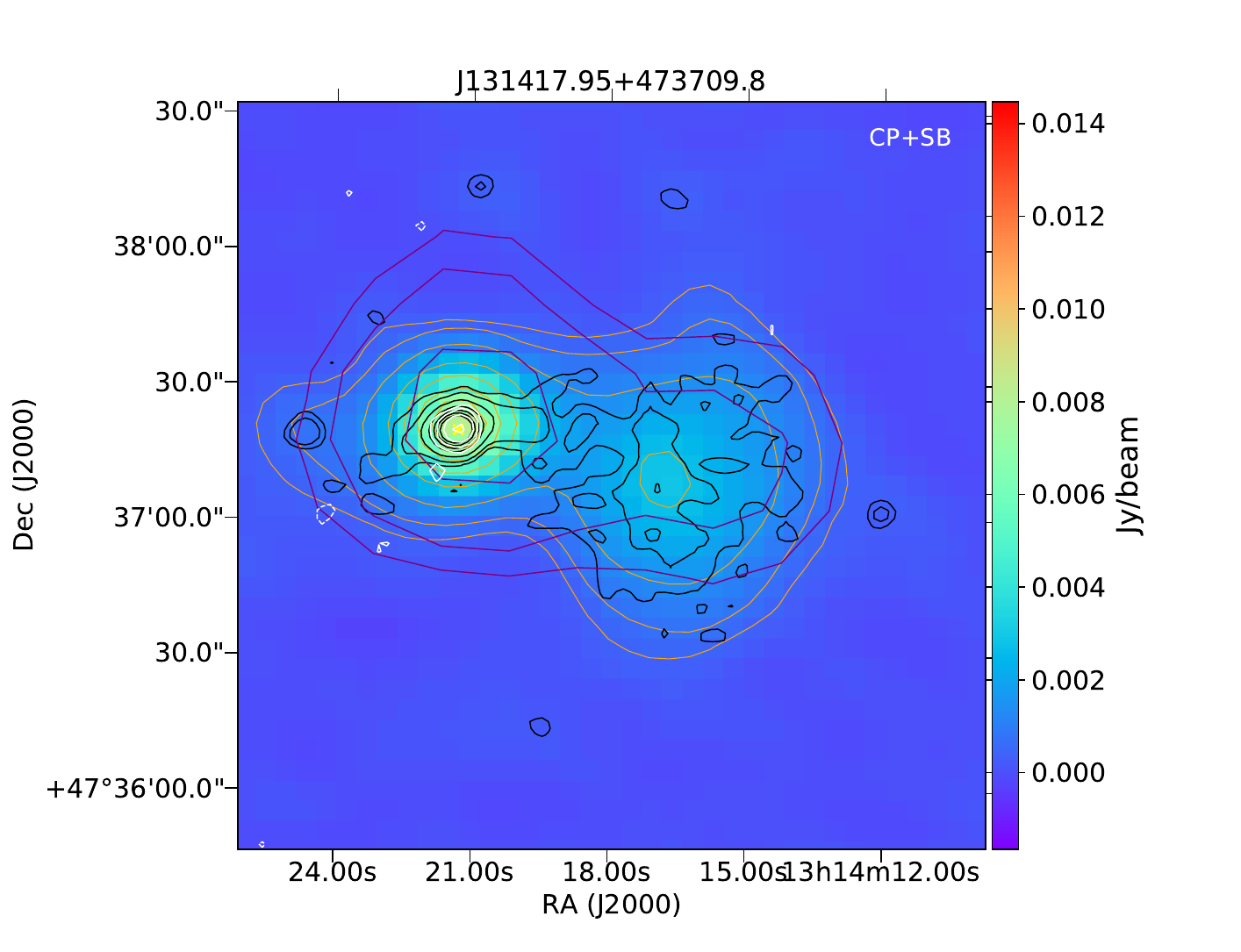}            
}
\caption{\footnotesize{Images of 69 candidate restarted galaxies selected based on
high radio $\mathrm{CP_{1.4GHz}}$ combined with low SB of extended emission, a steep spectrum of the core, and USS extended emission coupled with a bright core and summarised in Tables.~\ref{list of sources} and ~\ref{Core flux densities}. 
Radio contours from VLA FIRST maps (white, 5$\arcsec$), LOFAR high-resolution maps (black, 6$\arcsec$), and NVSS maps (purple, 45$\arcsec$) are overlaid on the LOFAR low-resolution resolution maps (orange, 20$\arcsec$).
The contouring of all the maps is made at $\,\sigma_\mathrm{local}\times(-3,3,5,10,20,30,40,50,100,150,200)$ levels, with $\sigma_\mathrm{local}$ representing the local RMS noise of the corresponding maps.
The host galaxy position is marked with a yellow cross. 
}}
\label{fig:3}
\end{figure*}

\begin{figure*}[ht!]
\centerline{\includegraphics[width=0.3\textwidth]{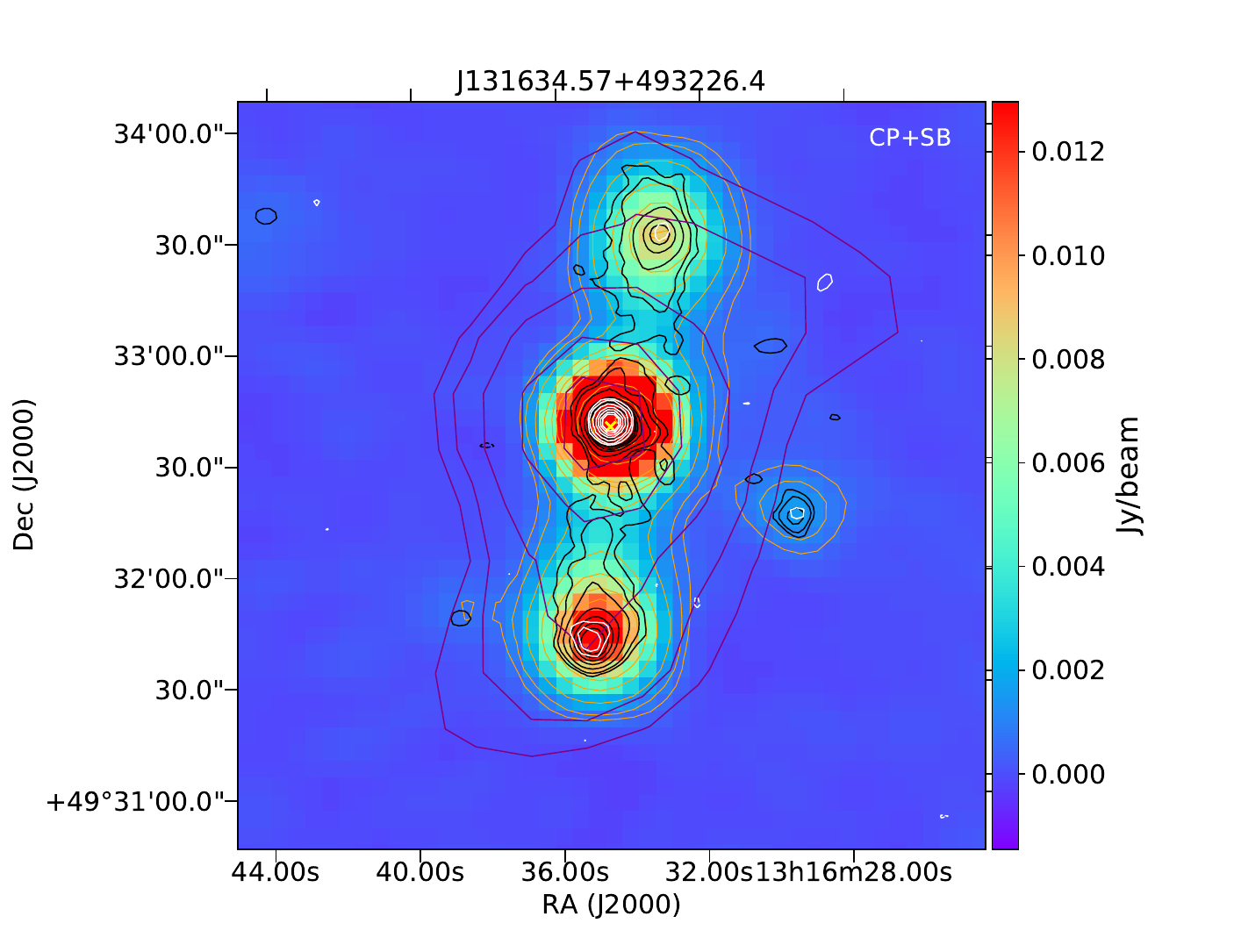}
            \includegraphics[width=0.3\textwidth]{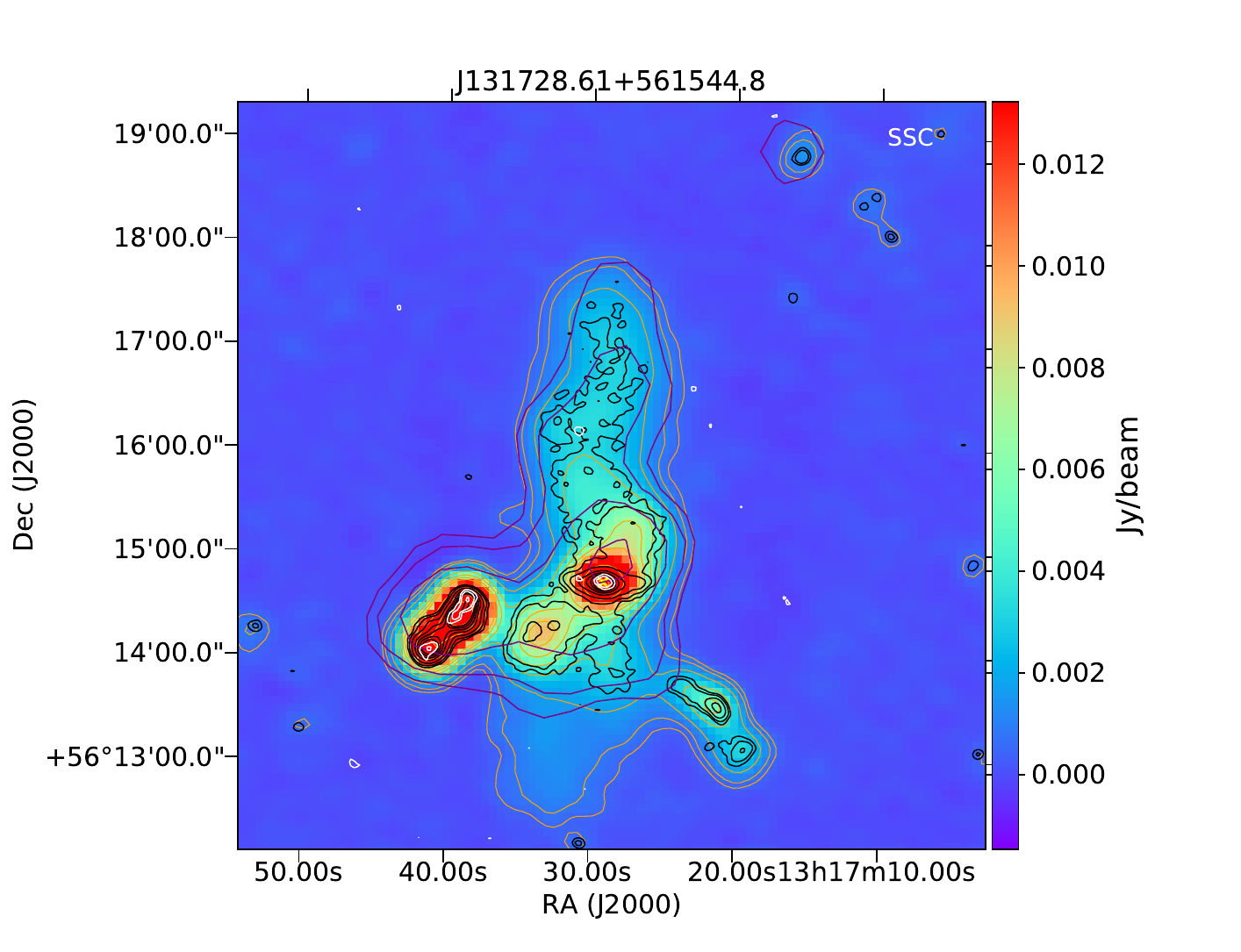}
            \includegraphics[width=0.3\textwidth]{Maps_final/tplot_72_3sig.pdf}
            }        
\centerline{\includegraphics[width=0.3\textwidth]{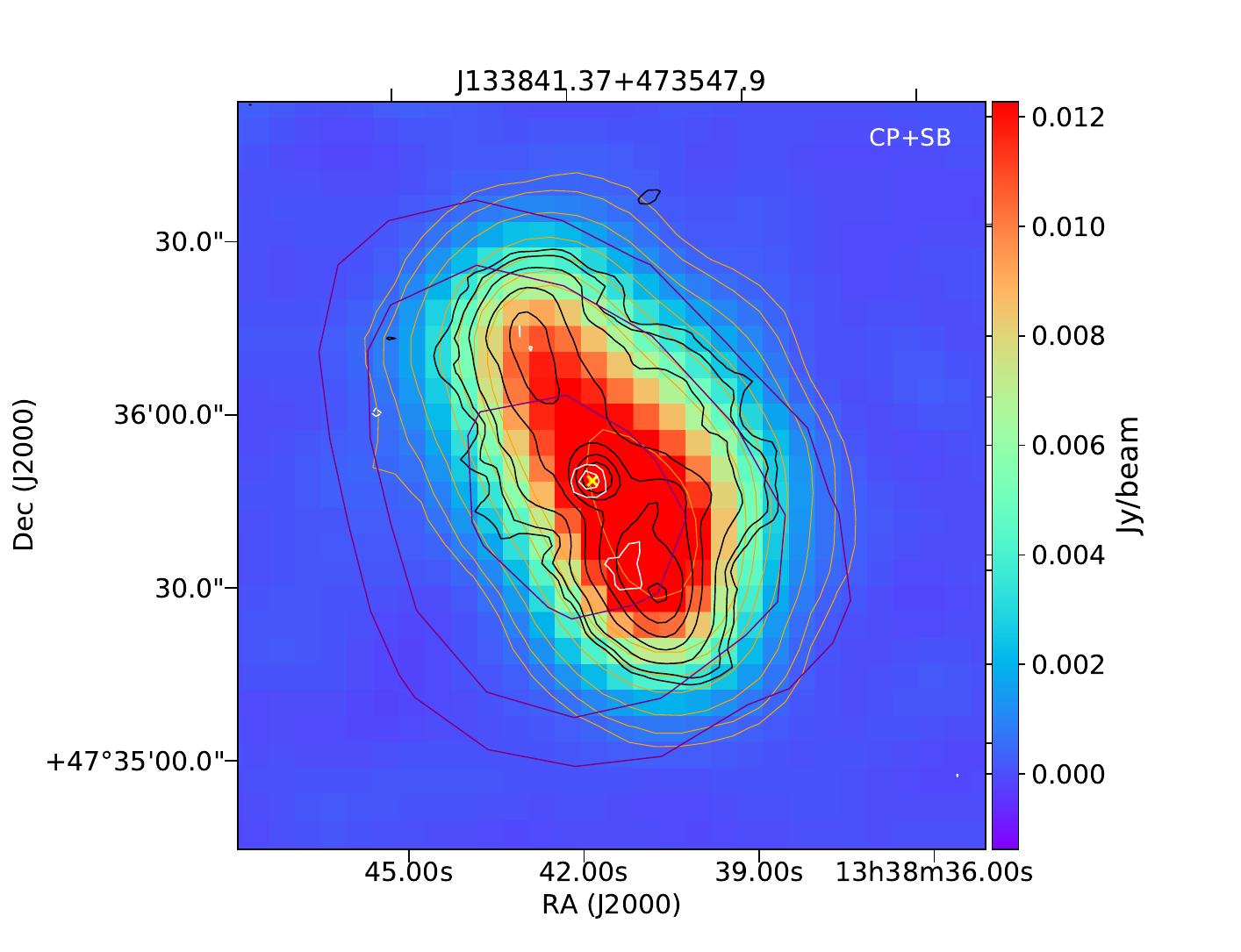}
            \includegraphics[width=0.3\textwidth]{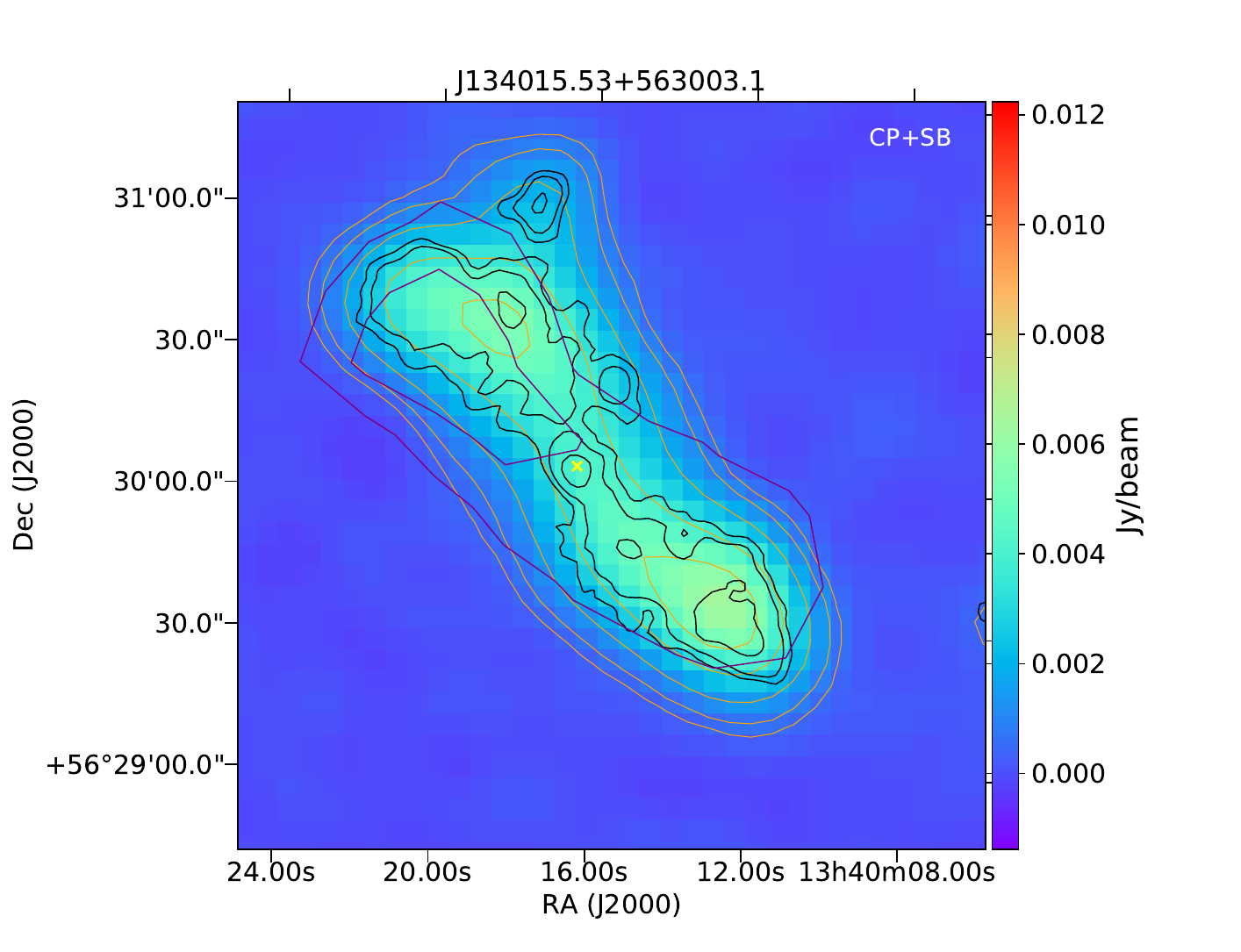}
            \includegraphics[width=0.3\textwidth]{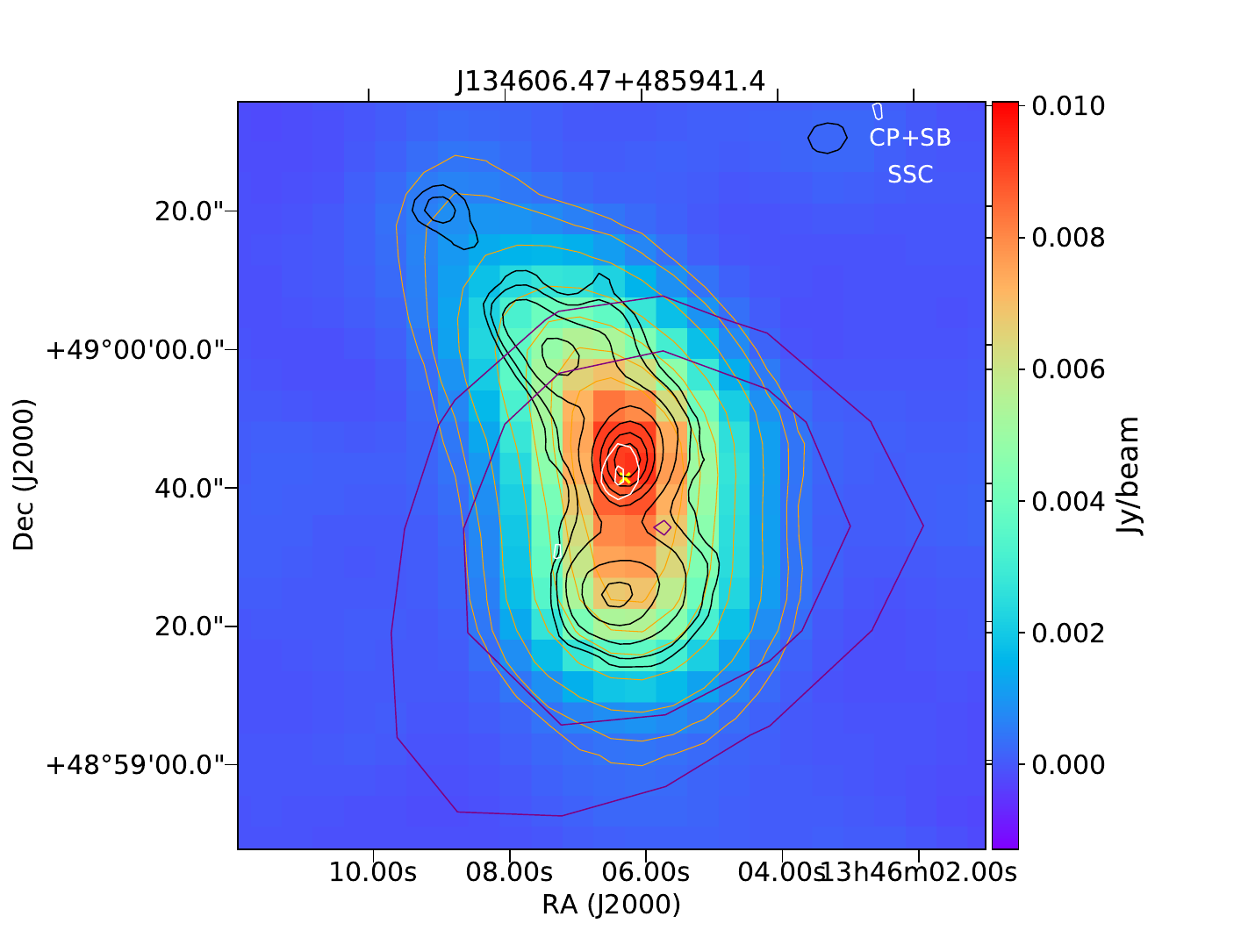}          
           }         
\centerline{\includegraphics[width=0.3\textwidth]{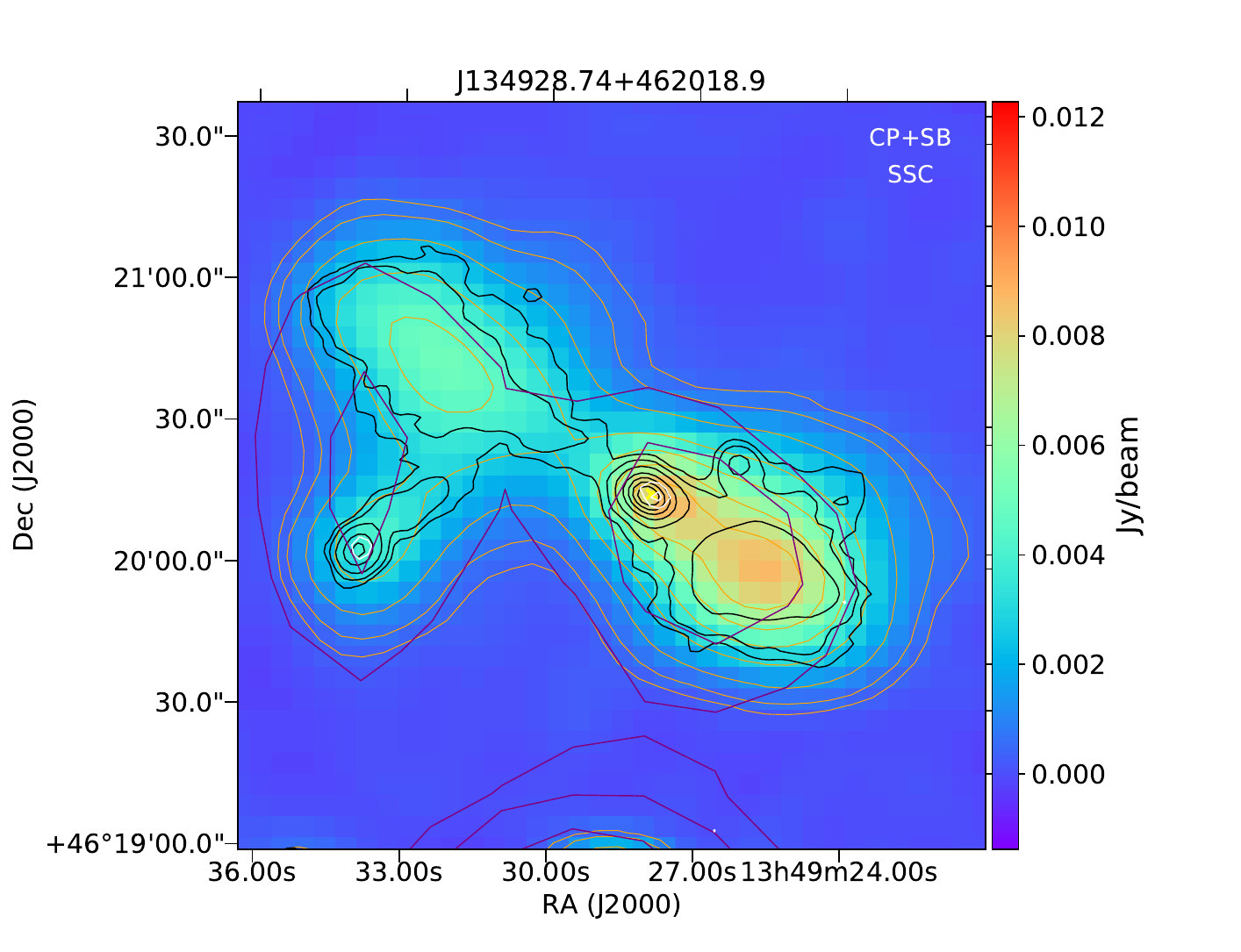}
            \includegraphics[width=0.3\textwidth]{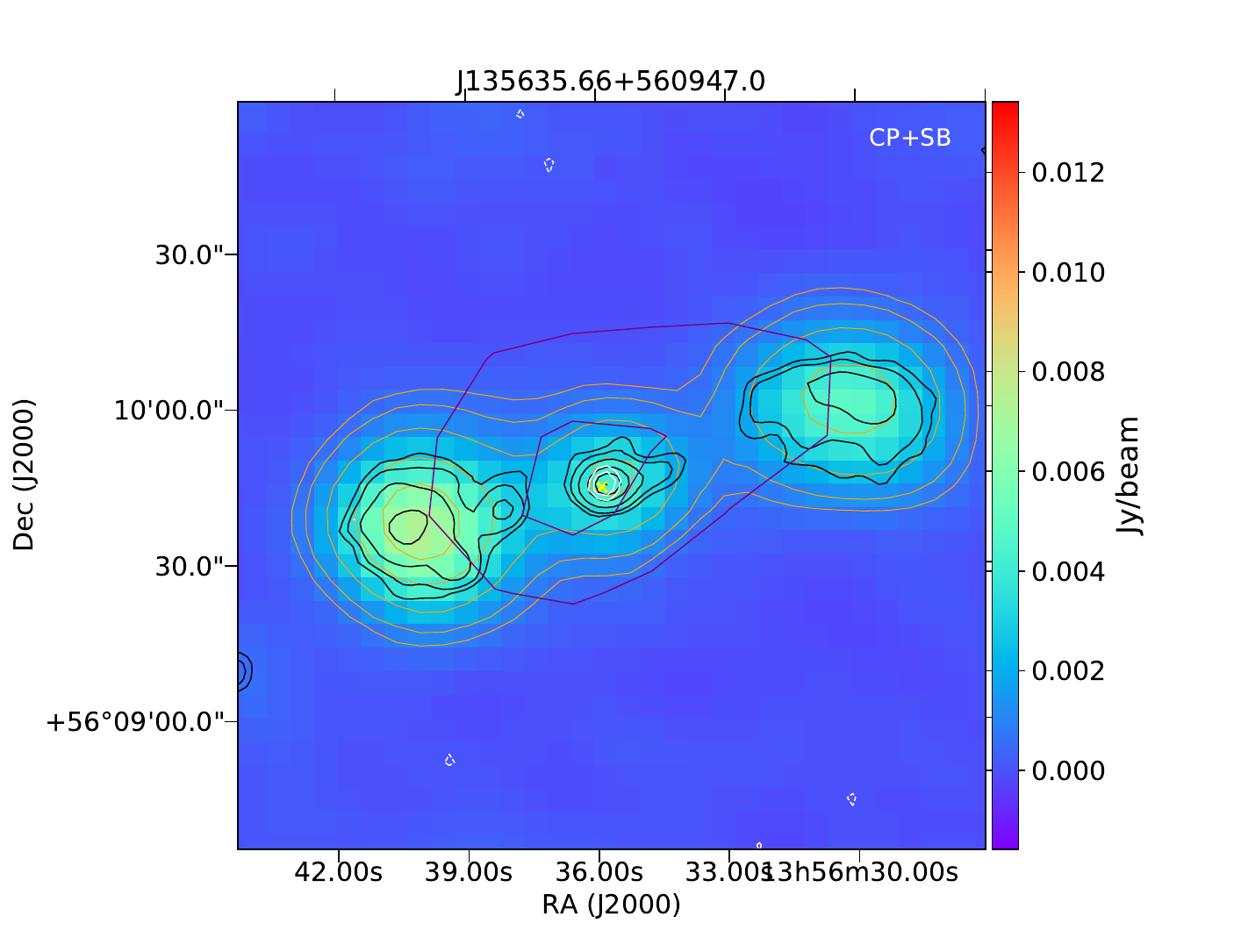}
            \includegraphics[width=0.3\textwidth]{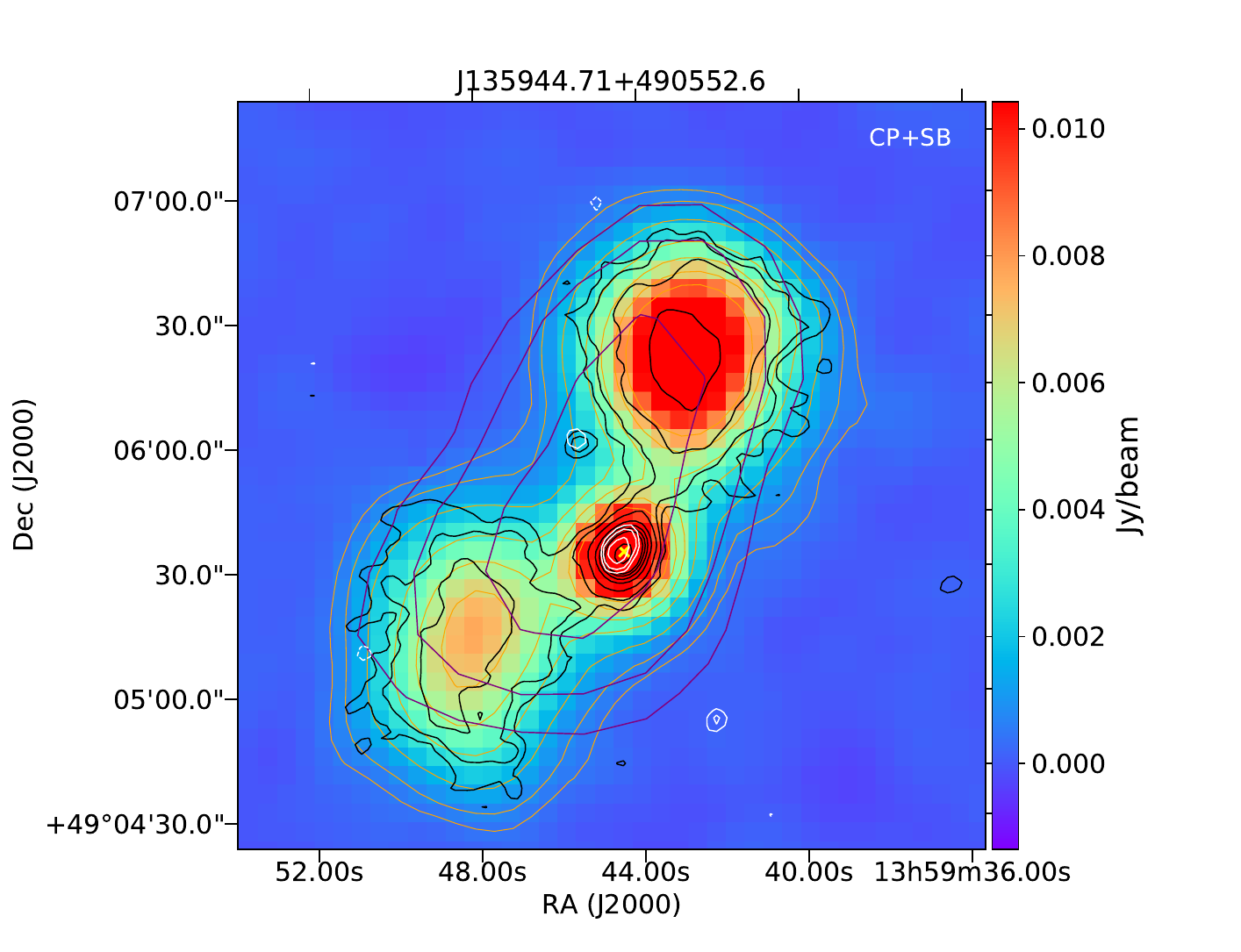}            
}
\centerline{\includegraphics[width=0.3\textwidth]{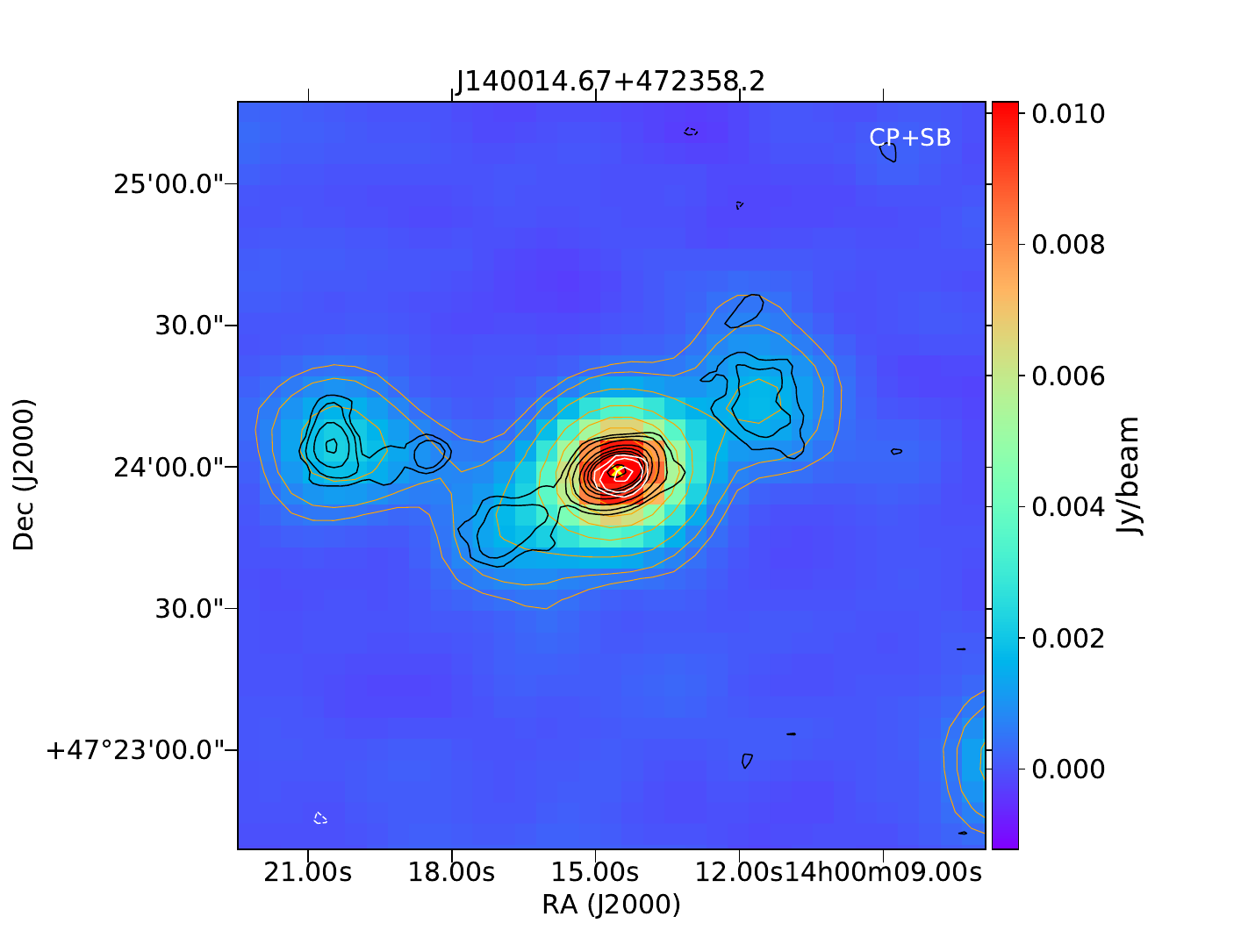}
            \includegraphics[width=0.3\textwidth]{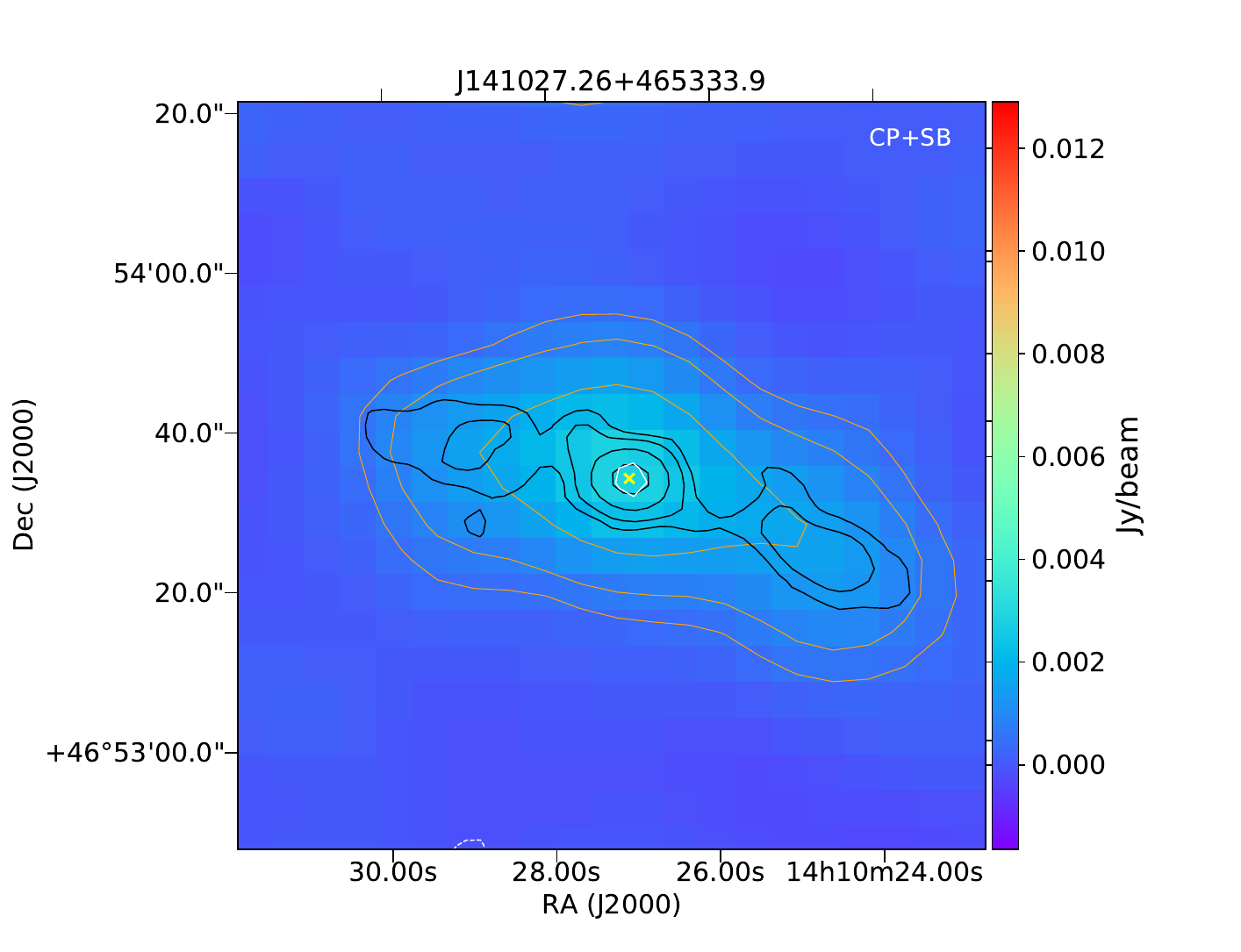}
            \includegraphics[width=0.3\textwidth]{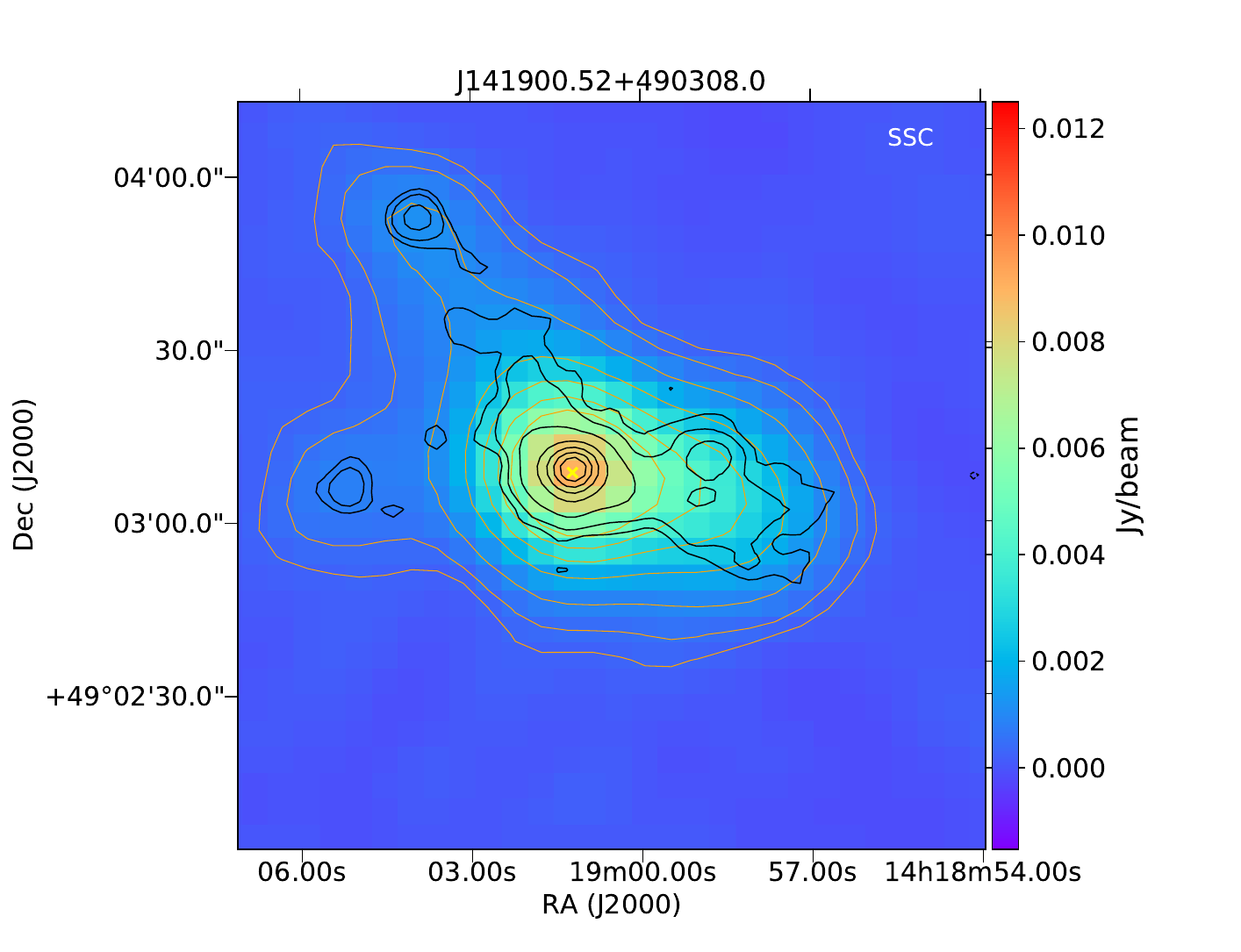}            
}
\centerline{\includegraphics[width=0.3\textwidth]{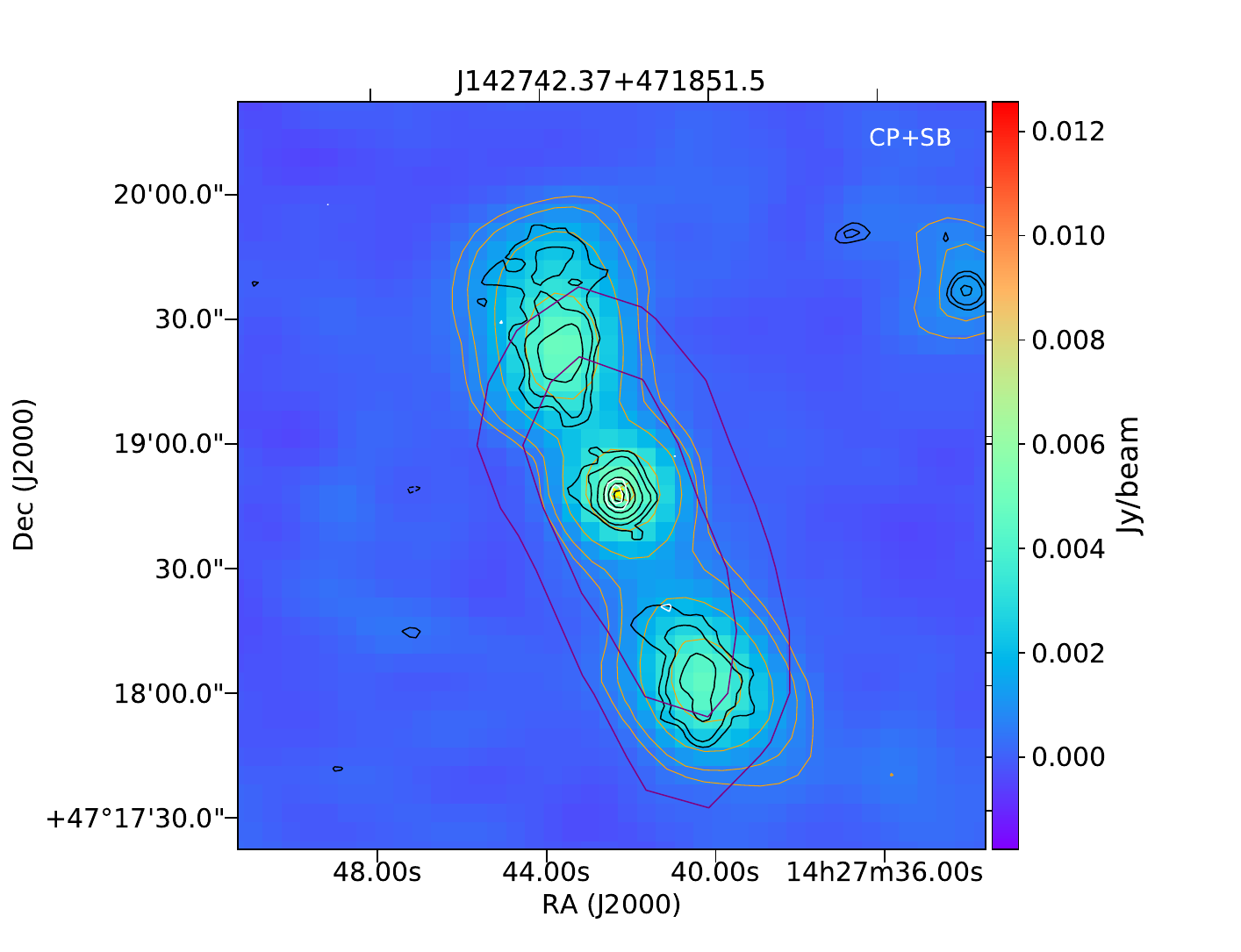}
            \includegraphics[width=0.3\textwidth]{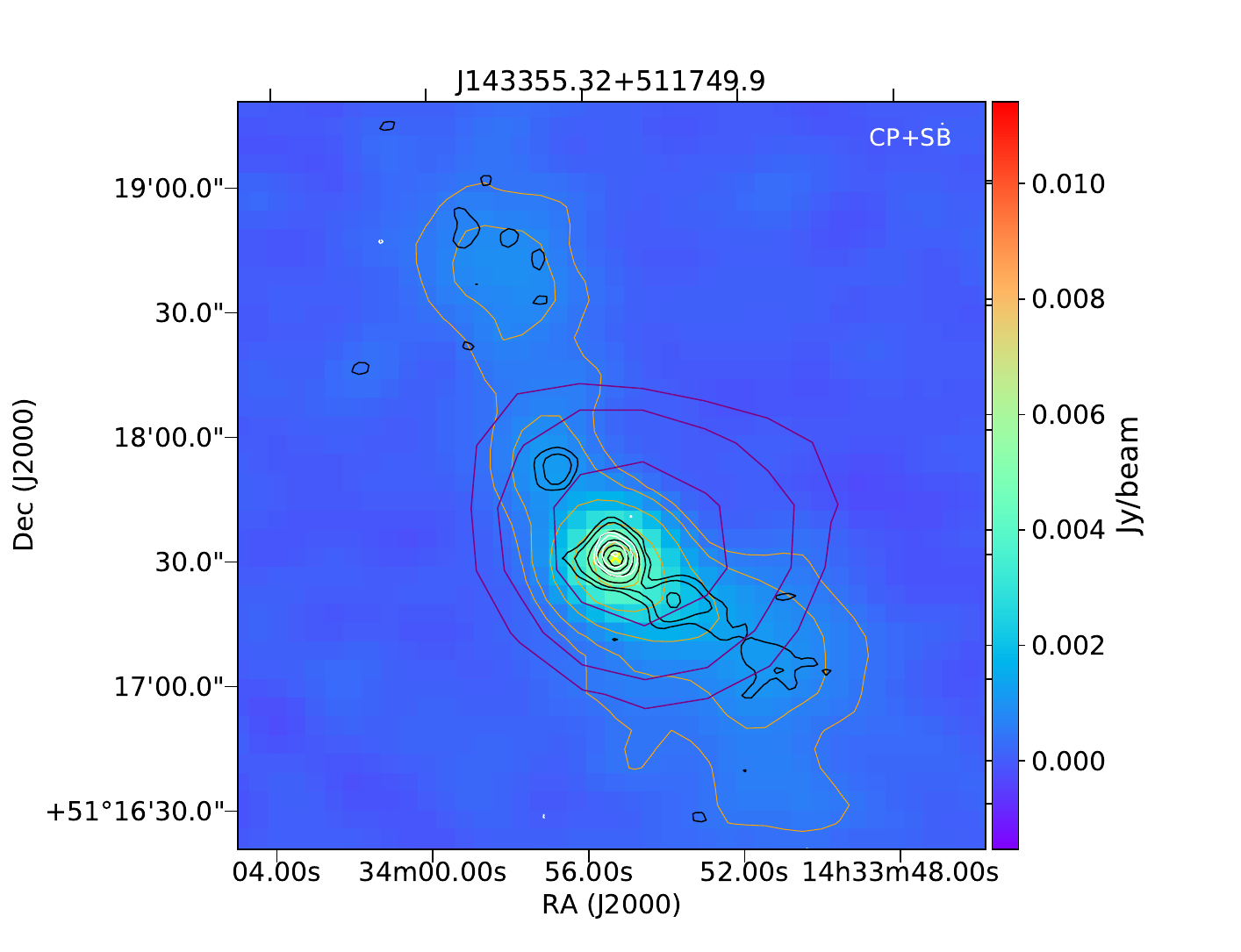}
            \includegraphics[width=0.3\textwidth]{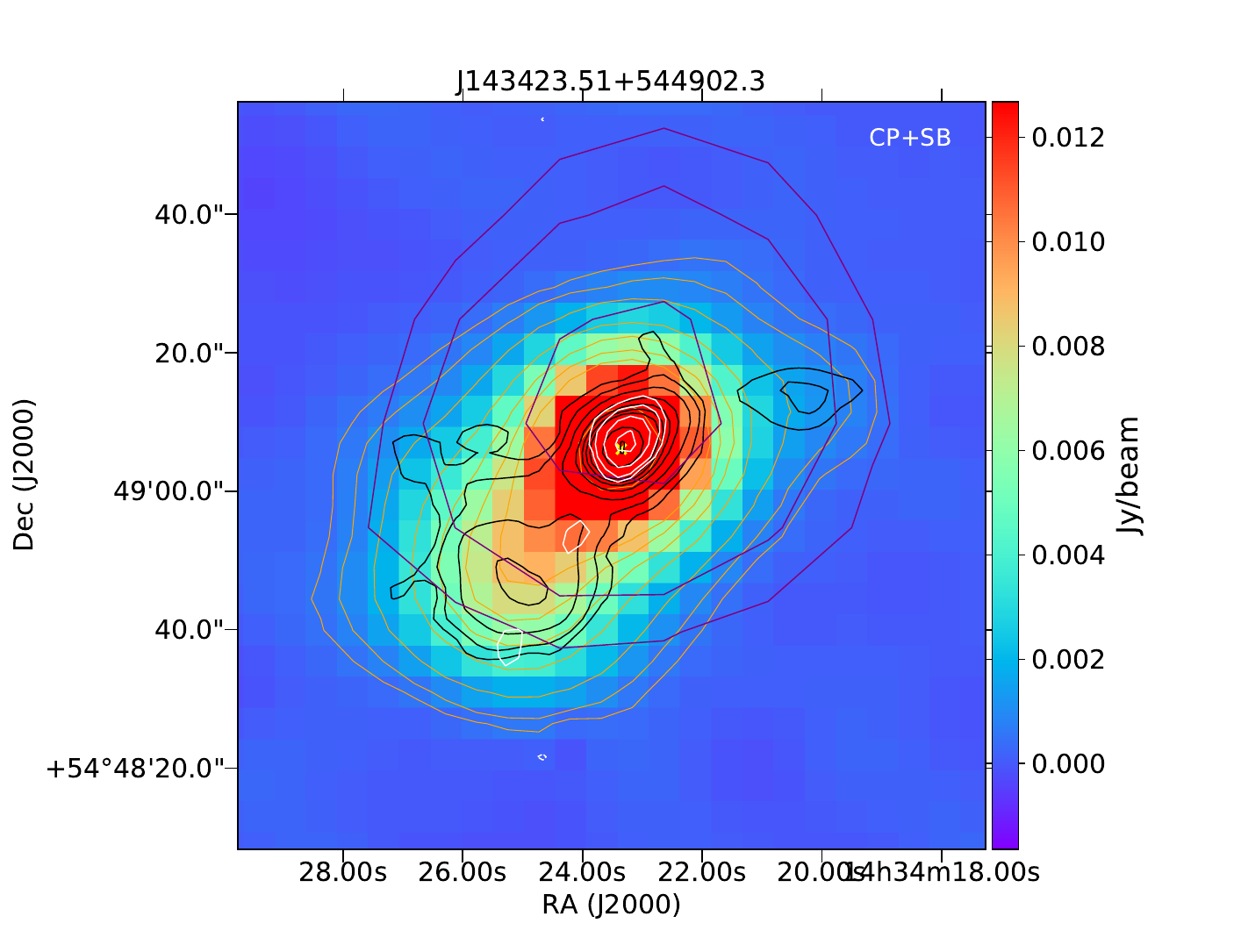}            
}
\caption{\footnotesize{Images of 69 candidate restarted galaxies selected based on
high radio $\mathrm{CP_{1.4GHz}}$ combined with low SB of extended emission, a steep spectrum of the core, and USS extended emission coupled with a bright core and summarised in Tables.~\ref{list of sources} and ~\ref{Core flux densities}. 
Radio contours from VLA FIRST maps (white, 5$\arcsec$), LOFAR high-resolution maps (black, 6$\arcsec$), and NVSS maps (purple, 45$\arcsec$) are overlaid on the LOFAR low-resolution resolution maps (orange, 20$\arcsec$).
The contouring of all the maps is made at $\,\sigma_\mathrm{local}\times(-3,3,5,10,20,30,40,50,100,150,200)$ levels, with $\sigma_\mathrm{local}$ representing the local RMS noise of the corresponding maps.
The host galaxy position is marked with a yellow cross.
}}
\label{fig:4}
\end{figure*}

\begin{figure*}[ht!]
\centerline{\includegraphics[width=0.3\textwidth]{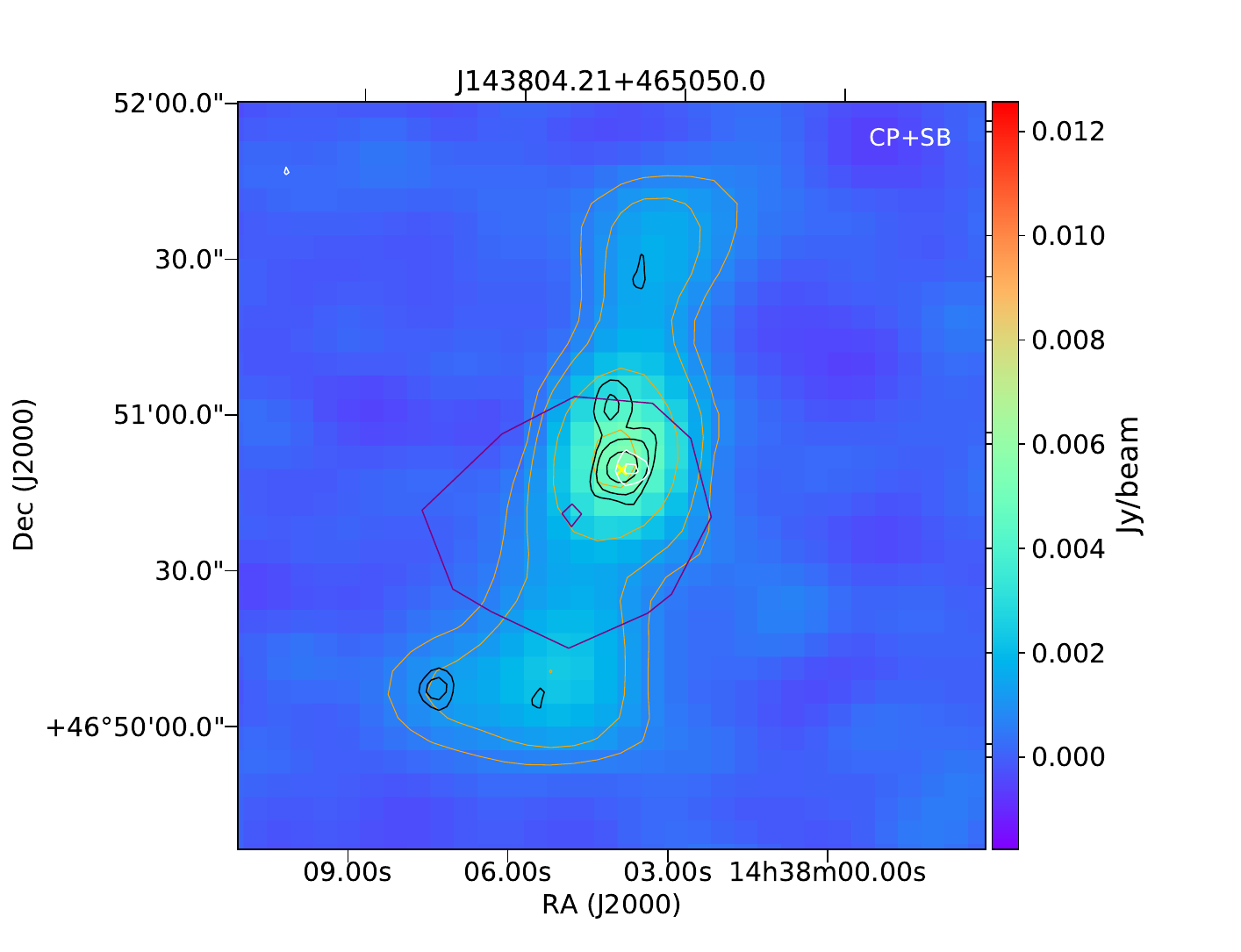}
            \includegraphics[width=0.3\textwidth]{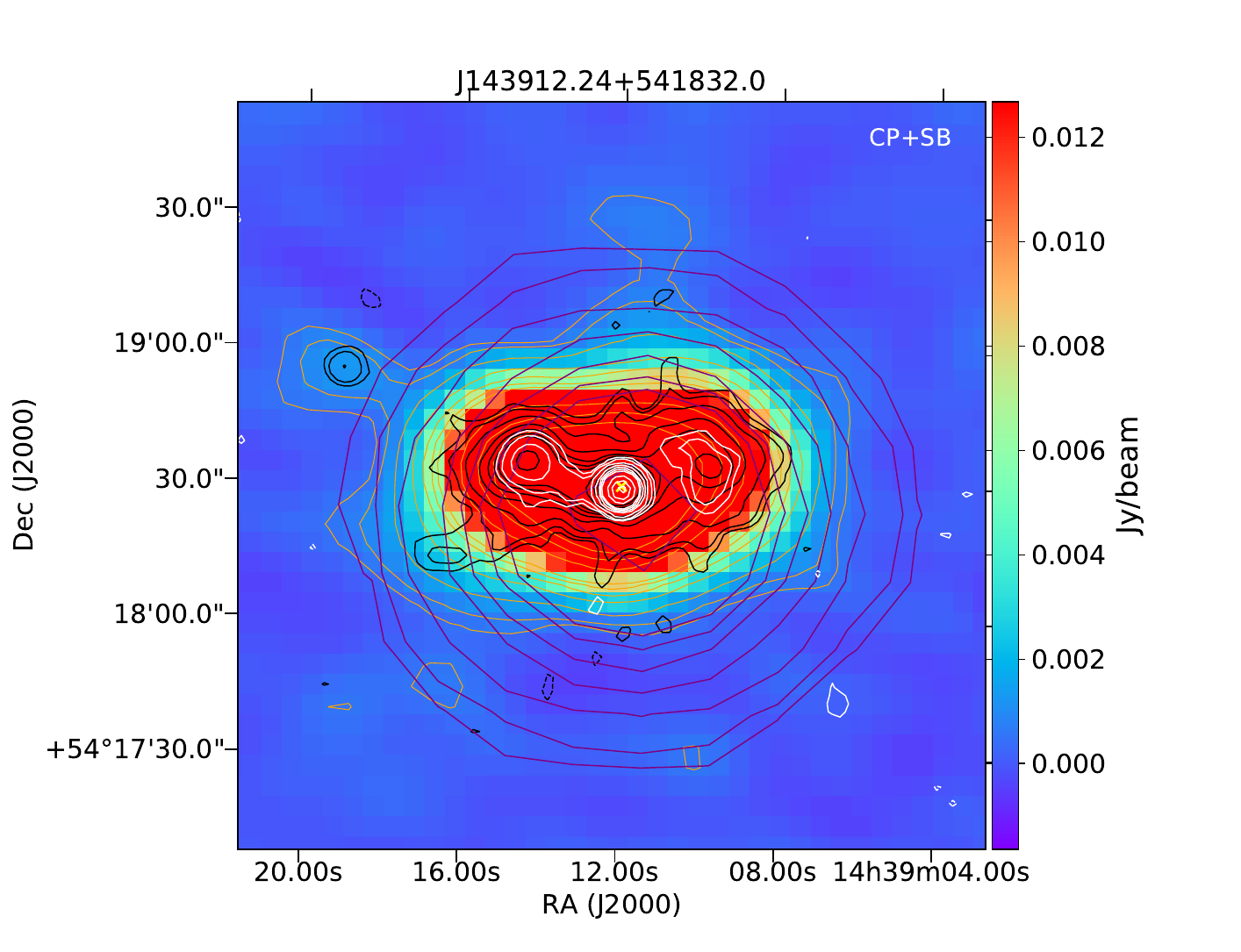}
            \includegraphics[width=0.3\textwidth]{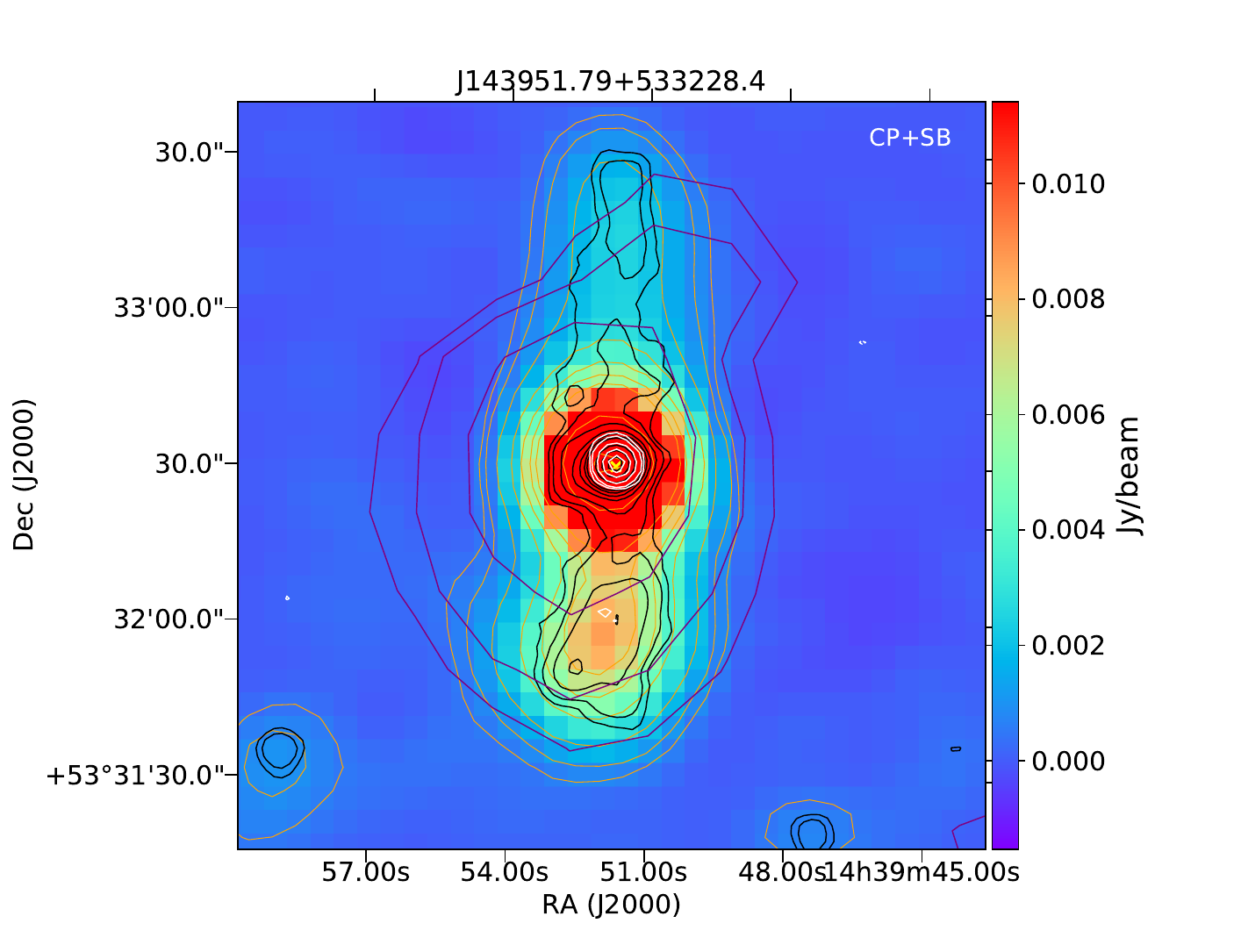}                   }        
\centerline{\includegraphics[width=0.3\textwidth]{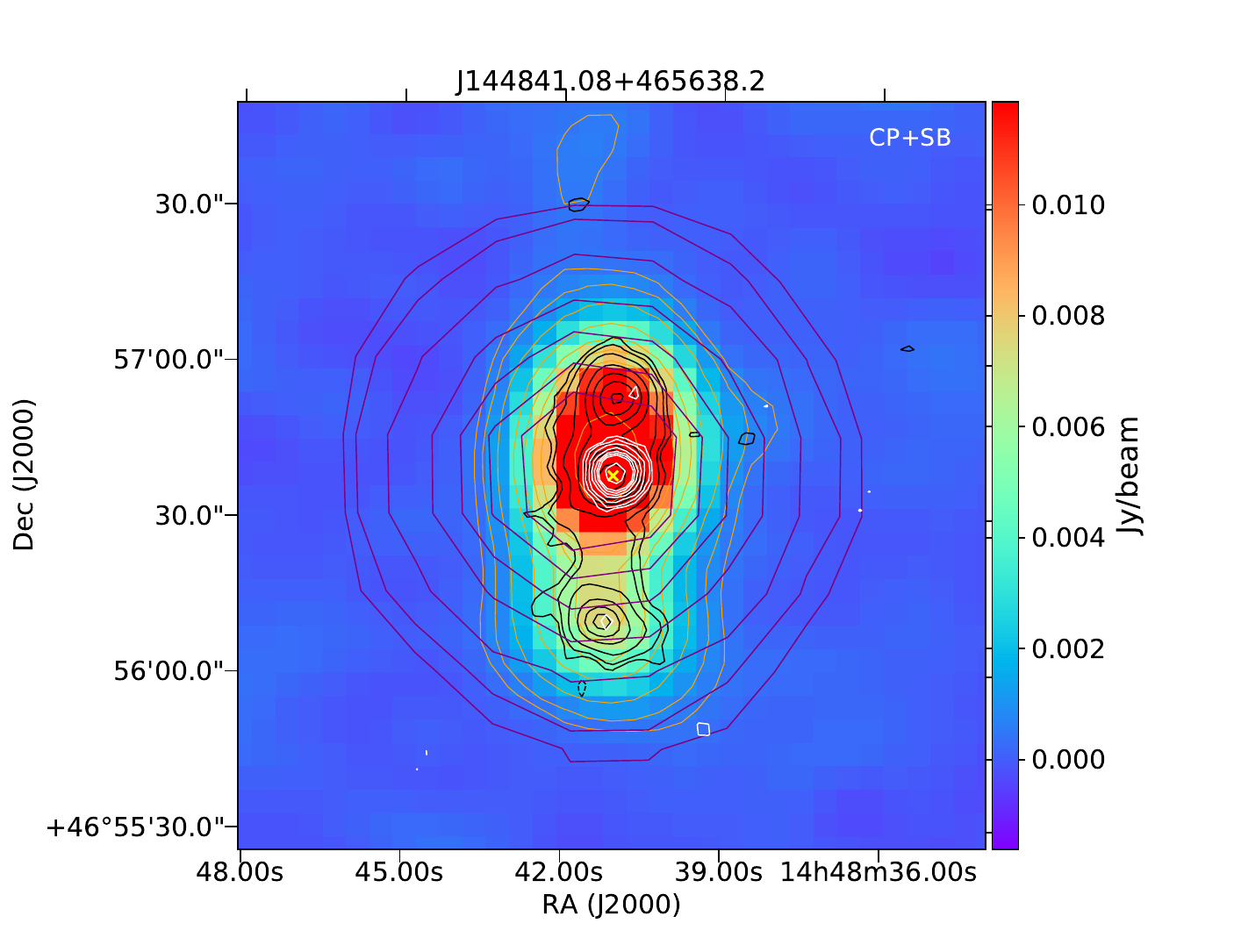}
            \includegraphics[width=0.3\textwidth]{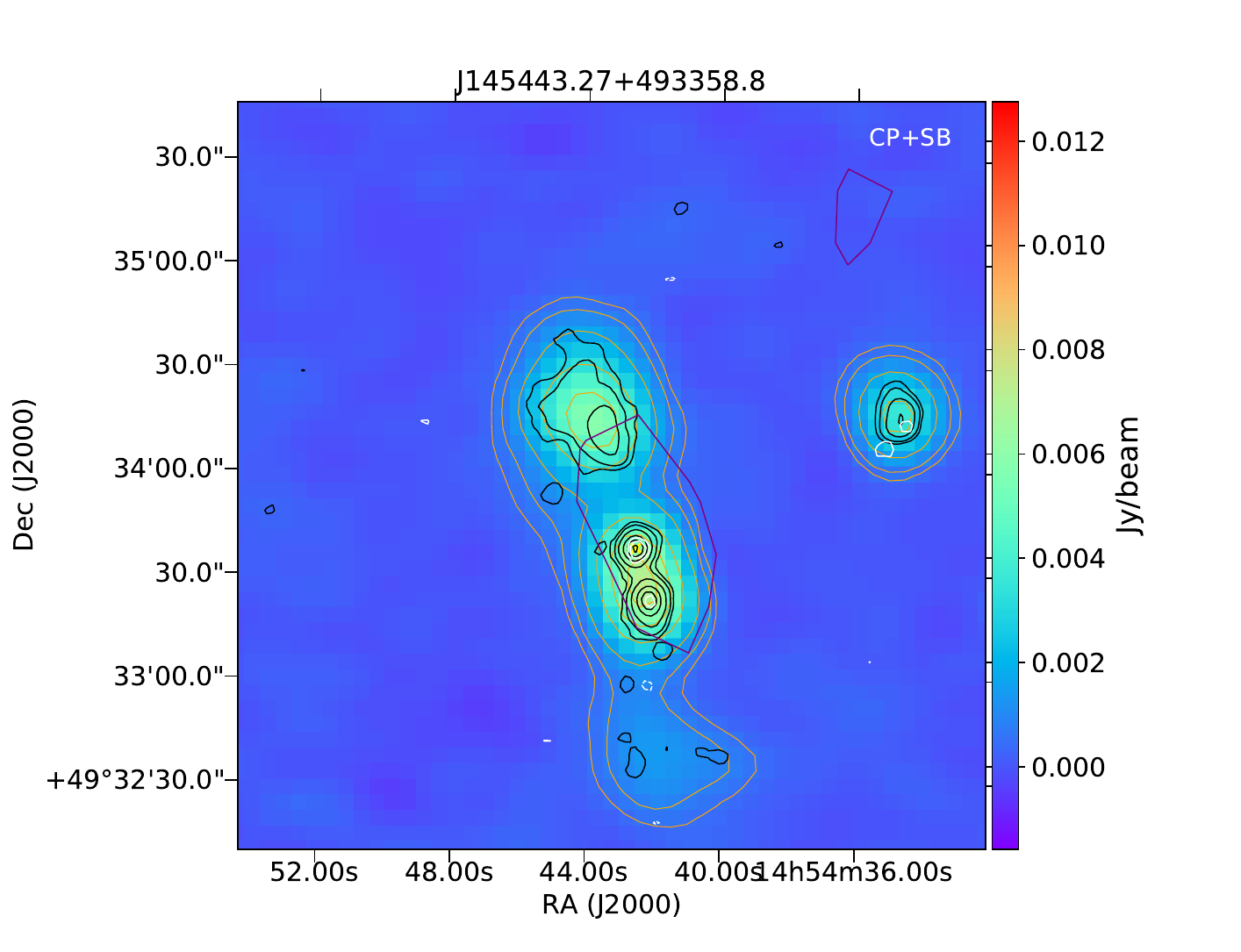}
            \includegraphics[width=0.3\textwidth]{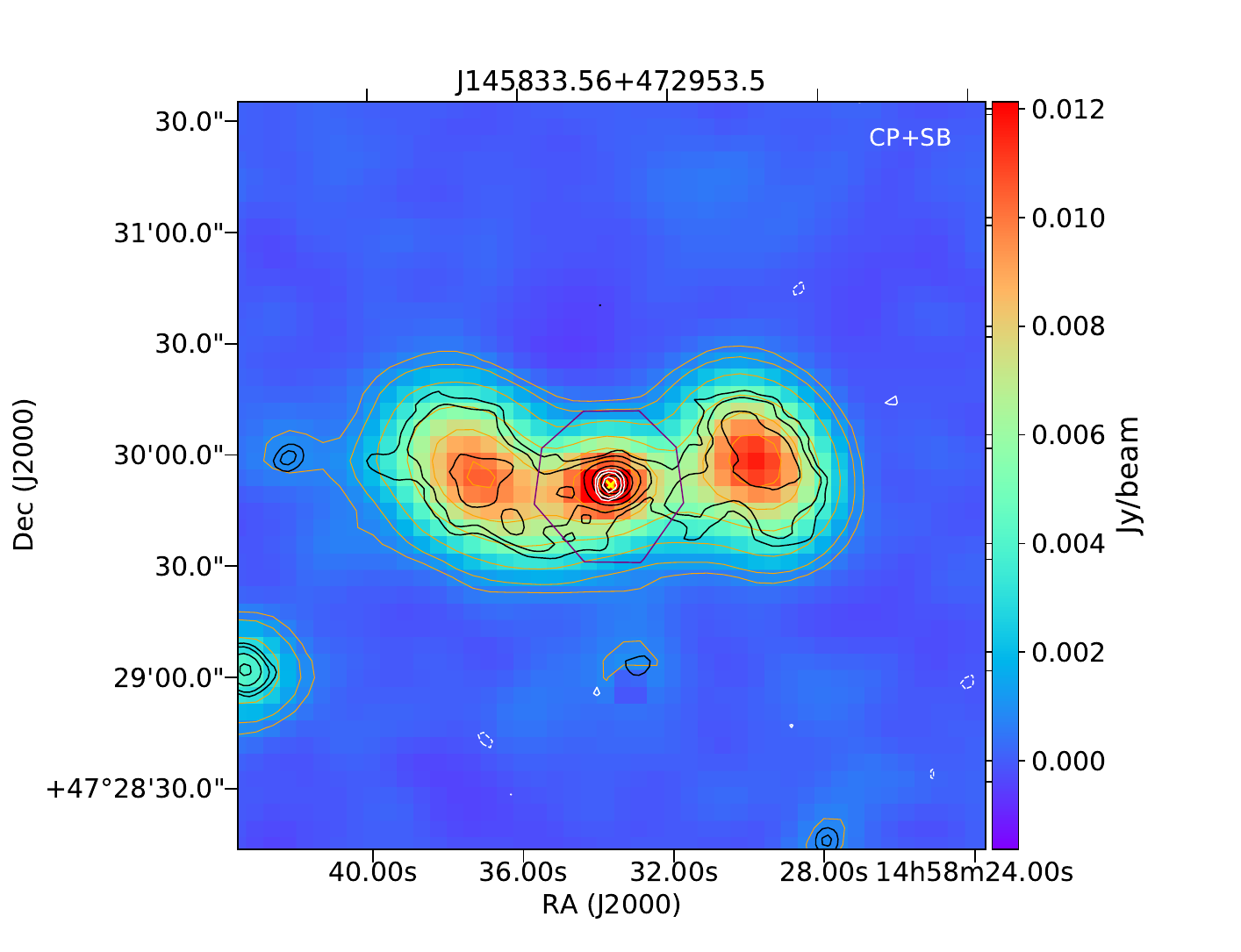}          
           }         
\centerline{\includegraphics[width=0.3\textwidth]{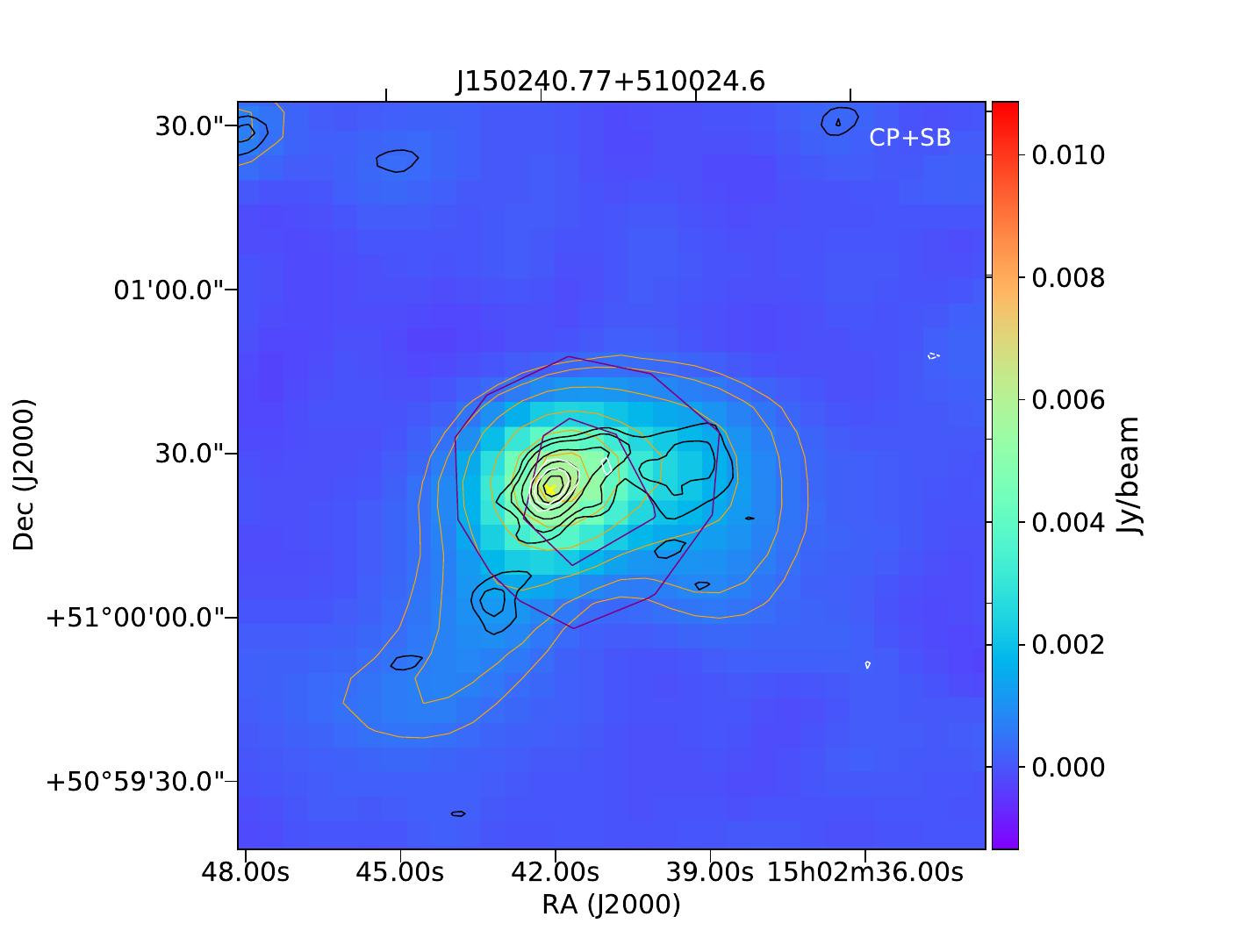}
            \includegraphics[width=0.3\textwidth]{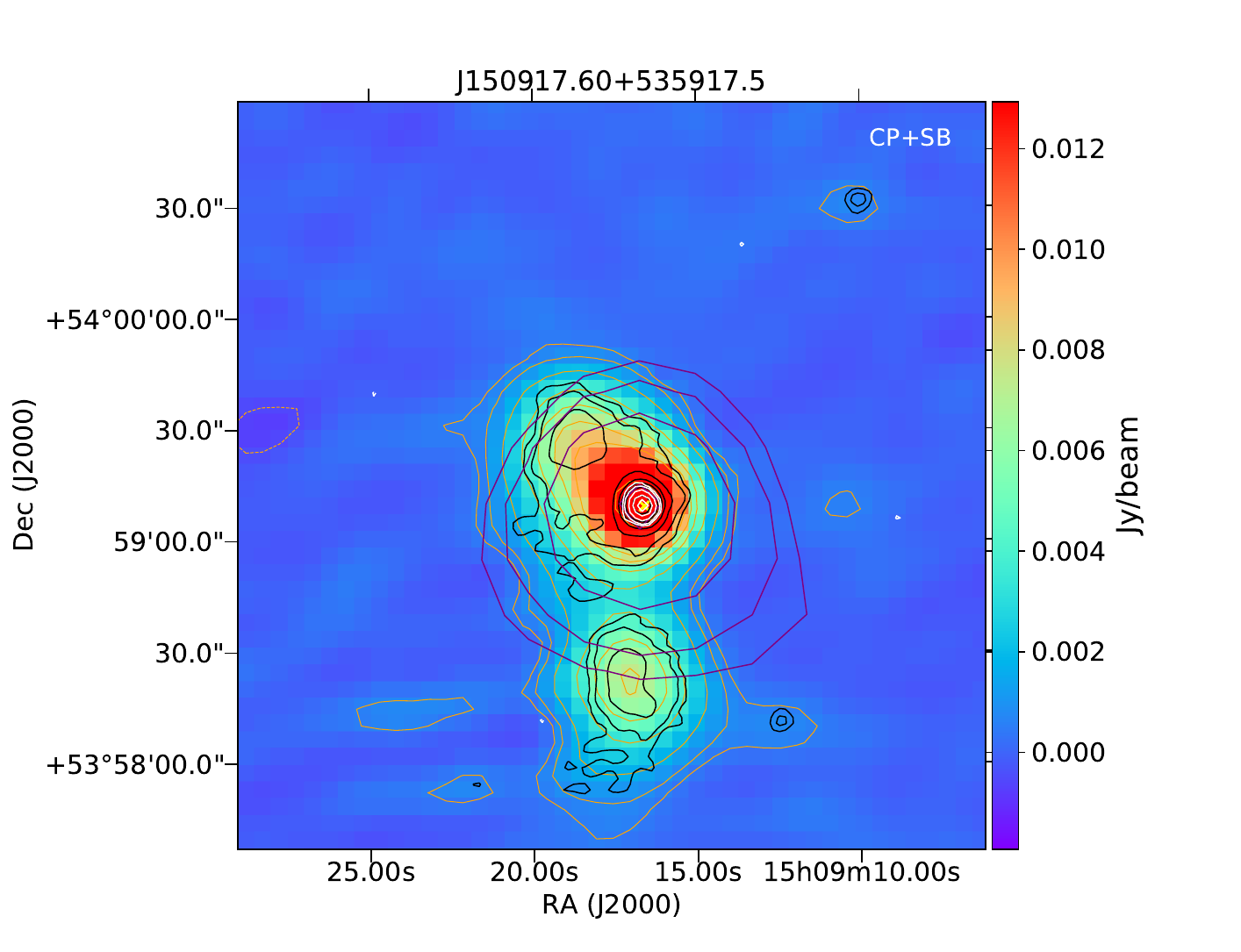}
            \includegraphics[width=0.3\textwidth]{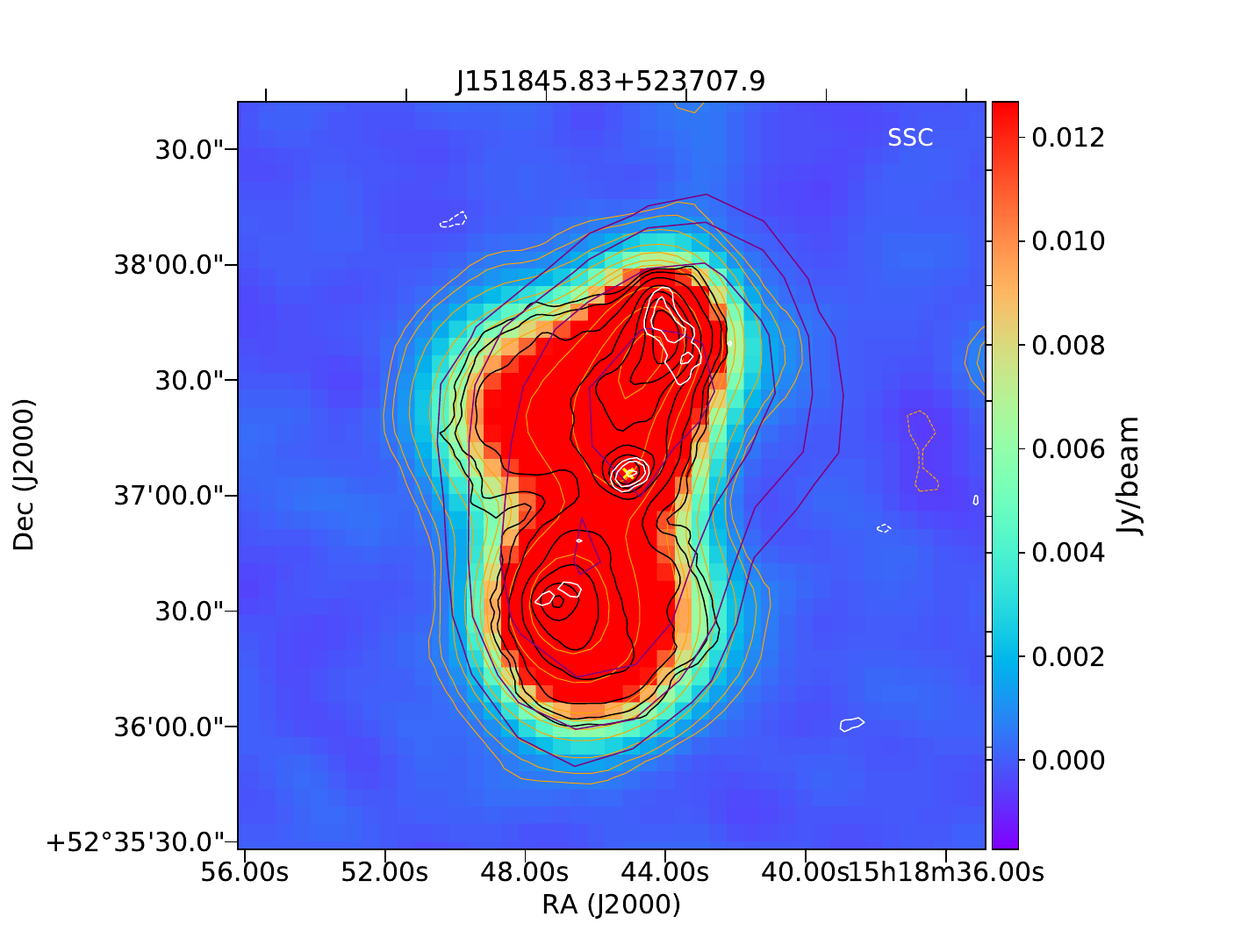}            
}
\caption{\footnotesize{Images of 69 candidate restarted galaxies selected based on
high radio $\mathrm{CP_{1400}}$ combined with low SB of extended emission, a steep spectrum of the core, and USS extended emission coupled with a bright core and summarised in Tables.~\ref{list of sources} and ~\ref{Core flux densities}.
Radio contours from VLA FIRST maps (white, 5$\arcsec$), LOFAR high-resolution maps (black, 6$\arcsec$), and NVSS maps (purple, 45$\arcsec$) are overlaid on the LOFAR low-resolution resolution maps (orange, 20$\arcsec$).
The contouring of all the maps is made at $\,\sigma_\mathrm{local}\times(-3,3,5,10,20,30,40,50,100,150,200)$ levels, with $\sigma_\mathrm{local}$ representing the local RMS noise of the corresponding maps.
The host galaxy position is marked with a yellow cross.
}}
\label{fig:5}
\end{figure*}

\begin{figure*}[ht!]
\centerline{\includegraphics[width=0.3\textwidth]{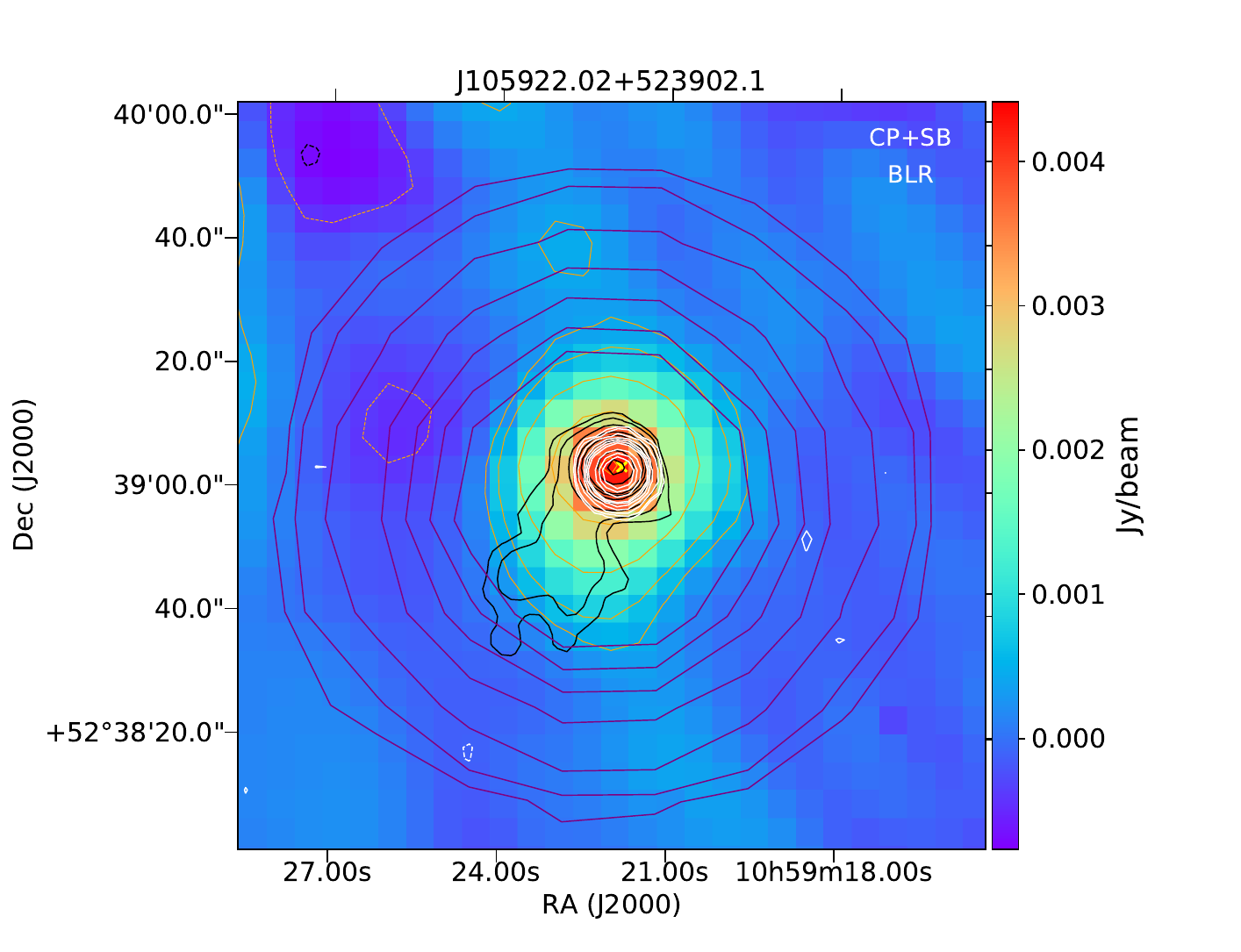}
            \includegraphics[width=0.3\textwidth]{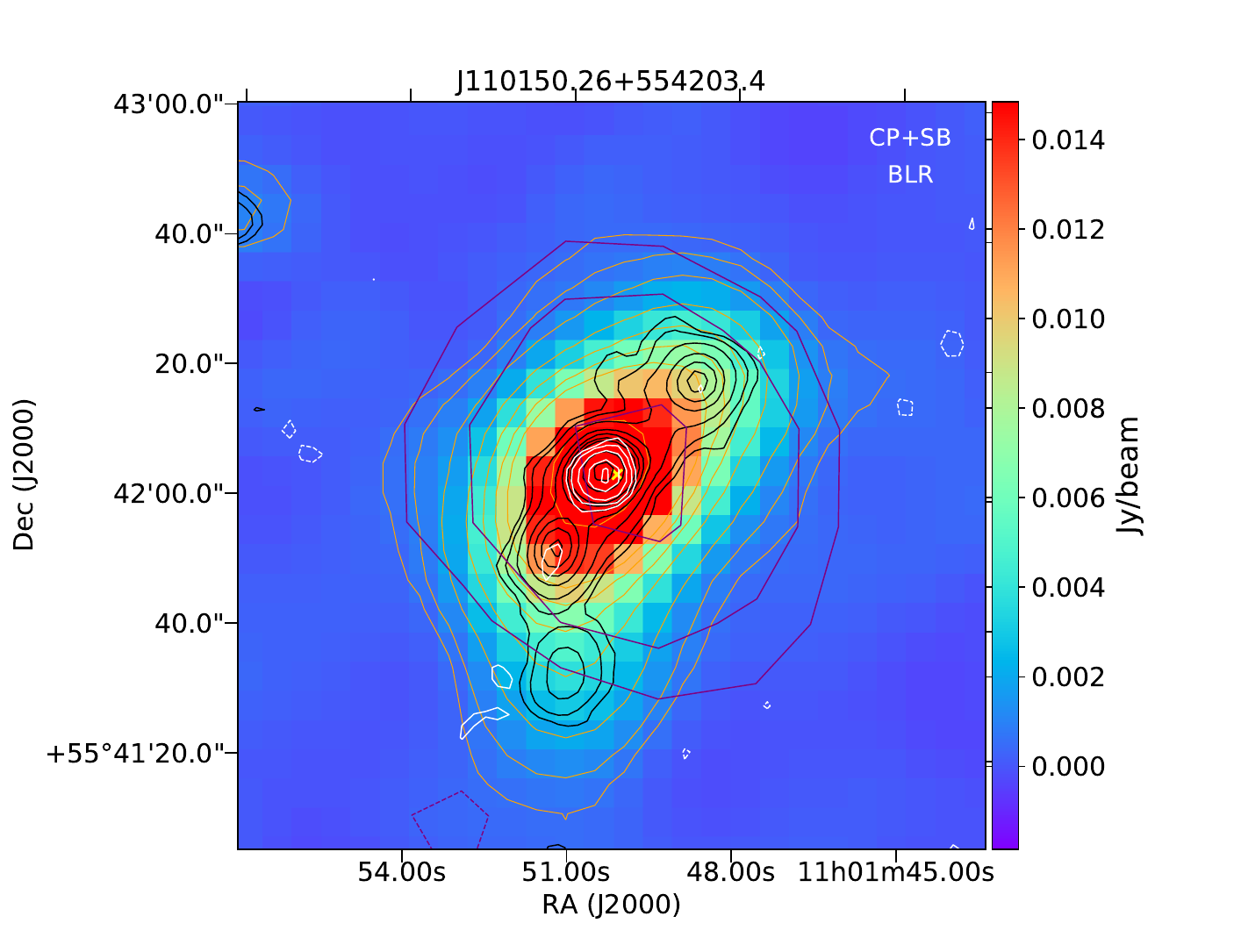}
            \includegraphics[width=0.3\textwidth]{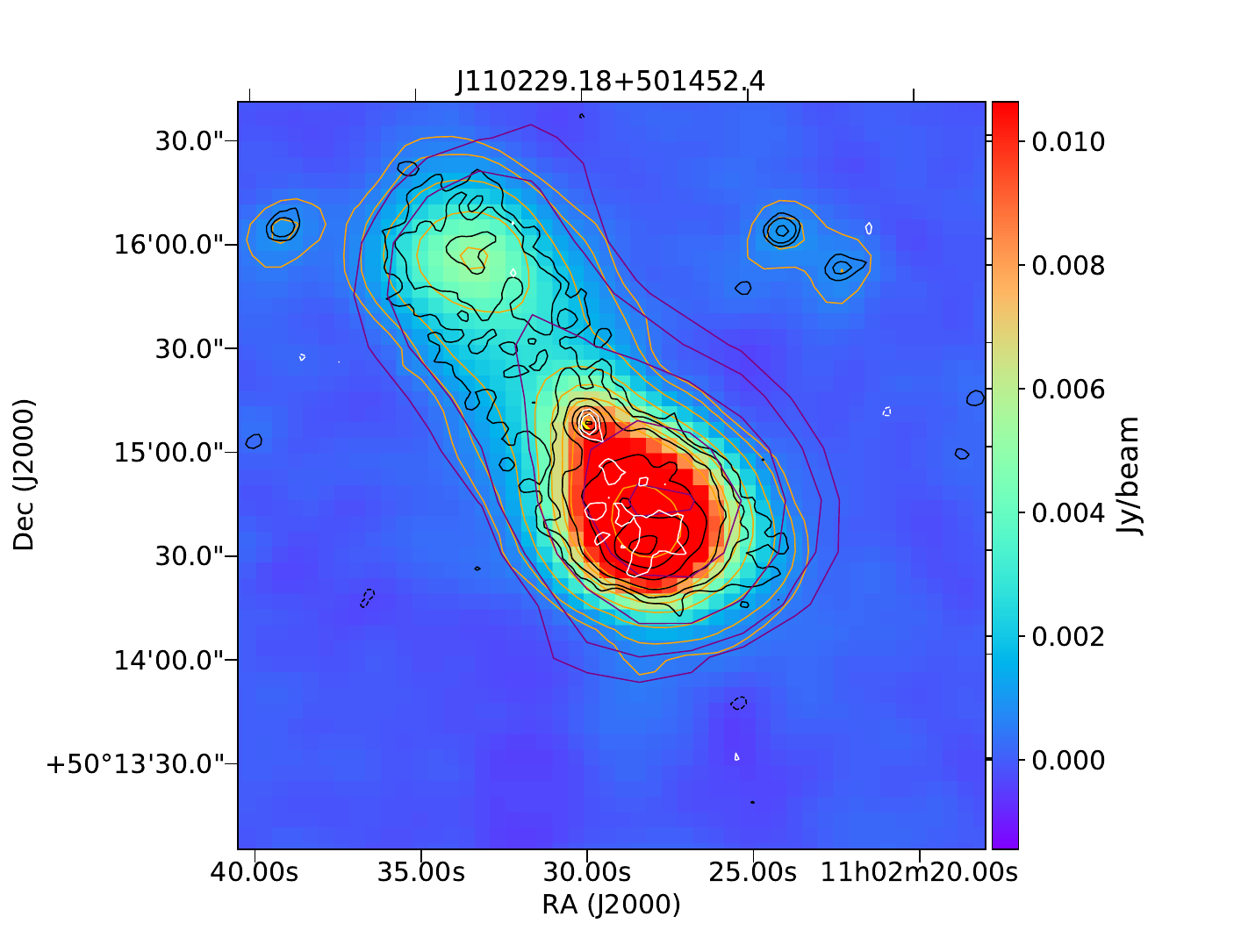}
            }        
\centerline{\includegraphics[width=0.3\textwidth]{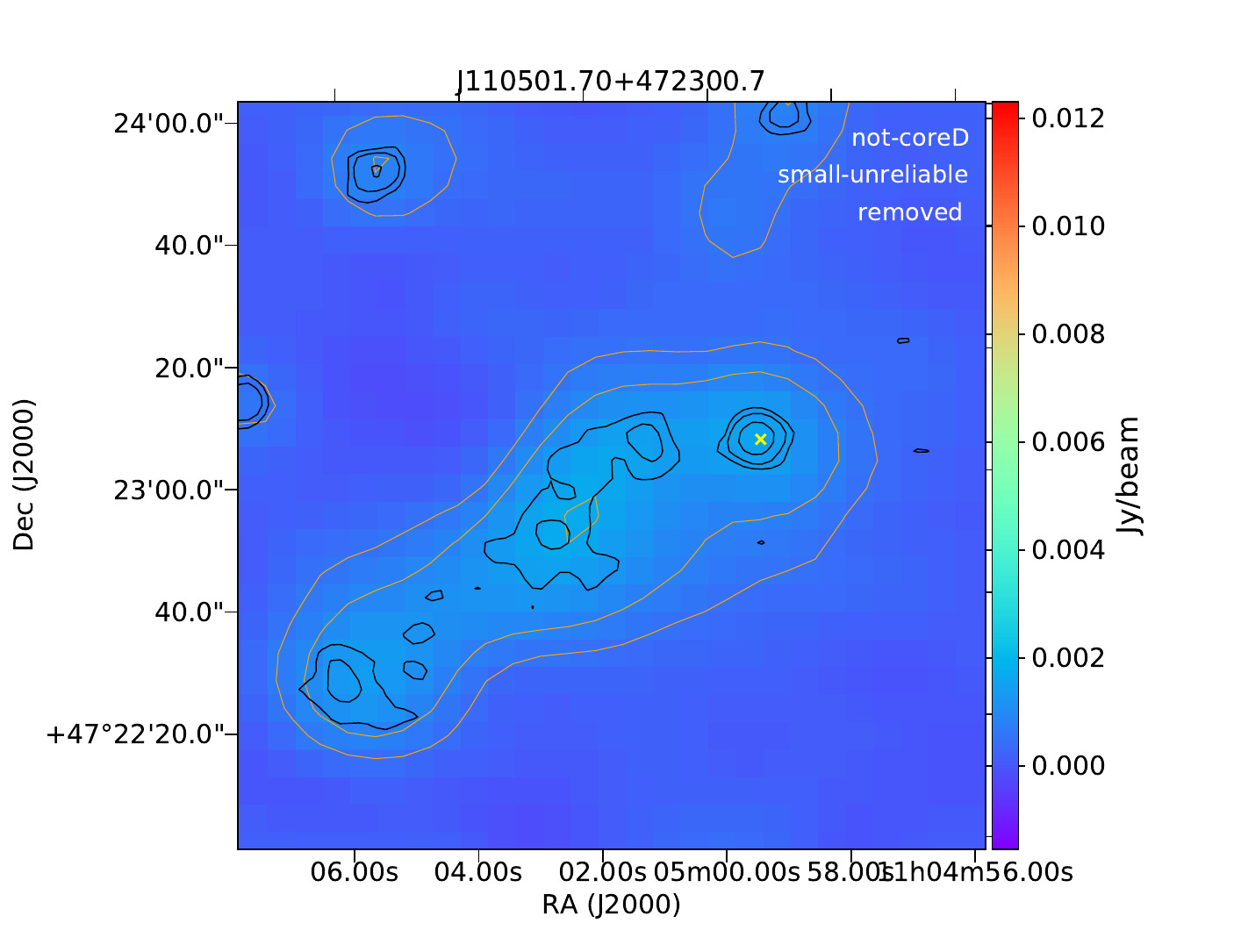}
            \includegraphics[width=0.3\textwidth]{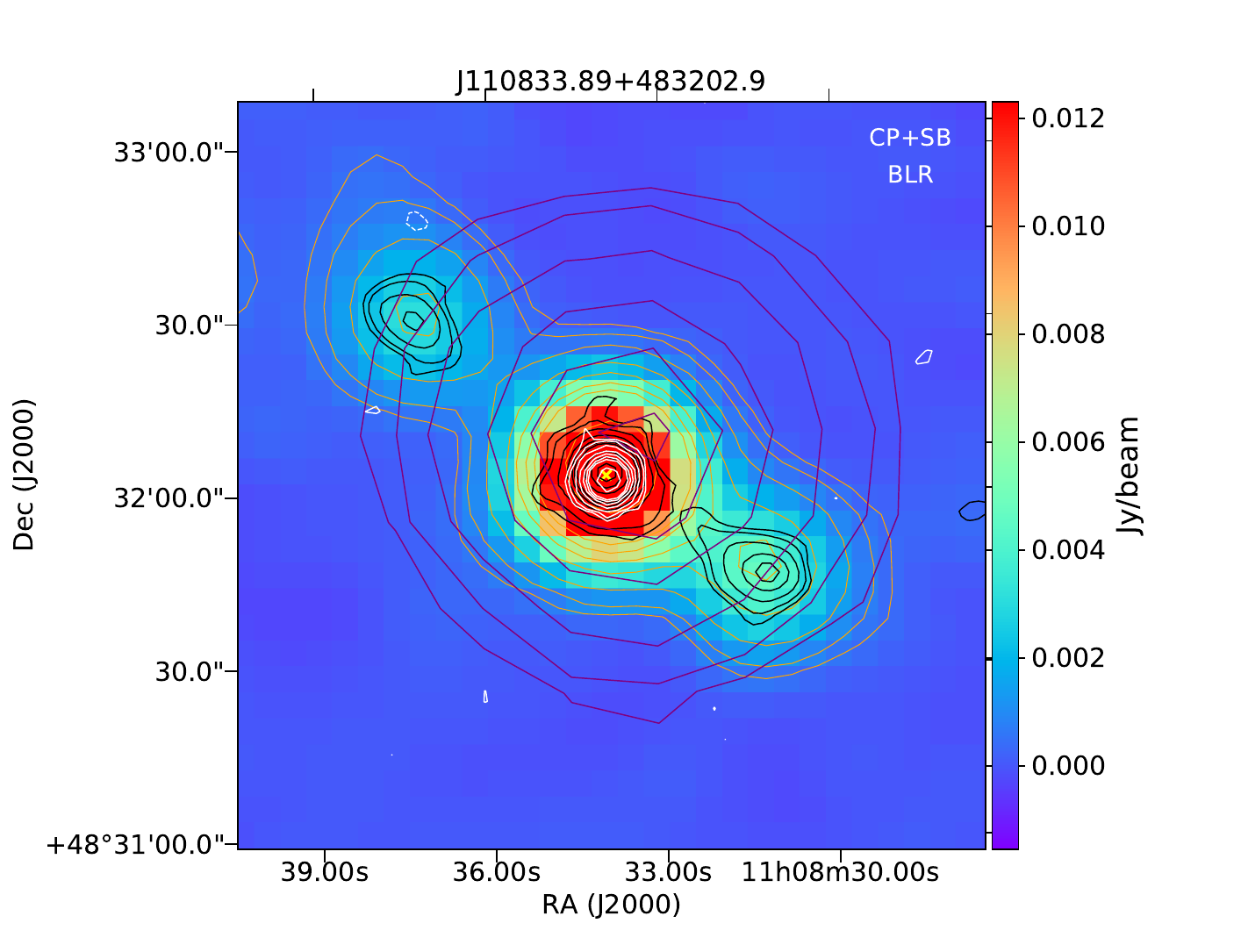}
            \includegraphics[width=0.3\textwidth]{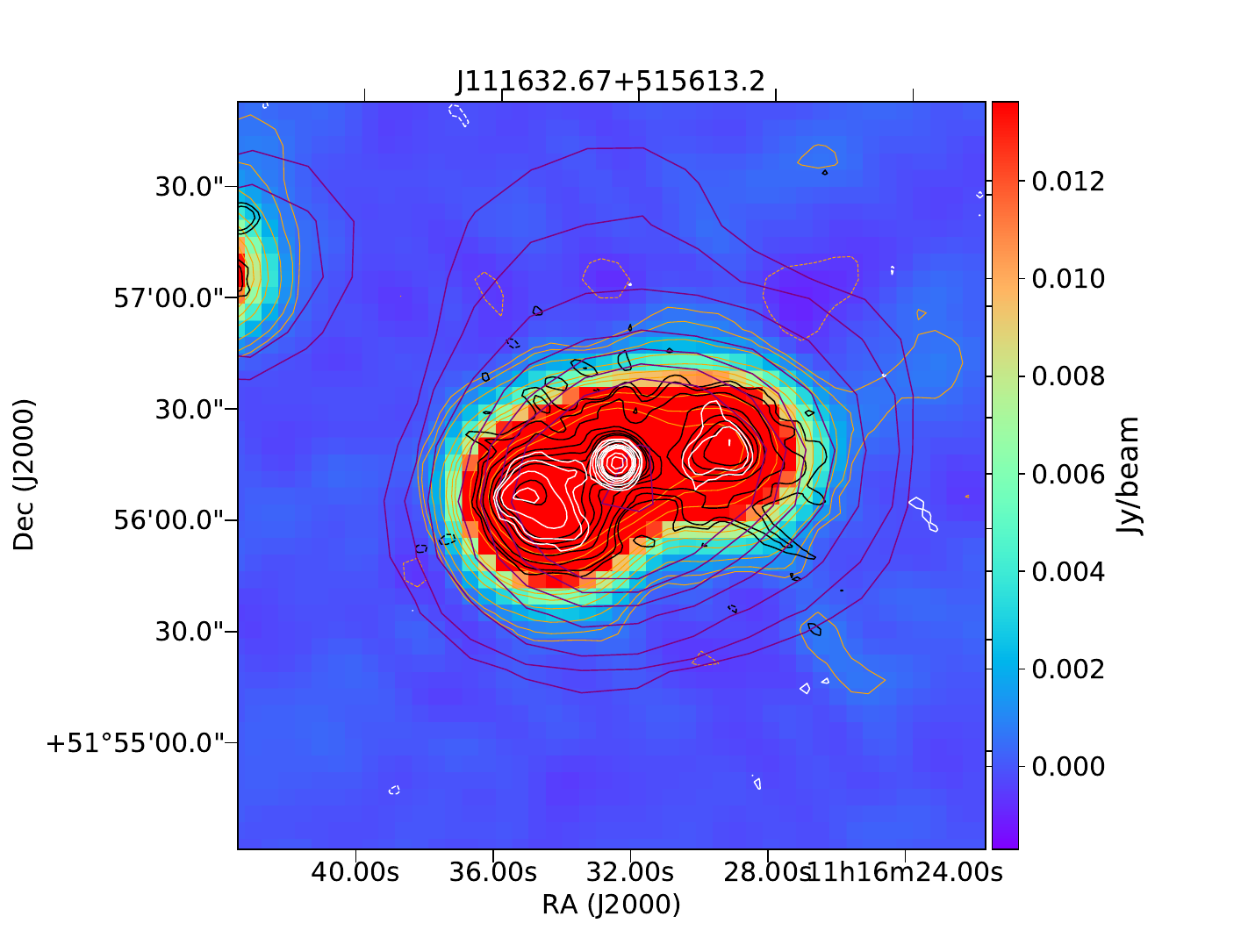}          
           }         
\centerline{\includegraphics[width=0.3\textwidth]{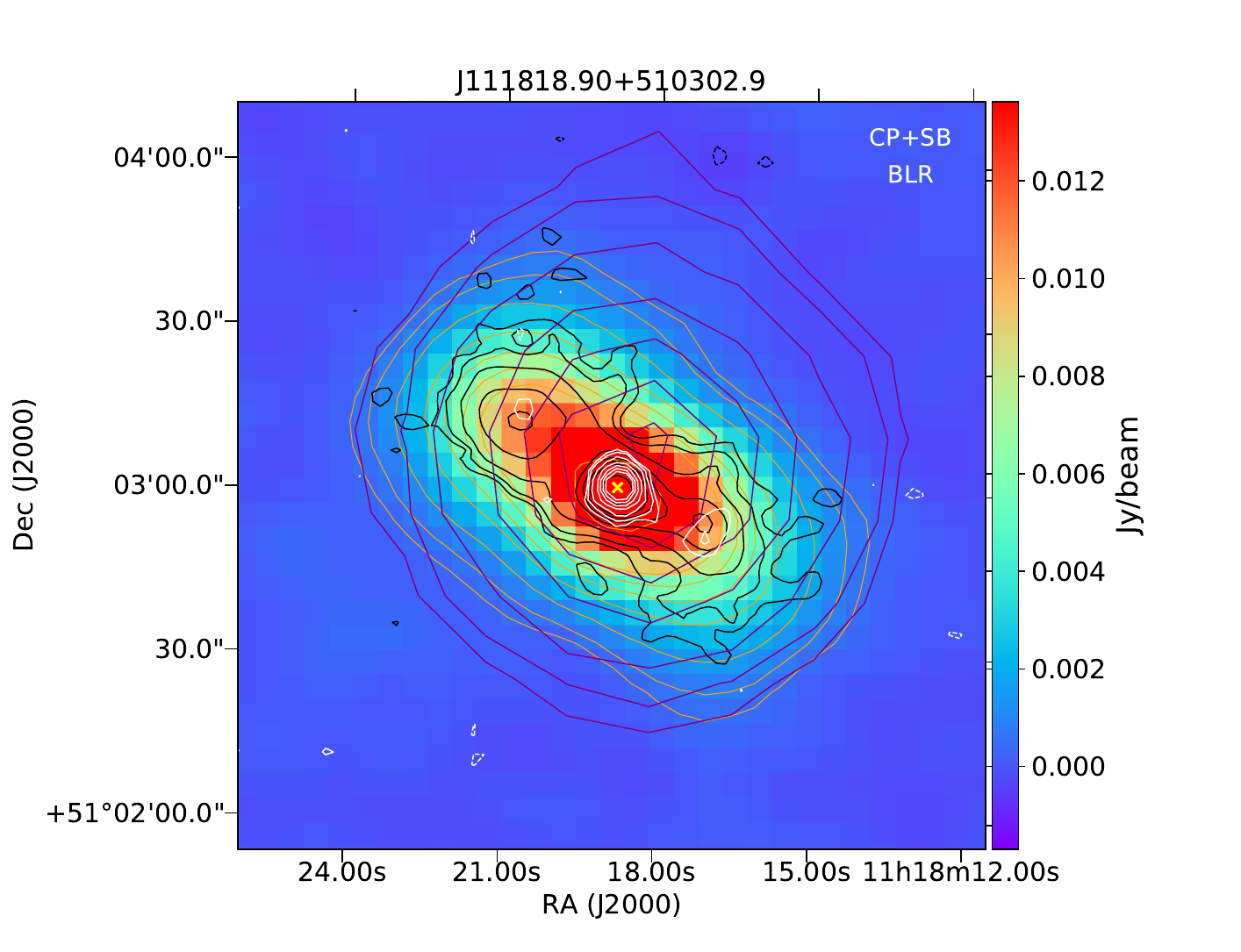}
            \includegraphics[width=0.3\textwidth]{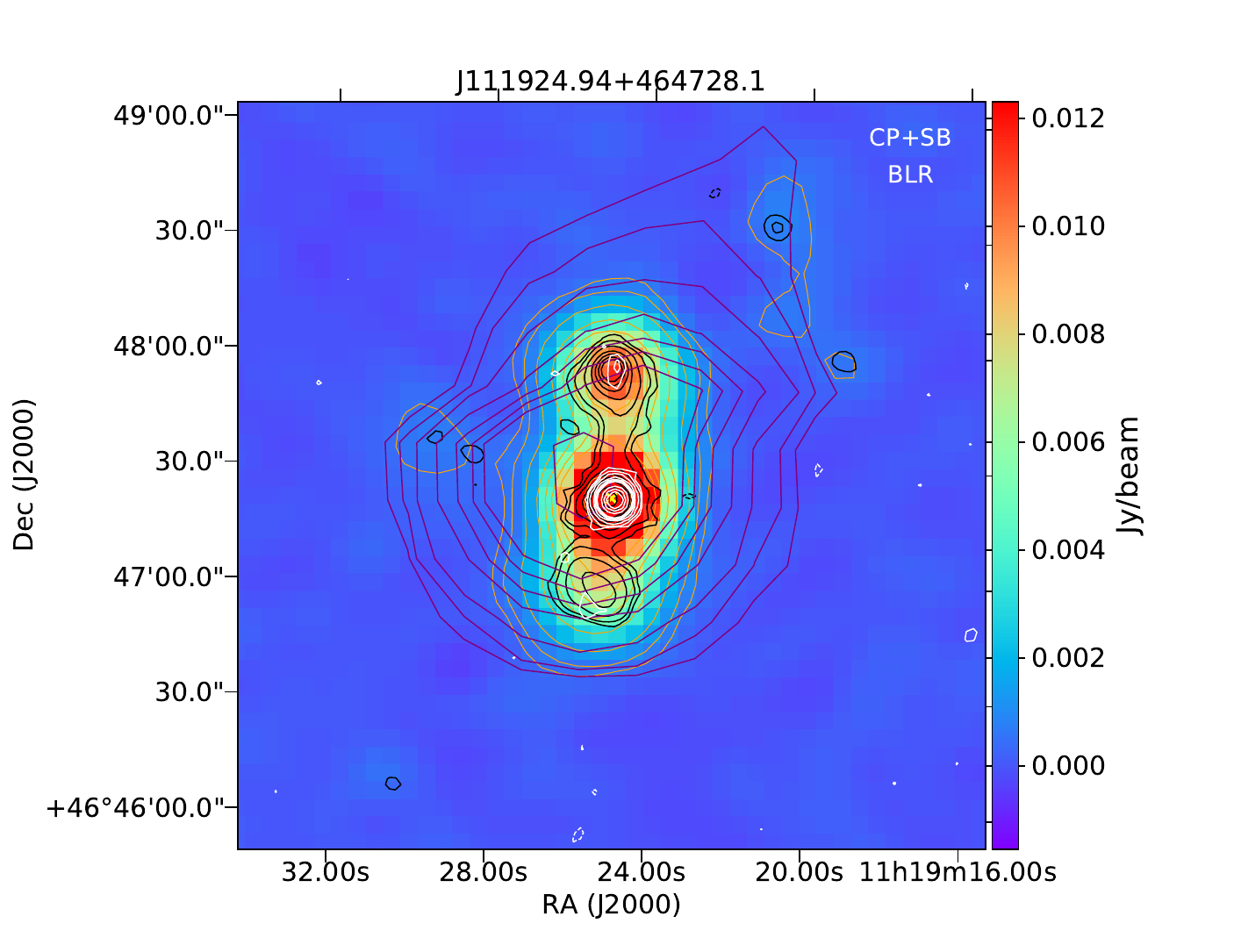}
            \includegraphics[width=0.3\textwidth]{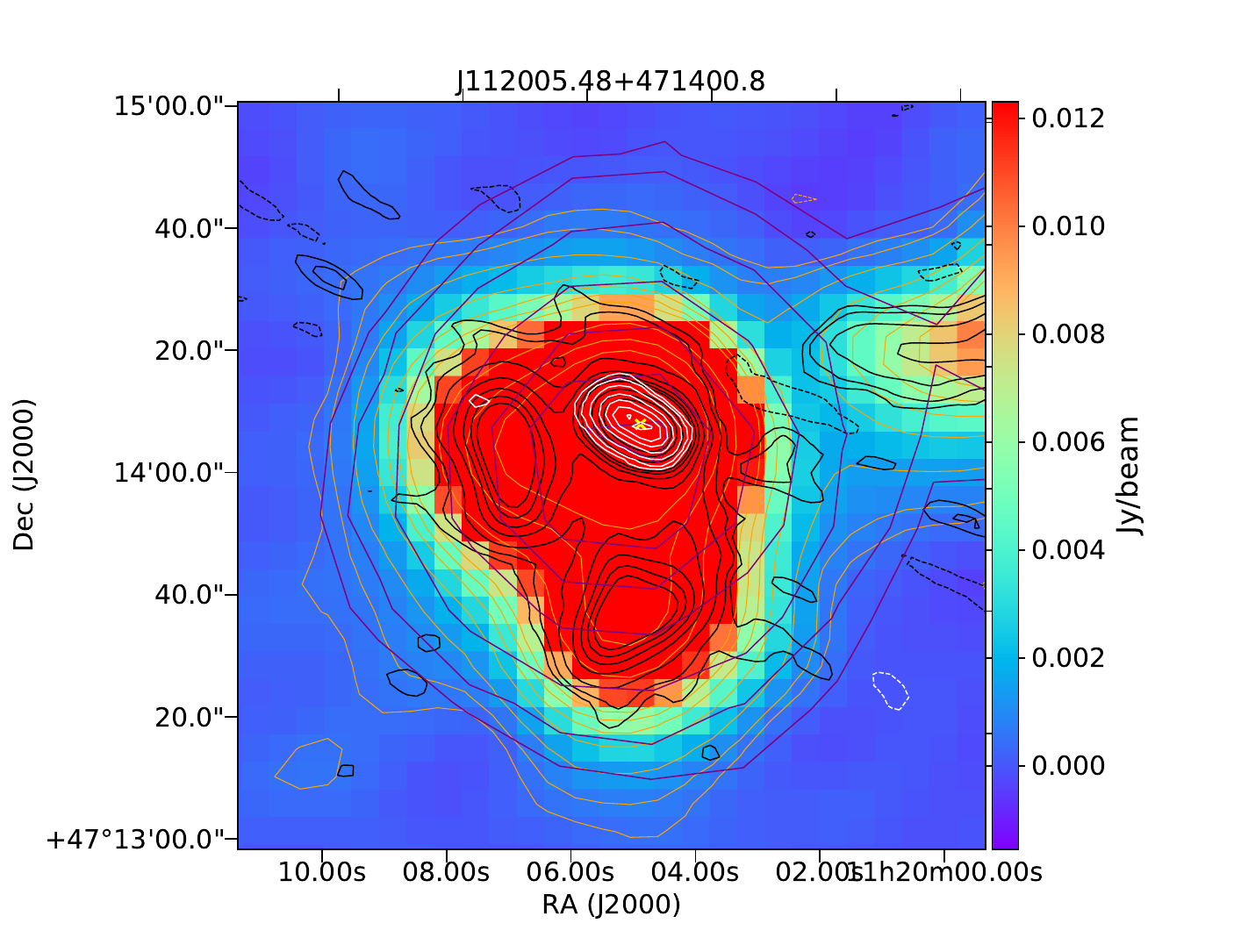}            
}
\centerline{\includegraphics[width=0.3\textwidth]{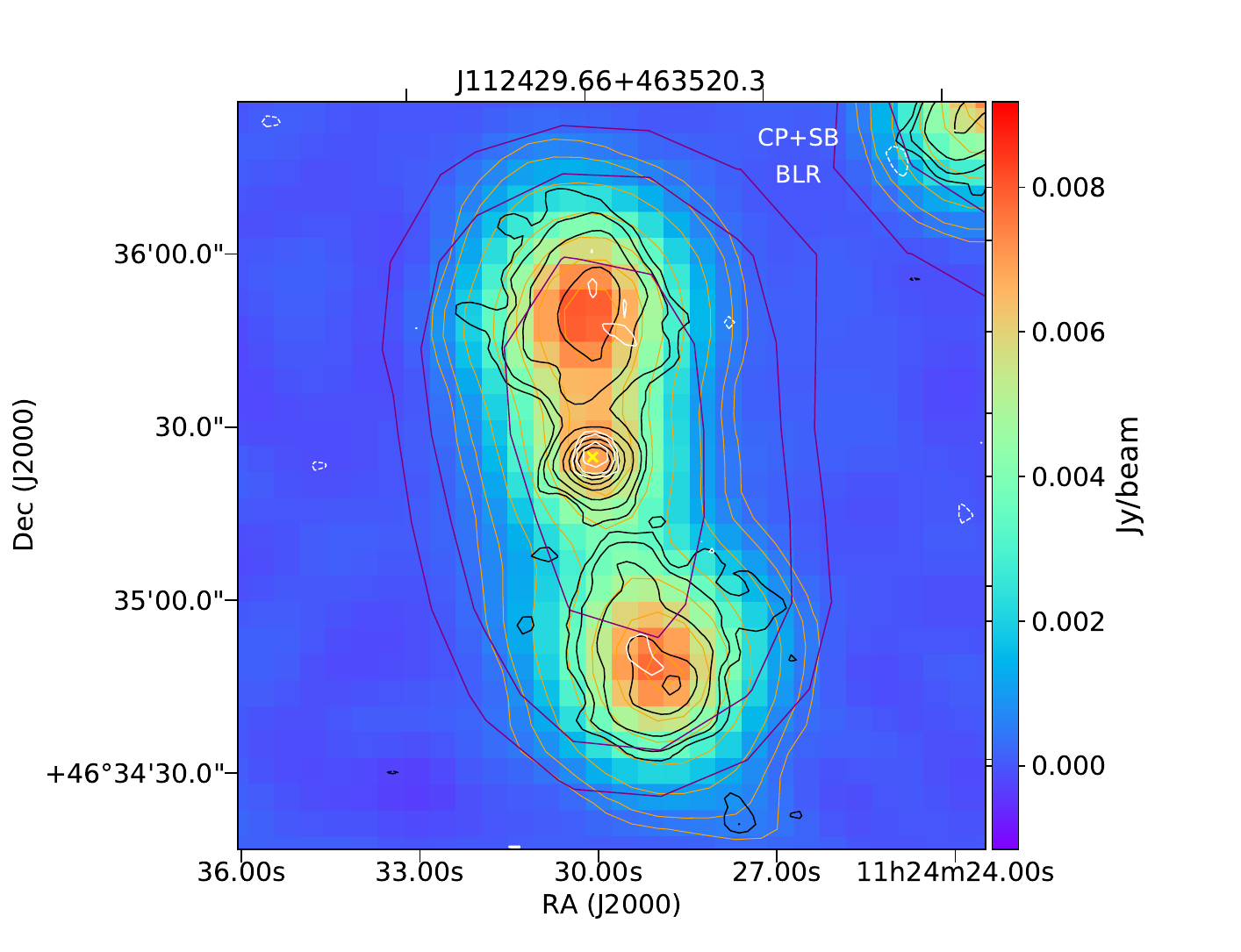}
            \includegraphics[width=0.3\textwidth]{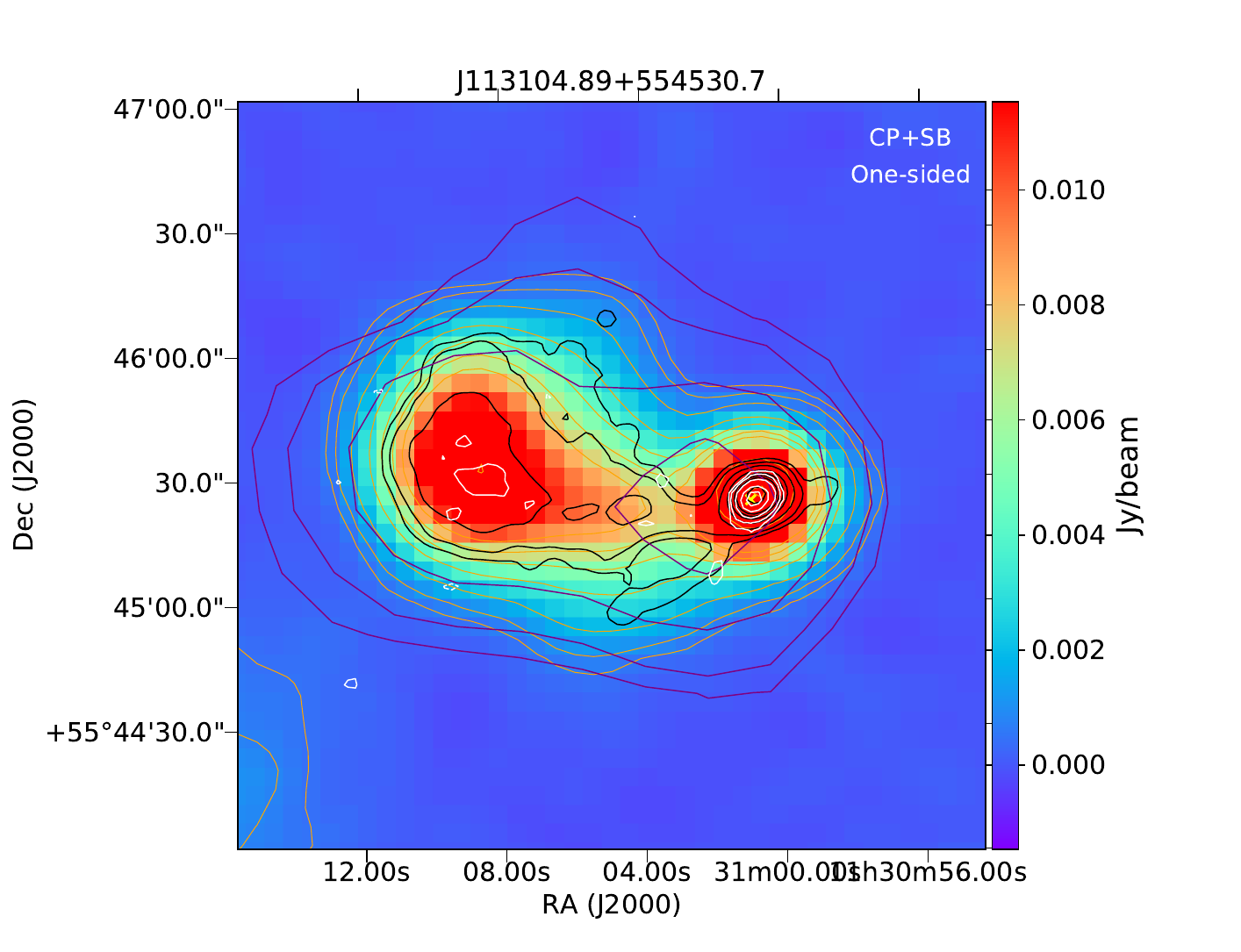}
            \includegraphics[width=0.3\textwidth]{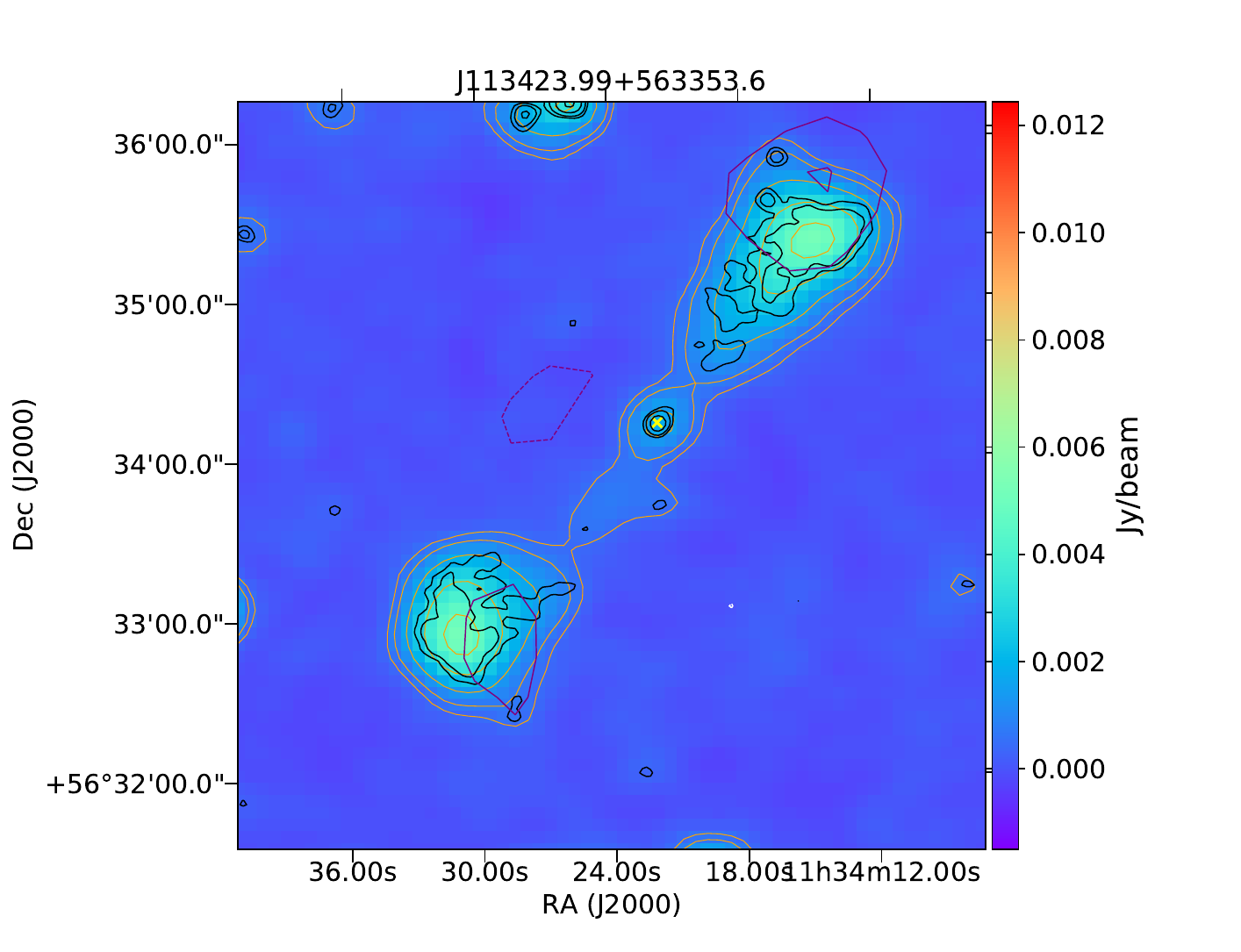}            
}
\centerline{\includegraphics[width=0.3\textwidth]{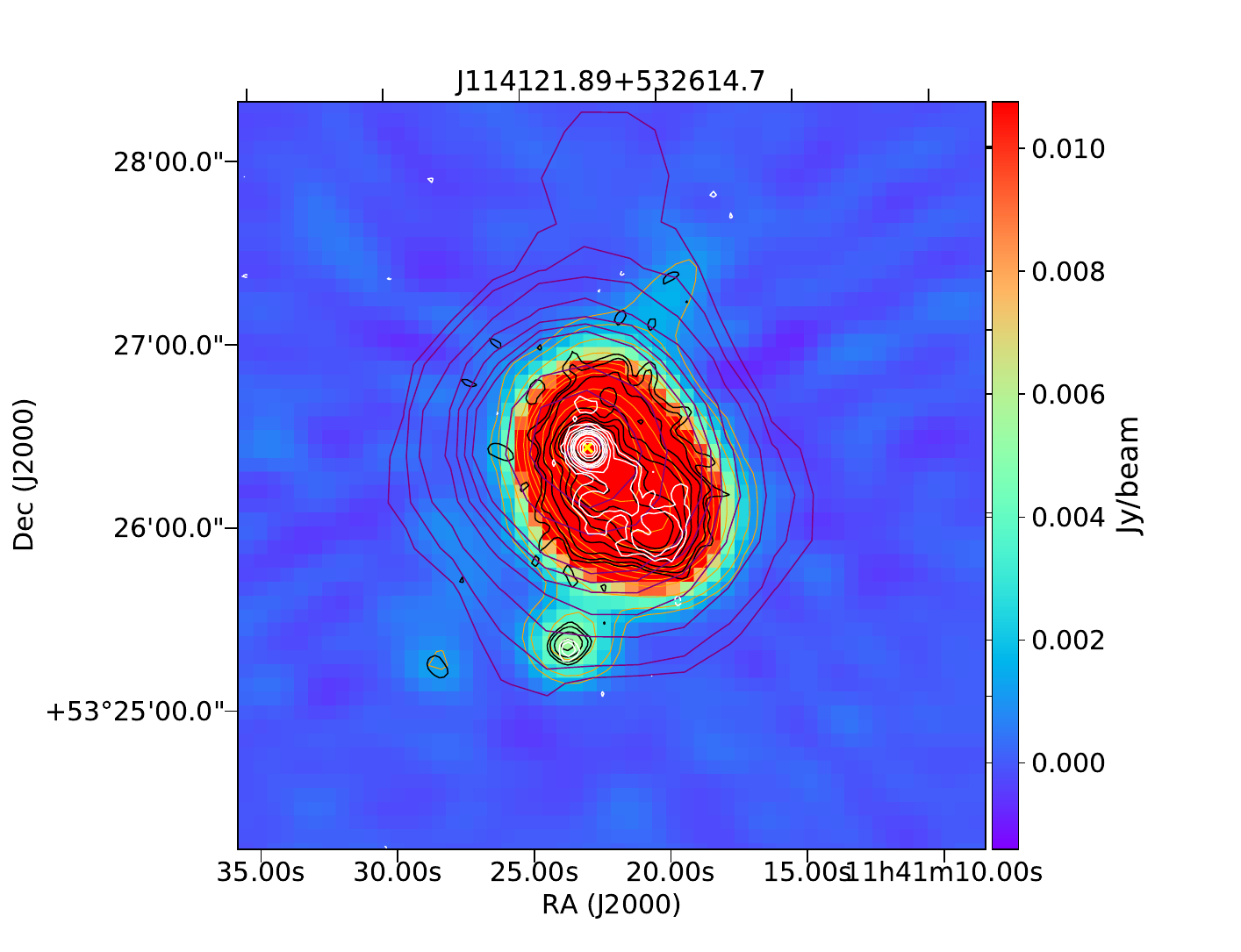}
            \includegraphics[width=0.3\textwidth]{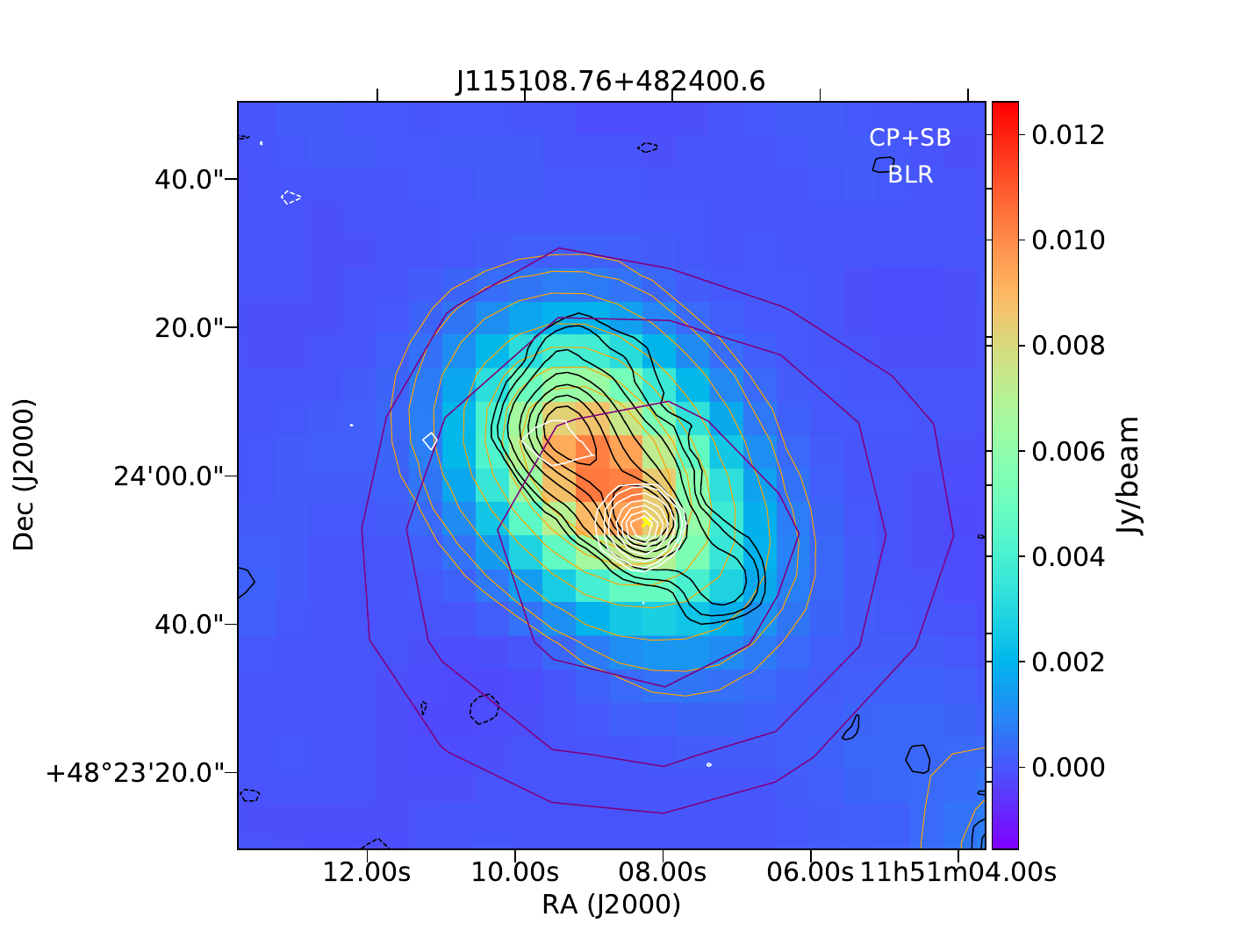}
            \includegraphics[width=0.3\textwidth]{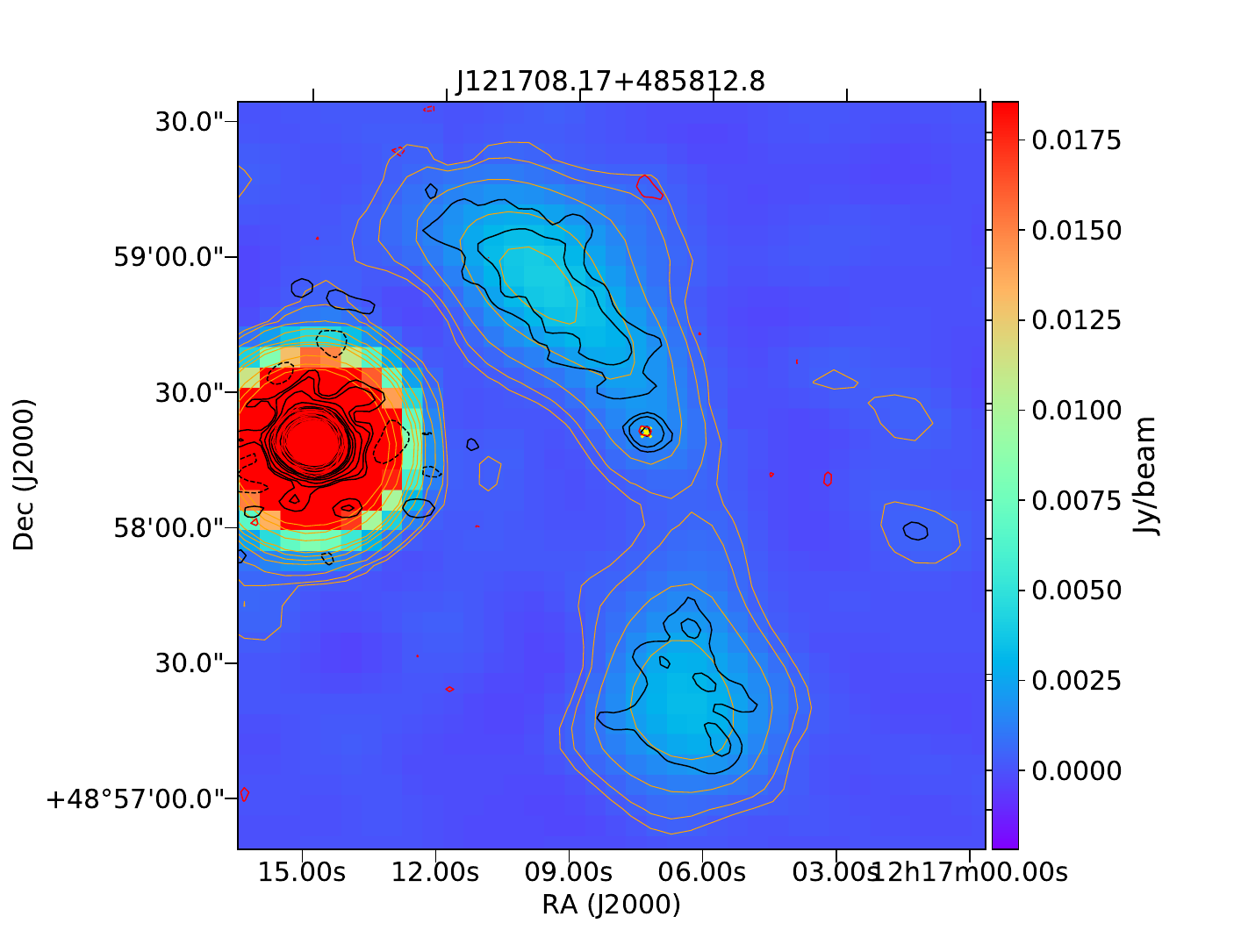}            
}
\caption{\footnotesize{
Images of sources excluded from the sample of restarted candidates following the criteria discussed in Sect.~\ref{High radio core prominence and low surface brightness}, Sect.~\ref{Steep spectral index of the core} and Sect.~\ref{Spectral index of the extended emission} and summarised in Tables.~\ref{list of sources-rejected} and ~\ref{Core flux densities-rejected}.
Radio contours from VLA FIRST maps (white, 5$\arcsec$), LOFAR high-resolution maps (black, 6$\arcsec$), and NVSS maps (purple, 45$\arcsec$) are overlaid on the LOFAR low-resolution resolution maps (orange, 20$\arcsec$).
The contouring of all the maps is made at $\,\sigma_\mathrm{local}\times(-3,3,5,10,20,30,40,50,100,150,200)$ levels, with $\sigma_\mathrm{local}$ representing the local RMS noise of the corresponding maps.
The host galaxy position is marked with a yellow cross.
}}
\label{fig:6}
\end{figure*}

\begin{figure*}[ht!]
\centerline{\includegraphics[width=0.3\textwidth]{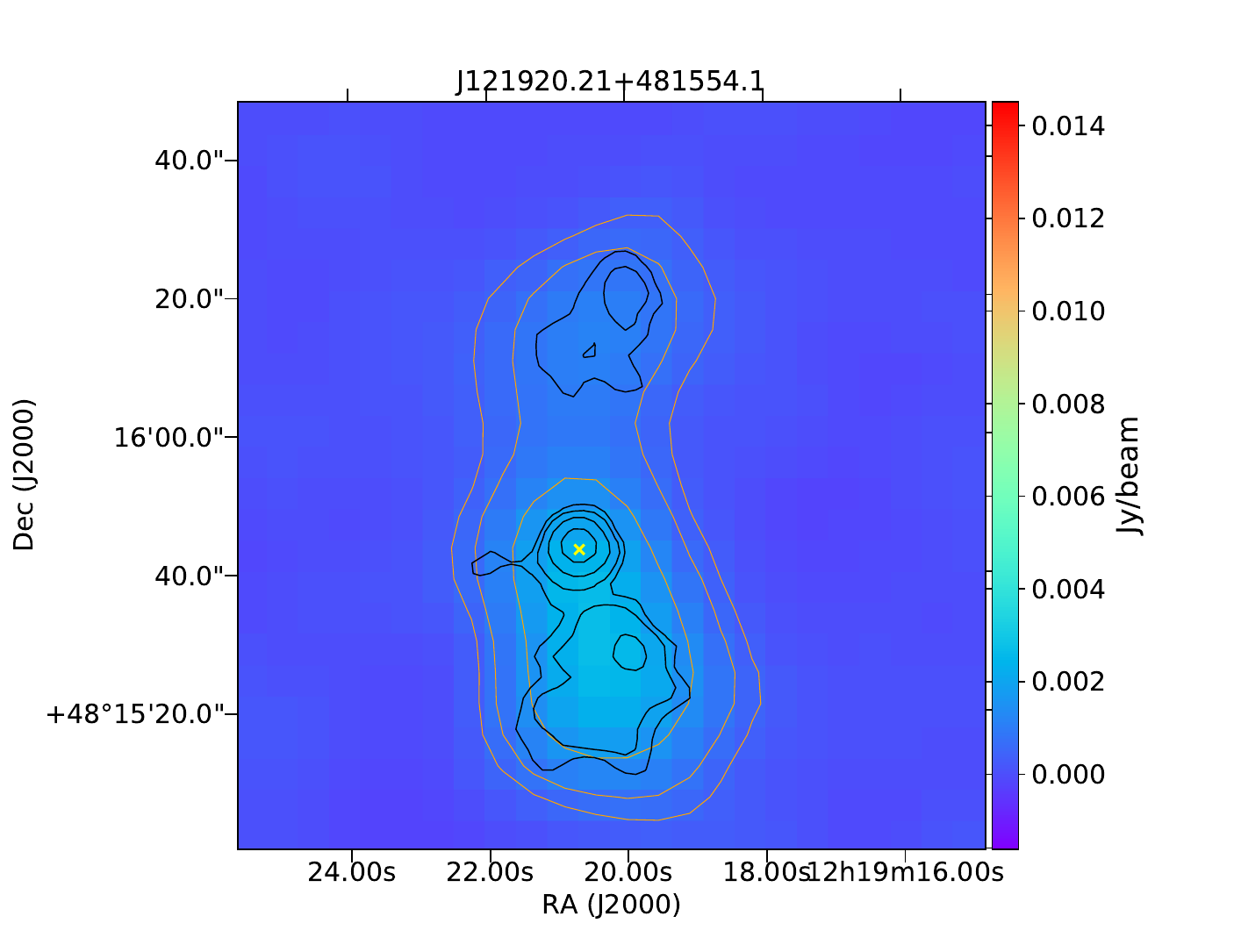}
            \includegraphics[width=0.3\textwidth]{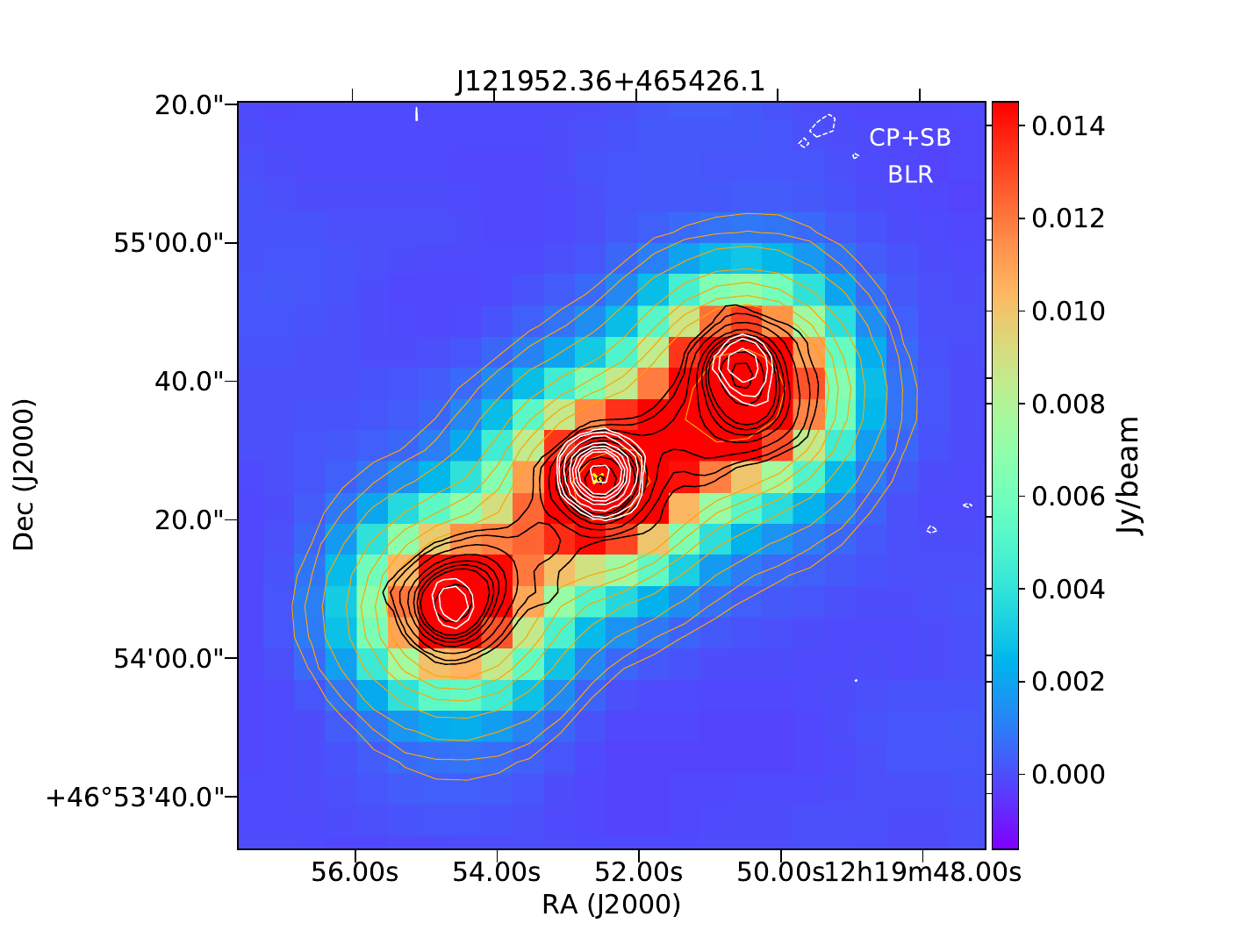}
            \includegraphics[width=0.3\textwidth]{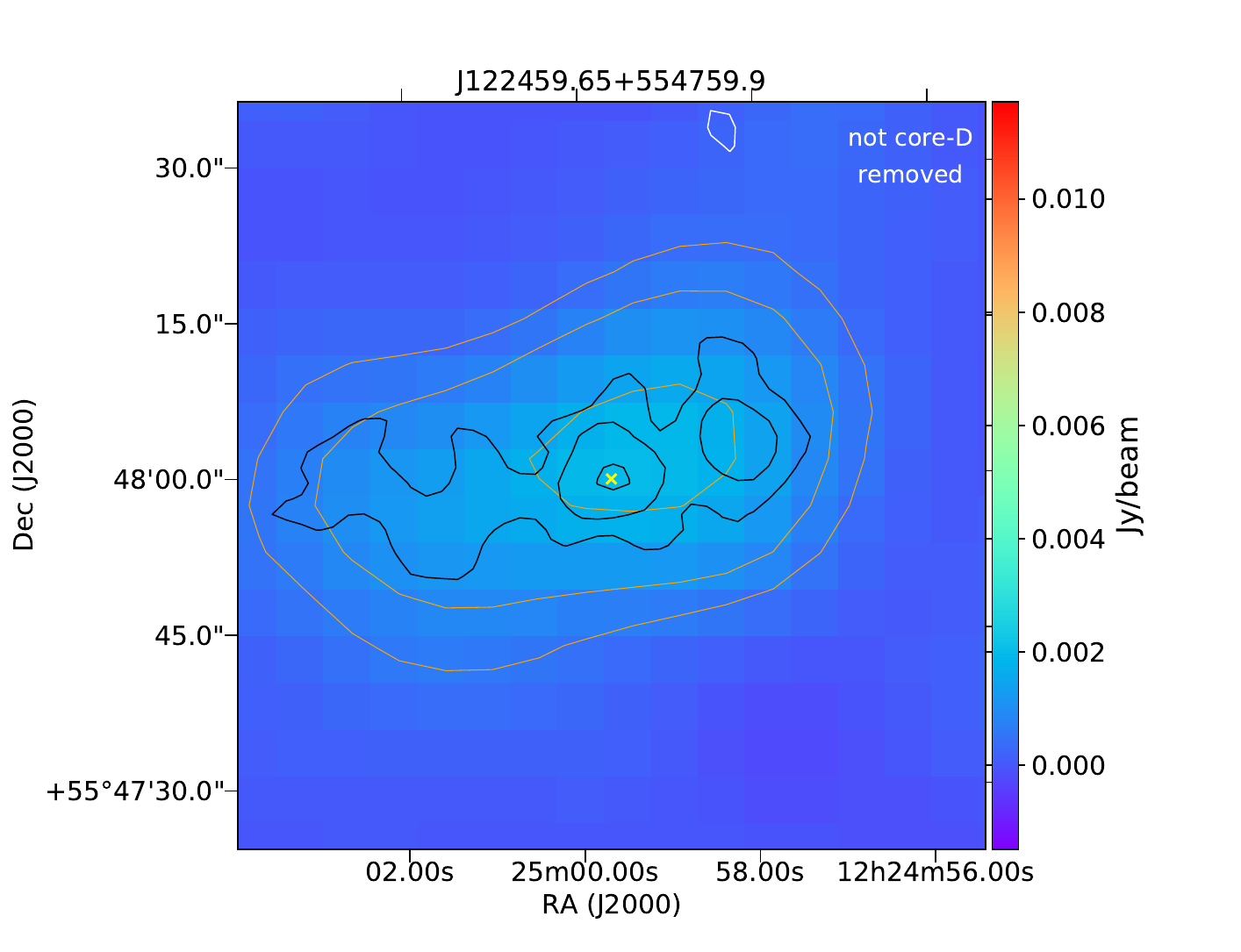}                   }        
\centerline{\includegraphics[width=0.3\textwidth]{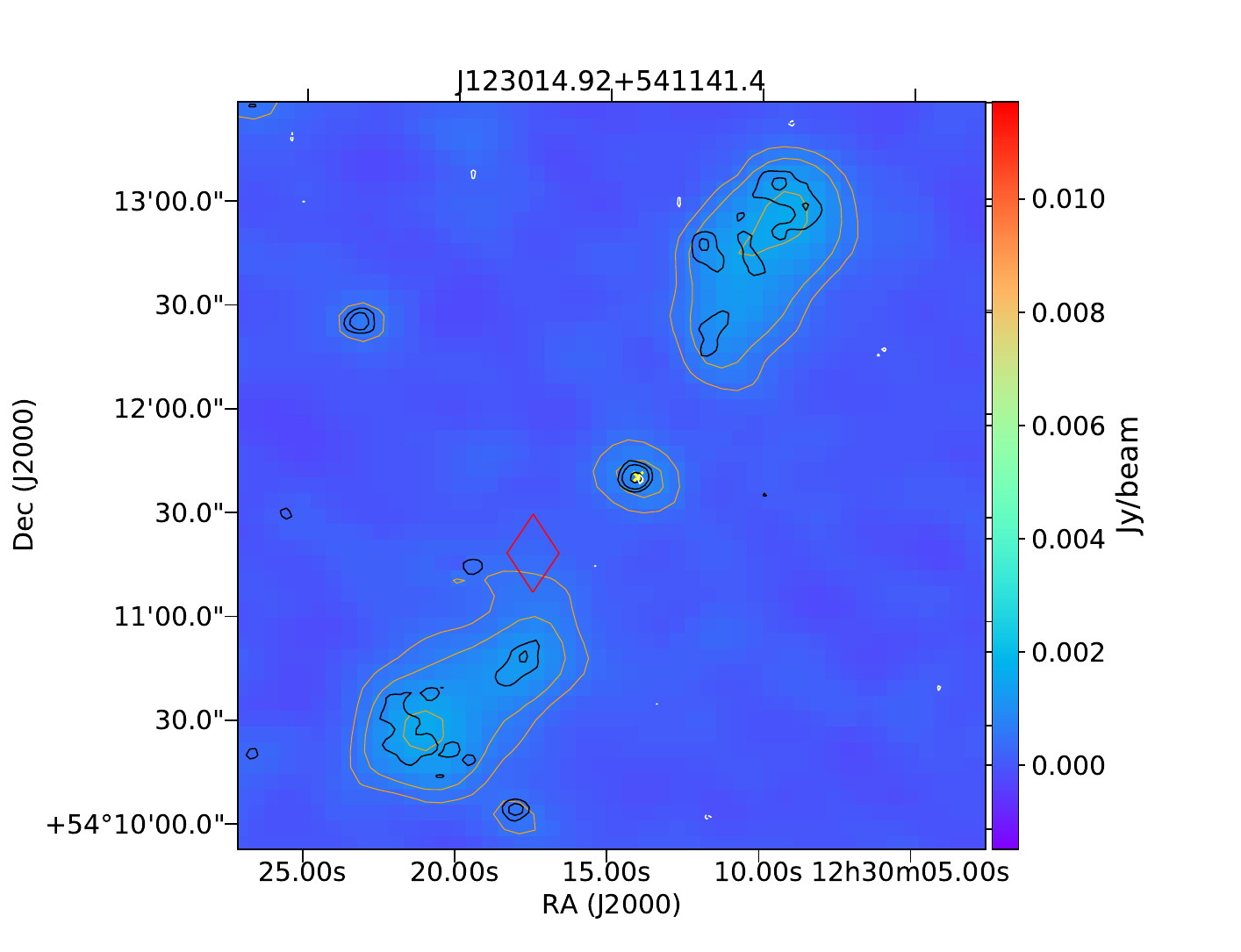}
            \includegraphics[width=0.3\textwidth]{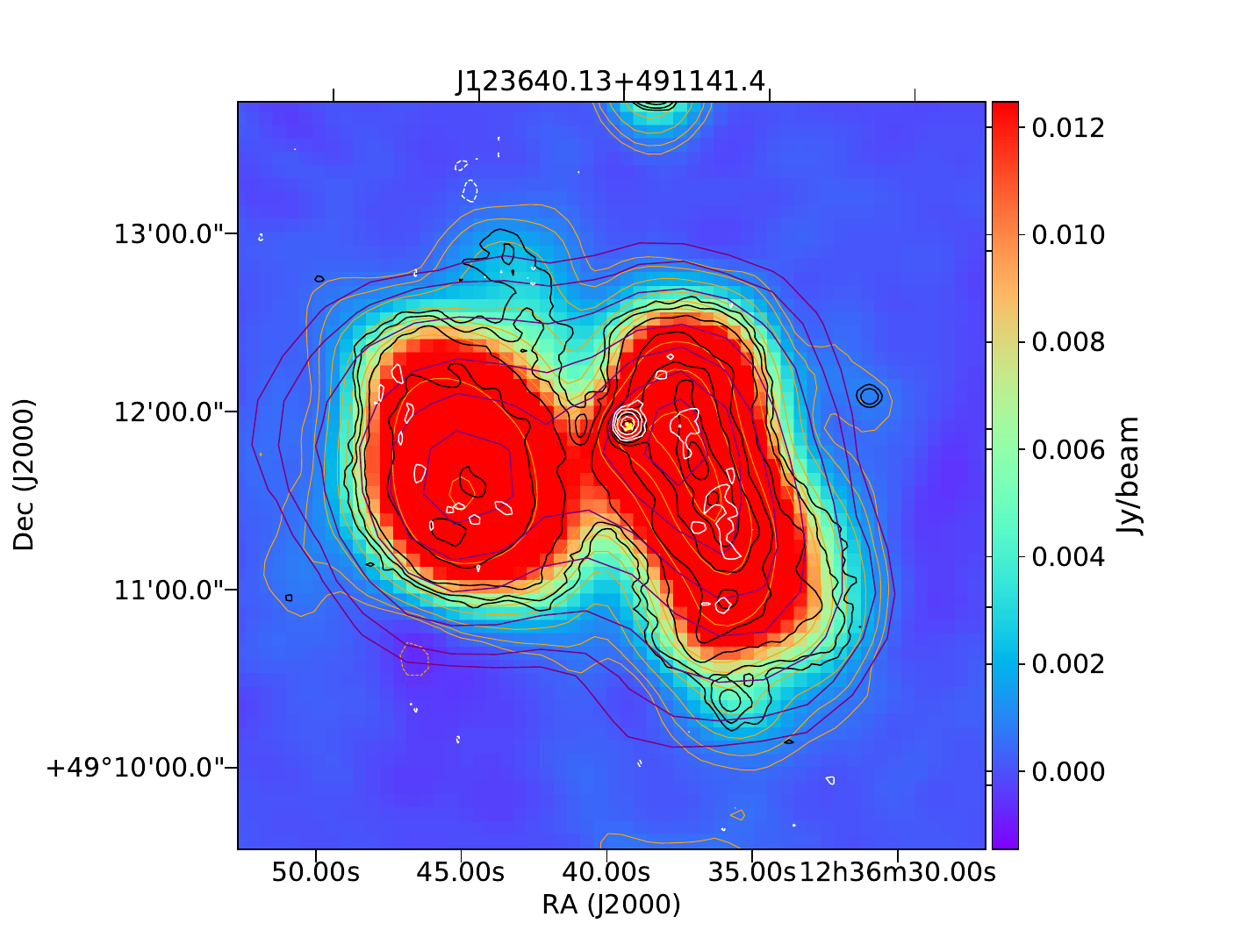}
            \includegraphics[width=0.3\textwidth]{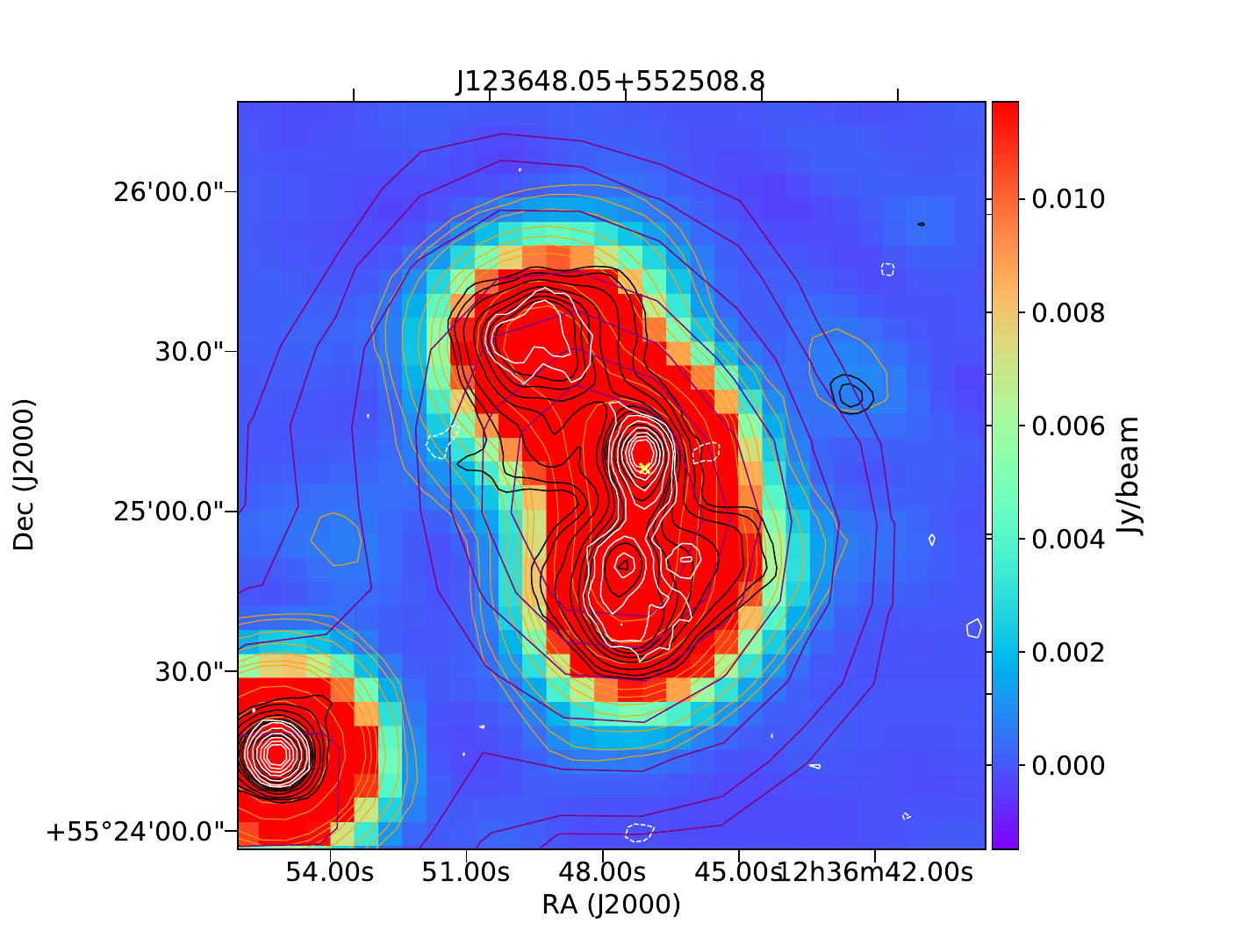}          
           }         
\centerline{\includegraphics[width=0.3\textwidth]{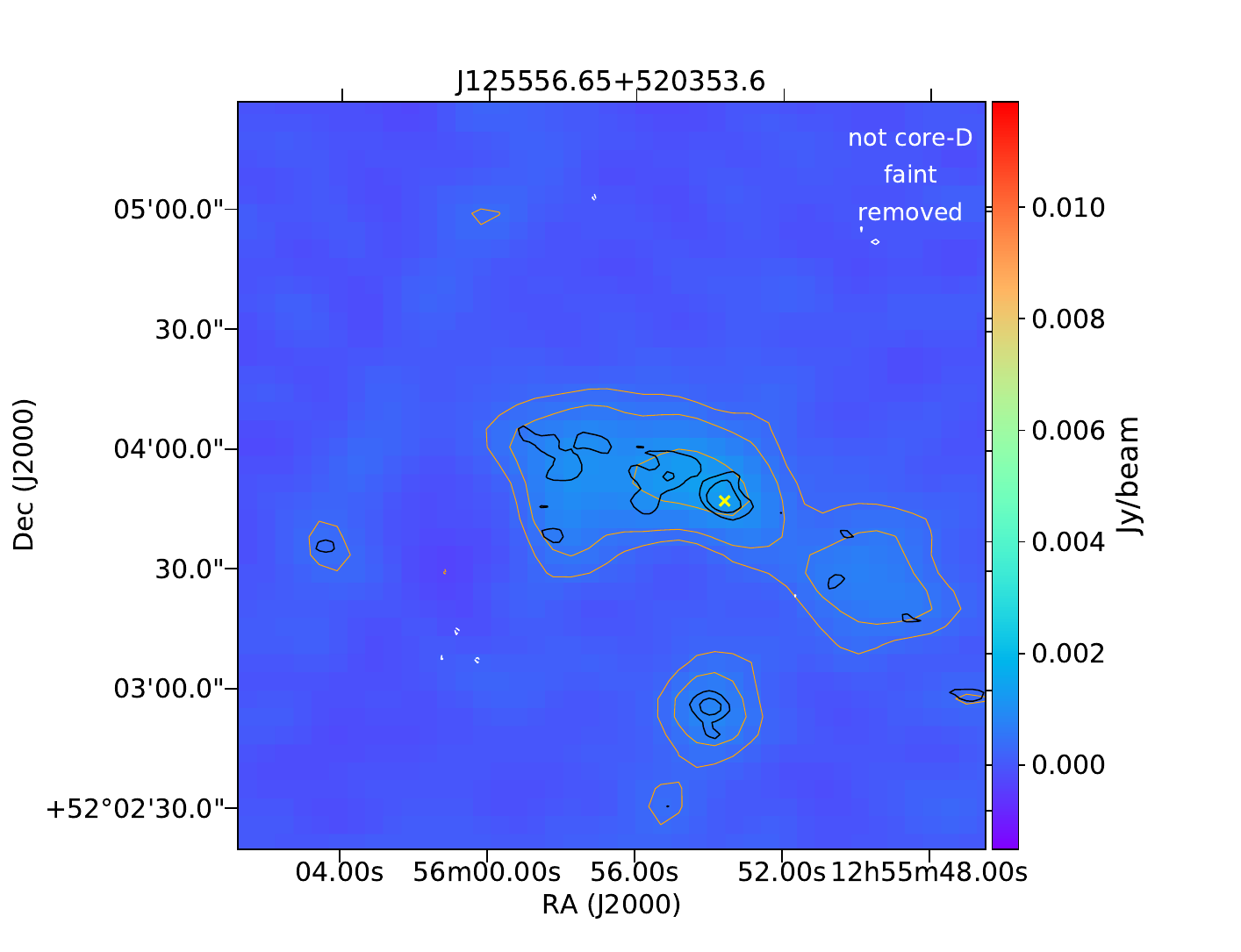}
            \includegraphics[width=0.3\textwidth]{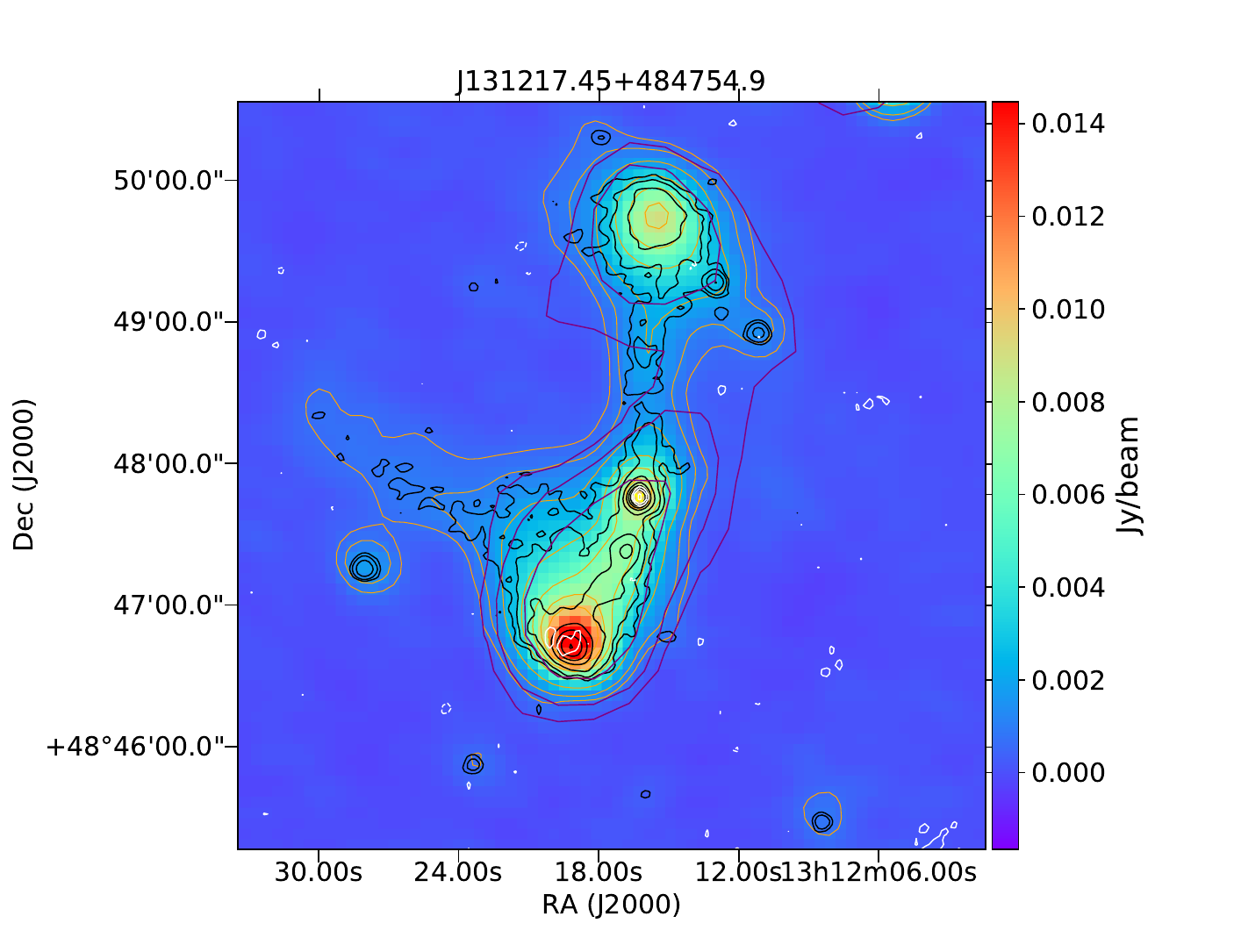}
            \includegraphics[width=0.3\textwidth]{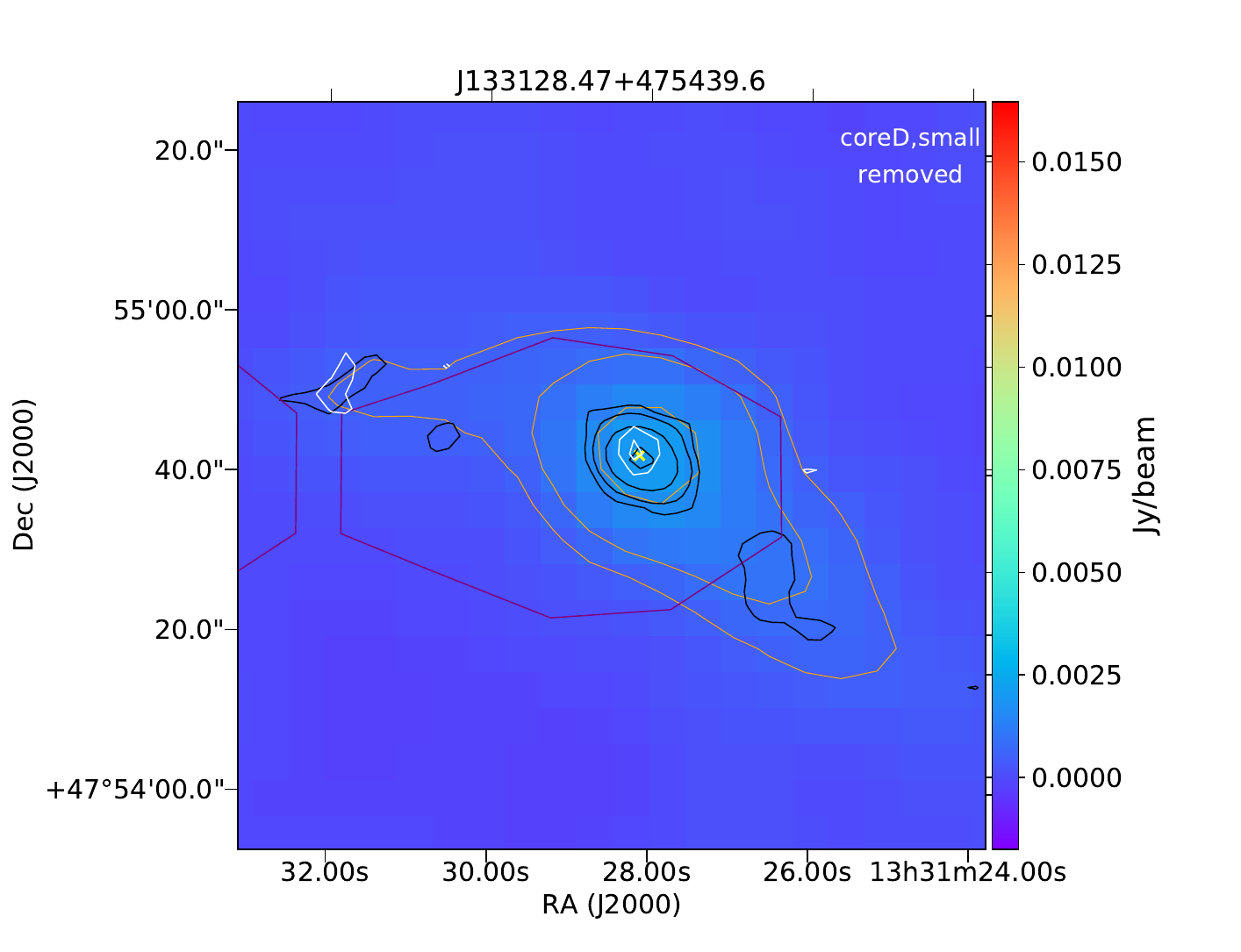}            
}
\centerline{\includegraphics[width=0.3\textwidth]{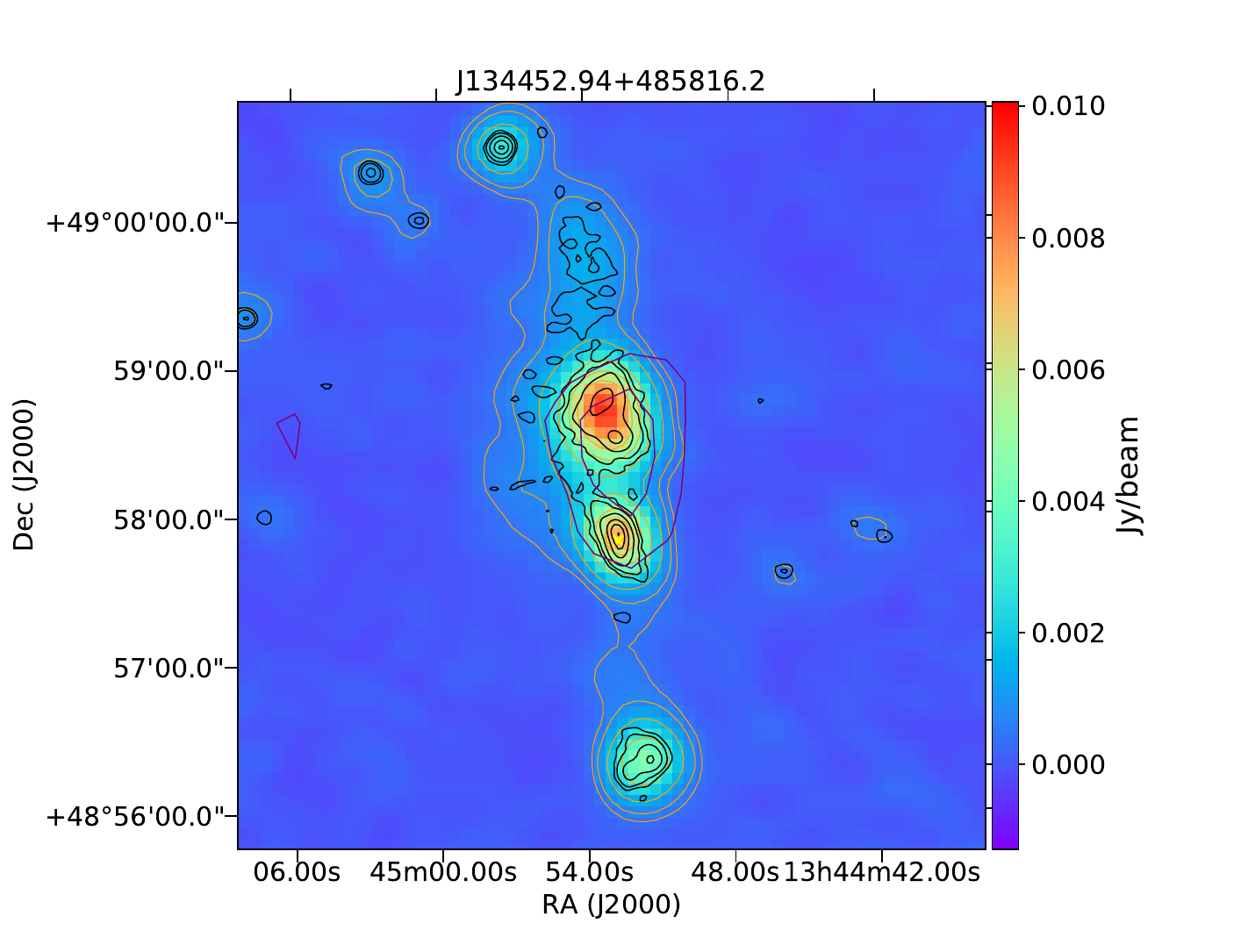}
            \includegraphics[width=0.3\textwidth]{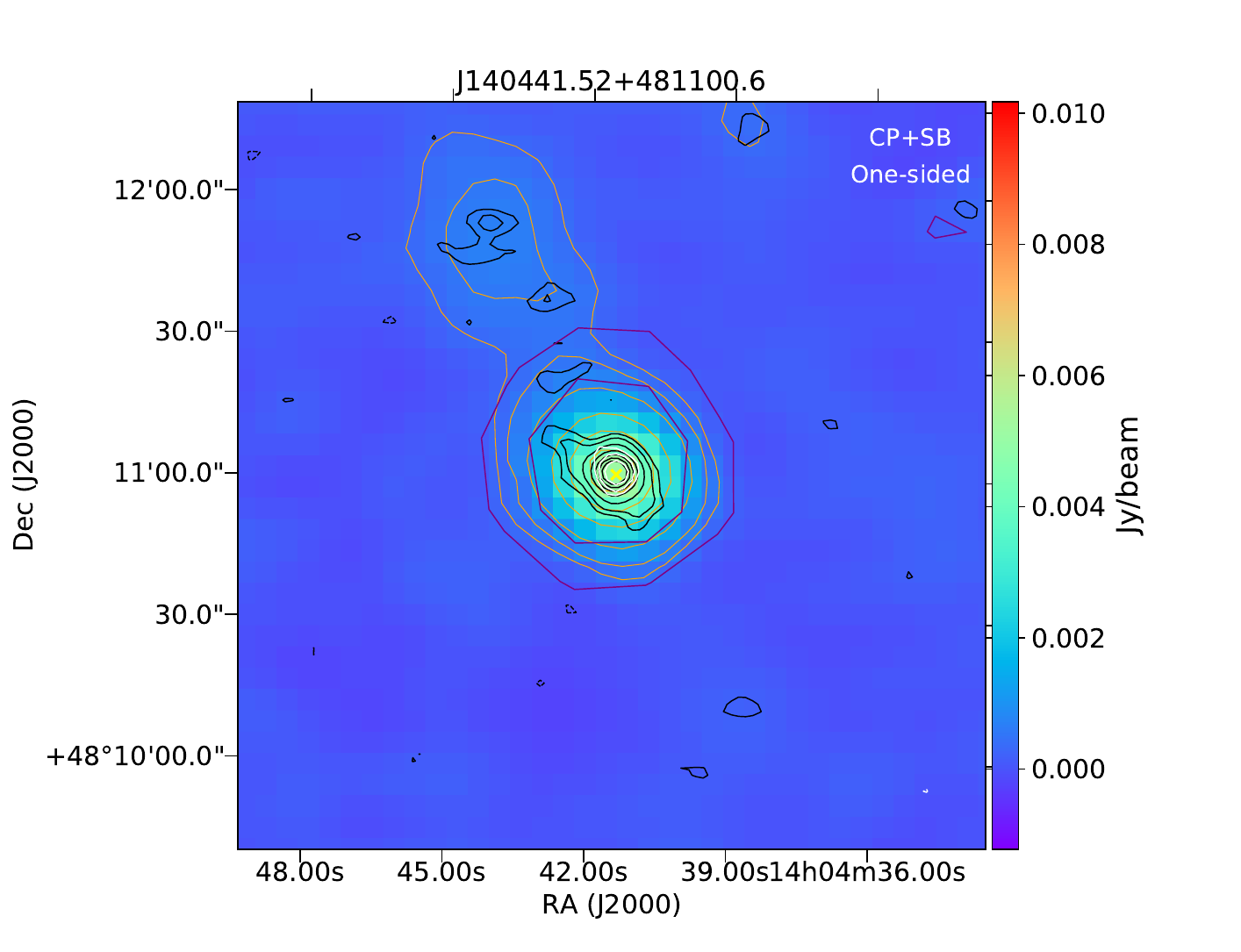}
            \includegraphics[width=0.3\textwidth]{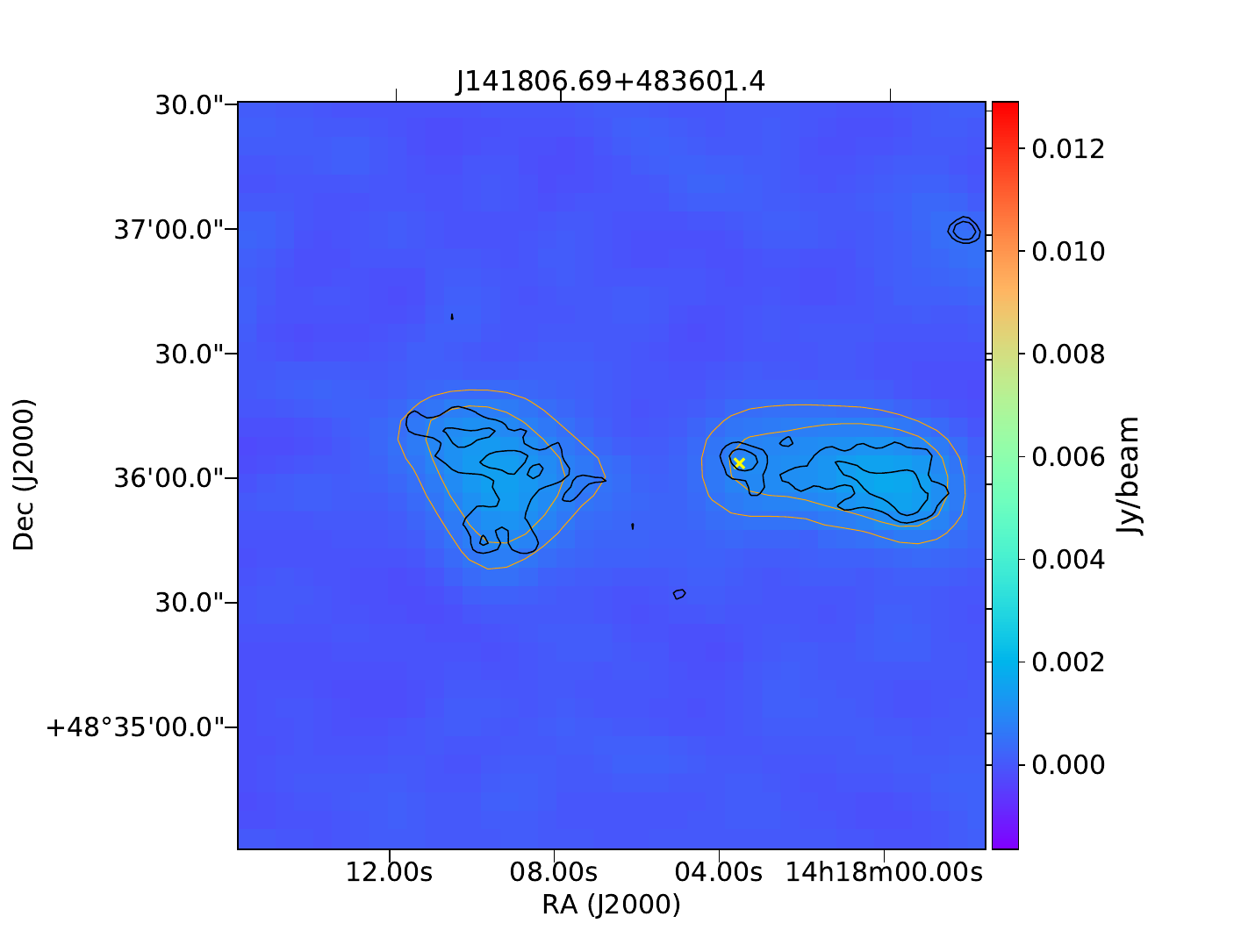}            
}
\centerline{\includegraphics[width=0.3\textwidth]{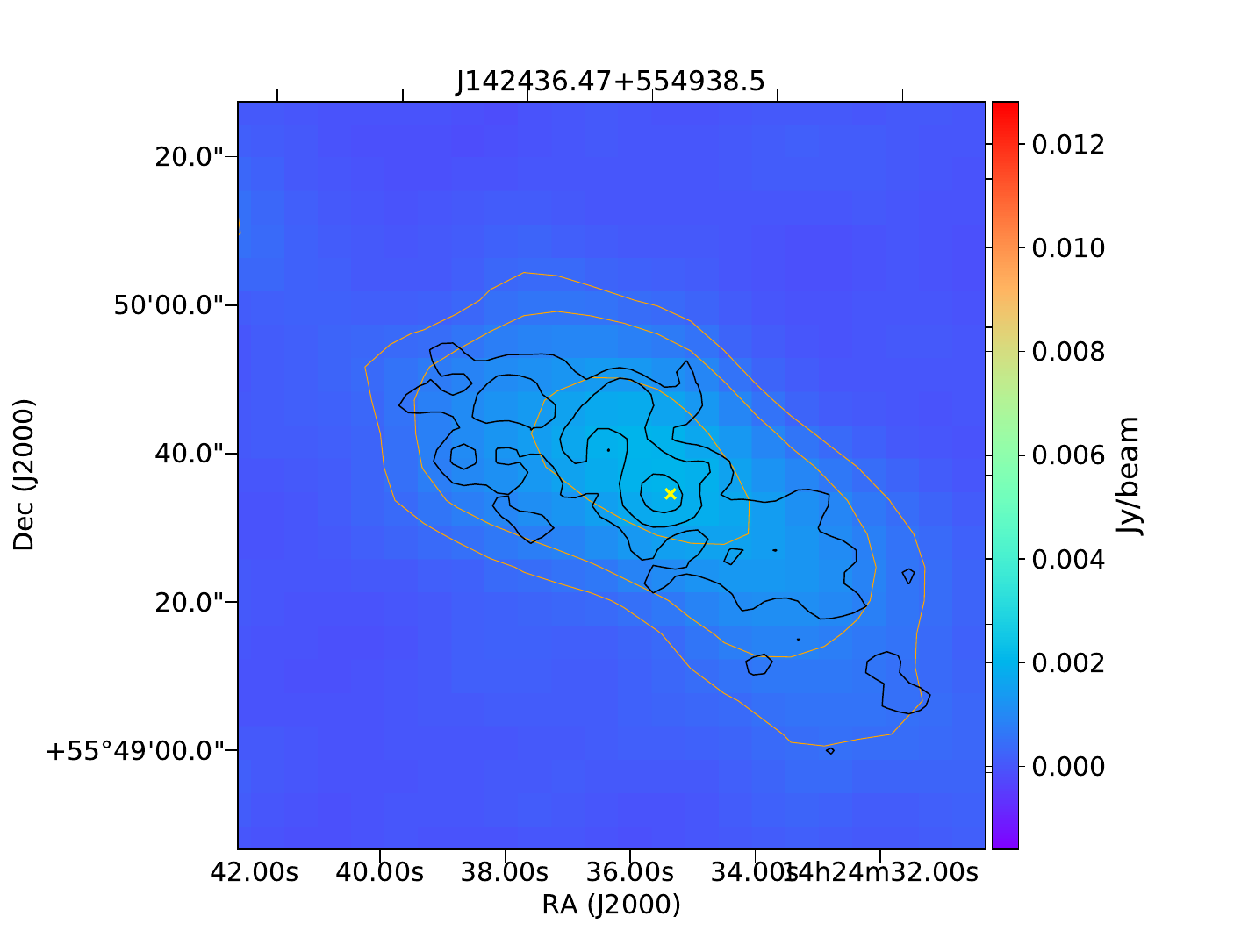}
            \includegraphics[width=0.3\textwidth]{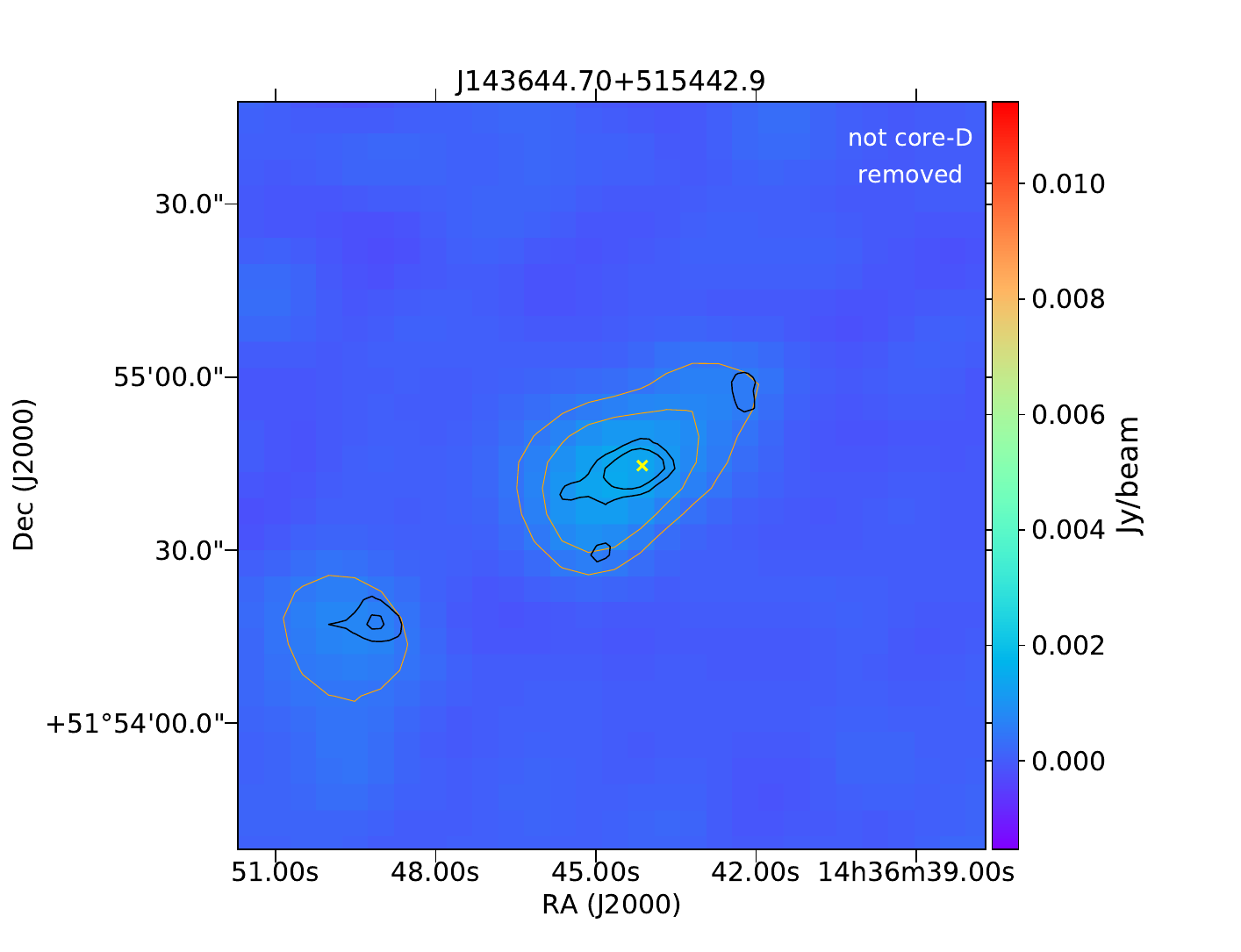}
            \includegraphics[width=0.3\textwidth]{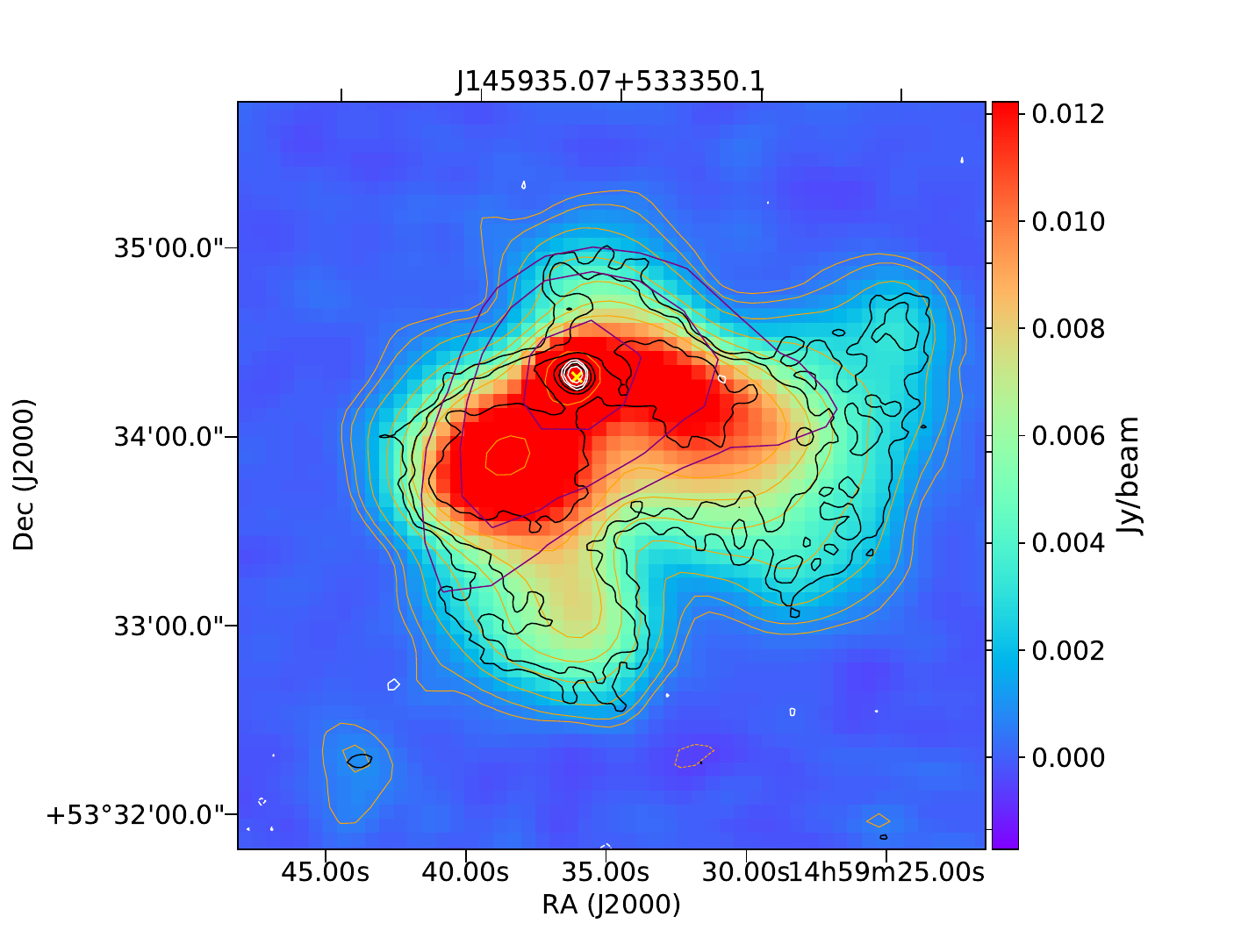}            
}
\caption{\footnotesize{
Images of sources excluded from the sample of restarted candidates following the criteria discussed in Sect.~\ref{High radio core prominence and low surface brightness}, Sect.~\ref{Steep spectral index of the core} and Sect.~\ref{Spectral index of the extended emission} and summarised in Tables.~\ref{list of sources-rejected} and ~\ref{Core flux densities-rejected}.
Radio contours from VLA FIRST maps (white, 5$\arcsec$), LOFAR high-resolution maps (black, 6$\arcsec$), and NVSS maps (purple, 45$\arcsec$) are overlaid on the LOFAR low-resolution resolution maps (orange, 20$\arcsec$).
The contouring of all the maps is made at $\,\sigma_\mathrm{local}\times(-3,3,5,10,20,30,40,50,100,150,200)$ levels, with $\sigma_\mathrm{local}$ representing the local RMS noise of the corresponding maps.
The host galaxy position is marked with a yellow cross.
}}
\label{fig:7}
\end{figure*}

\end{appendix}
%%%%%%%%%%%%%%%%%%%%%%%%%%%%%%%%%%%%%%%%%%%%%%%%%%%%%%%%%%%%%%%%%%
\twocolumn
\end{onecolumn}

\end{document}